\documentclass[11pt,final,twoside,fleqn,dvips,a4paper,ps2pdf]{book}

\usepackage{phdthesis}

\usepackage[T1]{fontenc}
\usepackage[latin1]{inputenc}
\usepackage[english]{babel}

\usepackage{mathptmx}
\usepackage[scaled=.90]{helvet}
\usepackage{courier}
\usepackage{color}

\usepackage{setspace}
\usepackage{fancyhdr}
\usepackage{mcite}

\usepackage{upgreek}

\usepackage{indentfirst}

\usepackage{amsmath}
\usepackage{graphics}
\usepackage{graphicx}
\usepackage{verbatim}
\usepackage{subfigure}
\usepackage{caption2}
\usepackage{psfrag}
\usepackage{slashed}

\setcaptionwidth{0.8\textwidth}

\usepackage{array}

\usepackage{cite}
\usepackage{comment}
\usepackage[toc,page]{appendix}

\usepackage{Titulobook}

\begin{document}

\title{Flavour Changing at Colliders in the Effective Theory Approach}
\author{Renato Batista Guedes Júnior}
\date{2008}
\maketitle

\chapter*{Acknowledgments}\label{cap:acknowledgments}
\addcontentsline{toc}{chapter}{Acknowledgments}

I would like to express my gratitude to all those that have accompanied me in the elaboration of this thesis, in particular, my supervisors, Doctor Rui Alberto Serra Ribeiro dos Santos and Professor Augusto Manuel Albuquerque Barroso. Without their guidance, support, interest and friendship, I would never have been able to pursue this thesis. I would also like to thank Professor Pedro Miguel Martins Ferreira for the assistance he provided at all levels of the research project.

I also thank Professor Orlando Oliveira, Rita Coimbra and Miguel Won for the assistance they provided of the research Project. I must acknowledge Professor António Onofre, Professor Arhrib Abdesslam and Professor Pedro Texeira-Dias for receive me in their research groups and give me all the support. A very special thanks to Steven Warren, Ana Catarina and António Paço to help me in theirs free time in linguistic review.

Finally, I would like to send my thanks to my friends Ana Rajado, António Paço, Filipa Lopes, Filipe Veloso, Inês de Castro, Nuno Castro, Raquel Varela and Renato Texeira. These individuals always helped me to keep my life in context. I am grateful to my office colleagues as well as to the Interdisciplinary Complex of the University of Lisbon and the Center for Theoretical and Computational Physics (CFTC) of the University of Lisbon, where this thesis was carried out, for the use of the physical resources available. This thesis is supported by Fundação para a Ciência e Tecnologia under Contract No. SFRH/BD/19781/2004.

\chapter*{Resumo}\label{cap:resumo}
\addcontentsline{toc}{chapter}{Resumo}

{\bf Palavras-chave:} Quark top, correntes neutrais, violação de sabor, modelos para além do modelo padrão, propriedade dos leptões, flavor symmetries

\vspace{2cm}

Nesta tese discutiremos a parametrização de efeitos relacionados com nova física em altas energias, isto é, física para além do modelo padrão (MP) das interacções fortes e electrofracas. Tal parametrização será feita com recurso a um Lagrangiano efectivo deduzido com o pressuposto de que estes novos efeitos só se manifestam para escalas de energia da ordem do $TeV$ e, por imposição, este Lagrangiano terá as mesmas invariâncias do MP. Expandiremos este Lagrangiano em ordem ao inverso da escala de energia. Desta forma, ele será composto por uma serie de termos infinitos de acordo com a escala de energia. Posto que cada termo terá uma dimensão de massa específica, podemos alternativamente identificar cada termo pela sua dimensão. Cada termo da expansão é, por sua vez, composto por inúmeras componentes ou operadores (operadores efectivos) dos quais nos serviremos para agrupar ou identificar os mesmos termos. Veremos ainda que o primeiro termo tem dimensão quatro e assumiremos como sendo o Lagrangiano do MP. Assim, aceitamos que a física descrita pelo Lagrangiano efectivo seja, em primeira aproximação, a mesma do MP sendo as correcções descritas pelos restantes termos.

Passaremos a uma fase seguinte onde nos propomos usar o método anteriormente descrito para parametrizar os efeitos de correntes neutras com mudança de sabor ({\it flavour changing neutral current} -- FCNC) na produção de {\it single} quark top (ou apenas {\it single} top). De uma forma mais precisa, sabemos que o quark top, sendo a partícula elementar mais massiva da natureza até aqui conhecida, decai quase exclusivamente em $bW$ (onde $b$ é o quark botton e $W$ o bosão vectorial de mesmo nome) antes de se hadronizar. Dizemos quase exclusivamente porque as excepções, $dW$ e $sW,$ são extremamente suprimidas pelos elementos fora da diagonal da matriz de Cabibbo-Kobayashi-Maskawa (CKM). Também, de acordo com o MP, decaimentos neutros do quark top, ou seja, decaimentos em $qg,$ $q\gamma$ e $qZ$ (onde $q$ pode ser o quark up (u) ou ocharm (c), $g$ o gluão, $\gamma$ o fotão e $Z$ o bosão vectorial de mesmo nome) são impossíveis a nível árvore -- para ajustar a nomenclatura, chamaremos decaimento forte, no caso do gluão e electrofraco no caso do fotão ou do bosão $Z.$ Assim, aliando mecanismo de supressão da matriz CKM com a impossibilidade de existência de correntes neutras com mudança de sabor, o MP prevê uma secção eficaz na produção de {\it single} top através de FCNC bastante suprimida. E se, ao contrário das previsões do MP, FCNC ocorre na natureza? Qual seria o impacto dessas correntes neutras na produção de {\it single} top? Independentemente da resposta positiva ou negativa para a existência de tais correntes -- que não conhecemos, naturalmente -- é sempre possível parametrizar esta hipótese através do Lagrangiano efectivo atrás discutido. No nosso estudo, expandiremos este Lagrangiano até a dimensão seis. Esta escolha prende-se com o facto de procurarmos a primeira contribuição para FCNC, ou seja, a primeira contribuição após a do MP. Os termos de dimensão cinco, a primeira contribuição a seguir à do MP, violam o número leptônico e bariônico para além de não contribuírem com qualquer operador para este estudo. Assim, imposto a conservação destas duas quantidades, devemos excluir os operadores de dimensão cinco e avançar para os de dimensão seis. Acordada a dimensão máxima do Lagrangiano efectivo, torna-se necessário encontrar os operadores que possam contribuir para decaimento do quark top com FCNC ($t\rightarrow qg,$ $t\rightarrow q\gamma$ e $t\rightarrow qZ$) e então, derivar as regras de Feynman desses operadores. Assim, seremos capazes de parametrizar as contribuições desses processos para o decaimento do quark top. Mostraremos ainda que estes decaimentos não são independentes: devido à imposição de invariância de gauge, os decaimentos com FCNC electrofracos estão relacionados, ou seja, o decaimento do top em quark+fotão tem impacto no decaimento em quark+Z e vice-versa. Seremos também capazes de relacionar as taxas de decaimento com FCNC com as secções eficazes de produção de {\it single} top ou mesmo das produções parciais: produção de $t+\gamma,$ $t+Z$ e $t+q$ (neste último caso $q$ pode ser um quark qualquer, excepto o quark top). Estamos interessados em particular no efeito combinado (forte e electrofraco) da produção com FCNC do quark top. Uma forma de abordar as FCNC é pela parametrização do espaço onde a sua existência seria observada (ou não). Um dos objectivos centrais desta tese é precisamente confrontar a hipótese de FCNC na produção do quark top com os valores ou limites que serão originados no LHC. Enquanto aguardamos estes dados, é possível confrontar valores obtidos através de parâmetros gerados aleatoriamente com as características previstas para o LHC. Podemos ainda, fixada a taxa de decaimento com FCNC do quark top, estudar como este decaimento influencia os processos de produção de {\it single} top, ou seja, definir regiões onde esperamos registar as contribuições para a produção do quark top. Nalguns casos, esta contribuição pode ser registada ou ``vista'' experimentalmente no LHC. Mesmo em casos onde não é possível registar a taxa de decaimento devido à sua pequena dimensão, encontraremos regiões onde o seu efeito na secção eficaz de produção de single top pode ser registado. Finalmente, faremos algumas considerações sobre a contribuição para produção de $t\bar t$ devida a FCNC e mostraremos que esta contribuição muito dificilmente poderá ser visto no LHC.

Finalmente, focaremos a nossa atenção no sector leptónico. Seguindo os procedimentos descritos e aplicados no estudo de FCNC, abordaremos o problema relacionado com violação de sabor entre os leptões ({\it lepton flavour violation} -- LFV), ou seja, faremos uso de operadores de dimensão seis para descrever LFV. Ainda é mister reconhecer que a violação de sabor no sector leptónico é absolutamente proibida pelo MP em todas as ordens. Veremos também que os limites experimentais para tais processos são extremamente fortes (comparativamente com os limites para os processos bariónico envolvendo FCNC). Assim como fizemos no estudo de FCNC, estudaremos os decaimentos envolvendo LFV, descritos pelo decaimento de um leptão mais pesado em três mais leves, tanto através de vértices do tipo $l_h\,l_l\,\gamma$ e $l_h\, l_lZ,$ onde $l_h$ e $l_l$ são os leptões mais pesado e mais leve, respectivamente, como através de operadores de quatro-fermiões, já aqui referidos. Procederemos da mesma forma que no sector bariónico para estabelecer relações entre a taxa de decaimento e as secções eficazes dos processos que envolvem LFV, nomeadamente os processos que podem ocorrer no ILC, a colisão entre electrão e positrão resultando em $\mu^-\,e^+,$ $\tau^-\,e^+$ e $\tau^-\,\mu^+$ bem como os processo conjugados de carga. Veremos que de acordo com os parâmetros do ILC, tais processos podem ser visto e, inclusive, o seu estudo é bastante facilitado recorrendo a alguns simples procedimentos para a análise do sinal, tais como cortes simples do momento transverso. Por fim, chamamos a atenção para um aspecto muito importante referente às nossas conclusões: todas as relações e parametrizações referenciadas anteriormente entre taxas de decaimento e secções eficazes não se esgotam com este trabalho, antes pelo contrário. Será necessário proceder à referida parametrização de acordo com os dados experimentais que nos chegarão do LHC num futuro bem próximo, ou ainda do ILC (neste caso num futuro cada vez mais distante), mas que hoje mesmo abundam provindo de outros aceleradores. Portanto, há muitos aspectos a serem estudados e muitas simulações a serem feitas. Dito isto, um fio condutor de toda esta tese será a apresentação de todas as expressões analíticas (decaimentos e secções eficazes anómalas), para que os nossos colegas experimentalistas possam proceder a simulações com os seus geradores de Monde-Carlo e definir as possíveis regiões onde poderemos assinalar nova física ligada a processos com FCNC e LFV.

\chapter*{Abstract}\label{cap:resume}
\addcontentsline{toc}{chapter}{Abstract}

{\bf Keywords:} Top quarks, neutral currents, flavour violation, models beyond the standard models, leptons properties, flavor symmetries

\vspace{2cm}

In this thesis we will discuss the parameterisation of effects related with new physics in high energies, that is, physics beyond the standard model (SM) of the strong and electroweak interactions. Such parameterization will be made with resource to an effective Lagrangian deduced considering that these new effects can only be seen energy scale of the order of TeV and, we force, this Lagrangian to be invariancies under the same symmetries. We will expand this Lagrangian to the inverse of the energy scale. This way, it will be composed of a series of infinite terms in accordance with the energy scale. Since each term will have a specific mass dimension, we can alternatively identify each term by its dimension. Each term of the expansion is, in turn, made of innumerable components or operators (effective operators) which we will use to group or identify the same terms. We will also see that the first term has dimension four and will assume it as being the Lagrangian of the SM. Thus, we accept that the physics described by the effective Lagrangian is in a first approach, the same of the SM, while the corrections are described by the remaining terms.

We will then proceed to the next phase where we propose to use the above mentioned method to parameterize the effects of flavour-changing neutral current (FCNC) in the production of single top quark (or just single top). We know that the top quark, being nature's most massive elementary particle known so far, decays almost exclusively to $bW$ (where $b$ is the bottom quark and $W$ the weak boson) before it hadronizes. We say almost exclusively because the exceptions, $dW$ e $ sW$, are extremely suppressed by the off-diagonal elements of the Cabibbo-Kobayashi-Maskawa matrix (CKM). Also, in accordance with the SM, top quark neutral decays, meaning decays in $qg, $ $q\gamma$ and $qZ$ (where $q$ may be the up (u) or charm (c) quark, $g$ the gluon, $\gamma$ the photon e $Z$ the vector boson) are impossible at tree level. To adjust the nomenclature, we will name it strong decay in the case of the gluon, and electroweak in the case of the photon or the $Z$ boson. Therefore, the CKM matrix suppression mechanism together with the impossibility of FCNC, makes a tiny prediction for values of the cross section in the SM. But what if, contradicting the SM forecasts, FCNC happens in nature more than what the SM forecast? Which would be the impact of these neutral currents in the production of single top? Whether the answer to these questions is positive or negative - which, of course, we do not know -- it is always possible to parameterize this hypothesis through the above mentioned effective Lagrangian. In our study we will expand this Lagrangian to dimension six. This choice is due to the fact that we are looking for a first contribution for the FCNC, that is, the first contribution after the SM. The terms of dimension five, which would be the first contribution after the SM, violate the leptonic and barionic number. In addition they do not contribute with any operator to this study. Thus, imposing the conservation of these quantum number, we must exclude dimension five operators and proceed to dimension six operators. Having established the maximum dimension of the effective Lagrangian, it is necessary to find those operators that might contribute to the top quark decay with FCNC ($t\rightarrow q\, g, $ $t\rightarrow q\,\gamma$ e $t\rightarrow q\, Z$) and then, to derive the corresponding Feynman rules. Thus, we will be able to parameterize the contributions of these processes to the top quark decay. We will also show that these decays are not independent - due to the imposition of gauge invariance, electroweak FCNC are related, that is, top quark decay in quark+photon has an impact in quark+$Z$ quark decay and vice versa. We will also be able to relate FCNC decay with the cross sections of production of single top or even the partial productions: production of $t\,+\,\gamma, $ $t\,+\,Z$ e $t\,+\,q$ (in this last case q quark can be any quark exception made to the top quark). We are particularly interested in the combined effect (strong and electroweak) of top quark production with FCNC. One way to approach FCNC is through parametrization of the space where its existence would be observed (or not). One of the central aims of this thesis is to confront the hypothesis of FCNC in the production of top quark with the values or limits that will be originated in the LHC. While we wait for these data, it is possible to collate values gotten through parameters randomly generated with the characteristics foreseen for the LHC. Having established the top quark FCNC branching ratio, we still can study how this decay influences the processes of production of single top, that is, to define regions where we hope to register the contributions for the production of top quark. In some cases, this contribution can experimentally be registered or "seen" in the LHC. Even in cases where one cannot register the decay due to its smallest, we will still find regions where its effect in the cross section of single top production can be seen. Finally, we will make some considerations on the contribution of $t \bar t$ production with FCNC and we will show that this contribution could hardly be seen in the LHC.

Finally, we will focus our attention in the leptonic sector. Following the procedures that we described and applied in the FCNC study, we will approach the problem related with lepton flavour violation (LFV), that is, we will use dimension six operators to describe LFV. We assume that flavour violation in the lepton sector is absolutely forbidden by the SM of all orders. We will also see that the experimental limits for such processes are extremely strong (as compared with the limits for baryonic processes involving FCNC). As in the FCNC study, we will study decays involving LFV, described by a heavier lepton decaing into three lighter leptons, through vertices of the type $l_h\,l_l\,\gamma$ e $l_h\, l_lZ,$ where $l_h$ e $l_l$ are the heavier and lighter leptons and through four-fermion operators, as described. We will proceed in the same way as in the barionic sector to establish relations between the branching ratios and the cross sections of processes involving LFV, namely the processes that can occur in the ILC or electron-positron collision resulting in $\mu^- \, e^+,$ $ \tau^- \, e^+$ and $\tau^- \, \mu^+,$ as well as, the charge conjugate processes. We will see that according to the ILC parameters, such processes can be seen and its study is quite facilitated when we appeal to some simple procedures for the analysis of the signal, such as simple cuts of the transverse moment. Finally, we note that this work does not put an end to all before mentioned relations and parameterizations between branching ratio and cross section. Quite the opposite: it will be necessary to parameterize the experimental data coming from the LHC in the near future, and from the ILC (in a more distant future). Therefore, there are many aspects to be studied and many simulations to be done. This thesis will follow the line of presenting all the analytical expressions (decays and anomalous cross sections), so that our experimentalist colleagues can proceed with the simulation in their Monte-Carlo generators and define the possible regions where we will be able to find new physics in processes with FCNC and LFV.

\chapter*{Preface}\label{cap:preface}
\addcontentsline{toc}{chapter}{Preface}

The research included in this thesis has been carried out at the Interdisciplinary Complex of the University of Lisbon and the Center for Theoretical and Computational Physics (CFTC) of the University of Lisbon in the Physics Department of the Faculty of Sciences of the University of Lisbon. The list of works published and under review included in this thesis in given below:
\begin{enumerate}
\item P. M. Ferreira, R. B. Guedes, and R. Santos, {\it Lepton flavour violating processes at the international linear collider,} Phys. Rev. {\bf D75,} 055015 (2007), hep-ph/0611222~\cite{Ferreira:2006dg};
\item P. M. Ferreira, R. B. Guedes, and R. Santos, {\it Combined effects of strong and electroweak FCNC effective operators in top quark physics at the LHC,} Phys. Rev. {\bf D77,} 114008 (2008), hep-ph/0611222~\cite{Ferreira:2008cj};
\item R. A. Coimbra,P. M. Ferreira, R. B. Guedes, O. Oliveira, R. Santos and Miguel Won,
(preparation)~\cite{Ferreira:2008xx}.
\end{enumerate}

\tableofcontents


\listoffigures

\listoftables

\pagenumbering{arabic} \setcounter{page}{1}

\chapter{Introduction}

The Large Hadron Collider LHC will soon begin operating. The number of top quarks that will
produce is of the order of millions per year. Such large
statistics will enable precision studies of top quark physics -- this
being the least well-know elementary particle discovered so far. The
study of flavour changing neutral current (FCNC) interactions of the
top quark is of particular interest. In fact, the FCNC decays of the
top -- decays to a quark of a different flavour and a gauge boson, or
a Higgs scalar -- have branching ratios which can vary immensely from
model to model -- from the extremely small values expected within the
standard model (SM) to values measurable at the LHC in
certain SM extensions.

The use of anomalous couplings to study possible new top physics at
the LHC and at the Tevatron has been the subject of many
works~\cite{Malkawi:1995dm,*Han:1996ce,*Han:1996ep,*Whisnant:1997qu,*Hosch:1997gz,*Han:1998tp,*Tait:2000sh,*Carlson:1994bg,*Rizzo:1995uv,*Tait:1996dv,*Espriu:2001vj,Hikasa:1998wx}.
The cross section for those processes were calculated in a recent
series of
papers~\cite{Ferreira:2005dr,Ferreira:2006xe,Ferreira:2006in} where
 FCNC interactions associated with the strong
interaction were considered -- decays of the type $t\rightarrow u\,g$ or
$t\rightarrow c\,g$ -- describing them using the most general
dimension six FCNC Lagrangian emerging from the effective operator
formalism~\cite{Buchmuller:1985jz}. The FCNC vertices originating
from that Lagrangian also give substantial contributions to
production processes of the top quark, such as the associated production of a
single top quark alongside a jet, a Higgs boson or an electroweak
gauge boson. As we will see, the study of
refs.~\cite{Ferreira:2005dr,Ferreira:2006xe,Ferreira:2006in}
concluded that, for large values of $BR(t\rightarrow q\,g)$, with $q
= u, c$, these processes of single top production might be
observable at the LHC.

Following the treatment of those articles~\cite{Ferreira:2005dr,Ferreira:2006xe,Ferreira:2006in}, the next logical step is to use the same treatment for the electroweak sector, by considering FCNC interactions leading to decays of the form $t\rightarrow q\,\gamma$ or $t\rightarrow q\,Z.$  In some extensions of the SM these branching ratios can be as large as, if not larger than, those of the strong FCNC interactions involving gluons. In this thesis we extend the analysis of previous works and consider the most general dimension six FCNC lagrangian in the effective operator formalism which leads to $t\rightarrow q\,\gamma$ and $t\rightarrow q\,Z$ decays. We will study the effects of these new electroweak FCNC interactions in the decays of the top quark and its expected production at the LHC. We will study in detail processes such as $t\,+\,\gamma,$ $t\,+\,Z$ and $t\,+\,j$ production, for which both strong and electroweak FCNC interactions contribute. The automatic gauge invariance of the effective operator formalism will allow us to detect correlations among several FCNC observables.

As we said, we expect many exciting discoveries to arise from the LHC experiments.
However, the LHC is a hadronic machine, and as such precision
measurements will be quite hard to undertake there. Also, the
existence of immense backgrounds at the LHC may hinder discoveries
of new physical phenomena already possible with the energies that this
accelerator will achieve. Thus it has been proposed to build a new
electron-positron collider, the International Linear Collider [1].
This would be a collider with energies on the TeV range, with
extremely high luminosities. The potential for new physics with such
a machine is immense. Here, we will focus on a specific
sector: the possibility of processes which violate lepton flavour.

We now know that the solar and atmospheric neutrino
problems~\cite{Bahcall,*Mohapatra,*Fogli:2005cq} arise not from
shortcomings of solar models but from particle physics. Namely, the
recent findings by the SNO
collaboration~\cite{Aharmim:2007nv,*Aharmim:2005gt,*Ahmad:2002jz}
have shown beyond doubt that neutrinos oscillate between families as
they propagate over long distances. Leptonic flavour violation (LFV)
is therefore an established experimental fact. The simplest
explanation for neutrino oscillations is that neutrinos have masses
that differ from zero -- extremely low masses, but nonzero nonetheless.
Oscillations with zero neutrino masses are possible, but only in
esoteric models~\cite{Barroso:1994sg}. With nonzero neutrino masses,
flavour violation in the charged leptonic sector becomes a reality
whereas with massless neutrinos, it is not allowed in the SM.
This is a sector of particle physics for which we
already have many experimental results~\cite{Yao:2006px}, which set
stringent limits on the extent of flavour violation that may occur.
Nevertheless, as we will show, even with all known experimental
constraints it is possible that signals of LFV may be observed at
the ILC, taking advantage of the large luminosities planned for that
machine. There has been much attention devoted to this subject. For
instance, in
refs.~\cite{Delepine:2001di,*Illana:2000ic,*FloresTlalpa:2001sp,*Perez:2003ad}
effective operators were used to describe LFV decays of the Z boson.
LFV decays of the Z boson were also studied in many extensions of
the
SM~\cite{Frank:1996xt,*Ghosal:2001ep,*Iltan:2001au,*Yue:2002pk,*Cao:2003zv,*Yue:2004bs,*Iltan:2005jp}.
The authors of
refs.~\cite{Deppisch:2003wt,*Sun:2004cx,*Cannoni:2003my,Cannoni:2005gy}
performed a detailed study of LFV at future linear colliders,
originating from supersymmetric models. Finally, a detailed study of
the four-fermion operators in the framework of LFV is performed
in~\cite{Ibarra:2004pe}. In that work the exact number of
independent four-fermion operators is determined. Gauge invariance
is then used to constrain LFV processes which are poorly measured,
or not measured at all. In this work we carry out a
model-independent analysis of all possible LFV interactions which
might arise in extensions of the SM.\label{par:lep1}

We will follow refs.~\cite{Ferreira:2008cj,Ferreira:2008xx,Ferreira:2006dg}; as in the articles~\cite{Ferreira:2005dr,Ferreira:2006xe,Ferreira:2006in} our methodology will differ from that of previous work in this area; whenever possible, we will present full analytical expressions that our colleagues at the Tevatron or LHC may use in their Monte Carlo generators, to study the sensitivity of the experiments to this new physics.
This thesis is organized as follows: in chapter~\ref{cap:capI} we review the theory obtained through an effective Lagrangian. We expand this Lagrangian in series of $ 1 / \Lambda $ where $ \Lambda $ is the energy scale where we suppose our Lagrangian is valid. This allows us to truncate the Lagrangian and make the computation of each term separately following the order that interests us. In our case, we truncate the Lagrangian in the third term, i.e. $\Lambda^4,$ or dimension six, according to their mass dimension. We justify the assumption that the first term or zero order term matches the SM Lagrangian and, finally, that the terms of order one or dimension five must  be ignored for the study of flavour violation. We review the theoretical predictions for the production of $ t\bar t $ and single top quark associated with the FCNC processes, the experimental limits available from the CDF and D0 collaborations of Fermilab as well as the theoretical predictions for LHC. Then, we summarize the experimental limits of the LFV in the leptons colliders.

In chapter~\ref{cap:capII} we introduce our FCNC operators, explaining what are the physical
criteria behind their choice.We also present the Feynman rules
for the new anomalous top quark interactions which will be the base
of all the work that follows. We use those Feynman rules to
compute and analyze the branching ratios of the top quark FCNC
decays, with particular emphasis on the relationship between
$Br(t\rightarrow q\gamma$ and $Br(t\rightarrow qZ).$ Then
we study the cross section for production, at the LHC, of a single
top and a photon or a $Z$ boson, with all FCNC interactions, both
strong and electroweak, included. We also investigate whether or note it
would be possible to conclude, from the data, that any FCNC
phenomena observed would have at their roots the strong or the
electroweak sectors. Finally, we present a general discussion of the
results and draw some conclusions.

In chapter~\ref{cha:lepton} we present the effective operator
formalism and list the operators which contribute to
lepton-violating processes with gauge bosons interactions
and four-fermion contact terms. We use the existing experimental
bounds on decays such as $\mu\rightarrow e\gamma$ to exclude several
of the operators which could {\it a priori} give a contribution to the
processes that we will be considering. We also analyze the role that the
equations of motion of the fields play in further simplifying our
calculations. Having chosen a set of effective operators, we proceed
to calculate their impact on LFV decays of leptons, deducing
analytical expressions for those quantities. Likewise, we will
present analytical results for the cross sections and asymmetries of
several LFV processes which might occur at the ILC. Finally, we
analyze these results performing a scan over a wide range of values
for the anomalous couplings and consider the
possible observability of these effects at the ILC.

\begin{savequote}[9pc]
\linespread {1.0}
 \sffamily
Like good musicians, good physicists know which scale are relevant
for which compositions. \qauthor{C. P. Burgess}
\end{savequote}

\chapter{Phenomenological Lagrangian and effective operators}\label{cap:capI}

We are interested in parameterizing new physics related to top quark physics due to flavour changing neutral current (FCNC) as well as to lepton flavour violation (LFV) in the charged sector.  We will discuss this parametrization through an effective theory, that is, using terms of an effective Lagrangian with vertices with flavour violation whose strength must be determined by experimental data.

The experimental data does not provide us with information that allow us to probe new physics beyond the SM. This situation is ripe to change with the introduction of the Large Hadron Collider (LHC) and the  construction of the International Linear Collider (ILC). In the first case, we will be searching for new physics through the possibility of FCNC in top quark production at the $TeV$ scale. In the second one, we will be looking for LVF.  In this chapter, we review the effective Lagrangian technique. After that we explore the limits of these two new manifestations of physics within the SM framework and review the relevant experimental data.

\section{Effective Lagrangian}

The SM has a great success in explaining the most important phenomena at the fundamental level. Nevertheless, it does not have the trademark of a fundamental theory; it has too many parameters and has no prediction regarding the number of particles. The phenomenological success of the SM in the low energy range (the SM was successfully tested in the $W$ mass range) provides a fundamental constraint to explore physics in the higher energy range.    We can suppose that there is a more general theory of which the SM is its low energy limit. The problem is how to describe such a theory.

In general, when we study a system for which we do
not have enough experimental information, or for which the theory does not
give us enough information about some observation, we can proceed
in two ways: a model dependent way (as in supersymmetry, dynamical
symmetry breaking technicolor model, etc.) or a model independent way.
By model independent we mean the effective Lagrangian
approach. In other words, we can parameterize the
unknown effects by introducing new terms in the
Lagrangian, whose coefficients must be experimentally determined~\cite{Manohar:1996cq,Young:1997,Wudka:1999ax,Rothstein:2003mp,Burgess:2007,Han:2008es}. We
must establish the specific way to parametrize these effects
according to the specific problems that we have to solve. In this
study of FCNC and LFV we do not know this general theory so we can try to describe it
through an effective theory i.e. we can try to write an effective FCNC and LFV Lagrangian of the general theory.

When we handle quantum field theory we automatically limit the role
that a higher-energy scale (Plank scale) plays in the description of
low-energy process. In this sense, the identification of how the scale
enters in the calculation provides us with an important way to analyze
systems with different scales~\cite{Burgess:2007}. The effective
theory supplies a tool for exploiting the simplification that arises
from systems presenting a large hierarchy of scale. For example, if
one assumes that some physical phenomenon is not observed below a
certain energy scale $\Lambda,$  all Fourier components of fields
above a scale $\Lambda$ are not directly observable and the
Lagrangian of this theory  must be obtained by integration over the variable
observable at an energy larger then $\Lambda.$ In this case, a real
field -- just to simplify -- $\phi$ can be split in two pieces
($\phi=\phi_0+\phi_1$) according to the energy scale such that
\cite{Wudka:1999ax}:
\begin{equation}
\phi_0(\vec k):|\vec k|<\Lambda \:\:\:\:\: \phi_1(\vec
k):\Lambda<|\vec k|<\Lambda_1,
\end{equation}
where $\Lambda$ is the energy scale and $\vec k$ the momentum. Then
by definition:
\begin{equation}
e^{iS^{\textrm{eff}}} = \int [d\phi_1] e^{iS[\phi_0,\phi_1]}\;\;\;
\textrm{where} \;\;\; S^{\textrm{eff}}=\int  dx\;{\cal
L}^{\textrm{eff}}.
\end{equation}
$S^{\textrm{eff}}$ and ${\cal L}^{\textrm{eff}}$ are the effective
action and effective Lagrangian respectively.
$\cal{L}^{\textrm{eff}}$ can be obtained by the expansion of
$S^{\textrm{eff}}$ in powers of $1/\Lambda.$

To build the effective Lagrangian, one should choose the  degrees of freedom to include  and the respective
symmetries which restrict the possible terms in the effective
Lagrangian. The
effective Lagrangian has an infinite number of terms each with constant
coefficients to be determined experimentally. The effective current-current interaction introduced
by Fermi in order to explain the weak interaction is a well-known example of an effective theory.

We have assumed that there is a general theory that we do not know and,
that the SM is its low-energy limit. In other words, we
must assume that physics beyond the SM is not observed below a
certain energy scale $\Lambda.$  The
 effective Lagrangian can be expanded in powers of
$1/\Lambda:$
\begin{equation}
{\cal L}^\textrm{eff} = {\cal L}^\textrm{SM} +
\frac{1}{\Lambda}{\cal L}^\textrm{(5)} + \frac{1}{\Lambda^2} {\cal
L}^\textrm{(6)} + \cdots ,\label{eq:expand-1}
\end{equation}
where ${\cal L}^\textrm{SM}$ is the term of order zero and matches the Lagrangian of SM which is not sensitive to the
energy scale and has mass-like dimension four.
 The term ${\cal
L}^\textrm{(5)}$ is the order one and the mass dimension
five and so on. This expansion is
convenient because each term is limited by a power of $1/\Lambda.$ We truncate the above series according to the degree of accuracy we
demand. This approach is appropriate for one last reason: it allows us to
focus on the phenomenology common to all new
physics models \cite{Cheng:2007}. We truncate this
series in order two or mass dimension six, ${\cal L}^{(6)}.$ Let us
write ${\cal L}^{(i)}$ as a linear combination of effective
operators of dimension $i$
\begin{equation}
{\cal L}^{(i)} = \sum_j {\cal O}^{(i)}_j =\sum_j\alpha_j
O^{(i)}_j\label{eq:coeff}.
\end{equation}
where $\alpha_j$ are unknown parameters which represent the
coupling strengths and the subscript $i$ -- $i=5,6\cdots$ -- denotes
the dimension of the Lagrangian term. From this procedure an
infinite group of effective operators with the same dimension arises. Finally, we don't know exactly
what the scale is but this is not important for the calculation because we
can parameterize the new physics and including $\Lambda$ in
the unknown coefficients (see eq.~\ref{eq:coeff}).
\label{par:cap1-ener}

Such an approach has been used by several authors, as can be
seen in the following references~\cite{Malkawi:1995dm,*Han:1996ce,*Han:1996ep,*Whisnant:1997qu,*Hosch:1997gz,*Han:1998tp,*Tait:2000sh,*Carlson:1994bg,*Rizzo:1995uv,*Tait:1996dv,*Espriu:2001vj,Hikasa:1998wx}.

\subsection{The standard Lagrangian ${\cal L}^\textrm{SM}$}

Now the task is to build all effective operators of a certain dimension respecting the imposed symmetries. The first term is the SM Lagrangian. We will now briefly review the SM Lagrangian.

The standard model of strong and electroweak interaction can be
described as a Yang-Mills theory, i.e. the Lagrangian
$\cal{L}^{\textrm{SM}}$ is locally $SU(3)\times SU(2)\times U(1)$
invariant \cite{Yang:1954ek,Abers:1973qs}. The symmetry uses
12 gauge vector fields, the gauge bosons.\label{par:tira4}
One important aspect of the SM that should be accounted for is that the
weak bosons have mass. One can not insert this
term into the Lagrangian by hand because such a Lagrangian would not
be gauge invariant. A possible mechanism for
giving mass to the gauge field was found by Higgs who introduced a
complex scalar doublet field $\phi$ with a non-vanishing vacuum
expectation value (vev) -- we will use
$v=<\phi>=246/\sqrt{2}\,GeV$ -- in the Lagrangian. After spontaneous
breaking of the local symmetry the weak bosons acquire mass and a new
scalar boson appears, the Higgs boson. This is known as the Higgs
mechanism~\cite{Higgs:1964pj,*Higgs:1966ev}.\label{par:tira5}

A realistic model of the electroweak interactions was proposed by
Glashow, Salam and Weinberg \cite{Weinberg:1967tq,*Salam1968}.
It uses two non-equivalent representations for the fermions: the
left-handed particles are $SU(2)$ doublets and the right-handed
particles are singlets. In this scheme, the Lagrangian must be
$SU(2)\times U(1)\footnote{We remind that the vector fields $(B_\mu$
and $W^i_\mu$ with $i=1,2,3)$ are not the physical fields. The physical fields
$A_\mu$, $Z_\mu$ and $W^\pm_\mu$ related to the later by the Weinberg angle.}$ invariant. After
spontaneous symmetry breaking, the gauge fields of the weak
interaction become massive. Like the boson fields, we cannot insert
the fermion mass terms by hand as it is not
$SU(2)\times U(1)$ invariant as well. The masses of the fermions are generated
by a mechanism similar to that of the bosons in the well-known Yukawa Lagrangian~\footnote{ The strong
interaction between the quarks, known as quantum chromodynamics
(QCD), appears when one includes $SU(3)$
invariance~\cite{Fritzsch:1973pi}.}.

In order to establish the notation, we will describe the Lagrangian
of the SM. First, the fields: the left-handed lepton doublet and
right-handed charged lepton are represented by $\ell_L$ and $e_R,$
respectively; the left-handed quark doublets by $q_L$ and
right-handed quarks by  $u_R$ and $d_R$; finally, the Higgs
boson doublet by $\phi.$

The gauge fields are:\\
%
\begin{align}
\textrm{gluons}: &\;  G_\mu^a,\:\:\: a=1...8,\nonumber\\
& \; G_{\mu\nu}^a=\partial_\mu G_\nu^a - \partial_\nu G_\mu^a+g_S f_{abc} G_\mu^b  G_\nu^c;\label{eq:gboson}\nonumber\\
W\;\;\textrm{ bosons}: & \;  W_\mu^I,\:\:\: I=1...3,\nonumber\\
&\; W_{\mu\nu}^I=\partial_\mu W_\nu^I - \partial_\nu W_\mu^I+g \epsilon_{IKJ} W_\mu^K  W_\nu^J;\nonumber\\
B\;\; \textrm{ bosons}: &\;   B_\mu,\nonumber\\
&\;  B_{\mu\nu}=\partial_\mu B_\nu - \partial_\nu B_\mu.
\end{align}
where $G_{\mu\nu}^a,$ $W_{\mu\nu}^I$ and $B_{\mu\nu}$ are the field strengths of the $SU(3),$ $SU(2)$ and $U(1)$  interactions respectively.
The Lagrangian is given by
\begin{align}
{\cal L}^{SM} &=  -\frac{1}{4} G^a_{\mu\nu} G^{a\mu\nu} - \frac{1}{4} W^I_{\mu\nu} W^{I\mu\nu} - \frac{1}{4} B_{\mu\nu} B^{\mu\nu}\nonumber\\
&+  (D_\mu\phi)^\dag (D^\mu\phi) + m^2 \phi^\dag \phi -\frac{1}{2}
\lambda(\phi^\dag \phi)^2\nonumber\\
&+  i \bar\ell_L \slashed{D} \ell_L + i \bar e_R \slashed{D} e_R + i
\bar q_L \slashed{D} q_L + i \bar u_R \slashed{D} u_R + i \bar d_R
\slashed{D} d_R\nonumber\\
&+  \Gamma_e \bar\ell_L e_R\phi + \Gamma_u \bar q_L u_R\tilde\phi +
\Gamma_d \bar q_L d_R\phi + h.c.,\label{eq:smlagran}
\end{align}
where
\begin{equation}
D_\mu = \partial_\mu - i g_s\frac{\lambda^a}{2} G^a_{\mu} -
ig\frac{\tau^I}{2}W^I_\mu-i g'YB_\mu \label{eq:der-cov}
\end{equation}
is the covariant derivative and $\Gamma_{\ell,q_u,q_d}$ are Yukawa
couplings for the leptons, up quarks and down quarks, respectively.
$\lambda^a$ acts on colour or $SU(3)$ indices, $\tau^I$ on $SU(2)$
indices and $Y$ is the hypercharge with value assigned as follows:
$Y(\ell_L)=-\frac{1}{2},$ $Y(e_r)=-1,$ $Y(q_L)=\frac{1}{6},$
$Y(u_R)=\frac{2}{3},$ $Y(d_R)=-\frac{1}{3}$ and
$Y(\phi)=\frac{1}{2}.$

A final feature of the SM is that it is a renormalizable theory
\cite{Hooft:1971fh}: a Yang-Mills theory with
spontaneous symmetry breaking is a renormalizable theory if the mass
dimension of the Lagrangian is less than or equal to four, in four space-time dimensions.

\subsection{Dimension five and six  effective operators}

While the choice of the symmetry limits the number of possible terms in the Lagrangian for each order it is obvious that this does not determine which terms are responsible for a particular manifestation of new physic. For example, the term ${\cal L}^\textrm{(6)}$ has 80 effective operators (plus hermitic conjugate) i.e. there are 80 different ways to link together the ``SM fields'' in an $SU(3)\times SU(2)\times U(1)$ invariant operator with mass dimension six and subject to the same broken symmetry as the SM. Some of these operators have only vector bosons. It is obvious  that these operators are not important if we are studying LFV because it is not possible to obtain any vertices with flavour violation. Thus, the choice of operators to include in the Lagrangian depends on the problem we are studying.

We will see that the term ${\cal L}^\textrm{(5)}$ has no interest to us. However, its simplicity may help us illustratring the general construction of the operators.

The ${\cal L}^\textrm{(5)}$ operators are, in principle, all possible combinations of the fields and derivatives building a scalar singlet with mass dimension five. It is not possible to create such an operator, for example, with a fermionic field because such an operator would have at least dimension six. Also, it is not possible to make a dimension five operator with scalar only because we would need five of such fields and it is not possible to put together five doublets in a singlet. In the same way we can eliminate all operators made up by vectors only, fermions and vectors, scalar and vectors and, finally, vectors, fermion and scalar together.

One possible candidate would be to have two fermions and two scalars. If we choose $\phi$ e $\phi^*$ as scalar, the hypercharge must be zero. This is only possible if we use the fermion and its hermitian conjugate. But this is not a gauge invariant scalar. As an alternative, we can use two fields: $\phi$ and its conjugate field defined as $\tilde\phi=i\,\sigma_2\, \phi^*$ in an $SU(2)$ triplet.  For fermions, we must have two doublets of $SU(2)$ to form a scalar. This can be achieved using a fermion field $\Psi$ and its conjugate $\Psi^c.$ Its components $\psi^c$ transform according to~\footnote{The relation is the same well known relation between the plane wave solution to the Dirac positive energy solution $u$ and negative one $v:$ $v=i\, \gamma_2\, u^*.$}
\begin{equation}
\psi^c = i \, \gamma_2 \, \psi^*.
\end{equation}
The conjugate doublet $\Psi^c,$ like $\Psi,$ is a helicity state so both transform in the same way under $SU(2).$ From the chiral or helicity projections $\gamma_{L,R}$ definition:
\begin{equation}
\gamma_{R,L}=\frac{1\pm \gamma_5}{2};
\end{equation}
and its commutation relation with Dirac matrix, we have
\begin{align}
(\Psi_L)^c & = (\Psi)^c_R, \nonumber\\
(\Psi_R)^c & = (\Psi)^c_L.
\end{align}
Following this, only one operator is left for the leptons and
quarks. That is
\begin{align}
{\cal L}^{(1)} & = \bar \ell^c_R \, \phi \, \tilde\phi \, \ell_L \,
+ \, h.c.,
\end{align}
where $\ell^c$ is the lepton conjugate doublet and is similar to the baryon one. We note
that this operator is written as a Majorana mass term $\bar \ell^c_R \, \ell_L \,.$

Finally, ${\cal L}^{(5)}$ breaks baryon and lepton quantum number so we do not need to worry about it as we demand the conservation of these quantum nunbers. Even in
the neutral sector, through seesaw mechanism, the neutrino mass
(Dirac mass) $m_\nu\sim 10^{-2}\; GeV$ requires a Majorana mass
$m_R\sim 10^{16}\; GeV$ for a scale of $1\;
GeV$~\cite{Paschos:2007pi}. This is a typical scale of grand unified
theories but it is clearly out of the reach of the next colliders. We are interested in flavour-violation with lepton and baryon number conservation.

The next term is the dimension six ${\cal L}^{(6)}.$ It is possible
to build 80 (plus hermitian conjugate) of such effective operators
and the corresponding list can be obtained in
\cite{Buchmuller:1985jz}. Now we describe any possible kind of operator according to the fields they contain and we exemplify each class with an effective operator:
\begin{description}
\item[Vectors Only:] there are four such operators and all are make up of three $G$ or three $W$ vector
bosons. For example, $O_G  =  f_{abc} G_\mu^{a\nu} G_\nu^{b\lambda}
G_\lambda^{c\mu}.$ The vector arise in the operator either though their field strengths or covariant derivative;

\item[Fermions Only:] these are four-fermion operators and all of them satisfy fermion number conservation. There are four different groups of operators
and they are of the form $\bar LL\bar LL,$ $\bar RR\bar RR,$ $\bar
LR\bar RL$ and $\bar LR\bar LR,$ where $L$ and $R$ are left-handed
and right-handed fields, respectively. Below, is an example for each
group (we use lepton and baryon fields indiscriminately):
    \begin{description}
    \item[$\bar LL\bar LL:$] $O^{(1)}_{\ell\ell}  =
    \frac{1}{2}(\bar\ell_L\gamma_\mu\ell_L)(\bar\ell_L\gamma^\mu\ell_L).$

    \item[$\bar RR\bar RR:$] $O_{ee}  =
    \frac{1}{2}(\bar e_R\gamma_\mu e_R)(\bar e_R\gamma^\mu e_R).$

    \item[$\bar LR\bar RL:$] $O_{\ell e}  =
    (\bar\ell_L e_R)(\bar e_R\ell_L).$

    \item[$\bar LR\bar LR:$] $O^{(1)}_{qq}  =
    (\bar q_L u_R)(\bar q_L d_R).$
    \end{description}
For our study, this family of operators has special interest because
they are responsible for the four-fermion operator with flavour
change;

\item[Scalars Only:]
these operators have either six bosons or four bosons and two
derivatives. An example of both is: $O_{\phi}  =
\frac{1}{3}(\phi^\dag\phi)^3$ and $O_{\partial\phi} =
\frac{1}{2}\partial_\mu(\phi^\dag\phi)\partial^\mu(\phi^\dag\phi);$

\item[Fermions and vectors:]
these operators require two fermions and three other powers of mass
that can come via a covariant derivative and one field strength.
Here is an exemple: $O_{\ell W} = i\bar\ell_L
\tau^I\gamma_\mu D_\nu\ell_lW^{I\mu\nu};$
%

\item[Scalars and vectors:]
the $\phi$ and $\phi^\dag$ must come in equal numbers in order to
ensure $SU(2)\times U(1)$ invariance. With just one of each
scalar field we can use two fields strengths, one fields strengths
and two covariant derivatives or four covariant
derivatives\footnote{In the last case we can use the equation of
motion and split this operators in fermion fields only. The
equation of motion to $\phi^\dag$ is $D^2\phi-\Gamma^\dag_e \bar e
\ell +i \Gamma_u\bar q\sigma_2 u-\Gamma^\dag_d\bar d q.$}. In the
case of two of each scalar field one can have two covariant
derivatives with act ont two diferents fields -- $(
D^\mu\phi^\dag)\phi+\phi^\dag D^\mu\phi.$ Applying the equation of
motion one obtain operators like $O_{\phi G} =
\frac{1}{2}(\phi^\dag\phi)G^a_{\mu\nu} G^{a\mu\nu};$

\item[Fermions and scalars:]
in this case one must have two fermions and three bosons or two
bosons and one derivative which acts on a gauge invariant quantity.
These
operators are for example $O_{e \phi} = (\phi^\dag\phi)(\ell_L e_R\phi);$

\item[Vectors, fermions and scalars:]
one can separate two situations: two  fermions and one scalar or two
scalar. In the first case, we can have two covariant derivatives or
one field strength. The two covariant derivative can act both on
scalar field, one on the scalar field and one on a fermion or both
on the fermions. An example of this kind of operators is: $O_{D e} =
(\bar\ell_L D_\mu e_R)D^\mu\phi.$

In the case with two scalar fields one must have one covariant
derivative. From the hypercharge assignments the only possibility is
the derivative act in the scalar. An example of these fields:
$O_{\phi\ell} = i(\phi^\dag D_\mu\phi)(\bar\ell\gamma^\mu \ell).$
\end{description}

We must identify amongst the 80 operators with dimension six
all those which are relevant for our research. After this, it is necessary to check if all are linearly
independent, that is if they are not connected by the equation of
motion or some other operation. If that is the case, we can perform some simplifications.
Once the operators are defined, the next step is to determine the
Feynman rules (in Appendix~\ref{app:cap1} we review the derivation of a Feynman rule) and calculate the processes of interest. So far we have looked at the method from a general point of view. For our work, as mentioned before, we are interested in those processes with FCNC in
$t\bar t$ and single top quark production as well as LFV in the charged sector. Also, we can conclude that in the construction of our effective operators we have to deal with SM terms according to the equation~\ref{eq:smlagran} and terms of dimension six FCNC and LFV Lagrangian.

We will point out top quark properties and experimental limits of FCNC
and LFV, as well as the theoretical framing of the production and
expected experimental limits of top quark in the LHC and
experimental limits of LFV in the next ILC according SM. The point
of this discussion lies on the fact that one can constrain the space
of valid solution for flavour violation and identify the conditions
for physics beyond SM in the LHC and the ILC.

\section{FCNC in the top physics}

The LHC is a proton-proton collider being constructed at CERN in a
tunnel about $100\; m$ below the ground and with $27\; Km$ in circumference.
It's center-of-mass energy is $14\; TeV$ and the expected luminosity is $10^{34}\;
cm^{-2}s^{-1}.$  Some of its main goals are the search for the Higgs
boson, the search for new phenomena such as supersymmetry, extra
dimensions, mini black holes and to perform precision measurement of the
SM. Particularly relevant to this thesis are the precision
measurements related for the production and decay of the top quark.

According to the SM, top quark can be created in pairs via the strong
force or singly (single top quark production) via the electroweak
interaction.

\subsection{Top quark-antiquark pairs production}

 The top quark was discovered at Fermilab
in 1995~\cite{Abachi:1995iq,*Abe:1995hr} in the mass range predicted
by SM $\sim 170.9 \pm 1.8\,GeV$~\cite{:2007bxa}. Its large Yukawa coupling in the symmetry breaking sector (due
to its big mass) offers the possibility to look for new physics. The
top quark, unlike the other quark, decays almost exclusively in
$t\rightarrow bW$ before its hadronization due to its extremely short
lifetime of $\sim 4\times 10^{-25}\; s.$ The ratio between the decay
time scale and the hadronization time scale is about one order of
magnitude. The next decays are $t\rightarrow sW$ and
$t\rightarrow dW,$ both suppressed by the square of the CKM matrix
elements. Taking $|V_{ts}|\sim 0.04$ and $|V_{td}|\sim 0.01,$ we
obtain $Br(t\rightarrow sW)\sim 1.6\times 10^{-3}$ and
$Br(t\rightarrow dW)\sim 1.6\times 10^{-4}$ \cite{Beneke:2000hk}.

\begin{figure}[!htbp]
  \begin{center}
    \includegraphics[scale=1.0]{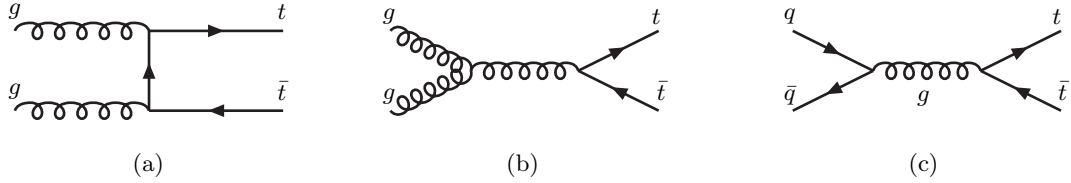}
    \caption{Feynman diagrams for $t\bar t$ production: (a) and (b) by gluon fusion and (c) by $q\bar q$ annihilation.}
    \label{fig:ttbars}
  \end{center}
\end{figure}

At the Tevatron, with a center of mass energy $\sqrt{s} = 1.96\; TeV,$ top
quarks are produced predominantly in pairs via the strong interaction
-- $85\%$ by $q\bar q$ annihilation and $15\%$ by gluon-gluon
fusion. The corresponding SM cross section in NLO+NNLL\footnote{Next-to-leading order + next-to-next-to-leading logarithmic.} is $6.77\pm
0.42\;pb$ for a top quark mass of $175 GeV$~\cite{Kidonakis:2003qe} and the Feynman diagrams to this
process in the SM are shown in fig.~\ref{fig:ttbars}.

Most of top quarks produced at the LHC -- about $833\pm 100 pb$
\cite{Bonciani:1998vc,Beneke:2000hk} -- will be quark-antiquark pairs. Of these,
approximately $83\%$ will be produced by gluon-gluon fusion and
$17\%$ by $q\bar q$ annihilation.

\subsection{Single top production}\label{sec:singletop}

Studying single-top production at hadron
colliders is important for a number of reasons: it provides a direct measurement of the CKM matrix element $|V_{tb}|^2;$
 it measures the spin polarization of top quarks;  lastly, the presence of various new non-SM phenomena
may be inferred by observing deviations from the predicted rate of
the single-top signal~\cite{Ciobanu:2007an}. We are particularly interested in this last reason.

\begin{figure}[!htbp]
  \begin{center}
    \includegraphics[scale=1.0]{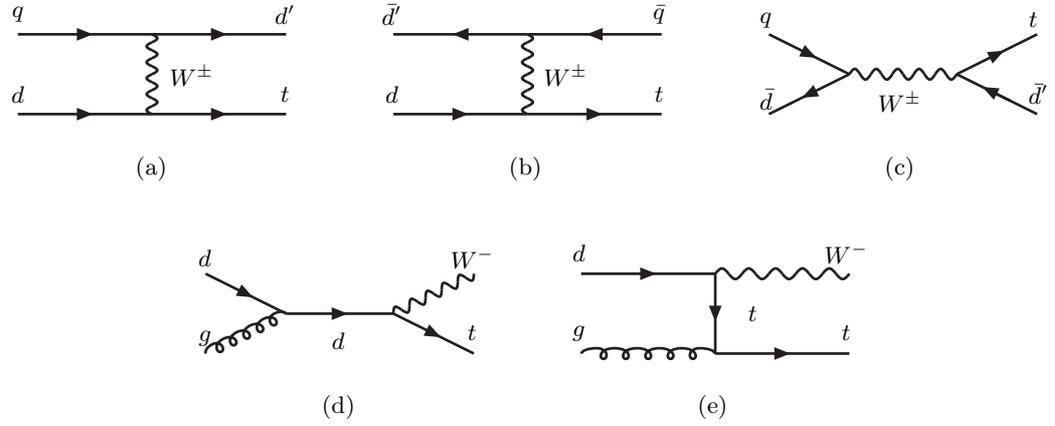}
    \caption{Feynman diagrams for electroweak single top production: (a) and (b) t-channel, (c) s-channel, (d) s-channel associated $tW$ production and (e) t-channel associated $tW$ production.}
    \label{fig:single-t}
  \end{center}
\end{figure}

\begin{figure}[!htbp]
  \begin{center}
    \includegraphics[scale=1.4]{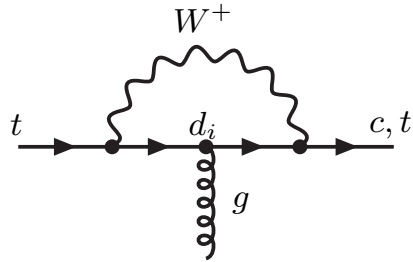}
    \caption{Top quark flavour change to one loop.}
    \label{fig:one-loop}
  \end{center}
\end{figure}

There are three electroweak production
mechanisms for single top quarks in the SM: $t$-channel
$(qd\rightarrow t d')$ and $(d\bar d'\rightarrow t \bar q)$ as we see in Fig.~\ref{fig:single-t}-(a) and (b);
$s$-channel $(q\bar{d}\rightarrow t\bar d')$~\ref{fig:single-t}-(c);
associated $tW$ production $(gd\rightarrow
tW),$~\ref{fig:single-t}-(d) and (e).
The theoretical single top quark production cross section at the
Tevatron is $\sim 2.9\; pb.$ Evidence of a single top quark production
with a significance of $3.4$ standard deviation was reported by the D0
Collaboration \cite{Abazov:2006gd}. The LHC will be able to measure
the assumed SM cross section of single top events at NLO to be $\sim 245\pm 12\; pb$~\cite{Sullivan:2004ie}, $\sim 11\pm 1\; pb$~\cite{Campbell:2005bb} and $\sim 66\pm 2\; pb$~\cite{Sullivan:2004ie} for those that occur through the $t$~-channel, $s$~-channel and associated $tW$ production, respectively. The neutral coupling preserves flavour; this implies that FCNC are absent at the tree level. In
principle, the top production could occur at
one loop (see Fig.\ref{fig:one-loop} for example).
Nevertheless, because of Glashow-Iliopous-Maiani (GIM) cancelation~\cite{Glashow:1970gm,Glashow:1976nt,*Paschos:1976ay} the flavour changing related to
radiative corrections is suppressed. As such, the branching ratios
of these rare top decays are immensely suppressed but can be much
larger in extensions of the model. Essentially, in a different
theory  the existence of new particles will give new contributions
to the top rare decays. There can be differences of as
much as thirteen orders of magnitude between the SM branching ratios
and those in some models, as may be observed in Tab.~\ref{tab:br}.
\begin{table}[!htb]
\vspace*{0.2cm}
\begin{center}
\begin{small}
\begin{tabular}{lccccc}
\hline  \hline
 Process  & SM & QS & 2HDM & MSSM & $R \!\!\!\!\!\!  \not \quad$ SUSY
   \\
\hline  \\
$t \to u Z$ & $8 \times 10^{-17}$ & $1.1 \times 10^{-4}$
  & $-$
  & $2 \times 10^{-6}$ & $3 \times 10^{-5}$ \\
$t \to u \gamma$ & $3.7 \times 10^{-16}$ & $7.5 \times 10^{-9}$
 & $-$
 & $2 \times  10^{-6}$ & $1 \times 10^{-6}$ \\
$t \to u g$ & $3.7 \times 10^{-14}$ & $1.5 \times 10^{-7}$
  & $-$
  & $8 \times 10^{-5}$ & $2 \times 10^{-4}$ \\
%
%
$t \to c Z$ & $1 \times 10^{-14}$ & $1.1 \times 10^{-4}$
  & $\sim 10^{-7}$
  & $2 \times 10^{-6}$ & $3 \times 10^{-5}$ \\
$t \to c \gamma$ & $4.6 \times 10^{-14}$ & $7.5 \times 10^{-9}$
 & $\sim 10^{-6}$
 & $2 \times 10^{-6}$ & $1 \times 10^{-6}$ \\
$t \to c g$ & $4.6 \times 10^{-12}$ & $1.5 \times 10^{-7}$
  & $\sim 10^{-4}$
  & $8 \times 10^{-5}$ & $2 \times 10^{-4}$ \\
\hline \hline
\end{tabular}
\end{small}
\end{center}
\caption{Branching ratios for FCNC decays of the top quark in the SM
and several possible extensions: the quark-singlet model (QS), the
two-higgs doublet model (2HDM), the minimal supersymmetric model
(MSSM) and SUSY with R-parity violation. See ref.
\protect\cite{AguilarSaavedra:2004wm,Luke:1993cy,*Atwood:1996vj,*Yang:1997dk,*Guasch:1999jp,*Delepine:2004hr,*Liu:2004qw,Eilam:2006rb,*Cao:2007dk,*Arhrib:2006pm,*Arhrib:2006sg,*Guasch:2006hf,*LopezVal:2007rc,*AguilarSaavedra:2002kr,*delAguila:1998tp,*Cheng:1987rs,*Bejar:2000ub,*Li:1993mg,*deDivitiis:1997sh,*Lopez:1997xv,*Eilam:2001dh}
for details.} \label{tab:br}
\end{table}
In the SM these type of decays are so rare that they will never be observed in
experiments due to lack of sensitivity (the sensitivity of the experiments at
the LHC is of $O(10^{-5})$ at best~\cite{delAguila:2008iz}).

As we have said, we can parameterize the effects of FCNC through an effective Lagrangian. It is therefore important to use the experimental data to limit the coefficients of operators related with all FCNC. Indirect
bounds~\cite{Fox:2007in,Larios:2004mx,*Peccei:1990uv,*Han:1995pk,*Martinez:1994my,*Han:1996ep}
come from electroweak precision physics and from B and K physics.
The strongest bounds so far are the ones in \cite{Fox:2007in} where
invariance under $SU(2)_L$ is required for the set of operators
chosen. Top and bottom physics are related and B physics
can be used to set limits on operators that involve top and bottom
quarks through gauge invariance. Regarding $Br (t \,\rightarrow\,
q\, Z)$ and $Br (t\,\rightarrow \,q\, \gamma)$, the only direct
bounds available to date are the ones from the Tevatron (CDF). The
CDF collaboration has searched its data for signatures of $t
\,\rightarrow \,q \, \gamma$ and $t\, \rightarrow \,q \, Z$ (where
$q\,=\,u,c$). Both analyses use $p\bar{p}\, \rightarrow \,
t\,\bar{t} $ data and assume that one of the top decays according
to the SM into $W\,b$. The results are presented in Table
\ref{tab:limits}.
\begin{table}[!hbp]
\begin{center}
  \begin{tabular}{ | l | c | c | c |}
    \hline
     & LEP & HERA & Tevatron  \\
     \hline \hline\\[-0.4cm]
    $Br(t \rightarrow q \, Z)$      & $ < \, 7.8 \% \,$\cite{Heister:2002xv,*Abdallah:2003wf,*Abbiendi:2001wk,*Achard:2002vv} & $  < \, 49\% \,$ \cite{Chekanov:2003yt}
    & $ < \, 10.6 \% \,^d$ \cite{Aaltonen:2008aa}\protect\footnote{{\bf [VER SE A REFERÊNCIA \cite{Aaltonen:2008aa} ACESCENTA ALGO DE
NOVO]}}  \\ \hline
    $Br(t \rightarrow q \, \gamma)$ & $ < \, 2.4 \% \,$~\cite{Heister:2002xv,*Abdallah:2003wf,*Abbiendi:2001wk,*Achard:2002vv} & $ < \, 0.75 \% \,$~\cite{Chekanov:2003yt}
    & $ < \, 3.2 \% \,^d$ \cite{Abe:1997fz}  \\ \hline
    $Br(t \rightarrow q \, g)$      & $ < \, 17 \% \,$ \cite{Beneke:2000hk}  & $ < \, 13 \% \,$\cite{Ashimova:2006zc,*Aktas:2003yd,Chekanov:2003yt}
    & $ < \, O (0.1 - 1 \%) \,$ \cite{Abazov:2007ev,Cheng:2007} \\
    \hline
  \end{tabular}
\end{center}
\caption{Current experimental bounds on FCNC branching ratios. The
upperscript ``d'' refers to bounds obtained from direct
measurements, as is explained in the text.} \label{tab:limits}
\end{table}
 The bounds on the branching ratios from LEP and ZEUS are bounds on
the cross section that were then translated into bounds on the
branching ratios through the anomalous couplings. The LEP bounds use
the same anomalous coupling for the $u$ and $c$ quarks and the ZEUS
bound is only for the process involving a $u$ quark. The bounds on
$Br (t \rightarrow \,q\, g)$ are all from cross sections translated
into branching ratios. Usually only one operator is considered, the
chromomagnetic one, which makes the translation straightforward. The
same searches are being prepared for the LHC. A detailed discussion
with all present bounds on FCNC and the predictions for the LHC can
be found in \cite{Carvalho:2007yi,delAguila:2008iz,Ball:2007zza}.
With a luminosity of $100\; fb^{-1}$ and in the absence of signal,
the 95\% confidence level expected bounds on the branching ratios
are $Br(t \,\rightarrow\, q\, Z)\,\sim\,10^{-5}$, $Br(t
\,\rightarrow\, q\, \gamma)\,\sim\,10^{-5}$ and $Br(t
\,\rightarrow\, q\, g)\,\sim\,10^{-4}$. In this thesis, we assume a $10^{-2}$ upper bond of those FCNC branching ratios in except when otherwise mentioned.

\section{Linear Collider and LFV}\label{sec1:lepton}

\begin{table}[t]
\begin{tabular}{cc}
 Process & Upper bound
\\
\hline $\tau\,\rightarrow\,e\,e\,e$ & $2.0 \times 10^{-7}$    \\
$\tau\,\rightarrow\,e\,\mu\,\mu$ & $2.0 \times 10^{-7}$    \\
$\tau\,\rightarrow\,\mu\,e\,e$ & $1.1 \times 10^{-7}$   \\
$\tau\,\rightarrow\,\mu\,\mu\,\mu$ & $1.9 \times 10^{-7}$     \\
$\mu\,\rightarrow\,e\,e\,e$ & $1.0 \times 10^{-12}$   \vspace{0.2cm}  \\
  \hline
$Z\,\rightarrow\,e\,\mu$ & $1.7 \times 10^{-6}$   \\
$Z\,\rightarrow\,e\,\tau$ & $9.8 \times 10^{-6}$
 \\
$Z\,\rightarrow\,\tau\,\mu$ & $1.2 \times 10^{-5}$  \\
\end{tabular}
\caption{Experimental constraints on flavour-violating decay
branching ratios \protect\cite{Yao:2006px}.} \label{tab:dec}
\end{table}

In chapter~\ref{cha:lepton} we will parameterize the possibility of LFV in the charged sector through effective operators. Whereas with massless neutrinos LFV is
not allowed in the SM, in the analysis of the signal and
background this is a major simplification when compared to the analysis of
FCNC. On the other hand, this is a sector of particle physics for
which we already have many experimental results~\cite{Yao:2006px},
all of which set stringent limits on the extent of flavour violation
that may occur. Nevertheless, as we will show in this thesis, even with all known
experimental constraints, signals of LFV may be
observed. The LHC is a hadronic machine, and as such precision
measurements will be quite hard to undertake there. Also, the
existence of immense backgrounds at the LHC may hinder discoveries
of new physical phenomena already possible at the energies that this
accelerator will achieve. Thus it has been proposed to build a new
electron-positron collider, the International Linear Collider (ILC)
\cite{Battaglia:2006bv}. This would be a collider with  a
center-of-mass energy of ground $1\; TeV$ and a planned
integrated luminosity of $1\; ab^{-1}.$  The potential
for new physics with such a machine is immense.

Finally, in Table \ref{tab:dec} we present the experimental limits from the branching ratios of lepton decay  with flavour-violation. In the final part of the chapter~\ref{cha:lepton} we will discuss the possible improvement of these limits in the existing colliders.

\newpage
\addcontentsline{toc}{section}{{\bf Appendix 2}}

\begin{subappendices}

\section{Effective operators and Feynman rules}\label{app:cap1}

We will give an example of the derivation of the Feynman rules on a dimension six operator~\cite{Ryder:1996}
\begin{align}
{\cal O}_{qG} & = \sum_{i,j}  \frac{i}{\Lambda^2}
\alpha_{ij} \bar{q}^i \gamma^\mu \lambda^a D^\nu q^j G^a_{\mu\nu},\label{eq:ane1}
\end{align}
where $i,j=1,2,3$ correspond to three quark families. As we are interested in flavour violation, we must get  $i\neq j$ and
\begin{align}
{\cal O}_{qG} & = \sum_{i\neq j} \frac{i}{\Lambda^2}
\alpha_{ij} \bar{q}^i \gamma^\mu\lambda^a D^\nu q^j G^a_{\mu\nu}.\label{eq:ane2}
\end{align}
Taking a particular $i$ and $j$
\begin{align}
\tilde{\cal O}_{qG} & =  \frac{i}{\Lambda^2} \ \alpha_{ij}
\bar{q}^i \gamma^\mu\lambda^a D^\nu q^j G^a_{\mu\nu} + 
\frac{i}{\Lambda^2} \alpha_{ji} \bar{q}^j \gamma^\mu\lambda^a D^\nu
q^i G^a_{\mu\nu}.\label{eq:ane3}
\end{align}
The hermitian conjugate of the second term in the right side is:
\begin{align}
\tilde{\cal O}^\dag_{qG} & =  \frac{-i}{\Lambda^2}
\alpha^*_{ji} \bar{q}^i \gamma^\mu \overleftarrow{D}^\nu\lambda^a q^j
G^a_{\mu\nu}.\label{eq:ane4}
\end{align}
Now, let's add this term to the first one in the right side of eq.~\ref{eq:ane1} and we will designate result by ${\cal O}_{qG}$ again
\begin{align}
{\cal O}_{qG} & =  \frac{i}{\Lambda^2}  \bar{q}^i
\gamma^\mu \lambda^a(\alpha_{ij} \overrightarrow{D}^\nu-\alpha^*_{ji}
\overleftarrow{D}^\nu) q^j G^a_{\mu\nu}.\label{eq:ane5}
\end{align}
According with eg.~\ref{eq:der-cov}, the covariant derivative has a partial derivative whose term generates a triple vertex and the other one is comprised by gauge fields and originates a four vertex. Let's focus in the triple vertices
\begin{align}
{\cal O}_{qG} & =  \frac{i}{\Lambda^2} \alpha_{ij}
\bar{q}^i \gamma^\mu \lambda^a(\alpha_{ij}
\overrightarrow{\partial}^\nu-\alpha^*_{ji}
\overleftarrow{\partial}^\nu) q^j (\partial_\mu
G^a_{\nu}-\partial_\nu G^a_{\mu}).\label{eq:ane6}
\end{align}
Introducing the identity $1=\gamma_L + \gamma_R$ and some manipulation, we get
\begin{align}
{\cal O}_{qG} & = \frac{i}{\Lambda^2} \left[\bar{u}^i_L
\gamma^\mu \lambda^a(\alpha_{ij} \overrightarrow{\partial}^\nu-\alpha^*_{ji}
\overleftarrow{\partial}^\nu) u^j_L
+ \bar{d}^i_L \gamma^\mu\lambda^a
(\overrightarrow{\partial}^\nu-\overleftarrow{\partial}^\nu)
d^j_L\right]
(\partial_\mu G^a_{\nu}-\partial_\nu G^a_{\mu})\nonumber\\
& +  \frac{i}{\Lambda^2}  \bar{u}^i_R \gamma^\mu\lambda^a
(\alpha_{ij} \overrightarrow{\partial}^\nu-\alpha^*_{ji}
\overleftarrow{\partial}^\nu) u^j_R (\partial_\mu
G^a_{\nu}-\partial_\nu G^a_{\mu}).\label{eq:ane7}
\end{align}
We can concentrate in the right sector (the left one is similar) so, from eq.~\ref{eq:ane7}
\begin{align}
{\cal O}^R_{qG} & = \frac{i}{\Lambda^2} \bar{u_R}^i
\gamma^\mu \gamma_R \lambda^a(\alpha_{ij}
\overrightarrow{\partial}^\nu-\alpha^*_{ji}
\overleftarrow{\partial}^\nu) u^j (\partial_\mu
G^a_{\nu}-\partial_\nu G^a_{\mu}).\label{eq:ane8}
\end{align}

\begin{figure}[h]
  \begin{center}
    \includegraphics[scale=0.8]{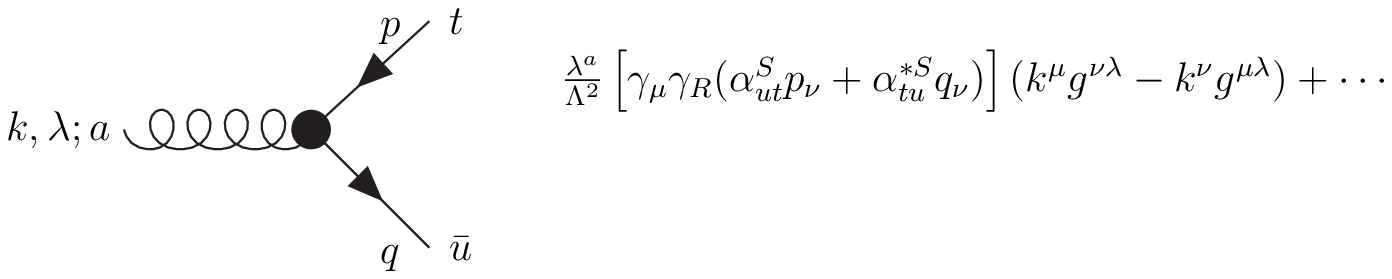}
    \caption{Feynman rules for the $gt\bar u.$}
    \label{fig:ttbarsm}
  \end{center}
\end{figure}

The functional $\Gamma[\phi]$ generates the n-point vertex function $\Gamma^n(p)$ and is defined by the Legendre transformation\cite{Ryder:1996}
\begin{equation}
W[J]=\Gamma[\phi]+\int dx\, J(x)\,\phi(x),\label{eq:ane9}
\end{equation}
where $J(x)$ is the source, $\phi$ is a generic field of the theory and $W[J]$ is the generating functional which generates only connected Green's functions. The relation between $\Gamma[\phi]$ and the Lagrangian is given by
\begin{align}
\Gamma[\phi] & = \int d^4x\, {\cal L}_{int}[\phi].\label{eq:ane10}
\end{align}
Using the eqs.~\ref{eq:ane9} and~\ref{eq:ane10} the 3-point vertex function in spacial coordinates are given by
\begin{align}
\Gamma^{(3)}_\lambda(x_1,x_2,x_3) & = \frac{\delta^3 \Gamma}{\delta
G^b_\lambda(x_3)  \delta \bar{u}^k(x_2) \delta u^l(x_1)}\nonumber\\
& = i \frac{\lambda^a}{\Lambda^2} \int d^4x \delta^{il}
\delta^4(x-x_2) \gamma^\mu \gamma_R (\alpha_{ij}
\overrightarrow{\partial}^\nu-\alpha^*_{ji}
\overleftarrow{\partial}^\nu)
\delta^4(x-x_1) \delta^{jk}\nonumber\\
& \times \left(g_{\nu\lambda}\partial_\mu
-g_{\mu\lambda}\partial_\nu \right) \delta^4(x-x_3) \delta^{ab}.\label{eq:ane11}
\end{align}
We use the Fourier transform to get the 3-point vertex function in the momentum space (we use a convention where the $u$ and $G$ momenta incoming and the $\bar u$ momenta is outcoming). So
\begin{align}
TF & = (2\pi)^4 \delta^4(p-q+k) \Gamma_\lambda(p,q,k)\nonumber\\
& = i \frac{\lambda^a}{\Lambda^2} \int d^4x d^4x_1 d^4x_2 d^4x_3
e^{-i(p x_1 - q x_2 + k x_3)}
 \delta^4(x-x_2) \gamma^\mu \gamma_R
(\alpha_{ij} \overrightarrow{\partial}^\nu-\alpha^*_{ji}
\overleftarrow{\partial}^\nu)
\delta^4(x-x_1)\nonumber\\
& \times \left(g_{\nu\lambda}\partial_\mu
-g_{\mu\lambda}\partial_\nu \right) \delta^4(x-x_3)\nonumber\\
& = i \frac{\lambda^a}{\Lambda^2} \int d^4x d^4x_2 d^4x_3 e^{-i(p x
- q x_2 + k x_3)} \gamma^\mu \gamma_R \delta^4(x-x_2)(\alpha_{ij} i
p^\nu-\alpha^*_{ji} \overleftarrow{\partial}^\nu)
\nonumber\\
& \times \left(g_{\nu\lambda}\partial_\mu
-g_{\mu\lambda}\partial_\nu \right) \delta^4(x-x_3).\label{eq:ane12}
\end{align}
Recall that
\begin{align}
\int d^4x \partial^\mu \; \delta^4(x) & = \int d^4x \; d^4p \;
\partial^\mu e^{i p^\nu x_\nu}\nonumber\\
& = i \; \int d^4x \; d^4p \;  p^\nu \; g_{\mu \nu} \; e^{i p^\nu x_\nu}\nonumber\\
& = i \; \int d^4p \;  p_\mu  \; \int d^4x \; e^{i p \cdot x}\nonumber\\
& = i \; \int d^4p \;  p_\mu  \; \delta^4(p)\nonumber\\
& = i \; p_\mu
\end{align}
and
\begin{align}
TF & = i \frac{\lambda^a}{\Lambda^2} \int d^4x d^4x_3 e^{-i\left((p
- q) x + k x_3\right)} \gamma^\mu \gamma_R (i) (\alpha_{ij} p^\nu +
\alpha^*_{ji} q^\nu) \left(g_{\nu\lambda}\partial_\mu
-g_{\mu\lambda}\partial_\nu \right) \delta^4(x-x_3)\nonumber\\
& = i \frac{\lambda^a}{\Lambda^2} \int d^4x  e^{-i(p - q + k )x}
\gamma^\mu \gamma_R (i)(\alpha_{ij} p^\nu + \alpha^*_{ji} q^\nu) (i)
\left(g_{\nu\lambda} (p_\mu + q_\mu) -g_{\mu\lambda} (p_\nu + q_\nu)
\right).\nonumber
\end{align}
Momentum conservation in the vertex implies
\begin{align}
TF & = i \frac{\lambda^a}{\Lambda^2} \int d^4x  e^{-i(p - q + k )x}
\gamma^\mu \gamma_R (\alpha_{ij} p^\nu + \alpha^*_{ji} q^\nu)
\left(k_\mu g_{\nu\lambda}  - k_\nu g_{\mu\lambda}
\right)\nonumber\\
& = i \frac{\lambda^a}{\Lambda^2} \gamma^\mu \gamma_R (\alpha_{ij}
p^\nu + \alpha^*_{ji} q^\nu) \left(k_\mu g_{\nu\lambda}  - k_\nu
g_{\mu\lambda} \right)\int d^4x  e^{-i(p - q + k )x}\nonumber\\
& = i \frac{\lambda^a}{\Lambda^2} \gamma^\mu \gamma_R (\alpha_{ij}
p^\nu + \alpha^*_{ji} q^\nu) \left(k_\mu g_{\nu\lambda}  - k_\nu
g_{\mu\lambda} \right) (2\pi)^4 \delta^4(p-q+k)\label{eq:doisvinte}
\end{align}
and replacing~\ref{eq:ane12} in the left side of~\ref{eq:doisvinte}
\begin{align}
\Gamma_\lambda(p,q,k) & = i \frac{\lambda^a}{\Lambda^2} \gamma^\mu
\gamma_R (\alpha_{ij} p^\nu + \alpha^*_{ji} q^\nu) \left(k_\mu
g_{\nu\lambda} - k_\nu g_{\mu\lambda} \right).\label{eq:vertex}
\end{align}
We represent this vertex by a black circle in the Feynman diagrams. Thus, eq.~\ref{eq:vertex} is represented by the fig~\ref{fig:ttbarsm}
where the ellipsis recall that there should be more terms that must be include in this vertex (see fig.~\ref{fig:feyngamma}).

Now, we discuss some points that will be useful for this work. According with our convention, the momentum of the boson vector is always entering in the vertex while with the fermions, the momentum is entering in the particle case and leaving in the anti-particle. We have also computed a rule where the boson is represented by a gluon -- so this means a strong interaction. The same treatment can be applied to the photon and to the $Z$ boson.  We will discuss the effective operators responsible for these vertices in the next two chapters.

\end{subappendices}

\chapter{Top quark production and decay in the effective Lagrangian
approach}\label{cap:capII}
	
According to the previous chapter, the effective Lagrangian can be expanded as
\begin{equation}
{\cal L}^\textrm{eff} = {\cal L}^\textrm{SM} +
\frac{1}{\Lambda}{\cal L}^\textrm{(5)} + \frac{1}{\Lambda^2} {\cal
L}^\textrm{(6)} + \cdots .\label{eq:expand-1}
\end{equation}
We truncate this Lagrangian in the order $\Lambda^{-2}$ -- or dimension six -- and ignore the dimension five term. The contributions of order $\Lambda^{-2}$ and $\Lambda^{-4}$ to the single top quark and $t\bar t$ production with FCNC are summarized in Tab.~\ref{tab:sec}.
%
\begin{table}[!h]
\begin{center}
  \begin{tabular}{|l|l|l|}
    \hline
    1 & direct production & $p\; p\rightarrow (g\;q)\rightarrow t\; $ \\
     \hline
    2 & top + jet production & $p\; p\rightarrow (g\;g)\rightarrow \bar q\; t\; +\; X$ \\
     & & $p\; p\rightarrow (g\;q)\rightarrow g\; t\; +\; X$ \\
     & & $p\; p\rightarrow (\bar q\;q)\rightarrow \bar q\; t\; +\; X$ \\
     & & (including 4-fermion interactions) \\
    \hline
    3 & top + anti-top production & $p\; p\rightarrow (g\;g)\rightarrow \bar t\; t\; +\; X$ \\
     & & $p\; p\rightarrow (\bar q\;q)\rightarrow \bar t\; t\; +\; X$ \\
     \hline
    4 & top + gauge boson production & $p\; p\rightarrow (g\;q)\rightarrow \gamma\; t\; +\; X$ \\
     & & $p\; p\rightarrow (g\;q)\rightarrow Z\; t\; +\; X$ \\
     & & $p\; p\rightarrow (g\;q)\rightarrow W\; t\; +\; X$ \\
     \hline
     5 & top + Higgs production & $p\; p\rightarrow (g\;p)\rightarrow H\; t\; +\; X$ \\
     \hline
  \end{tabular}
\end{center}
\caption{Contributions of order $\Lambda^{-2}$ and $\Lambda^{-4}$ to the cross section of
top production~\protect\cite{delAguila:2008iz}.} \label{tab:sec}
\end{table}
The FCNC processes $pp \rightarrow t \, Z$ and $pp \rightarrow t \, \gamma$ were studied in great detail for the Tevatron in~\cite{delAguila:1999ac} and for the LHC in~\cite{delAguila:1999ec}. The authors of~\cite{Ferreira:2005dr,Ferreira:2006xe,Ferreira:2006in} have presented a complete study of those processes in the strong sector; they have presented a full analytical expression of all processes listed in that table as well as the conditions and limits for their observation at the LHC. The process label in the Tab.~\ref{tab:sec} by 1 and 2 was treated in the article~\cite{Ferreira:2005dr,Ferreira:2006xe} and 3, 4 and 5 in article~\cite{ Ferreira:2006in}. Now, our propose is, following refs.~\cite{Ferreira:2008cj,Ferreira:2008xx}, to present the same treatment for the electroweak sector i.e. the contribution due the effective FCNC electroweak operators to top quark production; finally we study the combined effects of strong and electroweak effective FCNC operators in top production.

\section{FCNC effective operators}
\label{sec:eff2}

We are interested in effective operators of dimension six that
contribute to flavour-changing interactions of the top quark in the
weak sector. As we have said before we do not consider ${\cal L}^{(5)}$ in our analysis. We follow
refs.~\cite{Ferreira:2005dr,Ferreira:2006xe,Ferreira:2006in}, where
the authors considered FCNC top effective operators which affect the
strong sector. Namely, operators which, amongst other things,
contribute to FCNC decays of the form $t\, \rightarrow \,u\,g$ or
$t\, \rightarrow \,c\,g$. The operators they considered were
expressed as
\begin{align}
{\cal O}_{tG} &= i \frac{\alpha^S_{it}}{\Lambda^2}\,
\left(\bar{u}^i_R \, \lambda^{a} \, \gamma_{\mu}  D_{\nu} t_R\right)
\, G^{a \mu \nu} \;\;\;,\;\;\;{\cal
O}_{tG\phi}=\frac{\beta^{S}_{it}}{\Lambda^2}\,\left(\bar{q}^i_L\, \,
\lambda^{a} \, \sigma^{\mu\nu}\,t_R\right)\, \tilde{\phi}
\,G^{a}_{\mu\nu}\;\;\; , \label{eq:opst}
\end{align}
where the coefficients $\alpha^S_{it}$ and $\beta^{S}_{it}$ are
complex dimensionless couplings. The fields $u^i_R$ and $q^i_L$
represent the right-handed up-type quark and left-handed quark
doublet of the first and second generation -- this way FCNC occurs.
$G^a_{\mu\nu}$ is the gluonic field tensor. There are also
operators, with couplings $\alpha^S_{ti}$ and $\beta^{S}_{ti}$,
where the positions of the top and $u^i$, $q^i$ spinors are
exchanged in the expressions above. Also, the hermitian conjugates
of all of these operators are obviously included in the lagrangian.
These operators contribute to FCNC vertices of the form
$g\,t\,\bar{u_i}$ (with $u_i \,=\,u\,,\,c$). The operators with
$\alpha^S$ couplings, due to their gauge structure (namely, the
covariant derivative acting on a quark spinor), also contribute to
quartic vertices of the form $g\,g\,t\,\bar{u_i}$,
$g\,\gamma\,t\,\bar{u_i}$ and $g\,Z\,t\,\bar{u_i}$.

Their criteria in choosing these operators was that they contributed
only to FCNC top physics, not affecting low energy physics. In that
sense, operators that contributed to top quark phenomenology but
which also affected bottom quark physics (in the notation of
ref.~\cite{Buchmuller:1985jz}, operators ${\cal O}_{qG}$) were not
considered. Recently, a study based on constraints from B
physics~\cite{Fox:2007in} using the predictions for the
LHC~\cite{Carvalho:2007yi,delAguila:2008iz,Ball:2007zza}, has showed
that, in fact, some of the constraints on dimension 6 operators
stemming from low energy physics are already stronger than some of
the predictions for the LHC. This is true for the operators denoted
in~\cite{Fox:2007in} by $LL$, which are the ones built with two
$SU(2)$ doublets that have left out in
refs.~\cite{Ferreira:2005dr,Ferreira:2006xe,Ferreira:2006in}.
Obviously the gauge structure is felt more strongly in the left-left
(LL) type of operators than in the right-right type. Hence, they
concluded that the LL operators will not be probed at the LHC
because they are already constrained beyond the expected bounds
obtained for a luminosity of 100 $fb^{-1}$. Limits on LR and RL
operators are close to those experimental bounds and RR operators
are the ones that will definitely be probed at the LHC. Moreover,
since more results will come from the B factories and the Tevatron,
the constraints will be even stronger by the time the LHC starts to
analyse data. Therefore the criteria in the choice of operators is
well founded so we will also not consider LL operators in the
electroweak sector.

\subsection{Effective operators contributing to electroweak FCNC top decays}
\noindent

We will now consider all possible dimension six effective
operators which contribute to top decays of the form
$t\,\rightarrow\,u_i\,\gamma$ and $t\,\rightarrow\,u_i\,Z$. First we
have the operators analogous to those of eq.~\eqref{eq:opst} in the
electroweak sector,
\begin{align}
{\cal O}_{tB} &= i
\frac{\alpha^B_{it}}{\Lambda^2}\,\left(\bar{u}^i_R \, \,
\gamma_{\mu} D_{\nu} t_R \right) \, B^{\mu \nu}\;\;\; , &
\;\;\;{\cal O}_{tB\phi} =
\;\;\frac{\beta^{B}_{it}}{\Lambda^2}\,\left(\bar{q}^i_L\,
\sigma^{\mu\nu}\,t_R\right)\, \tilde{\phi} \,B_{\mu\nu} \;\;\; , \nonumber \\
{\cal O}_{tW\phi}
&=\frac{\beta^{W}_{it}}{\Lambda^2}\,\left(\bar{q}^i_L\, \, \tau_{I}
\, \sigma^{\mu\nu}\,t_R\right)\, \tilde{\phi}\,W^{I}_{\mu\nu}\;\;\;
, & \label{eq:cap2-op4}
\end{align}
where $\alpha^B_{ti}$, $\beta^{B}_{ti}$ and $\beta^{W}_{ti}$  are
complex dimensionless couplings, and $B^{\mu \nu}$ and
$W^{I}_{\mu\nu}$ are the $U(1)_Y$ and $SU(2)_L$ field tensors,
respectively. As before, we also consider the operators with
exchanged quark spinors, corresponding to couplings $\alpha^B_{ti}$,
$\beta^{B}_{ti}$ and $\beta^{W}_{ti}$, and the hermitian conjugates
of all of these terms.

The electroweak tensors ``contain'' both the photon and $Z$ boson
fields, through the well-known Weinberg rotation. Thus they
contribute simultaneously to vertices of the form $Z \,\bar{t} \,
u_i$ and $\gamma\, \bar{t} \, u_i$ when we consider the partial
derivative of $D^\mu$ in the equations~\eqref{eq:cap2-op4}, or when
we replace the Higgs field $\phi$ by its vev $v$ in them. We will
isolate the contributions to FCNC photon and $Z$ interactions in
these operators defining new effective couplings
$\{\alpha^\gamma\,,\,\beta^{\gamma}\}$ and
$\{\alpha^Z\,,\,\beta^Z\}$. These are related to the initial
couplings via the Weinberg angle $\theta_W$ by
\begin{equation}
\alpha^{\gamma}\;=\;\cos\theta_W \, \alpha^{B} \qquad \; , \; \qquad
\alpha^{Z}\;=\; - \sin\theta_W \, \alpha^{B} \label{eq:alf}
\end{equation}
and
\begin{equation}
\left\{
\begin{array}{c}
   \beta^{\gamma} \, = \, \sin\theta_W \beta^{W} + \cos \theta_W \beta^{B}\\
  \beta^{Z} \, = \, \cos\theta_W \beta^{W} - \sin \theta_W \beta^{B}  \\
\end{array}
\right. . \label{eq:bet}
\end{equation}
As we will see, these Weinberg rotations will introduce a certain
correlation between FCNC processes involving the photon or the $Z$.

Because the Higgs field is electrically neutral but has weak
interactions, there are more effective operators which will only
contribute to new $Z$ FCNC interactions given by
\begin{align}
{\cal O}_{D_t} &=\frac{\eta_{it}}{\Lambda^2}\,\left(\bar{q}^i_L\,
D^{\mu}\,t_R\right)\, D_{\mu} \tilde{\phi} \, \;\;\;,\;\;\; {\cal
O}_{\bar{D}_t}=\frac{\bar{\eta}_{it}}{\Lambda^2}\,\left( D^{\mu}
\bar{q}^i_L\, \,t_R\right)\, D_{\mu} \tilde{\phi}
\label{eq:cap2-op1}
\end{align}
and
\begin{equation}
{\cal O}_{\phi_t}
 \, = \, \theta_{it} \, (\phi^{\dagger} D_{\mu} \phi) \, (\bar{u^i_R} \gamma^{\mu} t_R)  \;\;\;
 ,
\label{eq:cap2-op2}
\end{equation}
and another operator with coupling $\theta_{ti}$ with the position
of the $u^i$ and $t$ spinors exchanged. As before, the coefficients
$\eta_{it}$, $\bar{\eta}_{it}$ and $\theta_{it}$ are complex
dimensionless couplings.

\subsection{Feynman rules for top FCNC weak interactions}

\begin{figure}[ht]
\begin{minipage}[t]{1.0\linewidth}
  \begin{center}
    \includegraphics[scale=0.7]{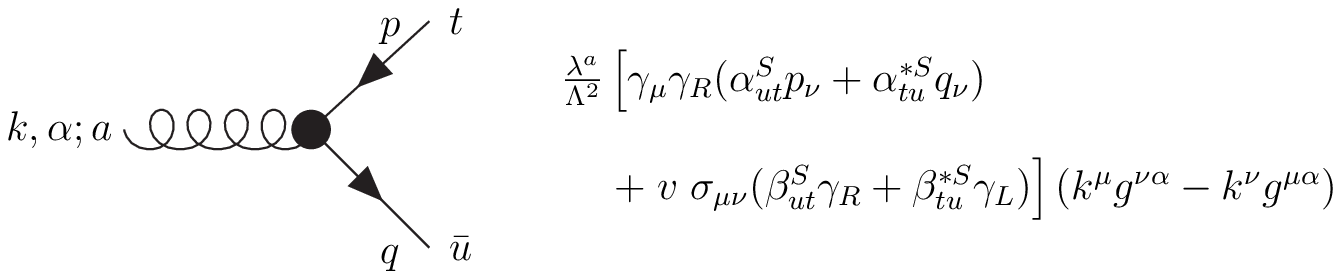}
    \caption{Feynman rules for the anomalous vertex $g \, t\, \bar{u} $.}
    \label{fig:feyngluon}
  \end{center}
\end{minipage}
\vspace{0.5cm}
\begin{minipage}[c]{1.0\linewidth}
  \begin{center}
    \includegraphics[scale=0.7]{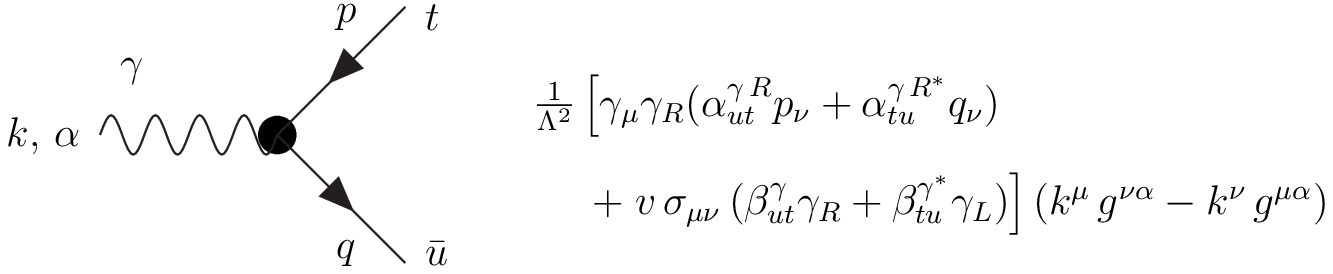}
    \caption{Feynman rules for the anomalous vertex $\gamma \, t\, \bar{u} $.}
    \label{fig:feyngamma}
  \end{center}
\end{minipage}
\vspace{0.5cm}
\begin{minipage}[b]{1.0\linewidth}
  \begin{center}
    \includegraphics[scale=0.7]{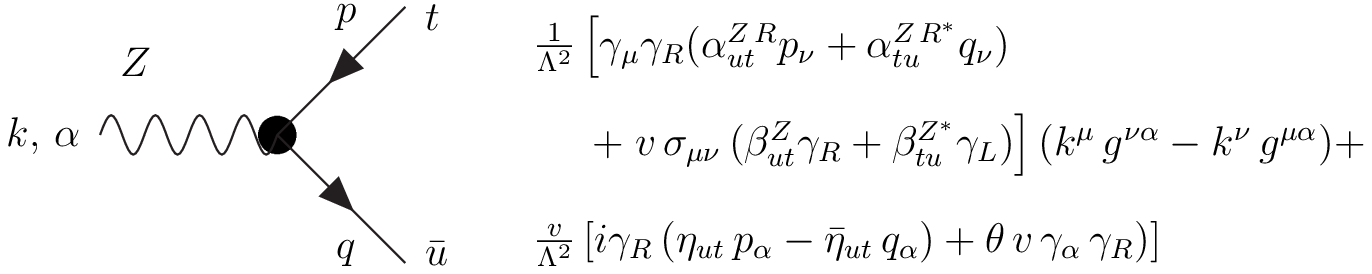}
    \caption{Feynman rules for the anomalous vertex $Z \, t\, \bar{u} $.}
    \label{fig:feynZ}
  \end{center}
\end{minipage}
\end{figure}

\begin{figure}[htbp]
\begin{minipage}[t]{1.0\linewidth}
  \begin{center}
    \includegraphics[scale=0.7]{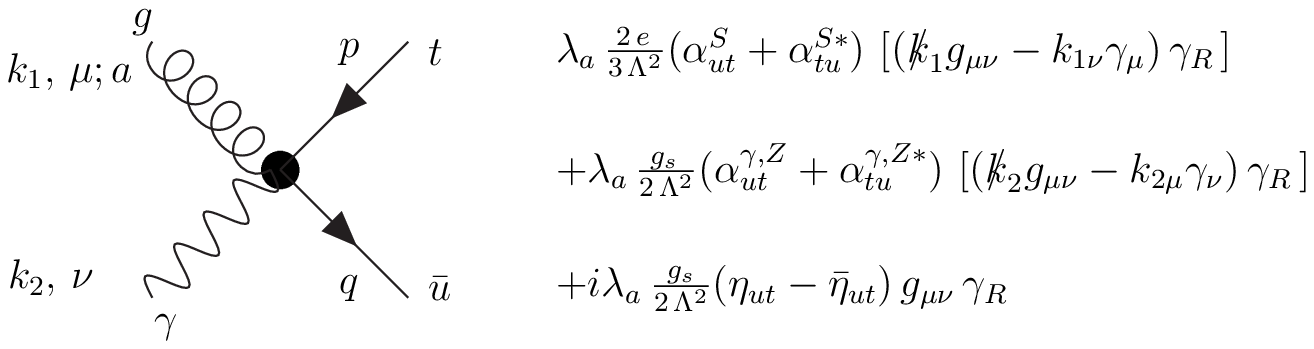}
    \caption{Feynman rules for the anomalous quartic vertex $\gamma \, g\, t \bar{u} $.}
    \label{fig:feyn4gamma}
  \end{center}
\end{minipage}
\vspace{0.5cm}
\begin{minipage}[c]{1.0\linewidth}
  \begin{center}
    \includegraphics[scale=0.7]{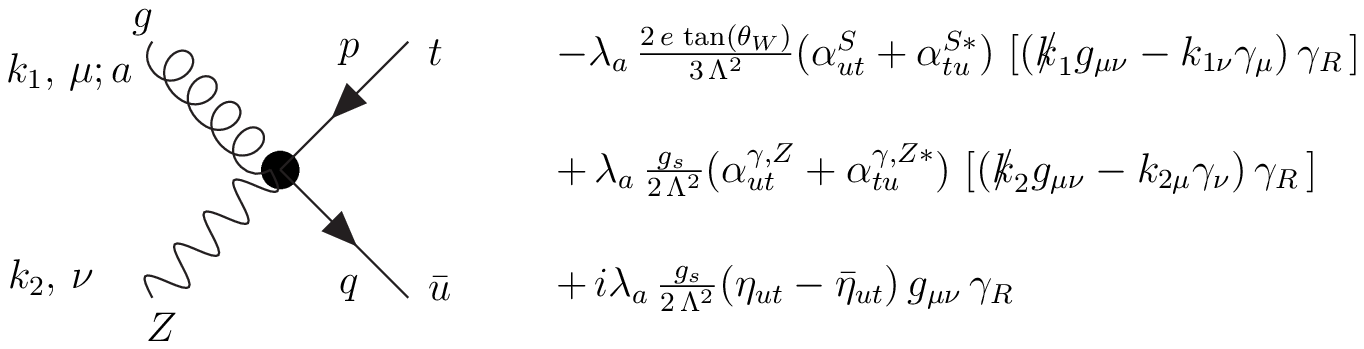}
    \caption{Feynman rules for the anomalous quartic vertex $Z \, g\, t \bar{u} $.}
    \label{fig:feyn4Z}
  \end{center}
  \end{minipage}
\end{figure}

The complete effective lagrangian can now be written as a function
of the operators defined in the previous section,
\begin{align}
{\cal L} &  = i \frac{\alpha^S_{it}}{\Lambda^2}\left( \bar{u}^i_R
\lambda^a\gamma^{\mu} D^{\nu} t_R \right)  G_{\mu \nu}^a + i
\frac{\alpha^S_{ti}}{\Lambda^2}\left( \bar{t}_R  \lambda^a
\gamma^{\mu}
D^{\nu} u^i_R \right)  G_{\mu \nu}^a \nonumber \vspace{0.3cm}\\
& + \frac{\beta^{S}_{it}}{\Lambda^2}\left(\bar{q}^i_L \lambda^{a}
\sigma^{\mu\nu}t_R\right) \phi G^{a}_{\mu\nu} +
\frac{\beta^{S}_{ti}}{\Lambda^2}\left(\bar{t}_L  \lambda^{a}
\sigma^{\mu\nu}u^i_R\right) \tilde{\phi}G^{a}_{\mu\nu} \nonumber  \vspace{0.3cm} \\
&  + i \frac{\alpha^B_{it}}{\Lambda^2}\left( \bar{u}^i_R
\gamma^{\mu} D^{\nu} t_R \right)  B_{\mu \nu}+i
\frac{\alpha^B_{ti}}{\Lambda^2}\left( \bar{t}_R   \gamma^{\mu}
D^{\nu} u^i_R \right)  B_{\mu \nu} \nonumber \vspace{0.3cm} \\
 & + \frac{\beta^{W}_{it}}{\Lambda^2}\left(\bar{q}^i_L
\tau_{I}  \sigma^{\mu\nu}t_R\right) \phi W^{I}_{\mu\nu} +
\frac{\beta^{W}_{ti}}{\Lambda^2}\left(\bar{t}_L  \tau_{I}
\sigma^{\mu\nu}u^i_R\right) \tilde{\phi}W^{I}_{\mu\nu} \nonumber  \vspace{0.3cm} \\
 & + \frac{\beta^{B}_{it}}{\Lambda^2}\left(\bar{q}^i_L
\sigma^{\mu\nu}t_R\right) \tilde{\phi}B_{\mu\nu} +
\frac{\beta^{B}_{ti}}{\Lambda^2}\left(\bar{t}_L
\sigma^{\mu\nu}u^i_R\right) \phi B_{\mu\nu} \nonumber \vspace{0.3cm} \\
 & + \frac{\eta_{it}}{\Lambda^2}\left(\bar{q}^i_L
D^{\mu}t_R\right) D_{\mu} \tilde{\phi} +
\frac{\bar{\eta}_{it}}{\Lambda^2}\left( D^{\mu}
\bar{q}^i_L t_R\right) D_{\mu} \tilde{\phi} \nonumber \vspace{0.3cm} \\
 & + \frac{\theta_{it}}{\Lambda^2}  (\phi^{\dagger} D_{\mu} \phi)  (\bar{u^i_R} \gamma^{\mu}
 t_R)  +  \frac{\theta_{ti}}{\Lambda^2}  (\phi^{\dagger} D_{\mu} \phi)  (\bar{t_R} \gamma^{\mu}
 u^i_R)  + \mbox{h.c.}  .
 \label{eq:lag}
\end{align}
This lagrangian describes new vertices of the form $g \,\bar{u}\,
t$, $Z \,\bar{u}\, t,$ $\gamma \,\bar{u}\, t,$ $Z \,\bar{u}\, t,$
$\bar{u} \,t \,\gamma \, g$ and $\bar{u} \,t \, Z \, g$ (and many
others) and their charge conjugate vertices. For simplicity we
redefine the $\eta$ and $\theta$ couplings as $ \eta \rightarrow
(\sin (2 \theta_W)/e) \, \eta$ and $ \theta \rightarrow (\sin (2
\theta_W)/e) \, (\theta_{it} \,-\,\theta^*_{ti})$. The Feynman rules
for the FCNC triple vertices are shown in
figures~\ref{fig:feyngamma} and~\ref{fig:feynZ}~\footnote{The
Feynman rules for the charge-conjugate vertices are obtained by
simple complex conjugation as was seen before. The exception is the
$\theta$ term, which due to our definition of the $\theta$ coupling
in eq.~\eqref{eq:cap2-op2}, will become $-\theta^*$ for the vertex
$Z \, u\, \bar{t}$.}.
The gauge structure of the terms in eq. \eqref{eq:lag} gives rise to
new quartic vertices. Most of the couplings which contribute to the
triple vertices of figs. \ref{fig:feyngluon}, \ref{fig:feyngamma} and \ref{fig:feynZ}
also contribute to the quartic ones. The Feynman rules for the
quartic vertices we will need are shown in figures
\ref{fig:feyn4gamma} and \ref{fig:feyn4Z}. We see
that these quartic interactions receive contributions from both the
strong and electroweak effective operators. Their presence is
mandatory because of gauge invariance and they will be of great
importance to obtain several elegant results which we present in
section \ref{sec:sig}.

For comparison, the FCNC lagrangian considered by the authors of
ref.~\cite{delAguila:1999ec} consisted in
\begin{align}
\cal{L} & =
\frac{g}{2\cos\theta_W}\,\bar{t}\,\gamma_\mu\,(X_{tq}^L\,\gamma_L\,+\,
X_{tq}^R\,\gamma_R)\,q\,Z^\mu\;+\;\displaystyle{\frac{g}{2\cos\theta_W}}\,\bar{t}\,
(k_{tq}^{(1)}\,-\,i\,k_{tq}^{(2)}\gamma_5)\,\displaystyle{\frac{i\sigma_{\mu\nu}q^\nu}{m_t}}
\,q\,Z^\mu \nonumber \\
 & +\; e\,\bar{t}\,(\lambda_{tq}^{(1)}\,-\,i\,\lambda_{tq}^{(2)}\gamma_5)\,
 \displaystyle{\frac{i\sigma_{\mu\nu}q^\nu}{m_t}}\,q\,A^\mu\;+\;
g_S\,\bar{t}\,(\zeta_{tq}^{(1)}\,-\,i\,\zeta_{tq}^{(2)}\gamma_5)\,
 \displaystyle{\frac{i\sigma_{\mu\nu}q^\nu}{m_t}}\,T^a\,q\,G^{a \mu}\;+\;h.c.
 \label{eq:lagj}
 \end{align}
Notice that whereas we consider a generic scale $\Lambda$ for new
physics, these authors set $\Lambda\,=\,m_t$. Also, it is easy to
recognize several of our couplings in the lagrangian above; for
instance, we have
\begin{align}
\frac{v^2}{\Lambda^2}\,\theta & =
\frac{g}{2\cos\theta_W}\,X_{tq}^R , \nonumber\\
\frac{v}{\Lambda^2}\,\beta_{qt}^Z & =
 \frac{g}{4\cos\theta_W
m_t}\,\left(k_{tq}^{(1)}\,-\,i\,k_{tq}^{(2)}\right), \nonumber\\
\frac{v}{\Lambda^2} \, \beta_{qt}^\gamma & = \frac{\,e}{2 \, m_t}
(\lambda_{tq}^{(1)} \,-\,i \,\lambda_{tq}^{(2)}),
\nonumber\\
\frac{v}{\Lambda^2} \,\beta_{qt}^S & =
 \frac{g_S}{4 \, m_t} (\xi_{tq}^{(1)}-i \,
\xi_{tq}^{(2)}).
 \label{eq:bet}
\end{align}
Notice that due to our choice of effective operators the couplings
of the form $\beta_{qt}$ and $\beta_{tq}$, and others, are treated
as independent -- meaning, the lagrangian~\eqref{eq:lagj} does not
contain our couplings $\beta_{tq}$.  Also, couplings of the form
$\{\alpha\,,\,\eta\}$ are not present in~\eqref{eq:lagj}, and the
photon and $Z$ couplings therein presented are taken to be
completely independent, unlike what we considered. Their $X_{tq}^L$
coupling hasn't got an equivalent in our formulation. We could
obtain it through a $\theta$-like effective operator, namely,
\begin{equation}
(\phi^{\dagger} D_{\mu} \phi) \, (\bar{q^i_L} \gamma^{\mu} q^j_L)
\;\;\; ,
\end{equation}
where one of the quark doublets $q^i$, $q^j$ would contain the top
quark. It is easy to see, though, that this operator would have a
direct contribution to bottom quark physics, thus violating one of
our selection criteria for the anomalous top interactions. One
important remark: the authors of ref.~\cite{delAguila:1999ec} do not
consider the quartic vertices of figs.~\ref{fig:feyn4gamma}
and~\ref{fig:feyn4Z}  in their calculations of cross sections for
$t\,+\,\gamma$ and $t\,+\,Z$ production. That's entirely correct,
since their analysis does not involve couplings like
$\{\alpha\,,\,\eta\}$, the only ones who contribute to those quartic
vertices.

\section{FCNC branching ratios of the top}
\label{sec:brs}

In this section we compute all FCNC decay widths of the top quark.
The decay width for $t\,\rightarrow\,u\,g$ is
given by
\begin{align}
\Gamma (t \rightarrow u g) & =  \frac{m^3_t}{12
\pi\Lambda^4}\,\Bigg\{ m^2_t \,\left|\alpha_{tu}^S  +
(\alpha^S_{ut})^* \right|^2 \,+\, 16 \,v^2\, \left(\left|
\beta_{tu}^S \right|^2 + \left| \beta_{ut}^S \right|^2 \right) \nonumber \\
 & + \,8\, v\, m_t\,\mbox{Im}\left[ (\alpha_{ut}^S  + (\alpha^S_{tu})^*)
\, \beta_{tu}^S \right] \Bigg\} \label{eq:widS}\;\;\; ,
\end{align}
with an analogous expression for $\Gamma (t \rightarrow c g)$, with
different couplings. The electroweak sector operators we discussed
in the previous section contribute to new FCNC decays, namely,
$t\,\rightarrow \,u\,\gamma$ (and $t\,\rightarrow \,c\,\gamma$, with
{\em a priori} different couplings), for which we obtain a width
given by the following expression:
\begin{align}
\Gamma (t \rightarrow u \gamma) & = \frac{m^3_t}{64
\pi\Lambda^4}\,\Bigg\{ m^2_t \,\left|\alpha_{tu}^{\gamma}  +
(\alpha^{\gamma}_{ut})^* \right|^2 \,+\, 16 \,v^2\, \left(\left|
\beta_{tu}^{\gamma} \right|^2 + \left| \beta_{ut}^{\gamma} \right|^2
\right)\nonumber \\
 & + \, 8\, v\, m_t\,\mbox{Im}\left[ (\alpha_{ut}^{\gamma}  + (\alpha^{\gamma}_{tu})^*)
\, \beta_{tu}^{\gamma} \right] \Bigg\} \label{eq:widW} \;\;\; .
\end{align}
Notice how similar this result is to eq.~\eqref{eq:widS}. We will
also have contributions from these operators to $t\,\rightarrow
\,u\,Z$ ($t\,\rightarrow \,c\,Z$), from which we obtain a width
given by
\begin{align}
 \Gamma(t\,\rightarrow \,u\,Z) & =  \frac{{\left( m_t^2 - m_Z^2 \right) }^2}{32\,m_t^3\,\pi
\,\Lambda^4}
\left[ K_1 \, \left| \alpha^Z_{ut} \right|^2 + K_2 \, \left|
\alpha^Z_{tu} \right|^2 + K_3 \, ( \left| \beta^Z_{ut} \right|^2 +
\left| \beta^Z_{tu} \right|^2) \right.
\nonumber \\
& +\, K_4 \, ( \left| \eta_{ut} \right|^2 + \left| \bar{\eta}_{ut}
\right|^2) + \, K_5 \, \left| \theta \right|^2 + K_6 \, Re \left[
\alpha^Z_{ut} \, \alpha^Z_{tu} \right] + K_7 \, Im \left[
\alpha^Z_{ut} \, \beta^Z_{tu} \right]
\nonumber \\
& + \, K_8 \, Im \left[ \alpha^{Z^*}_{tu} \, \beta^Z_{tu} \right] +
K_9 \, Re \left[ \alpha^Z_{ut} \theta^* \right]+ K_{10} \, Re \left[
\alpha^Z_{tu} \theta \right]
\nonumber \\
&   +\, \left. K_{11} \, Re \left[ \beta^Z_{ut}
(\eta_{ut}-\bar{\eta}_{ut})^* \right] + K_{12} \, Im \left[
\beta^Z_{tu} \, \theta \right] + K_{13} \, Re \left[ \eta_{ut}
\bar{\eta}_{ut}^* \right] \right]\;\;\; ,
\end{align}
where the coefficients $K_i$ are given by
\begin{align}
K_1 & =  \frac{1}{2} \, (m_t^4 + 4\,m_t^2\,m_Z^2 + m_Z^4) \qquad K_2
\, = \, \frac{1}{2} \,  (m_t^2 - m_Z^2)^2 \qquad K_3 \, = \, 4\,(
2\,m_t^2 + m_Z^2) \,v^2
 \nonumber  \\
K_4 & =  \frac{v^2}{4\,m_Z^2}(m_t^2 - m_Z^2)^2 \qquad K_5 \, = \,
\frac{v^4}{m_Z^2} ( m_t^2 + 2\,m_Z^2 ) \qquad K_6 \, = \, ( m_t^2 -
m_Z^2 ) \,( m_t^2 + m_Z^2)
\nonumber \\
K_7 & =  4\,m_t\, ( m_t^2 + 2\,m_Z^2 ) \,v \qquad K_8 \, = \,
4\,m_t\, ( m_t^2 - m_Z^2 ) \,v \qquad K_9 \, = \, -2\, ( 2\,m_t^2 +
m_Z^2 ) \,v^2
\nonumber \\
K_{10} & =  -2\, ( m_t^2 - m_Z^2) \,v^2 \qquad K_{11} \, = - K_{10}
\qquad K_{12} \, = \, -12 \,m_t\,v^3 \qquad K_{13} \, = \, \frac{-
v^2 }{m_Z^2} \, K_2 \;\;\;\ .
\end{align}

The anomalous couplings that describe the FCNC decays $t
\,\rightarrow\, q\, Z$ and $t \,\rightarrow\, q\, \gamma$ are not
entirely independent -- according to eqs.~\eqref{eq:alf}
and~\eqref{eq:bet} the couplings $\{\alpha^\gamma\,,\,\alpha^Z\}$
and $\{\beta^\gamma\,,\,\beta^Z\}$ are related to one another. This
will imply a correlation between the branching ratios for
these two decays. Then, gauge invariance imposes that one can
consider anomalous FCNC interactions that affect only the decay $t
\,\rightarrow\, q\, Z$, but any anomalous interactions which affect
$t \,\rightarrow\, q\, \gamma$ will necessarily have an impact on $t
\,\rightarrow\, q\, Z$. In particular, if one considers any sort of
theory for which $Br(t \,\rightarrow\, q\, \gamma)\,\neq\,0$, then
one will forcibly have $Br(t \,\rightarrow\, q\, Z)\,\neq\,0$. The
reverse of this statement is not necessarily true, since more
anomalous couplings contribute to the $Z$ interactions than do the
$\gamma$ ones.

\begin{figure}[htbp]
  \begin{center}
    \includegraphics[scale=0.7]{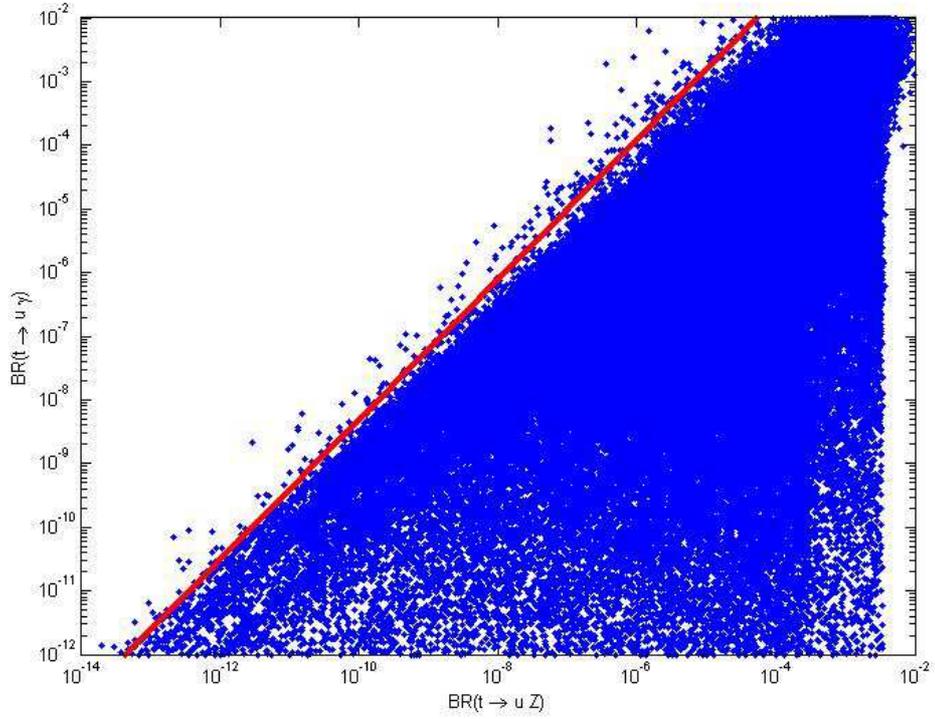}
    \caption{FCNC branching ratios for the decays $t \,\rightarrow\, u\, Z$ vs.
    $t \,\rightarrow\, u\, \gamma$.
    }
    \label{fig:brs}
  \end{center}
\end{figure}

If the couplings contributing to one of these branching ratios were
completely unrelated to those contributing to the other, then the
two branching ratios would be completely independent of one another.
As we see in figure~\ref{fig:brs} that is not the case. To obtain
this plot we considered that the total width of the top quark was
equal to $1.42$ GeV (a value which includes QCD corrections, and
taking
$V_{tb}\,\simeq\,1$~\cite{Beneke:2000hk,Denner:1990ns,*Eilam:1991iz,*Czarnecki:1998qc,*Chetyrkin:1999ju,*Oliveira:2001vw}),
set $\Lambda\,=\,1 TeV~\footnote{If one wishes to consider a
different scale for new physics, one will simply have to rescale the
values of the anomalous couplings.}$ and generated random complex
values of all the anomalous couplings, with magnitudes in the range
between $10^{-10}$ and $1$. We rejected those combinations of
parameters which resulted in branching ratios for $t \,\rightarrow\,
u\, Z$ and $t \,\rightarrow\, u\, \gamma$ larger than
$10^{-2}~\footnote{With all precision one should then add the
corresponding FCNC widths to the top total width quoted above.
However, the error we commit with this approximation is always
smaller than 2\%, and then only for the larger values of the
branching ratios considered.}.$ Regarding the $\{\alpha\,,\,\beta\}$
couplings, we first generated random values for
$\{\alpha_{ij}^B\,,\,\beta_{ij}^B\,,\,\beta_{ij}^W\}$ and then,
through eqs.~\eqref{eq:alf} and~\eqref{eq:bet} obtained
$\{\alpha^\gamma\,,\,\alpha^Z\}$ and
$\{\beta^\gamma\,,\,\beta^Z\}$.\label{par:top1}
%

With very few exceptions, we can even quote a rough
bound on the branching ratios by observing the straight line
drawn by us in the plot -- namely, that it is nearly impossible
to have $Br(t\,\rightarrow\, u\, \gamma)\,>\,500\,Br(t \,\rightarrow\,
u\,Z)^{1.1}$  -- but we remember that plot build as~\ref{fig:brs} gives us the parameter space of solution in an approximate way i.e. in the limit we must have a region with clear boundary. Again, if gauge invariance did not impose the conditions between
$\gamma$ and $Z$ couplings expressed in eqs.~\eqref{eq:alf}
and~\eqref{eq:bet}, what we would obtain in fig.~\ref{fig:brs}
would be a uniformly filled plot -- for a given value of $Br(t
\,\rightarrow\, u\, Z)$ one could have any value of $Br(t
\,\rightarrow\, u\, \gamma)$. If we take the point of view that any
theory beyond the SM will manifest itself at the TeV scale through
the effective operators of ref.~\cite{Buchmuller:1985jz} then this
relationship between these two FCNC branching ratios of the top is a
model-independent prediction. Finally, had we considered a more
limited set of anomalous couplings -- for instance, only $\alpha$ or
$\beta$ type couplings -- the plot in fig.~\ref{fig:brs} would be
considerably simpler. Due to the relationship between those
couplings, the plot would reduce to a band of values, not a wedge as
that shown. Identical results were obtained for the FCNC decays $t
\,\rightarrow\, c\, Z$ and $t \,\rightarrow\, c\, \gamma$.

\section{Strong vs. Electroweak FCNC contributions for cross sections of
associated single top production} \noindent \label{sec:sig}

\begin{figure}[htbp]
  \begin{center}
    \includegraphics[scale=0.5]{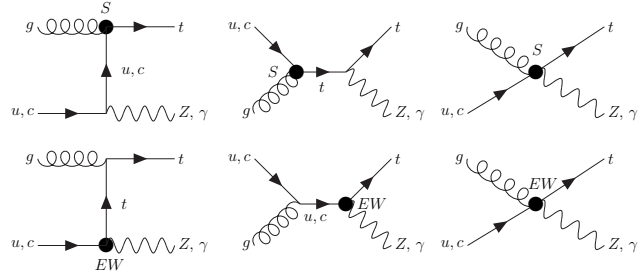}
    \caption{Feynman diagrams for $t \, Z$ and $t \, \gamma$ production
    with both strong and electroweak FCNC vertices.}
    \label{fig:gqtZgamma}
  \end{center}
\end{figure}

The anomalous operators that we have been considering contribute, not only
to FCNC decays of the top, but also to processes of single top
production. Namely to the associated production of a top quark
alongside a photon or a $Z$ boson, processes described by the
Feynman diagrams shown in fig.~\ref{fig:gqtZgamma}. The FCNC
vertices are represented by a solid dot, with the letter ``S''
standing for a strong FCNC anomalous interaction and a ``EW'' for
the electroweak one. Notice the four-legged diagrams, imposed by
gauge invariance.

\begin{figure}[!htbp]
  \begin{center}
    \includegraphics[scale=1.0]{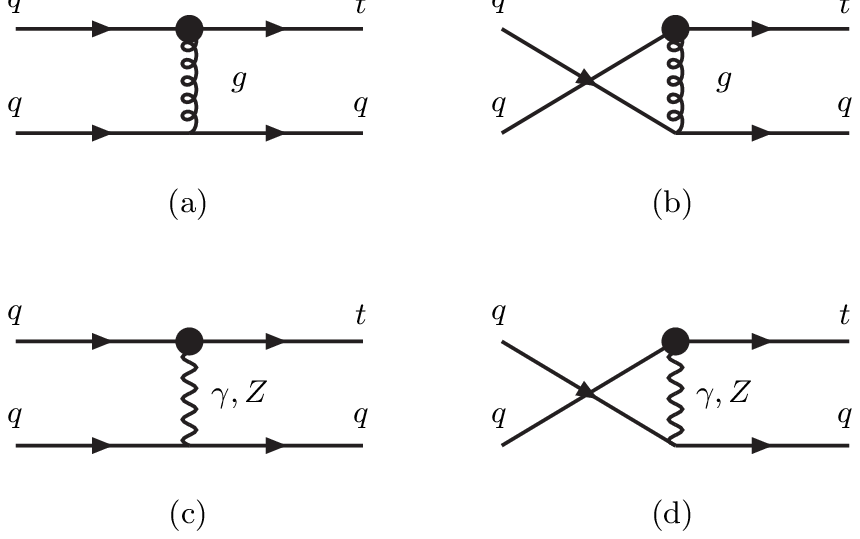}
    \caption{Feynman diagrams for $q\; q\rightarrow t\; q.$}
    \label{fig:single-t1}
  \end{center}
\end{figure}

\begin{figure}[!htbp]
  \begin{center}
    \includegraphics[scale=1.0]{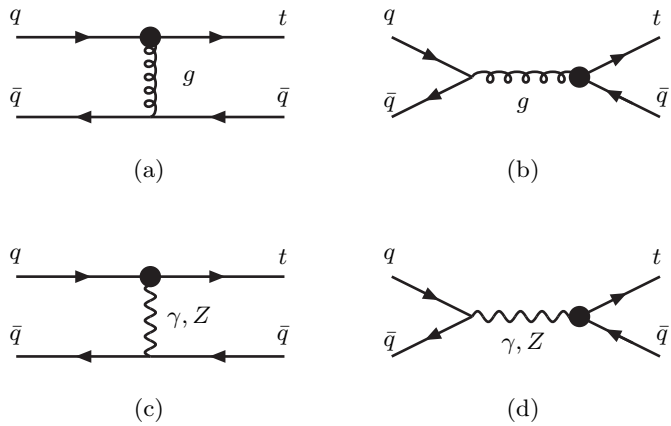}
    \caption{Feynman diagrams for $q\; \bar q\rightarrow t\; \bar q.$}
    \label{fig:single-t2}
  \end{center}
\end{figure}

Other possible processes of single top production involve quark-quark (or quark-antiquark) scattering ($tq$ production); we call them four-fermions processes. There are eight such processes which we list in the table~\ref{tab:ffchannel}.
\begin{table}[t]
\begin{center}
\begin{tabular}{ccc}\hline\hline \\
 Four-fermions process & Process number & Channel  \\
$uu\rightarrow tu$ & (1) & t,u (fig.~\ref{fig:single-t1})\\
$uc\rightarrow tc$ & (2) & t (fig.~\ref{fig:single-t1})\\
$u\bar u\rightarrow t\bar u$ & (3) & s,t (fig.~\ref{fig:single-t2})\\
$u\bar u\rightarrow t\bar c$ & (4) & s (fig.~\ref{fig:single-t2})\\
$u\bar c\rightarrow t\bar c$ & (5) & t (fig.~\ref{fig:single-t2})\\
$d\bar d\rightarrow t\bar u$ & (6) & s (fig.~\ref{fig:single-t2})\\
$ud\rightarrow td$ & (7) & t (fig.~\ref{fig:single-t1})\\
$u\bar d\rightarrow t\bar d$ & (8) & t (fig.~\ref{fig:single-t2})\\
\end{tabular}
\caption{List of single top production channel through quark-quark scattering.}\label{tab:ffchannel}
\end{center}
\end{table}
Here, we consider processes that involve only a single violation i.e. we do not consider processes like $s\bar d\rightarrow t\bar u$ for example. The resulting cross section has contributions from strong and electroweak operators like (3.2), (3.3) and (3.4) as well as from the four-fermions operators. However, we do not consider the contribution due to four-fermion operators.Figs.~\ref{fig:single-t1} and \ref{fig:single-t2}  show all Feynman diagrams in a generic way.
Finally, we can observe that  processes (1) to (5) in table~\ref{tab:ffchannel} have no correspondence at the tree level in the SM. This way, the first contribution to the cross section is of order $\Lambda^{-4}.$ This is not the case with the processes ((6) to (8)). Here, the processes have the SM contribution to the amplitude; they are represented by the Feynman diagrams~\ref{fig:single-t}-(a), \ref{fig:single-t}-(b) and \ref{fig:single-t}-(c) in page~\pageref{fig:single-t}. Thus, the first correction in these cross sections is the interference between SM and FCNC diagram with contribution of order $\Lambda^{-2}.$
The strong FCNC contribution to $t\gamma,$ $tZ$ and $tq$ production have already been considered in~\cite{Ferreira:2005dr, Ferreira:2006xe, Ferreira:2006in}. Our aim in this section is to investigate what is the combined influence of the strong and electroweak anomalous contributions to these processes. We note that the relation between strong and electroweak FCNC channel to the $tq$ production had been considered in~\cite{won:2008mt}.

\subsection{Cross section for $q \, g \rightarrow t \, \gamma$}\label{sec:tgamma}

\begin{figure}[htbp]
  \begin{center}
    \includegraphics[scale=0.7]{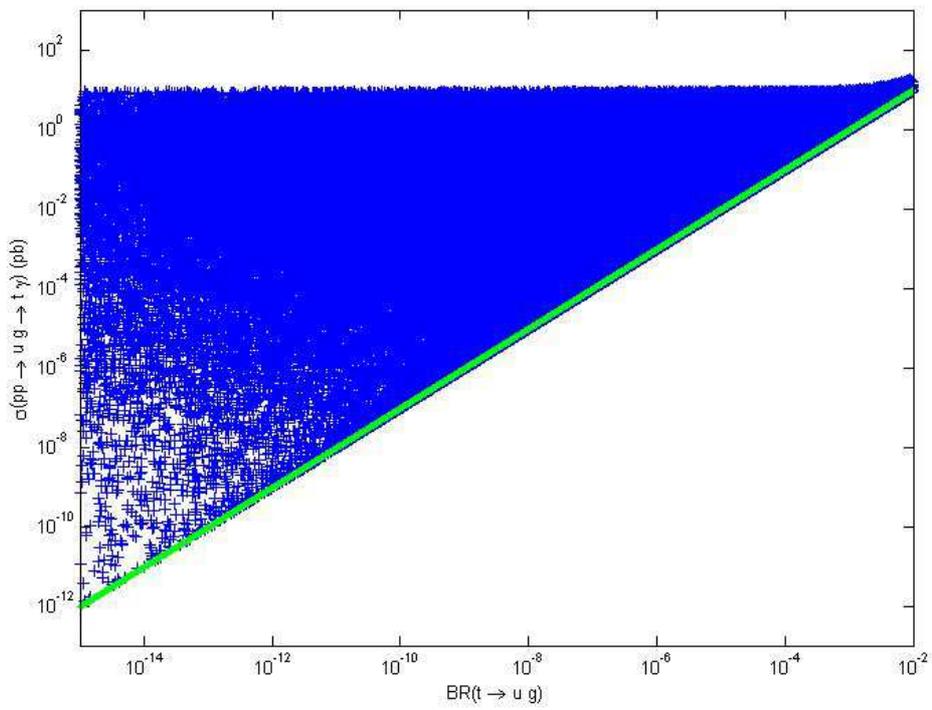}
    \caption{Total (blue crosses) and strong (grey) cross
    sections for the process $p\,p\rightarrow\,u\,g\rightarrow t\,\gamma$ versus the
    FCNC branching ratio for the decay $t\,\rightarrow\,u\,g$.}
    \label{fig:sigtga}
  \end{center}
\end{figure}

The total cross section for the associated FCNC production of a
single top quark and a photon including all the anomalous
interactions considered in section~\ref{sec:eff2} is given by
\begin{align}
\frac{d \,\sigma_{q \, g \rightarrow t \, \gamma}}{dt} & =
\frac{e^2}{18\,m_t^3\,s^2}\, F_{\gamma}(t,s) \,
\Gamma(t\,\rightarrow\,q\,g) \,+\, \frac{g_S^2}{6\,m_t^3\,s^2}\,
F_{\gamma}(s,t) \, \Gamma(t\,\rightarrow\,q\,\gamma) \;+\; \frac{e
\, g_S\,H_{\gamma}(t,s)}
  {96 \, \pi \, s^2 \, \Lambda^4}  \nonumber \\
& \times \,  \left\{ Re \left[ \left( \alpha^S_{it} +
(\alpha^S_{ti})^* \right) \left( \alpha^{\gamma}_{it} +
(\alpha^{\gamma}_{ti})^* \right) \right] + \frac{4v}{m_t} \, Im
\left[ ((\alpha^\gamma_{it})^* + \alpha^\gamma_{ti}) \, \beta^S_{ti}
+ (\alpha^S_{it} + ( \alpha^S_{ti})^*) \,
\beta^\gamma_{ti} \right] \right. \nonumber \\
& +\, \left. \frac{16v^2}{m_t^2} \, Re \left[ \beta^\gamma_{it}
(\beta^S_{it})^*  + \beta^\gamma_{ti} (\beta^S_{ti})^* \right]
\right\} \label{eq:tgam}
\end{align}
where we have defined the functions
\begin{align}
F_{\gamma}(t,s) & =  \frac{{m_t}^8 + 2\,s^2\,t\,\left( s + t \right)
-      {m_t}^6\,\left( s + 2\,t \right)  +
      {m_t}^4\,\left( s^2 + 4\,s\,t + t^2 \right)  -
      {m_t}^2\,s\,\left( s^2 + 6\,s\,t + 3\,t^2 \right)}{\left( {m_t}^2 - s \right)^2\,t}
\nonumber \\
H_{\gamma}(t,s)  & =   -\,\frac{2\, m_t^2}{3\,
    \left( m_t^2 - s \right) \,\left( m_t^2 - t \right) } \left(
3\,m_t^6 -
      4\,m_t^4\,\left( s + t \right)  - s\,t\,\left( s + t \right)  +
      m_t^2\,\left( s^2 + 3\,s\,t + t^2 \right)  \right) \;\;\; .
\end{align}
We used the couplings generated in the previous section for which we
computed the branching ratios presented in fig.~\ref{fig:brs}. We
also generated random complex values for the strong couplings
$\{\alpha_{ij}^S\,,\,\beta_{ij}^S\}$, once again requiring that
$Br(t\,\rightarrow\,u\,g)\,<\,10^{-2}$. To obtain the cross section
for the process $p\,p\,\rightarrow\,u\,g\,\rightarrow\,t\,\gamma$ at
the LHC we integrated the partonic cross section in
eq.~\eqref{eq:tgam} with the CTEQ6M partonic distribution
functions~\cite{Pumplin:2002vw}, with a factorization scale $\mu_F$
set equal to $m_t$. We also imposed a cut of 10 GeV on the $p_T$ of
the final state partons. In figure~\ref{fig:sigtga} we plot the
value of the cross section for this process against the branching
ratio of the FCNC decay of the top to a gluon.
We show both the
``strong'' cross section (in grey, corresponding to all couplings
but the strong ones set to zero) and the total cross section (in
blue crosses, including the effects of the strong couplings, the
electroweak ones and their interference). The most immediate
conclusion one can draw from fig.~\ref{fig:sigtga} is that the interference between the strong
and weak FCNC interactions is by and large
constructive. In fact, the vast majority of the points in fig.~\ref{fig:sigtga} which correspond to the total cross section lie above the line representing the contributions from the strong FCNC processes alone. For a small subset of points we may have $\sigma^{Total}(pp\rightarrow ug\rightarrow t\gamma)\; < \; \sigma^{S}(pp\rightarrow ug\rightarrow t\gamma)$, but in those cases the difference between both quantities is never superior to 1\%. Then, within an error of 1\%, the strong cross section $\sigma^{S}(pp\rightarrow ug\rightarrow t\gamma)$ (calculated in ref.~\cite{Ferreira:2006in}) is effectively a lower bound on the total cross section for this process.

Another interesting observation from fig.~\ref{fig:sigtga}: any
bound on $\sigma(p\,p\rightarrow u g\rightarrow t\,\gamma)$ (such as those which are
expected to come from the LHC results) immediately implies a bound
on $Br(t\,\rightarrow\,u\,g).$ However, a hypothetical direct determination of
$Br(t\,\rightarrow\,u\,g)$ would not determine the cross
section.\label{par:top3}
 Inversely, the discovery of the FCNC process $p\,p \rightarrow u g\rightarrow
t\,\gamma$ and obtention of a value for $\sigma(p\,p\rightarrow u
g\rightarrow t\,\gamma)$ would set an upper bound on
$Br(t\,\rightarrow\,u\,g)$, not fix its value.

\begin{figure}[htbp]
  \begin{center}
    \includegraphics[scale=0.7]{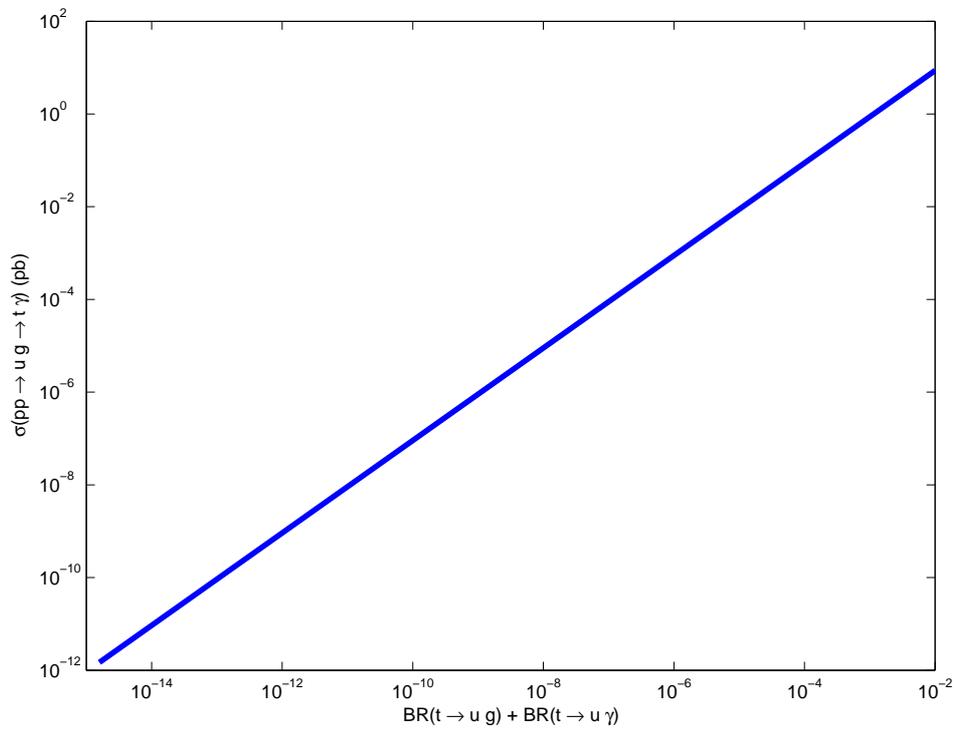}
    \caption{Total (electroweak and strong contributions) cross
    section for the process $p\,p\rightarrow\,u\,g\rightarrow t\,\gamma$ versus
    the sum of the FCNC branching ratios for the decays
    $t\,\rightarrow\,u\,\gamma$ and $t\,\rightarrow\,u\,g$.}
    \label{fig:sigbrs}
  \end{center}
\end{figure}

Had we plotted the electroweak cross section (the term proportional
to $\Gamma(t\,\rightarrow\,q\,\gamma)$ in eq.~\eqref{eq:tgam}) and
the total one versus $Br(t\,\rightarrow\,u\,\gamma)$, we would have
found a very similar picture to that of fig.~\ref{fig:sigtga}: a
straight line for the electroweak cross section and a wedge of
values lying mostly above it.\label{par:top4}
 We thus observe a great similarity in the behaviour
of the total cross sections with both FCNC branching ratios. In
fact, this is shown in quite an impressive manner in
fig.~\ref{fig:sigbrs}, where we
plot the total cross section against the {\em sum} of the FCNC
branching ratios. The ``line'' shown in this figure is actually a
very thin band, but this plot shows that, to good approximation, we
should expect a direct proportionality between the cross section for
the process $p\,p \rightarrow u\,g \rightarrow t\,\gamma$ and the
quantity
$Br(t\,\rightarrow\,u\,\gamma)\,+\,Br(t\,\rightarrow\,u\,g)$. In
fact we can even extract the proportionality constant from the plot
above, and obtain
\begin{equation}
\sigma(p\,p \rightarrow u\,g \rightarrow t\,\gamma)\;\simeq\;900\,
\left[Br(t\,\rightarrow\,u\,\gamma)\,+\,Br(t\,\rightarrow\,u\,g)\right]
\;\;\mbox{pb}\;\;\;,
\end{equation}
with a maximal deviation of about 9\%. Thus a measurement of this
cross section would determine the {\em sum} of the FCNC branching
ratios, but not each of them separately. Analogous results are
obtained for the processes involving the $c$ quark, the only
differences stemming from the parton density functions associated
with that particle. We obtain
\begin{equation}
\sigma(p\,p \rightarrow c\,g \rightarrow t\,\gamma)\;\simeq\;95\,
\left[Br(t\,\rightarrow\,c\,\gamma)\,+\,Br(t\,\rightarrow\,c\,g)\right]
\;\;\mbox{pb}\;\;\;,
\end{equation}
but the values of the cross section can now deviate as much as 19\%
from this formula. Notice that typical values of the cross section
for production of $t\,+\,Z$ via FCNC through a $c$ quark are roughly
ten times smaller than those of processes that go through a $u$
quark, which is of course due to the much smaller charm content of
the proton.

\begin{figure}[htbp]
  \begin{center}
    \includegraphics[scale=0.7]{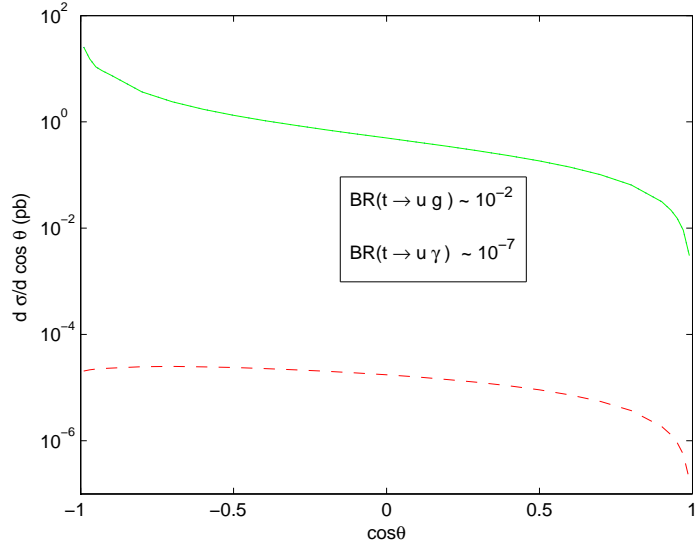}
    \caption{Differential cross section $p\,p\rightarrow\,u\,g\rightarrow t\,\gamma$ versus
    $\cos\theta$, for a typical choice of parameters with a branching ratio
    for $t\,\rightarrow\,u\,g$ much larger than $Br(t\,\rightarrow\,u\,\gamma)$. The strong
    contribution practically coincides with the total cross section (full line). The electroweak
    contribution is represented by the dashed line.}
    \label{fig:diff1}
  \end{center}
\end{figure}

Is there a way, then, to ascertain whether the main contribution to
$\sigma(p\,p\rightarrow u g\rightarrow t\,\gamma)$ stems from
anomalous strong interactions, or from weak ones? Indeed there is,
by analysing the differential cross section for this process. In
fig.~\ref{fig:diff1} we plot $d\sigma/d\cos\theta$ versus
$\cos\theta$, $\theta$ being the angle between the momentum of the
photon (or top) and the beam line. We show the strong and
electroweak contributions to this cross section, as well as its
total result. We chose a typical set of values for the anomalous
couplings producing a branching ratio for the FCNC decay
$t\,\rightarrow\,u\,g$ clearly superior to that of the decay
$t\,\rightarrow\,u\,\gamma$. As we see, the angular distribution of
the electroweak and strong cross sections is quite different. Since
the strong anomalous interactions are dominating over the
electroweak ones the total cross section mimics very closely the
strong one.

\begin{figure}[htbp]
  \begin{center}
    \includegraphics[scale=0.7]{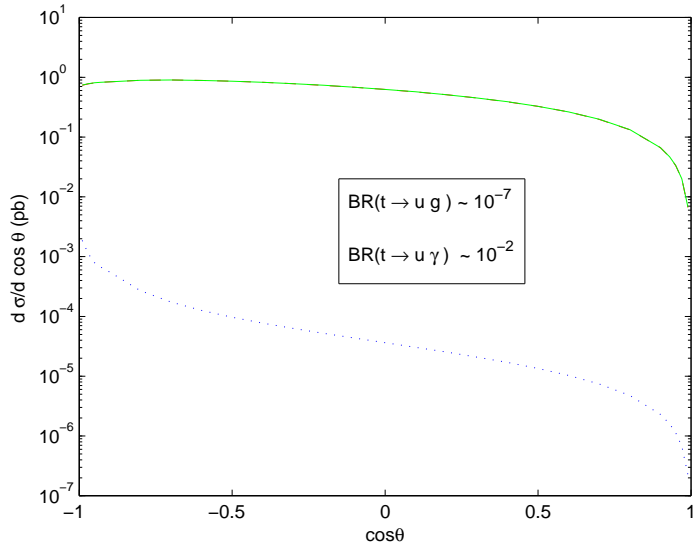}
    \caption{Differential cross section $p\,p\rightarrow\,u\,g\rightarrow t\,\gamma$ versus
    $\cos\theta$, for a typical choice of parameters with a branching ratio
    for $t\,\rightarrow\,u\,g$ much smaller than $Br(t\,\rightarrow\,u\,\gamma)$. The
    electroweak contribution practically coincides with the total cross section (full line). The
    strong contribution is represented by the dotted line.}
    \label{fig:diff2}
  \end{center}
\end{figure}
\begin{figure}[htbp]
  \begin{center}
    \includegraphics[scale=0.7]{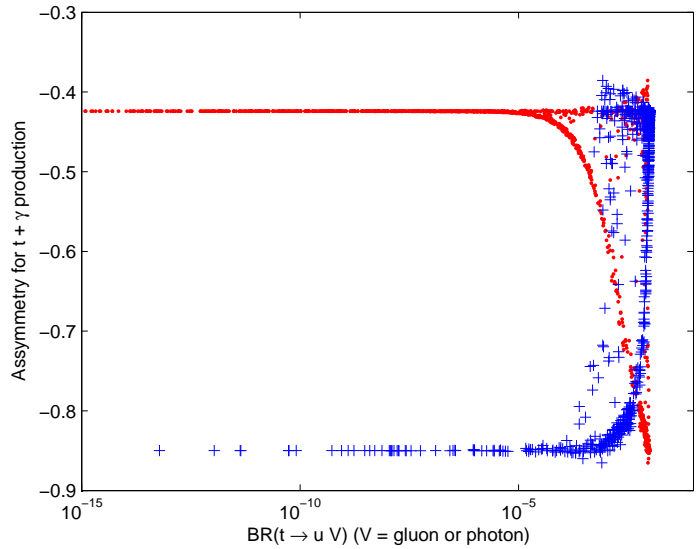}
    \caption{The angular asymmetry coefficient defined in eq.~\eqref{eq:assg}
    as a function of the branching ratios $Br(t\,\rightarrow\,u\,\gamma)$ (crosses)
    and $Br(t\,\rightarrow\,u\,g)$ (dots).}
    \label{fig:assg}
  \end{center}
\end{figure}

In fig.~\ref{fig:diff2} we show the inverse situation: a typical
set of values was chosen which gives us
$Br(t\,\rightarrow\,u\,\gamma)\,\sim\,10^{-2}$ and
$Br(t\,\rightarrow\,u\,g)\,\sim\,10^{-7}$, meaning a situation for
which the anomalous electroweak interactions are clearly dominant
over the strong ones. We see from the angular distribution of the
total cross section shown in fig.~\ref{fig:diff2} that it now
greatly resembles its electroweak component. Judging from
figs.~\ref{fig:diff1} and~\ref{fig:diff2}, the telltale sign of
dominance of strong FCNC interactions is a pronounced variation with
$\cos\theta$ in the cross section, whereas a dominance of
electroweak FCNC effects will produce a relatively ``flat'' cross
section. The Feynman diagrams of fig.~\ref{fig:gqtZgamma} help to
explain this difference in dependence with $\cos\theta$: the strong
cross section has a significant contribution from the $t$-channel
(since the $s$-channel diagram is suppressed by the top mass),
whereas the inverse happens for the electroweak cross section.
However, it should be pointed out that the four-legged diagrams
contributing to both cross sections will upset a clear $s$-or-$t$
channel dominance. Notice also that if FCNC produce branching ratios
of similar size in both sectors the difference in behaviour shown in
these plots will not be seen. In fact, we may get a better feel for
the different angular behavior of the strong and electroweak FCNC
interactions if we define an asymmetry coefficient for this cross
section,
\begin{equation}
A_{t + \gamma} \;=\;\frac{\sigma_{t + \gamma}(\cos\theta > 0)
\,-\,\sigma_{t + \gamma}(\cos\theta < 0)}{\sigma_{t +
\gamma}(\cos\theta > 0) \,+\,\sigma_{t + \gamma}(\cos\theta <
0)}\;\;\; . \label{eq:assg}
\end{equation}
To exemplify the relevance of this quantity, we generated a special
sample of anomalous couplings: random values of all strong and
electroweak couplings such that
$Br(t\,\rightarrow\,u\,\gamma)\,+\,Br(t\,\rightarrow\,u\,g) \sim
10^{-2}$. This will include the cases where one of the branching
ratios dominates over the other, and also the case where both of
them have similar magnitudes. We show the results in
fig.~\ref{fig:assg}, plotting the value of $A_{t + \gamma}$ in
terms of the two branching ratios whose sum is fixed to $10^{-2}$.
Looking at the far left of the plot we see that when the electroweak
FCNC interactions dominate over the strong ones $A_{t + \gamma}$
tends to a value of approximately $-0.85$, and in the reverse
situation we have $A_{t + \gamma}\,\sim\,-0.42$. However, when both
branching ratios have similar sizes, $A_{t + \gamma}$ can take any
value between those two limits.

\subsection{Cross section for $q \, g \rightarrow t \, Z$}
\label{sec:tz}

We can perform analysis similar to those of the previous section for
the associated production of a top and a Z boson. We computed an
analytical expression for the cross section of this process, which
is given by the sum of three terms,
\begin{equation}
\frac{d\sigma_{q g \rightarrow t Z}}{dt}\;=\;\frac{d\sigma^{EW}_{q g
\rightarrow t Z}}{dt} \;+\;\frac{d\sigma^{S}_{q g \rightarrow t
Z}}{dt} \;+\;\frac{d\sigma^{Int}_{q g \rightarrow t Z}}{dt} \;\;\;,
\label{eq:tz}
\end{equation}
with strong FCNC contributions ($\sigma^S$), electroweak ones
($\sigma^{EW}$) and interference terms between both sectors. The
expression for $d\sigma^{S}_{q g \rightarrow t Z}/dt$ was first
given in ref.~\cite{Ferreira:2006in}. The remaining formulae are
quite lengthy, involving many different combinations of anomalous
couplings with complicated coefficients. We present them
in Appendix~\ref{sec:apz} for completeness. To examine the values of these
cross sections at the LHC, we used the set of anomalous couplings
generated in the previous section, complemented with randomly
generated values for the $\eta$ and $\theta$
couplings~\footnote{Which, recall, do not contribute to FCNC
interactions involving the photon, only the Z.} and integrated the
expressions~\eqref{eq:tz} with the CTEQ6M pdf's. We chose
$\mu_F\,=\,m_t\,+\,m_Z$ and imposed a 10 GeV cut on the transverse
momentum of the particles in the final state.

\begin{figure}[htbp]
  \begin{center}
    \includegraphics[scale=0.7]{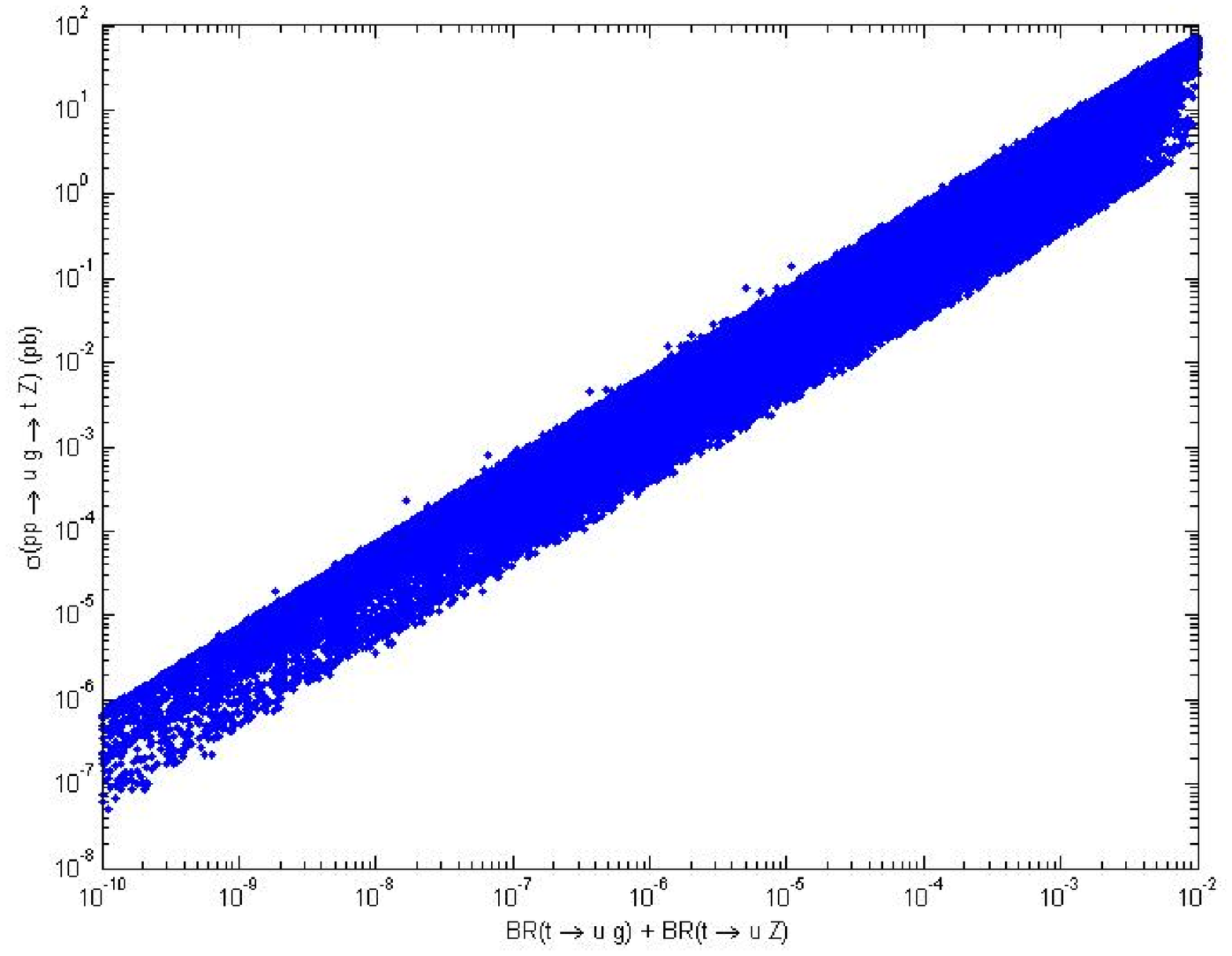}
    \caption{Total (electroweak and strong contributions) cross
    section for the process $p\,p\rightarrow\,u\,g\rightarrow t\,Z$ versus
    the sum of the FCNC branching ratios for the decays
    $t\,\rightarrow\,u\,Z$ and $t\,\rightarrow\,u\,g$.}
    \label{fig:tz}
  \end{center}
\end{figure}
\begin{figure}[htbp]
  \begin{center}
    \includegraphics[scale=0.7]{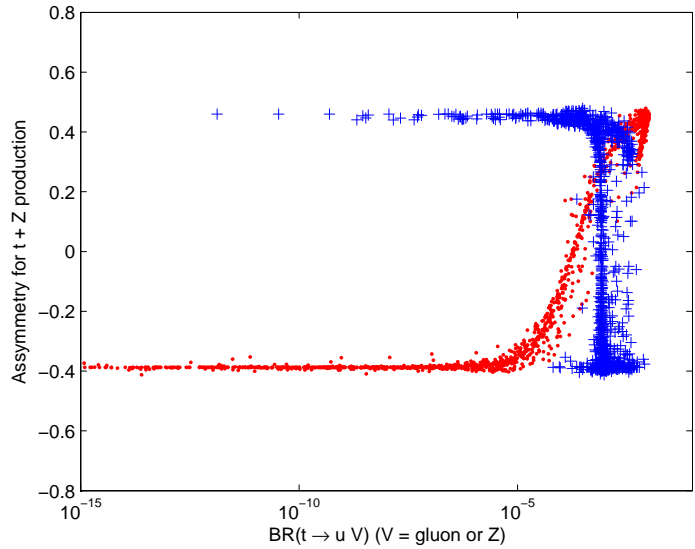}
    \caption{The angular asymmetry coefficient $A_{t + Z}$
    as a function of the branching ratios $Br(t\,\rightarrow\,u\,Z)$ (crosses)
    and $Br(t\,\rightarrow\,u\,g)$ (dots).}
    \label{fig:assz}
  \end{center}
\end{figure}

Unlike what was observed for the $t\,\gamma$ channel, there is no
direct proportionality between $\sigma^{EW}(pp\rightarrow u
g\rightarrow t Z)$ and $Br(t\,\rightarrow\,q\,Z)$ -- this is due to
the many different functions multiplying the several combinations of
anomalous couplings presented in Appendix~\ref{sec:apz}. Because the
functions $F_{1_Z}$ and $F_{2_Z}$ (eqs.~\eqref{eq:a12}) are very
similar, there is an {\em approximate} proportionality between the
branching ratio and $\sigma^S(pp\rightarrow u g\rightarrow t Z)$, as
was seen in ref.~\cite{Ferreira:2006in}. In fig.~\ref{fig:tz} we
plot the total cross section for this process against the sum
$Br(t\,\rightarrow\,u\,Z)\,+\,Br(t\,\rightarrow\,u\,g)$. We see,
from this plot, that the cross section for $t\,+\,Z$ production is
always contained between two straight lines, and it is easy to
obtain the following relation, valid for the overwhelming majority
of the points shown in fig.~\ref{fig:tz}:
\begin{equation}
200\, \left[Br(t \rightarrow u\,g)\,+\,Br(t \rightarrow
u\,Z)\right]\,<\, \sigma(p p \rightarrow u\,g \rightarrow t\,
Z)\,<\, 10^4\, \left[Br(t \rightarrow u\,g)\,+\,Br(t \rightarrow
u\,Z)\right] \,\mbox{(pb)}. \label{eq:rb}
\end{equation}
The thick band observed in this figure means any bounds obtained,
say, on the cross section, will translate into a less severe bound
on the sum of the branching ratios than what happened for the
$t\,+\,\gamma$ channel. For instance, in fig.~\ref{fig:sigbrs} an
upper bound on the cross section $\sigma(pp\rightarrow u
g\rightarrow t \gamma)$ of $10^{-2}$ implied
$Br(t\,\rightarrow\,u\,\gamma)\,+\,Br(t\,\rightarrow\,u\,g)\,<\,
10^{-5}$, whereas a similar bound on $\sigma(pp\rightarrow u
g\rightarrow t Z)$ gives us approximately, from the right-hand side
of the band in fig.~\ref{fig:tz},
$Br(t\,\rightarrow\,u\,Z)\,+\,Br(t\,\rightarrow\,u\,g)\,<\,
10^{-4}$. If we didn't have this band of values, but rather a line
corresponding to its left-hand side edge, the bound would be one
order of magnitude lower. As before, we obtain qualitatively
identical results for the processes involving the $c$ quark, and we
can quote rough bounds similar to those of eq.~\eqref{eq:rb},
\begin{equation}
30\, \left[Br(t \rightarrow c\,g)\,+\,Br(t \rightarrow
c\,Z)\right]\,<\, \sigma(p p \rightarrow c\,g \rightarrow t\,
Z)\,<\, 600\, \left[Br(t \rightarrow c\,g)\,+\,Br(t \rightarrow
c\,Z)\right] \,\mbox{(pb)}.
\end{equation}
\begin{figure}[htbp]
  \begin{center}
    \includegraphics[scale=0.7]{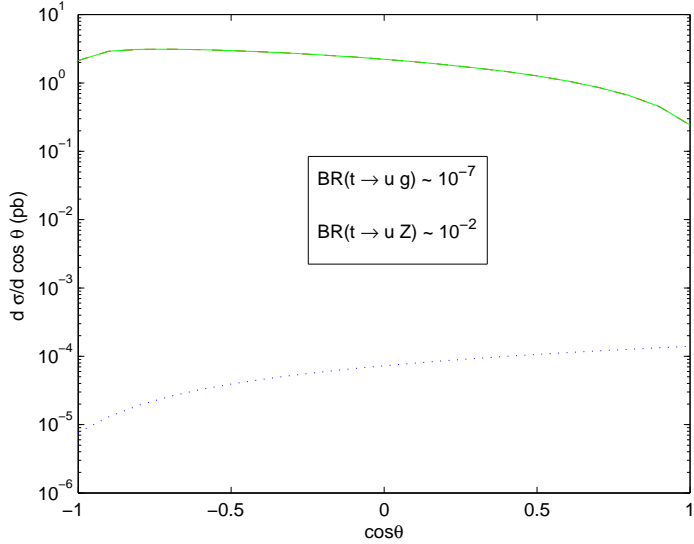}
    \caption{Differential cross sections for the process $p p \rightarrow u g
\rightarrow t Z$. Total (thick line), electroweak (dashed line) and
strong (dotted line) contributions. The electroweak contribution
practically coincides with the strong one.}
    \label{fig:difzf}
  \end{center}
\end{figure}
And again, we observe that the strong and electroweak cross sections
have different angular dependencies. In fig.~\ref{fig:difzf} we
plot the differential cross section for the process $p p \rightarrow
u g \rightarrow t Z$, both the strong and electroweak contributions,
for a typical choice of anomalous couplings for which the
electroweak FCNC interactions dominate over the strong ones. The
strong contributions increase with $\cos\theta$, whereas the
electroweak ones decrease. If the strong FCNC couplings dominate
over the electroweak ones, then the total cross section would very
closely mimic the angular dependence of the dotted line in
fig.~\ref{fig:difzf}. Once more, if the electroweak and strong
FCNC interactions have contributions of similar magnitudes, then it
will not be possible to distinguish them through this analysis. We
can define an asymmetry coefficient for the $t\,+\,Z$ process as
well, namely
\begin{equation}
A_{t + Z} \;=\;\frac{\sigma_{t + Z}(\cos\theta > 0) \,-\,\sigma_{t +
Z}(\cos\theta < 0)}{\sigma_{t + Z}(\cos\theta > 0) \,+\,\sigma_{t +
Z}(\cos\theta < 0)}\;\;\; .
\end{equation}
We will now use the set of anomalous couplings generated to produce
fig.~\ref{fig:assg} and plot the evolution of $A_{t + Z}$ with
both FCNC branching ratios in fig.~\ref{fig:assz}.
Again, we see a clear distinction between dominance of electroweak
FCNC interactions or strong FCNC ones. In the former case $A_{t +
Z}$ tends to a value of approximately $0.4$, and in the latter
situation we have $A_{t + Z}\,\sim\,-0.4$ -- this is particulary
interesting since the asymmetry changes signs, going from one regime
to the other. Once more, if both branching ratios have like sizes,
$A_{t + Z}$ may have any value between these two extreme.

\subsection{Cross section for the four-fermion channels}

In the following analysis we divide the four-fermions processes in two: those which don't have the SM contribution like processes (1) to (5) in table~\ref{tab:ffchannel} and processes with the SM contribution: (6) to (8).
The analytical expression for the cross section for the processes (1) to (5) is given by
\begin{equation}
\frac{d\sigma}{dt}=\frac{d\sigma^S}{dt}+\frac{d\sigma^{EW}}{dt}+\frac{d\sigma^{Int}}{dt},\label{eq:expressao1}
\end{equation}
where $\sigma^S,$ $\sigma^{EW}$ and $\sigma^{Int}$ refer to strong and electromagnetic FCNC contribution  and the interference between them, respectively. The electromagnetic term is the sum of the photon and $Z$ boson FCNC contribution:
\begin{equation}
\frac{d\sigma^{EW}}{dt}=\frac{d\sigma^\gamma}{dt}+\frac{d\sigma^Z}{dt};\label{eq:expressao2}
\end{equation}
and the interference term is given by
\begin{equation}
\frac{d\sigma^{Int}}{dt}=\frac{d\sigma^{S\gamma}}{dt}+\frac{d\sigma^{SZ}}{dt}+\frac{d\sigma^{\gamma Z}}{dt}.\label{eq:expressao3}
\end{equation}
where the superscript $S \gamma,$ $S Z$ and $\gamma Z$ relates to the interferences: strong+photon, strong+Z and photon+Z, respectively. We present the analytical expression for those processes in the appendix~\ref{sec:ap4f}. In the previous subsections we have generated values for all anomalous coupling that come out in the expression~\ref{eq:expressao1} and integrated it with CTEQ6M pdf's to compute the cross section in the LHC. Then, we again chose $\mu_F=m_t,$ imposed a $10 \; GeV$ cut on the transverse momentum and imposed that all branching ratio must be less than $10^{-2}\; pb.$ Now, we should add that the total cross section for $p\,p\rightarrow t\, +\, q$ the sum of both contributions: u quark and c quark.

\begin{figure}[htbp]
  \begin{center}
    \includegraphics[scale=0.7]{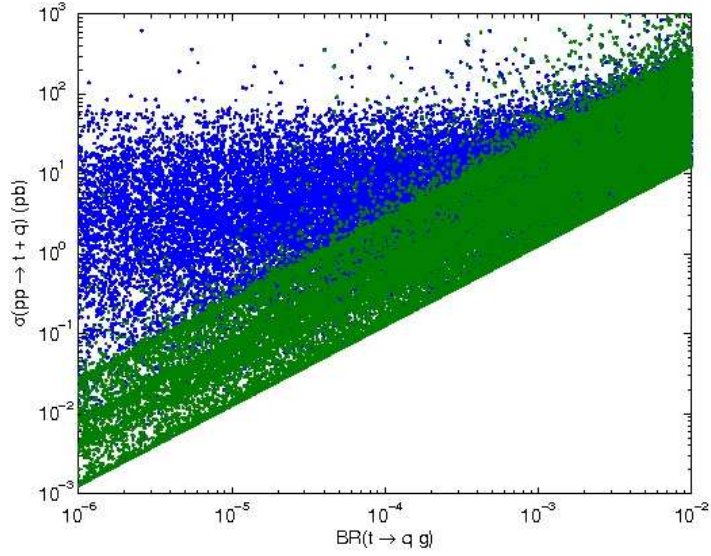}
    \caption{Total (blue) and strong (green) $p\,p\rightarrow t\, +\, q$ FCNC cross section as function of the branching ratio $Br(t\,\rightarrow\,q\,g).$}
    \label{fig:asstotal1}
  \end{center}
\end{figure}

In fig.~\ref{fig:asstotal1} we plot the value of the total and strong cross section for $t\,+\, q$ production against the strong FCNC branching ratio of the top quark.
Here, we register a difference between contribution from strong and electroweak decay for the $t\,+\, q$ FCNC production cross section and the contribution for the $t\,+\, \gamma$ and $t\,+\, Z$ one: the upper limit to the electroweak contribution is, in the best of cases of order of $\sim 10\, pb$ but the strong contribution is $\sim 10^{2}\, pb.$ As we saw in the previous sections ($t\,+\, \gamma$ and $t\,+\, Z$ production), the upper limit contribution of the strong and electroweak contribution were of the same order ($\sim 10^{2}\, pb$). Now, the upper limit of the strong and electroweak contribution is about $\sim 10\, pb$ to the strong FCNC branching ratio $\sim 10^{-3}$ and then it grow until $\sim 10^{2}\, pb$ to a branching ratio $10^{-2}.$ Thus, if we find a cross section bigger than $10^{1}\, pb,$  the main contribution  from the strong one but the reverse of this statement is not true.

In the previous sections we have used the asymmetry studies to have a criterion to distinguish the strong and electroweak FCNC contribution to $t\,+\,\gamma$ and $t\,+\,Z$ production. In $t\,+\, q$ FCNC production this is not possible because both the strong and the electroweak contribution have the same Feynman diagrams without any intermediate top quark.

\begin{figure}[htbp]
  \begin{center}
    \includegraphics[scale=0.7]{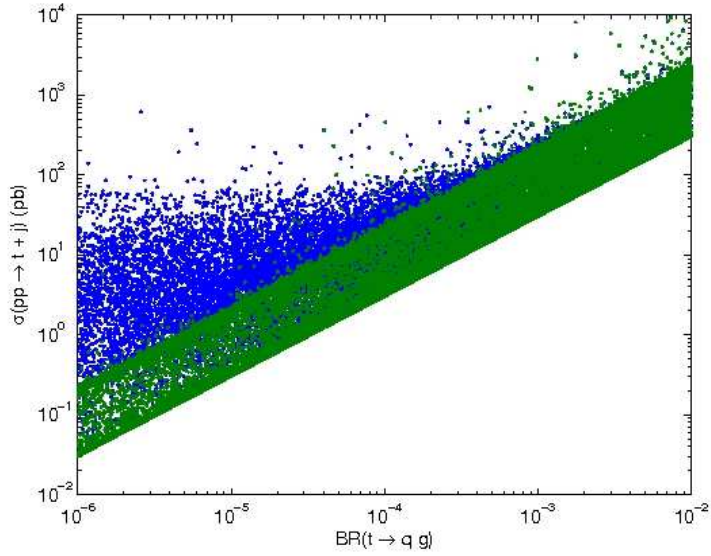}
    \caption{Total (blue) and strong (green) $p\,p\rightarrow t\, +\, j$ FCNC cross section as function of the branching ratio $Br(t\,\rightarrow\,q\,g).$}
    \label{fig:asstotal3}
  \end{center}
\end{figure}

The cross section for $t\,+\,j$ production is the sum of following cross sections: $p\, p\rightarrow g\, g\rightarrow t\,\bar q,$ $p\, p\rightarrow g\, q\rightarrow t\, g$~\cite{Ferreira:2006xe}, $p\, p\rightarrow g\, q\rightarrow t$ (direct top production~\cite{Ferreira:2005dr}) and $p\, p\rightarrow q\, q\rightarrow t\, q~\footnote{We didn't include contribution from processes (6) to (8) in table~\ref{tab:ffchannel}  because it is not expressive as we saw.}.$ In the SM, there are three possible processes to produce $t\,+\, j$ represented by the Feynman diagrams in the figs.~\ref{fig:single-t}(a), .\ref{fig:single-t}(b) and \ref{fig:single-t}(c). With a cross section $\sim 257\pm 12\, pb$\cite{Campbell:2005bb,Sullivan:2004ie}. In fig.~\ref{fig:asstotal3} we plot the FCNC cross section of $t\,+\, j$ production against the total FCNC branching ratio.
First, we note that the plot~\ref{fig:asstotal3} and \ref{fig:asstotal1} are very similar except that the contribution of strong FCNC cross section is bigger by a factor of ten. Thus, the upper limit to the cross section is of order of $\sim 10^{3}\, pb$ for a strong FCNC branching ratio equal to $10^{-2}$ and it is of order of $\sim 10^{2}\, pb$ for a strong FCNC branching ration about $10^{-4}.$ For large values of the strong branching ratio the number of FCNC events could be ten times larger than what is expected for the SM.

To conclude, we briefly comment processes (6) to (8) in the table~\ref{tab:ffchannel}. They have a contribution to the cross section of order $\Lambda^{-2}$ due to the interference with SM Feynman diagrams. The analytical expressions for these cross sections are given in appendix~\ref{sec:ap4f}. In the analysis of those processes we separate the final state in two: with a b quark (or $\bar b$) in the final state or without it. The reason for that is: first, the off-diagonal CKM matrix elements suppress final state with top quark and other quarks other than b one; finally, the b-tag technique allows one to identify a jet from the hadronization of a b quark in the final state. In fact, the contribution with a b quark is of order $\sim 10$ greater than the other contributions but even in this case the total contribution is less than one as was showed in the ref.~\cite{ won:2008mt}.

\section{Discussion and conclusions}
\label{sec:disc}

To summarize, we employed the effective operator formalism to parameterize the effects of any theory that might have as its low-energy limit the SM. The fact that we are working in a gauge invariant formalism allowed us to find many relations between couplings and quantities which, {\em a priori}, would not be related at all. In particular we found a near-proportionality between the cross section of associated top plus photon production at the LHC and the sum of the FCNC decays of the top to a photon and a gluon. We estimated the cross sections for $t\,+\,\gamma,$ $t\,+\,Z$ and $t\,+\,q$ production at the LHC and saw that, for large enough values of the top FCNC branching ratios, one might expect a significant number of events. We also concluded that, for these processes, the interplay between the strong and electroweak anomalous interactions tends to increase the values of the cross sections -- the interference between both FCNC sectors is mostly constructive. Also, if the cross section of $t\,+\,q$ production is of order of $10\, pb$ and bigger then we have a criterion to evaluate the dependence of this cross section and the strong FCNC branching ratio. Finally, we saw that the upper limit of the $t\,+\, j$ FCNC cross section is one order highest than the $t\,+\, q$. This was understood due the contribution of the gluonic processes.

Despite the advantage in using the cross section for $t+ j$ FCNC production due its large values, these processes do not offer a general criterion to study the individual contribution due to the strong or electroweak FCNC decay. In other words, we know the total cross section but it is not easy to distinguish the strong from the electroweak contribution. This is not matter case for the $t + \gamma$ or $t + Z$ production cross section where we can perform such studies using the FB asymmetry, as we saw. Thus, the general problem is: even if the top quark has indeed large FCNC branching ratios -- strong or electroweak ones --, which would lead to significant cross sections of associated single top production at the LHC, could those processes actually be observed? In other words, given the numerous backgrounds present at the LHC, is it possible to extract a meaningful FCNC signal from the expected data? The very thorough analysis of ref.~\cite{delAguila:1999ec} seems to indicate so. For instance, for $t\,+\,Z$ production they identify several possible channels available to identify the FCNC signal, summarised in table~\ref{tab:tZ}.
\begin{table}[t]
\begin{center}
\begin{tabular}{cccccc}\hline\hline \\
 Final State & \hspace{2cm} & Fraction (\%) & \hspace{2cm} &
 Backgrounds
\\ & & & \\ \hline \\
$tZ \rightarrow (bjj) \, (jj)$ &  & 22.2 & & $jjjjj$ &   \vspace{0.5cm} \\
$tZ \rightarrow (bjj) \, (\nu \bar{\nu})$ &  & 8.1 & & $t \bar{t}$, $Wt$, $Zjjj$ &   \vspace{0.5cm} \\
$tZ \rightarrow (bl \nu) \, (jj)$ &  & 7.5 & & $t \bar{t}$, $Wt$, $Wjjj$ & \vspace{0.5cm} \\
$tZ \rightarrow (bl \nu) \, (\nu \bar{\nu})$ &  & 2.7 & & $Wj$ &\vspace{0.5cm} \\
$tZ \rightarrow (bjj) \, (ll)$ &  & 2.3 & & $Zjjj$, $ZWj$ &  \vspace{0.5cm} \\
$tZ \rightarrow (bjj) \, (\bar{b} b)$ &  & 2.2 & & $b \bar{b} jjj$ &\vspace{0.5cm} \\
$tZ \rightarrow (bl \nu) \, (ll)$ &  & 0.8 &   & $ZWj$ & \vspace{0.5cm} \\
$tZ \rightarrow (bl \nu) \, (\bar{b} b)$ &  & 0.7 & & $t \bar{t}$,
$Wt$, $ZWj$, $W b \bar{b} j$& \\ & & &
\\\hline\hline
\end{tabular}
\caption{Possible final states in $t \, Z$ production, and main
backgrounds to each process \protect\cite{delAguila:1999ec}.}\label{tab:tZ}
\end{center}
\end{table}
For all of these processes, the processes $W Z j$, $t\bar{t}$ and single top production will also act as backgrounds. It is also likely, considering the immense QCD backgrounds, that only those processes with at least one lepton will be possible to observe at the LHC. To build this table, the top quark was considered to decay according to SM physics, $t \rightarrow b \, W$, and the several decay possibilities within the SM of the $W$ and $Z$ bosons give the possibilities listed therein. The fraction attributed to each channel corresponds to the percentages of each decay mode of the $W$ and $Z$ as well as a 90 \% tagging efficiency for lepton (electron or muon) tagging, and a 60 \% one for each b-jet. The most impressive result of ref.~\cite{delAguila:1999ec}, though, is the efficiency with which the FCNC signal is extracted from these backgrounds: they have shown that a battery of simple kinematical cuts on the observed particles is more than enough to obtain a very clear -- and statistically meaningful -- FCNC signal. For $t\,+\,Z$ production they conclude that the best channel would be $p\,p\rightarrow t\,Z\rightarrow l^+\,l^-\,l\,\nu\,b$. For $t\,+\,\gamma$ production the analysis is made simpler by the photon not having decay branching ratios, which aides the statistics obtained -- the best channel available would be $p\,p\rightarrow t\,\gamma\rightarrow \gamma\,l\,\nu\,b$. Clearly, only an analysis analogous to that of~\cite{delAguila:1999ec}, with the FCNC interactions considered in the present thesis included in an event generator, would be capable of reaching definite conclusions regarding which kinematical cuts would be better suited to obtain a clear FCNC signal. That study is beyond the scope of the present thesis, though a preliminary study of our strong FCNC interactions in the LHC environment, using the TopReX event generator~\cite{Slabospitsky:2002ag}, is about to be concluded~\cite{nuno:2008}. A word on higher-order QCD corrections: they are manifestly difficult to compute in the effective operator formalism, since the Lagrangian becomes non-renormalizable. A recent work using electroweak top FCNC couplings~\cite{Kidonakis:2003sc}, however, concluded that those corrections greatly reduce any dependence the results obtained at tree level might have on the scales of renormalization and factorization. These authors have also shown that the higher order corrections tend to slightly increase the leading order result. The analysis of the differential cross sections for $t\,+\,\gamma$ and $t\,+\,Z$ production will possibly allow the identification of the source of FCNC physics -- the strong or the electroweak sector.

Finally, we consider the process 3 in the table~\ref {tab:sec} i.e., $t\bar t$ production via FCNC operators. In ref.~\cite{ Ferreira:2006xe} the authors have calculated this cross section for the strong FCNC operator; in the appendix~\ref{sec:ttbar} we present the analytical expression for this process for the electroweak FCNC operator. To obtain the cross section for the process $p\,p\,\rightarrow\,q\,\bar q\,\rightarrow\,t\,\bar t$ at the LHC we integrated the partonic cross section in eq.~\eqref{eq:blabla} in the appendix~\ref{sec:ttbar} with the CTEQ6M pdf~\cite{Pumplin:2002vw}, with a factorization scale $\mu_F=2\,m_t$ .We also imposed a cut of 15 GeV on the $p_T$ of the final state partons. However, it is unlikely to register such FCNC process in the LHC. By a simple inspection of the coefficients,  supposing that all of coefficients are one with the same signal, the total cross section is about $20\, pb.$  The SM cross section to $t\bar t$ about $833\pm 100 pb$~\cite{Bonciani:1998vc,Beneke:2000hk} i. e., the expected experimental error is larger then all FCNC contribution.

\newpage


\addcontentsline{toc}{section}{{\bf Appendix 3}}

\begin{subappendices}

\section{Cross section expression for the process $q \, g \rightarrow
t \, Z$} \label{sec:apz}

As mentioned in section~\ref{sec:tz} the cross section for the
associated production of a top and a Z boson is given by three
terms, as in eq.~\eqref{eq:tz}. The strong FCNC contribution is
given by:
\begin{align}
\frac{d\, \sigma^S_{q \, g \rightarrow t \, Z}}{dt} & = \frac{e^2}
  {96 \, \pi \, s^2 \, \Lambda^4}
\left\{ F_{1_Z}(t,s) \, \left[ \left| \alpha^S_{qt} +
(\alpha^S_{tq})^* \right|^2 + \frac{8v}{m_t} \, Im \left[
(\alpha^S_{qt} + ( \alpha^S_{tq})^*) \, \beta^S_{tq}
\right]\right.\right.
 \nonumber \\
& +\left.\left. \frac{16v^2}{m_t^2} \, \left| \beta^S_{tq} \right|^2
\right]  + \, F_{2_Z}(t,s) \,  \frac{16v^2}{m_t^2} \, \left|
\beta^S_{qt} \right|^2 \right\}\;\;\;,
\end{align}
with coefficients
\begin{align}
F_{1_Z}(t,s) & =  \frac{- m_t^2}{72\, c_W^2\,m_Z^2\,{\left( m_t^2 -
s \right) }^2 \,s_W^2\,t^2}\nonumber \\
& \times\left[ 32\,m_t^8\,m_Z^2\, s_W^4\,\left( m_Z^2 - t \right)  +
32\,m_t^4\,m_Z^2\,s_W^4\, \left( m_Z^2 - t \right) \,\left( s^2 +
4\,s\,t + t^2\right)
\right.\nonumber \\
& + s^2\,t^2\,\left( 2\,m_Z^4\, \left( 9 - 24\,s_W^2 + 32\,s_W^4
\right)  + 9\,s\,t - 2\,m_Z^2\,\left( 9 - 24\,s_W^2 + 32\,s_W^4
\right)\,\left( s + t \right)  \right) \,\nonumber \\
& + m_t^2\,s\,t\,\left( -9\,s\,t^2 -64\,m_Z^4\,s_W^4\,\left( s + t
\right) + m_Z^2\,\left( 32\,s^2\,s_W^4 + 3\,s\,\left( 3 - 32\,s_W^2
+ 64\,s_W^4 \right) \,t \right.\right.
\nonumber \\
& \left.\left. + 96\,s_W^4\,t^2 \right) \right) - \left.
32\,m_t^6\,m_Z^2\,s_W^4\, \left( 2\,m_Z^2\,\left( s + t
\right)-t\,\left( s + 2\,t \right) \right)\right],
\nonumber  \\[0.35cm]
F_{2_Z}(t,s) & =  \frac{ m_t^2}{72 \, c_W^2\,m_Z^2\,{\left( m_t^2 -
s \right) }^2 \,s_W^2\,t^2} \left\{ - 2\,m_t^4\,m_Z^2\,{\left( 3 -
4\,s_W^2 \right) }^2\, \left( m_Z^2 - t \right) \,\left( s^2 +
4\,s\,t + t^2 \right)\right. \nonumber \\
& + s^2\,t^2\,\left( -2\,m_Z^4\, \left( 9 - 24\,s_W^2 + 32\,s_W^4
\right)  - 9\,s\,t + 2\,m_Z^2\,\left( 9 - 24\,s_W^2 + 32\,s_W^4
\right) \, \left( s + t \right)  \right)\nonumber \\
& + m_t^2\,s\,t\,\left[ 9\,s\,t^2 + 4\,m_Z^4\,{\left( 3 - 4\,s_W^2
\right) }^2\,\left( s + t \right)  + m_Z^2\,\left( -2\,s^2\,{\left(
3 - 4\,s_W^2 \right) }^2 \right.  \right.
\nonumber \\
& - \left. \left. 3\,s\,\left( 15 + 64\,\left( -s_W^2 + s_W^4
\right) \right) \,t - 6\,{\left( 3 - 4\,s_W^2 \right) }^2\,t^2
\right) \right] 2\,m_t^8\,m_Z^2\,{\left( 3 - 4\,s_W^2 \right) }^2\,
\left( -m_Z^2 + t \right)
\nonumber \\
& \left. + 2\,m_t^6\,m_Z^2\, {\left( 3 - 4\,s_W^2 \right) }^2\,
\left( 2\,m_Z^2\,\left( s + t \right)- t\,\left( s + 2\,t \right)
\right) \right\}\;\;\; . \label{eq:a12}
\end{align}
The electroweak FCNC contribution is given by the following
expression:
\begin{align}
\frac{d \, \sigma^{EW}_{q \, g \rightarrow t \, Z}}{dt} & =
\frac{g_s^2} {96 \, \pi \, s^2  \, \Lambda^4}
\left[ G_{1_Z}(t,s) \, \left| \alpha^Z_{qt} \right|^2 + G_{2_Z}(t,s)
\, \left| \alpha^Z_{tq} \right|^2 + G_{3_Z}(t,s) \, ( \left|
\beta^Z_{qt} \right|^2 + \left| \beta^Z_{tq} \right|^2) \right.
\nonumber \\
& + G_{4_Z}(t,s) \, ( \left| \eta_{qt} \right|^2 + \left|
\bar{\eta}_{qt} \right|^2) + \, G_{5_Z}(t,s) \, \left| \theta
\right|^2 + G_{6_Z}(t,s) \, Re \left[ \alpha^Z_{qt} \, \alpha^Z_{tq}
\right] \nonumber \\
& + G_{7_Z}(t,s) \, Im \left[ \alpha^Z_{qt} \, \beta^Z_{tq} \right]
+ \, G_{8_Z}(t,s) \, Im \left[ \alpha^{Z^*}_{tq} \, \beta^Z_{tq}
\right] + G_{9_Z}(t,s) \, Re \left[ \alpha^Z_{qt} \theta^* \right]
\nonumber \\
& + G_{10_Z}(t,s) \, Re \left[ \alpha^Z_{tq} \theta \right]  \left.
+ \, G_{11_Z}(t,s) \, Re \left[ \beta^Z_{qt}
(\eta_{qt}-\bar{\eta}_{qt})^* \right] \right.
\nonumber\\
& \left. + G_{12_Z}(t,s) \, Im \left[ \beta^Z_{tq} \, \theta \right]
+ G_{13_Z}(t,s) \, Re \left[ \eta_{qt} \bar{\eta}_{qt}^* \right]
\right]
\end{align}
where the $G_{i_Z}$ functions are given by
\begin{align}
G_{1_Z}(t,s) & =  \frac{1}{4\,s\,{\left( m_t^2 - t \right) }^2}\nonumber \\
& \times \left[ m_t^{10} + m_t^8\,\left( 2\,m_Z^2 - 2\,s - t \right)
+ m_t^6\,\left( -5\,m_Z^4 + s^2 + 4\,s\,t + t^2 - 4\,m_Z^2\,\left( s + t \right)\right)\right.\nonumber \\
& + m_t^2\,\left( 2\,m_Z^8 - 8\,m_Z^6\,t - 4\,m_Z^2\,t^2\,\left( s +
t \right)  + 2\,s\,t^2\,\left( s + t \right)  +
m_Z^4\,{\left( s + 3\,t \right) }^2 \right) \nonumber \\
& - m_Z^4\,t\,\left( 2\,m_Z^4 + s^2 + t^2 -
2\,m_Z^2\,\left( s + t \right)  \right) \nonumber \\
& \left. + m_t^4\,\left( 6\,m_Z^6 - 3\,m_Z^4\,t +
2\,m_Z^2\,t\,\left( s + 3\,t \right)  - t\,\left( 3\,s^2 + 6\,s\,t +
t^2 \right) \right) \right]\nonumber
\end{align}
\begin{align}
G_{2_Z}(t,s) & =  \frac{1}{4\,s\, {\left( m_t^2 - t \right) }^2}
\left[ m_t^{10} - m_t^8\,\left( 4\,m_Z^2 + 2\,s + t \right)  +
m_t^6\,\left( 7\,m_Z^4 + s^2 + 2\,m_Z^2\,t + 4\,s\,t +
t^2 \right) \right. \nonumber \\
& - m_Z^4\,t\, \left( 2\,m_Z^4 + s^2 + t^2 - 2\,m_Z^2\,\left( s + t
\right) \right)\nonumber \\
& - m_t^4\,\left( 6\,m_Z^6 + 3\,m_Z^4\,t - 2\,m_Z^2\,s\,\left( s +
3\,t \right)  + t\,\left( 3\,s^2 + 6\,s\,t + t^2 \right)
\right)\nonumber \\
& \left. + m_t^2\,\left( 2\,m_Z^8 + 4\,m_Z^6\,t + 2\,s\,t^2\,\left(
s + t \right)  + m_Z^4\,\left( s^2 - 6\,s\,t - 3\,t^2 \right)  +
m_Z^2\,\left( -2\,s^2\,t + 2\,t^3 \right)  \right) \right]\nonumber
\end{align}
\begin{align}
G_{3_Z}(t,s) & =  \frac{2\, v^2} {s\, {\left( m_t^2 - t \right) }^2}
\left[ 2\,m_t^8 -
m_t^6\,\left( 3\,m_Z^2 + 4\,s + 2\,t \right)\right. \nonumber \\
&  - t\,\left( 2\,m_Z^6 - 2\,m_Z^4\,\left( s + t \right)  -
4\,s\,t\,\left( s + t \right)  + m_Z^2\,{\left( s + t \right) }^2 \right)   \nonumber \\
& + m_t^4\,\left( 2\,m_Z^4 - m_Z^2\,\left( 2\,s + t \right)  +
2\,\left( s^2 + 4\,s\,t + t^2 \right)
\right)   \nonumber \\
& \left.  + m_t^2\,\left( 2\,m_Z^6 - 4\,m_Z^4\,t - 2\,t\,\left(
3\,s^2 + 6\,s\,t + t^2 \right)+ m_Z^2\,\left( s^2 + 6\,s\,t + 5\,t^2
\right)  \right) \right] \nonumber
\end{align}
\begin{align}
G_{4_Z}(t,s) & =  \frac{v^2}{8\, m_Z^2\,s\,{\left( m_t^2 - t \right)
}^2} \left[  m_t^{10} - m_t^8\, \left( 4\,m_Z^2 + 2\,s + t \right) +
m_t^6\,\left( 7\,m_Z^4 + 2\,m_Z^2\,t +
{\left( s + t \right) }^2 \right)  \right. \nonumber \\
& - m_Z^2\,t\,\left( 2\,m_Z^6 - 2\,m_Z^4\,\left( s + t \right)  -
4\,s\,t\,\left( s + t \right)  +
m_Z^2\,{\left( s + t \right) }^2 \right) \nonumber \\
&  - m_t^4\,\left( 6\,m_Z^6 + t\,{\left( s + t \right) }^2 +
m_Z^4\,\left( 2\,s + 3\,t \right)  - 2\,m_Z^2\,s\,\left( 3\,s + 5\,t \right) \right) \nonumber \\
&  \left. + m_t^2\,m_Z^2\, \left( 2\,m_Z^6 + 4\,m_Z^4\,t +
m_Z^2\,\left( s^2 - 2\,s\,t - 3\,t^2 \right)  + 2\,t\,\left( -5\,s^2
- 4\,s\,t + t^2 \right)  \right) \right]\nonumber
\end{align}
\begin{align}
G_{5_Z}(t,s) & =  \frac{v^4} {2\,m_Z^2\,s\, {\left( m_t^2 - t
\right) }^2}  \left[ m_t^8 - m_t^6\,\left( 2\,s + t \right)  -
2\,m_Z^2\,t\,\left( 2\,m_Z^4 + s^2 + t^2 -
2\,m_Z^2\,\left( s + t \right)  \right) \right. \nonumber \\
&   + m_t^4\,\left( -2\,m_Z^4 - 2\,m_Z^2\,t + {\left( s + t
\right) }^2 \right)  \nonumber \\
& \left. + m_t^2\,\left( 4\,m_Z^6 - 2\,m_Z^4\,t - t\,{\left( s + t
\right) }^2 + 2\,m_Z^2\,\left( s^2 - s\,t + 2\,t^2 \right) \right)
\right]\nonumber
\end{align}
\begin{align}
G_{6_Z}(t,s) & =  \frac{1}{2\,s\,{\left( m_t^2 - t \right) }^2}
\left[ m_t^{10} - m_t^8\,\left( 2\,m_Z^2 + 2\,s + t \right) +
m_t^6\,\left( m_Z^4 + s^2 + 4\,s\,t + t^2 \right) \right.
\nonumber \\
& + m_Z^4\,t\,\left( 2\,m_Z^4 + s^2 + t^2 - 2\,m_Z^2\,\left( s + t
\right)  \right) - m_t^2\,\left( 2\,m_Z^8 + m_Z^4\,{\left( s - t
\right) }^2 - 2\,s\,t^2\,\left( s + t \right)  \right)
\nonumber \\
&  \left. + m_t^4\,\left( 2\,m_Z^6 - m_Z^4\,t + 2\,m_Z^2\,t\,\left(
s + t \right)  - t\,\left( 3\,s^2 + 6\,s\,t + t^2 \right) \right)
\right]\nonumber
\end{align}
\begin{align}
G_{7_Z}(t,s) & =  \frac{2 \,m_t \,v}{s\, {\left( m_t^2 - t \right)
}^2} \left[ m_t^8 - m_t^6\,\left( 2\,s + t \right) + m_t^4\,\left(
-2\,m_Z^4 + s^2 + 4\,s\,t + t^2 - 2\,m_Z^2\,\left( s + t \right)
\right) \right.\nonumber \\
&        + 2\,t\,\left( -2\,m_Z^6 + 2\,m_Z^4\,\left( s + t \right) -
m_Z^2\,t\,\left( s + t \right) + s\,t\,\left( s + t \right)  \right)
\nonumber \\
&        \left.    + m_t^2\,\left( 4\,m_Z^6 - 2\,m_Z^4\,t +
2\,m_Z^2\,t\,\left( s + 2\,t \right)  - t\,\left( 3\,s^2 + 6\,s\,t +
t^2 \right)  \right) \right]\nonumber
\end{align}
\begin{align}
G_{8_Z}(t,s) & =  \frac{2 \,m_t \,v}{s\, {\left( m_t^2 - t \right)
}^2} \left[ m_t^8 - m_t^6\,\left( 3\,m_Z^2 + 2\,s + t \right)  +
m_t^4\,\left( 4\,m_Z^4 + s^2 + m_Z^2\,t + 4\,s\,t + t^2 \right)
\right.\nonumber \\
&          + t\,\left( 2\,m_Z^6 - 2\,m_Z^4\,\left( s + t \right)  +
2\,s\,t\,\left( s + t \right)  + m_Z^2\,\left( -s^2 + t^2 \right)
\right)\nonumber \\
&         \left.  - m_t^2\,\left( 2\,m_Z^6 + 2\,m_Z^4\,t -
m_Z^2\,\left( s^2 + 4\,s\,t + t^2 \right)  + t\,\left( 3\,s^2 +
6\,s\,t + t^2 \right)  \right) \right]\nonumber
\end{align}
\begin{align}
G_{9_Z}(t,s) & =  \frac{v^2}{s\,{\left( m_t^2 - t \right) }^2}
\left[ -2\,m_t^8 + m_t^4\, \left( -2\,m_Z^4 + m_Z^2\,t - 2\,t^2
\right) \right.\nonumber \\
&  +   m_t^6\,\left( 3\,m_Z^2 + 2\,\left( s + t \right) \right) +
m_Z^2\,t\,\left( 2\,m_Z^4 + s^2 + t^2 - 2\,m_Z^2\,\left( s + t
\right)  \right)\nonumber \\
&      \left. - m_t^2\,\left( 2\,m_Z^6 - 4\,m_Z^4\,t -
2\,t^2\,\left( s + t \right)  + m_Z^2\,\left( s^2 + 2\,s\,t + 5\,t^2
\right) \right)  \right]\nonumber
\end{align}
\begin{align}
G_{10_Z}(t,s) & =  \frac{-v^2}{s\, {\left( m_t^2 - t \right) }^2}
\left[ m_t^8 - m_t^6\,\left( 3\,m_Z^2 + t \right)  + m_t^4\,\left(
4\,m_Z^4 - s^2 + m_Z^2\,t - 2\,s\,t + t^2 \right) \right.
\nonumber \\
&  + m_Z^2\,t\, \left( 2\,m_Z^4 + s^2 + t^2 - 2\,m_Z^2\,\left( s + t
\right) \right)\nonumber\\
& \left. - m_t^2\, \left( 2\,m_Z^6 + 2\,m_Z^4\,t - s^2\,t + t^3 +
m_Z^2\,\left( s^2 - 4\,s\,t - t^2 \right)  \right) \right]\nonumber
\end{align}
\begin{align}
G_{11_Z}(t,s) & =  \frac{ v^2}{s\,{\left( m_t^2 - t \right) }^2}
\left[ m_t^8 - m_t^6\, \left( 3\,m_Z^2 - 2\,s + t \right)
\right.\nonumber \\
& + t\,\left( 2\,m_Z^6 - 2\,m_Z^4\,\left( s + t \right)  -
4\,s\,t\,\left( s + t \right) + m_Z^2\,{\left( s + t \right)
}^2\right)\nonumber\\
& + m_t^4\,\left( 4\,m_Z^4 - 3\,s^2 - 10\,s\,t + t^2 + m_Z^2\,\left(
2\,s + t \right)  \right)\nonumber \\
&         \left.      - m_t^2\,\left( 2\,m_Z^6 + 2\,m_Z^4\,t +
m_Z^2\,\left( s^2 - t^2 \right)  + t\,\left( -7\,s^2 - 10\,s\,t +
t^2 \right) \right)  \right]\nonumber
\end{align}
\begin{align}
G_{12_Z}(t,s) & =  \frac{-2\, m_t\, v^3}{s\, {\left( m_t^2 - t
\right) }^2} \,\left[ 3\,m_t^6 - m_t^4\,\left( 6\,m_Z^2 + 2\,s +
3\,t \right)  + m_t^2\,\left( 6\,m_Z^4 - s^2 - 2\,s\,t + 3\,t^2
\right) \right.\nonumber \\
&      + \left. t\,\left( -6\,m_Z^4 + s^2 - 2\,s\,t - 3\,t^2 +
6\,m_Z^2\,\left( s + t \right)  \right)  \right]\nonumber
\end{align}
\begin{align}
G_{13_Z}(t,s) & =  \frac{- v^2} {4\, m_Z^2\,s\,{\left( m_t^2 - t
\right) }^2}  \left[ m_t^{10} - m_t^8\,\left( 4\,m_Z^2 + 2\,s + t
\right)  + m_t^6\,\left( 7\,m_Z^4 + 2\,m_Z^2\,t + {\left( s + t
\right) }^2 \right) \right.\nonumber \\
&            - m_Z^2\,t\,\left( 2\,m_Z^6 - 2\,m_Z^4\,\left( s + t
\right)  - 4\,s\,t\,\left( s + t \right)  + m_Z^2\,{\left( s + t
\right) }^2 \right)\nonumber \\
&           - m_t^4\,\left( 6\,m_Z^6 + t\,{\left( s + t \right) }^2
+ m_Z^4\,\left( 2\,s + 3\,t \right)  - 2\,m_Z^2\,s\,\left( 3\,s +
5\,t \right)  \right)\nonumber \\
&       \left.    + m_t^2\,m_Z^2\, \left( 2\,m_Z^6 + 4\,m_Z^4\,t +
m_Z^2\,\left( s^2 - 2\,s\,t - 3\,t^2 \right)  + 2\,t\,\left( -5\,s^2
- 4\,s\,t + t^2 \right)  \right) \right]\;\;\; .
\end{align}
Finally, the strong-electroweak interference cross section is given
by
\begin{align}
\frac{d \, \sigma^{Int}_{q \, g \rightarrow t \, Z}}{dt} & = \frac{e
\, g_s} {96 \, \pi \, s^2  \, \Lambda^4}
\left[ H_{1_Z}(t,s) \, \left\{ Re \left[ \left( \alpha^S_{qt} +
\alpha^{S^*}_{tq} \right) \alpha^{Z^*}_{qt} \right] + \frac{4 \,
v}{mt} \, Im \left[ \beta^S_{tq} \, \alpha^Z_{qt} \right] \right\}
\right.
\nonumber \\
& + H_{2_Z}(t,s) \, \left\{ Re \left[ \left( \alpha^S_{qt} +
\alpha^{S^*}_{tq} \right) \alpha^{Z}_{tq} \right] + \frac{4 \,
v}{mt} \, Im \left[ \beta^S_{tq} \, \alpha^{Z^*}_{tq} \right]
\right\}
\nonumber \\
& + H_{3_Z}(t,s) \, \left\{ Im \left[ \left( \alpha^S_{qt} +
\alpha^{S^*}_{tq} \right) \beta^{Z}_{tq} \right] + \frac{4 \, v}{mt}
\, Re \left[ \beta^{S^*}_{tq} \, \beta^{Z}_{tq} \right] \right\}
\nonumber \\
& + H_{4_Z}(t,s) \, \left\{ Re \left[ \left( \alpha^S_{qt} +
\alpha^{S^*}_{tq} \right) \theta^* \right] + \frac{4 \, v}{mt} \, Im
\left[ \beta^{S}_{tq} \, \theta \right] \right\}+ H_{5_Z}(t,s) \, Re
\left[  \beta^S_{qt} \, \beta^{Z^*}_{qt} \right]
\nonumber \\
&   \left. + H_{6_Z}(t,s) \,  Re \left[  \beta^S_{qt} \, \left(
\eta_{qt} - \bar{\eta}_{qt} \right)^* \right] \right]
\end{align}
with $H_{i_Z}$ given by
\begin{align}
H_{1_Z}(t,s) & =  \frac{m_t^2}{6\,c_W\,\left( m_t^2 - s \right) \,
s_W\,\left( m_t^2 - t \right) \,t}
\nonumber \\
& \left[ 12\,m_t^6\,s_W^2\,t    - m_t^4\,\left( 4\,m_Z^4\,s_W^2 +
t\,\left( -3\,s + 16\,s\,s_W^2 + 16\,s_W^2\,t \right)  \right)
\right.
\nonumber \\
&       + t\,\left( 4\,m_Z^4\,s_W^2\,\left( -s + t \right)  +
s\,t\,\left( 3\,s - 4\,s\,s_W^2 - 4\,s_W^2\,t \right)  +
2\,m_Z^2\,s\,\left( 2\,s\,s_W^2 - 3\,t + 4\,s_W^2\,t \right) \right)
\nonumber \\
& + m_t^2\,\left( 4\,m_Z^4\,s\,s_W^2 + 2\,m_Z^2\,s\,\left( 3 -
8\,s_W^2 \right) \,t\right.
\nonumber \\
&  \left.\left. + t\,\left( s^2\,\left( -3 + 4\,s_W^2 \right)  +
3\,s\,\left( -1 + 4\,s_W^2 \right) \,t + 4\,s_W^2\,t^2 \right)
\right)  \right]\nonumber
\end{align}
\begin{align}
H_{2_Z}(t,s) & =  \frac{- m_t^2}{6\,c_W\, \left( m_t^2 - s \right)
\,s_W\,\left( m_t^2 - t \right) \,t}
\nonumber \\
&\left[   t\,\left( m_Z^2\,s\,\left( 4\,s\,s_W^2 - 3\,t \right)  +
4\,m_Z^4\,s_W^2\,\left( -s + t \right)  + s\,t\,\left( -3\,s +
4\,s\,s_W^2 + 4\,s_W^2\,t \right) \right)\right.
\nonumber \\
&         + 4\,m_t^6\,s_W^2\, \left( 2\,m_Z^2 - 3\,t \right)  +
m_t^4\,\left( -4\,m_Z^4\,s_W^2 - 8\,m_Z^2\,s\,s_W^2 + t\,\left(
-3\,s + 16\,s\,s_W^2 + 16\,s_W^2\,t \right)  \right)
\nonumber \\
&       + m_t^2\,\left( 4\,m_Z^4\,s\,s_W^2 + m_Z^2\,t\,\left( 3\,s -
8\,s_W^2\,t \right)\right.
\nonumber\\
&  \left.\left. + t\,\left( s^2\,\left( 3 - 4\,s_W^2 \right)  +
3\,s\,\left( 1 - 4\,s_W^2 \right) \,t - 4\,s_W^2\,t^2 \right)
\right) \right]\nonumber
\end{align}
\begin{align}
H_{3_Z}(t,s) & = \frac{m_t \, v}{3\,c_W\,\left( m_t^2 - s \right) \,
s_W\,\left( m_t^2 - t \right) \,t}\nonumber \\
& \left[ -8\,m_t^6\,s_W^2\, \left( m_Z^2 - 3\,t \right)  +
2\,m_t^4\,\left( 4\,m_Z^2\,s\,s_W^2 + t\,\left( 3\,s - 16\,s\,s_W^2
- 16\,s_W^2\,t \right) \right)\right.
\nonumber \\
&          + s\,t^2\,\left( m_Z^2\,\left( -3 + 8\,s_W^2 \right)  -
2\,\left( -3\,s + 4\,s\,s_W^2 + 4\,s_W^2\,t \right)  \right)
\nonumber \\
& + m_t^2\,t\,\left( m_Z^2\, \left( s\,\left( 3 - 16\,s_W^2 \right)
+ 8\,s_W^2\,t \right)\right.
\nonumber \\
& \left.\left.+ 2\,\left( s^2\,\left( -3 + 4\,s_W^2 \right) +
3\,s\,\left( -1 + 4\,s_W^2 \right) \,t + 4\,s_W^2\,t^2 \right)
\right) \right]\nonumber
\end{align}
\begin{align}
H_{4_Z}(t,s) & =  \frac{- m_t^2 \, v^2}{6\,c_W\,\left( m_t^2 - s
\right) \, s_W \,\left( m_t^2 - t \right) \,t} \left[
8\,m_t^6\,s_W^2 - 8\,m_t^4\,\left( m_Z^2 + s \right) \,s_W^2 \right.
\nonumber \\
&       + m_t^2\,\left( 8\,m_Z^2\,s\,s_W^2 + t\,\left( 9\,s -
16\,s\,s_W^2 - 8\,s_W^2\,t \right)  \right)
\nonumber \\
& \left.+ t\,\left( -8\,m_Z^2\,s_W^2\,\left( s - t \right)  +
s\,\left( 8\,s\,s_W^2 - 9\,t + 8\,s_W^2\,t \right)  \right)
\right]\nonumber
\end{align}
\begin{align}
H_{5_Z}(t,s) & =  \frac{4\, v^2}{3\, c_W \,\left( m_t^2 - s \right)
\, s_W\, \left( m_t^2 - t \right) \,t}
\nonumber \\
& \left[ s\,t^2\,\left( m_Z^2\,\left( -3 + 8\,s_W^2 \right)  -
2\,\left( 4\,s\,s_W^2 - 3\,t + 4\,s_W^2\,t \right) \right)
-2\,m_t^6\,\left( -3 + 4\,s_W^2 \right) \, \left( m_Z^2 - 3\,t
\right)\right.
\nonumber \\
& + 2\,m_t^4\,\left( m_Z^2\,s\, \left( -3 + 4\,s_W^2 \right) +
t\,\left( s\,\left( 9 - 16\,s_W^2 \right)  + 4\,\left( 3 - 4\,s_W^2
\right) \,t \right)  \right)
\nonumber \\
& + m_t^2\,t\,\left( m_Z^2\, \left( s\,\left( 9 - 16\,s_W^2 \right)
+ 2\,\left( -3 + 4\,s_W^2 \right) \,t \right)\right.
\nonumber \\
& \left.\left.+ 2\,\left( 4\,s^2\,s_W^2 + 6\,s\,\left( -1 + 2\,s_W^2
\right) \,t + \left( -3 + 4\,s_W^2 \right) \,t^2 \right)  \right)
\right]\nonumber
\end{align}
\begin{align}
H_{6_Z}(t,s) & = \frac{v^2}{3\,c_W\,\left( m_t^2 - s \right) \,
s_W\,\left( m_t^2 - t \right) \,t}
\nonumber \\
&\left[ - 2\,m_t^6\,\left( -3 + 4\,s_W^2 \right) \,\left( s + 3\,t
\right)  + m_t^4\,t\,\left( s\,\left( -15 + 16\,s_W^2 \right)  +
6\,\left( -3 + 4\,s_W^2 \right) \,t \right) \right.
\nonumber \\
&          + m_t^2\,t\,\left( -6\,s^2 + m_Z^2\,s\,\left( -3 +
8\,s_W^2 \right)  + s\,\left( 15 - 16\,s_W^2 \right) \,t + 2\,\left(
3 - 4\,s_W^2 \right) \,t^2 \right)
\nonumber \\
&     \left.      + 2\,m_t^8\,\left( -3 + 4\,s_W^2 \right)+
s\,t^2\,\left( m_Z^2\,\left( 3 - 8\,s_W^2 \right)  + 2\,\left(
4\,s\,s_W^2 + \left( -3 + 4\,s_W^2 \right) \,t \right) \right)
\right]\;\;\; .
\end{align}

\newpage

\section{Cross section expression for the four-fermion FCNC production process} \label{sec:ap4f}

The cross section for the four-fermion channel is given by the eq.~\eqref{eq:expressao1} and by the definition~\eqref{eq:expressao2} and~\eqref{eq:expressao3}. Now, we present each term according with the anomalous decay fot the process (1), (2), (3) and (6) in table~\ref{tab:ffchannel}. The strong FCNC contribution is given by:


{\bf Process (1)}

\begin{align}
\frac{d\sigma^S}{dt} & = \frac{g_3^2}{216\,\pi\, t\, u\,s^2\,\Lambda^4}\biggl\{F_1\times|\alpha^S_{ut}|^2+\,F_2\times|\alpha^S_{tu}|^2\, +\,F_3\times\Bigl[|\beta^S_{ut}|^2\,+\,|\beta^S_{tu}|^2\Bigr]\nonumber\\
& +\,F_4 \times Re(\alpha^S_{ut}\, \alpha^S_{tu})\,+\,F_5 \times Im(\alpha^S_{ut}\, \beta^{S}_{tu})\,+\,F_6 \times Im(\alpha^S_{tu}\, \beta^{S*}_{tu}) \biggr\}
\end{align}
with coefficients
\begin{align}
F_1 &= -3\, {m_t}^6\, \left(t + u \right) + 6\, {m_t}^4\, \left(t^2 + u^2 \right) +
  t\, u\, \left(7\, t^2 + 8\, t\, u + 7\, u^2 \right) \nonumber\\
  & - {m_t}^2\, \left(6\,
    t^3 + 13\, t^2\, u + 13\, t\, u^2 + 6\, u^3 \right)\nonumber\\
F_2 &= -3\,{m_t}^6\,\left( t + u \right)  + {m_t}^4\,\left( 6\,t^2 + 16\,t\,u + 6\,u^2 \right)  +
  t\,u\,\left( 7\,t^2 + 8\,t\,u + 7\,u^2 \right)\nonumber\\
  &  - {m_t}^2\,\left( 6\,t^3 + 17\,t^2\,u + 17\,t\,u^2 + 6\,u^3 \right)\nonumber\\
F_3 &= -16\,\left( 3\,{m_t}^4\,\left( t + u \right)  - 2\,{m_t}^2\,\left( 3\,t^2 + 4\,t\,u + 3\,u^2 \right)  +
    2\,\left( 3\,t^3 + 4\,t^2\,u + 4\,t\,u^2 + 3\,u^3 \right)  \right) \,v^2 \nonumber\\
F_4 &= -2\,\left( 3\,{m_t}^6\,\left( t + u \right)  - 2\,{m_t}^4\,\left( 3\,t^2 + 4\,t\,u + 3\,u^2 \right)  +
    t\,u\,\left( 7\,t^2 + 8\,t\,u + 7\,u^2 \right)\right. \nonumber\\
    & \left. + {m_t}^2\,\left( 6\,t^3 + t^2\,u + t\,u^2 + 6\,u^3 \right)  \right) \nonumber\\
F_5 &= -8\,m_t\,\left( 6\,t^3 + 7\,t^2\,u + 7\,t\,u^2 + 6\,u^3 + 3\,{m_t}^4\,\left( t + u \right)  -
    2\,{m_t}^2\,\left( 3\,t^2 + 2\,t\,u + 3\,u^2 \right)  \right) \,v \nonumber\\
F_6 &= 24\,m_t\,\left( t + u \right) \,\left( {m_t}^4 + 2\,t^2 + t\,u + 2\,u^2 - 2\,{m_t}^2\,\left( t + u \right)  \right) \,v
\end{align}
where $s, t,$ and $u$ are mandelstan variable and the $m_t$ the top quark mass.

{\bf Process (2)}

\begin{align}
\frac{d\sigma^S}{dt} & = \frac{g_3^2}{72\,\pi\, t\,s^2\,\Lambda^4}\biggl\{F_1\times|\alpha^S_{ut}|^2+\,F_2\times|\alpha^S_{tu}|^2\, +\,F_3\times\Bigl[|\beta^S_{ut}|^2\,+\,|\beta^S_{tu}|^2\Bigr]\nonumber\\
& +\,F_4 \times Re(\alpha^S_{ut}\, \alpha^S_{tu})\,+\,F_5 \times Im(\alpha^S_{ut}\, \beta^{S}_{tu})\,+\,F_6 \times Im(\alpha^S_{tu}\, \beta^{S*}_{tu}) \biggr\}
\end{align}
with coefficients
\begin{align}
F_1 &= -\left( \left( {m_t}^2 - t \right) \,
    \left( 4\,{m_t}^4 + s^2 + u^2 - 4\,{m_t}^2\,\left( s + u \right)  \right)  \right)\nonumber\\
F_2 &= -\left( \left( {m_t}^2 - t \right) \,\left( s^2 + u^2 \right)  \right)\nonumber\\
F_3 &=  -16\,\left( -2\,s\,u + {m_t}^2\,\left( s + u \right)  \right) \,v^2\nonumber\\
F_4 &= -2\,\left( {m_t}^2 + t \right) \,\left( s^2 + u^2 \right) \nonumber\\
F_5 &=  -8\,m_t\,\left( {m_t}^4 - t^2 - 2\,{m_t}^2\,u + 2\,t\,u + 2\,u^2 \right) \,v\nonumber\\
F_6 &= 8\,m_t\,\left( s^2 + u^2 \right) \,v
\end{align}

{\bf Process (3)}

\begin{align}
\frac{d\sigma^S}{dt} & = \frac{g_3^2}{216\,\pi\, t\, s^3\,\Lambda^4}\biggl\{F_1\times|\alpha^S_{ut}|^2+\,F_2\times|\alpha^S_{tu}|^2\, +\,F_3\times\Bigl[|\beta^S_{ut}|^2\,+\,|\beta^S_{tu}|^2\Bigr]\nonumber\\
& +\,F_4 \times Re(\alpha^S_{ut}\, \alpha^S_{tu})\,+\,F_5 \times Im(\alpha^S_{ut}\, \beta^{S}_{tu})\,+\,F_6 \times Im(\alpha^S_{tu}\, \beta^{S*}_{tu}) \biggr\}
\end{align}
with coefficients
\begin{align}
F_1 &= 6\,{m_t}^4\,{\left( s - t \right) }^2 - 3\,{m_t}^6\,\left( s + t \right)  +
  s\,t\,\left( 11\,s^2 + 16\,s\,t + 11\,t^2 \right) \nonumber\\
  & -  {m_t}^2\,\left( 6\,s^3 + 5\,s^2\,t + 5\,s\,t^2 + 6\,t^3 \right)\nonumber\\
F_2 &= -3\,{m_t}^6\,\left( s + t \right)  + {m_t}^4\,\left( 6\,s^2 + 20\,s\,t + 6\,t^2 \right)  +
  s\,t\,\left( 11\,s^2 + 16\,s\,t + 11\,t^2 \right) \nonumber\\
  & -  {m_t}^2\,\left( 6\,s^3 + 25\,s^2\,t + 25\,s\,t^2 + 6\,t^3 \right)\nonumber\\
F_3 &=  -16\,\left( 3\,{m_t}^6 - 3\,{m_t}^4\,\left( 2\,t + 3\,u \right)  -
    2\,u\,\left( 7\,t^2 + 7\,t\,u + 3\,u^2 \right)\right. \nonumber\\
    & \left. + 2\,{m_t}^2\,\left( 3\,t^2 + 10\,t\,u + 6\,u^2 \right) \right) \,v^2\nonumber\\
F_4 &=  -2\,\left( 3\,{m_t}^6\,\left( s + t \right)  -
    2\,{m_t}^4\,\left( 3\,s^2 + 2\,s\,t + 3\,t^2 \right)  +
    s\,t\,\left( 11\,s^2 + 16\,s\,t + 11\,t^2 \right)\right. \nonumber\\
    & \left. + {m_t}^2\,\left( 6\,s^3 - 7\,s^2\,t - 7\,s\,t^2 + 6\,t^3 \right)  \right)\nonumber\\
F_5 &=  -8\,m_t\,\left( 3\,{m_t}^6 - 3\,{m_t}^4\,\left( t + 3\,u \right)  -
    u\,\left( 19\,t^2 + 19\,t\,u + 6\,u^2 \right)\right. \nonumber\\
    & \left. + {m_t}^2\,\left( 3\,t^2 + 22\,t\,u + 12\,u^2 \right) \right) \,v\nonumber\\
F_6 &= 24\,m_t\,\left( {m_t}^2 - u \right) \,
  \left( {m_t}^4 + 3\,t^2 + 3\,t\,u + 2\,u^2 - {m_t}^2\,\left( 3\,t + 2\,u \right)  \right) \,v
\end{align}
%

The electroweak (photon) FCNC contribution is given by

{\bf Process (1)}

\begin{align}
\frac{d\sigma^\gamma}{dt} & = \frac{e^2}{216\,\pi\, t\, u\,s^2\,\Lambda^4}\biggl\{G_1\times|\alpha^\gamma_{ut}|^2+\,G_2\times|\alpha^\gamma_{tu}|^2\, +\,G_3\times\Bigl[|\beta^\gamma_{ut}|^2\,+\,|\beta^\gamma_{tu}|^2\Bigr]\nonumber\\
& +\,G_4 \times Re(\alpha^\gamma_{ut}\, \alpha^\gamma_{tu})\,+\,G_5 \times Im(\alpha^\gamma_{ut}\, \beta^{\gamma}_{tu})\,+\,G_6 \times Im(\alpha^\gamma_{tu}\, \beta^{\gamma *}_{tu}) \biggr\}
\end{align}
with coefficients
\begin{align}
G_1 &= 6\,{m_t}^4\,{\left( t - u \right) }^2 - 3\,{m_t}^6\,\left( t + u \right)  + t\,u\,\left( 11\,t^2 + 16\,t\,u + 11\,u^2 \right) \nonumber\\
& -  {m_t}^2\,\left( 6\,t^3 + 5\,t^2\,u + 5\,t\,u^2 + 6\,u^3 \right) \nonumber\\
G_2 &= -3\,{m_t}^6\,\left( t + u \right)  + {m_t}^4\,\left( 6\,t^2 + 20\,t\,u + 6\,u^2 \right)  +
  t\,u\,\left( 11\,t^2 + 16\,t\,u + 11\,u^2 \right) \nonumber\\
  & - {m_t}^2\,\left( 6\,t^3 + 25\,t^2\,u + 25\,t\,u^2 + 6\,u^3 \right) \nonumber\\
G_3 &= -16\,\left( 6\,t^3 + 4\,t^2\,u + 4\,t\,u^2 + 6\,u^3 + 3\,{m_t}^4\,\left( t + u \right)  -
    2\,{m_t}^2\,\left( 3\,t^2 + 2\,t\,u + 3\,u^2 \right)  \right) \,v^2 \nonumber\\
G_4 &= -2\,\left( 3\,{m_t}^6\,\left( t + u \right)  - 2\,{m_t}^4\,\left( 3\,t^2 + 2\,t\,u + 3\,u^2 \right)  +
    t\,u\,\left( 11\,t^2 + 16\,t\,u + 11\,u^2 \right)\right. \nonumber\\
    & \left. + {m_t}^2\,\left( 6\,t^3 - 7\,t^2\,u - 7\,t\,u^2 + 6\,u^3 \right)  \right) \nonumber\\
G_5 &= -8\,m_t\,\left( 6\,t^3 - t^2\,u - t\,u^2 + 6\,u^3 + 3\,{m_t}^4\,\left( t + u \right)  +
    {m_t}^2\,\left( -6\,t^2 + 4\,t\,u - 6\,u^2 \right)  \right) \,v \nonumber\\
G_6 &= 24\,m_t\,\left( t + u \right) \,\left( {m_t}^4 + 2\,t^2 + t\,u + 2\,u^2 - 2\,{m_t}^2\,\left( t + u \right)  \right) \,v
\end{align}

{\bf Process (2)}

\begin{align}
\frac{d\sigma^\gamma}{dt} & = \frac{e^2}{72\,\pi\, t\, s^2\,\Lambda^4}\biggl\{G_1\times|\alpha^\gamma_{ut}|^2+\,G_2\times|\alpha^\gamma_{tu}|^2\, +\,G_3\times\Bigl[|\beta^\gamma_{ut}|^2\,+\,|\beta^\gamma_{tu}|^2\Bigr]\nonumber\\
& +\,G_4 \times Re(\alpha^\gamma_{ut}\, \alpha^\gamma_{tu})\,+\,G_5 \times Im(\alpha^\gamma_{ut}\, \beta^{\gamma}_{tu})\,+\,G_6 \times Im(\alpha^\gamma_{tu}\, \beta^{\gamma *}_{tu}) \biggr\}
\end{align}
with coefficients
\begin{align}
G_1 &= -\left( \left( {m_t}^2 - t \right) \,
    \left( 4\,{m_t}^4 + s^2 + u^2 - 4\,{m_t}^2\,\left( s + u \right)  \right)  \right)\nonumber\\
G_2 &= -\left( \left( {m_t}^2 - t \right) \,\left( s^2 + u^2 \right)  \right)\nonumber\\
G_3 &=  -16\,\left( -2\,s\,u + {m_t}^2\,\left( s + u \right)  \right) \,v^2\nonumber\\
G_4 &=  -2\,\left( {m_t}^2 + t \right) \,\left( s^2 + u^2 \right)\nonumber\\
G_5 &=  -8\,m_t\,\left( {m_t}^4 - t^2 - 2\,{m_t}^2\,u + 2\,t\,u + 2\,u^2 \right) \,v\nonumber\\
G_6 &= 8\,m_t\,\left( s^2 + u^2 \right) \,v
\end{align}

{\bf Process (3)}

\begin{align}
\frac{d\sigma^\gamma}{dt} & = \frac{e^2}{216\,\pi\, t\, s^3\,\Lambda^4}\biggl\{G_1\times|\alpha^\gamma_{ut}|^2+\,G_2\times|\alpha^\gamma_{tu}|^2\, +\,G_3\times\Bigl[|\beta^\gamma_{ut}|^2\,+\,|\beta^\gamma_{tu}|^2\Bigr]\nonumber\\
& +\,G_4 \times Re(\alpha^\gamma_{ut}\, \alpha^\gamma_{tu})\,+\,G_5 \times Im(\alpha^\gamma_{ut}\, \beta^{\gamma}_{tu})\,+\,G_6 \times Im(\alpha^\gamma_{tu}\, \beta^{\gamma *}_{tu}) \biggr\}
\end{align}
with coefficients
\begin{align}
G_1 &= -3\,{m_t}^6\,\left( s + t \right)  + 6\,{m_t}^4\,\left( s^2 + t^2 \right)  +
  s\,t\,\left( 7\,s^2 + 8\,s\,t + 7\,t^2 \right) \nonumber\\
  & -  {m_t}^2\,\left( 6\,s^3 + 13\,s^2\,t + 13\,s\,t^2 + 6\,t^3 \right)\nonumber\\
G_2 &= -3\,{m_t}^6\,\left( s + t \right)  + {m_t}^4\,\left( 6\,s^2 + 16\,s\,t + 6\,t^2 \right)  +
  s\,t\,\left( 7\,s^2 + 8\,s\,t + 7\,t^2 \right) \nonumber\\
  & -  {m_t}^2\,\left( 6\,s^3 + 17\,s^2\,t + 17\,s\,t^2 + 6\,t^3 \right)\nonumber\\
G_3 &= -16\,\left( 3\,{m_t}^6 - 3\,{m_t}^4\,\left( 2\,t + 3\,u \right)  -
    2\,u\,\left( 5\,t^2 + 5\,t\,u + 3\,u^2 \right)\right. \nonumber\\
    & \left. + 2\,{m_t}^2\,\left( 3\,t^2 + 8\,t\,u + 6\,u^2 \right) \right) \,v^2 \nonumber\\
G_4 &= -2\,\left( 3\,{m_t}^6\,\left( s + t \right)  -
    2\,{m_t}^4\,\left( 3\,s^2 + 4\,s\,t + 3\,t^2 \right)  + s\,t\,\left( 7\,s^2 + 8\,s\,t + 7\,t^2 \right)\right. \nonumber\\
    & \left. +  {m_t}^2\,\left( 6\,s^3 + s^2\,t + s\,t^2 + 6\,t^3 \right)  \right) \nonumber\\
G_5 &= -8\,m_t\,\left( 3\,{m_t}^6 - 3\,{m_t}^4\,\left( t + 3\,u \right)  -
    u\,\left( 11\,t^2 + 11\,t\,u + 6\,u^2 \right)\right. \nonumber\\
    & \left. + {m_t}^2\,\left( 3\,t^2 + 14\,t\,u + 12\,u^2 \right)  \right) \,v \nonumber\\
G_6 &= 24\,m_t\,\left( {m_t}^2 - u \right) \,
  \left( {m_t}^4 + 3\,t^2 + 3\,t\,u + 2\,u^2 - {m_t}^2\,\left( 3\,t + 2\,u \right)  \right) \,v
\end{align}
%


The electroweak (Z boson) FCNC contribution is given by the following expression

{\bf Process (1)}

\begin{align}
\frac{d\sigma^Z}{dt} & = \frac{e^2}{62208\,\pi\, C_w\,S_w\,(m_Z^2-t)^2\,  (m_Z^2-u)^2\, s^2\,\Lambda^4}\biggl\{H_1\,|\alpha^Z_{ut}|^2+\,H_2\,|\alpha^Z_{tu}|^2\, +\,H_3\,|\beta^Z_{ut}|^2\,\Bigr]\nonumber\\
& +\,H_4 \, |\beta^Z_{tu}|^2\,+\,H_5 \,\Bigl[|\eta|^2\,+\,|\bar\eta|^2\,-2\,Re(\eta\, \bar\eta^*)\Bigr] \,+\,H_6 \, |\theta|^2 +\,H_7 \, Re(\alpha^Z_{ut}\, \alpha^Z_{tu}) \nonumber\\
& + H_8 \, Im(\alpha^Z_{ut}\, \beta^Z_{tu}) \, + \, H_9 \, Re(\alpha^Z_{ut}\, \theta^*) \, + \, H_{10} \, Im(\alpha^Z_{tu}\, \beta^{Z*}_{tu})  \, + \, H_{11} \, Re(\alpha^Z_{tu}\, \theta) \nonumber\\
& + \, H_{12} \Bigl[ Re(\beta^Z_{ut}\, \eta^*) \,- \,Re(\beta^Z_{ut}\, \bar \eta^*) \Bigr]\, + \, H_{13} \, Im(\beta^Z_{tu}\, \theta) \biggr\}
\end{align}
with coefficients
{\small
\begin{align}
H_1 &= -192\,{m_t}^4\,{S_w}^4\,{( t - u ) }^2\,( {m_Z}^4 - 2\,t\,u )  +
  t^2\,u^2\,( ( 243 - 432\,{S_w}^2 + 704\,{S_w}^4 ) \,t^2 )) + 1024\,{S_w}^4\,t\,u\nonumber\\
    &+  ( 243 - 432\,{S_w}^2 + 704\,{S_w}^4 ) \,u^2 )  -  192\,{m_t}^6\,{S_w}^4\,( -4\,{m_Z}^2\,t\,u + {m_Z}^4\,( t + u )  +  t\,u\,( t + u )  ) \nonumber\\ &
 - 2\,{m_Z}^2\,t\,u\,( 243\,t\,u\,( t + u )  - 432\,{S_w}^2\,t\,u\,( t + u )  + 64\,{S_w}^4\,( 4\,t^3 + 15\,t^2\,u + 15\,t\,u^2 + 4\,u^3 )  ) \nonumber\\
& +  2\,{m_Z}^4\,( 243\,t^2\,u^2 - 432\,{S_w}^2\,t^2\,u^2 + 32\,{S_w}^4\,( 3\,t^4 + 8\,t^3\,u + 16\,t^2\,u^2 + 8\,t\,u^3 + 3\,u^4 )  ) \nonumber\\
& + {m_t}^2\,( 2\,{m_Z}^2\,t\,u\,( ( 243 - 432\,{S_w}^2 + 64\,{S_w}^4 ) \,t^2 + 8\,( 243 - 432\,{S_w}^2 + 160\,{S_w}^4 ) \,t\,u \nonumber\\
& + ( 243 - 432\,{S_w}^2 + 64\,{S_w}^4 ) \,u^2 )  - t\,u\,( ( 243 - 432\,{S_w}^2 + 384\,{S_w}^4 ) \,t^3 \nonumber\\
& +4\,( 243 - 432\,{S_w}^2 + 80\,{S_w}^4 ) \,t^2\,u + 4\,( 243 - 432\,{S_w}^2 + 80\,{S_w}^4 ) \,t\,u^2 \nonumber\\
& + 3\,( 81 - 144\,{S_w}^2 + 128\,{S_w}^4 ) \,u^3 )  + {m_Z}^4\,( -1215\,t\,u\,( t + u )  + 2160\,{S_w}^2\,t\,u\,( t + u ) \nonumber\\
& + 64\,{S_w}^4\,( 3\,t^3 - 14\,t^2\,u - 14\,t\,u^2 + 3\,u^3 )  )  ) \nonumber\\
H_2 &= t^2\,u^2\,( ( 243 - 432\,{S_w}^2 + 704\,{S_w}^4 ) \,t^2 + 1024\,{S_w}^4\,t\,u + ( 243 - 432\,{S_w}^2 + 704\,{S_w}^4 ) \,u^2 ) \nonumber\\
& -  192\,{m_t}^6\,{S_w}^4\,u\,( {m_Z}^4 - 2\,{m_Z}^2\,t + t\,( t + u )  )  + 2\,{m_Z}^2\,t\,u\,( 256\,s^3\,{S_w}^4 - 192\,s^2\,{S_w}^4\,u - 192\,s\,{S_w}^4\,u^2 \nonumber\\
& +  27\,( -9 + 16\,{S_w}^2 ) \,t\,u\,( t + u )  )  +  64\,{m_t}^4\,{S_w}^4\,u\,( 3\,{m_Z}^4\,( 2\,s + 3\,u )  -     6\,{m_Z}^2\,t\,( 2\,s + 3\,u ) \nonumber\\
& + 2\,t\,( 3\,t^2 + 10\,t\,u + 3\,u^2 )  )  +  2\,{m_Z}^4\,( 96\,s^4\,{S_w}^4 + 128\,s^3\,{S_w}^4\,u + 320\,s^2\,{S_w}^4\,u^2 +     384\,s\,{S_w}^4\,u^3 \nonumber\\
& + 3\,u^2\,( ( 81 - 144\,{S_w}^2 ) \,t^2 + 64\,{S_w}^4\,u^2 )  )  -   {m_t}^2\,( t\,u\,( ( 243 - 432\,{S_w}^2 + 384\,{S_w}^4 ) \,t^3 +        1600\,{S_w}^4\,t^2\,u  \nonumber\\
& + 1600\,{S_w}^4\,t\,u^2 +        3\,( 81 - 144\,{S_w}^2 + 128\,{S_w}^4 ) \,u^3 )  +     {m_Z}^4\,( 192\,s^3\,{S_w}^4 + 448\,s^2\,{S_w}^4\,u + 960\,s\,{S_w}^4\,u^2 \nonumber\\
& + 3\,u\,( ( 81 - 144\,{S_w}^2 ) \,t^2 + 9\,( 9 - 16\,{S_w}^2 ) \,t\,u  +  256\,{S_w}^4\,u^2 )  )  - 2\,{m_Z}^2\,t\,u\, ( 320\,s^2\,{S_w}^4 + 768\,s\,{S_w}^4\,u \nonumber\\
& + 9\,( ( 27 - 48\,{S_w}^2 ) \,t^2 + ( 27 - 48\,{S_w}^2 + 64\,{S_w}^4 ) \,u^2 ) )  ) \nonumber
\end{align}
}

{\small
\begin{align}
H_3 &= -16\,( 192\,{m_t}^4\,{S_w}^4\,( -2\,{m_Z}^2\,t\,u + {m_Z}^4\,( t + u )   +       t\,u\,( t + u )  ) \nonumber\\
& + t\,u\,( -4\,{m_Z}^2\,s^2\,   ( 81 - 144\,{S_w}^2 + 160\,{S_w}^4 )  +   9\,s\,( -9 + 16\,{S_w}^2 ) \,( 3\,t^2 - 2\,t\,u + 3\,u^2 ) \nonumber\\
& +  128\,{S_w}^4\,( 3\,t^3 + 2\,t^2\,u + 2\,t\,u^2 + 3\,u^3 )  +   4\,{m_Z}^4\,( 9\,s\,( -9 + 16\,{S_w}^2 )  + 160\,{S_w}^4\,( t + u )  )  )  \nonumber\\
&  - {m_t}^2\,( -8\,{m_Z}^2\,s\,( -81 + 144\,{S_w}^2 + 32\,{S_w}^4 ) \,t\,u +   t\,u\,( 27\,s\,( -9 + 16\,{S_w}^2 ) \,( t + u ) \nonumber\\
& +  128\,{S_w}^4\,( 3\,t^2 + 2\,t\,u + 3\,u^2 )  )  + {m_Z}^4\,( 27\,s\,( -9 + 16\,{S_w}^2 ) \,( t + u ) \nonumber\\
& +  64\,{S_w}^4\,( 3\,t^2 + 10\,t\,u + 3\,u^2 )  )  )  ) \,v^2 \nonumber\\
H_4 &= -16\,( 192\,{m_t}^4\,{S_w}^4\,( -2\,{m_Z}^2\,t\,u + {m_Z}^4\,( t + u )  +  t\,u\,( t + u )  ) \nonumber\\
& - 64\,{m_t}^2\,{S_w}^4\, ( -4\,{m_Z}^2\,s\,t\,u + 2\,t\,u\,( 3\,t^2 + 2\,t\,u + 3\,u^2 )  +  {m_Z}^4\,( 3\,t^2 + 10\,t\,u + 3\,u^2 )  ) \nonumber\\
& + t\,u\,( ( 243 - 432\,{S_w}^2 + 384\,{S_w}^4 ) \,t^3 +    ( 243 - 432\,{S_w}^2 + 256\,{S_w}^4 ) \,t^2\,u \nonumber\\
& +  ( 243 - 432\,{S_w}^2 + 256\,{S_w}^4 ) \,t\,u^2 +     3\,( 81 - 144\,{S_w}^2 + 128\,{S_w}^4 ) \,u^3 \nonumber\\
& +  2\,{m_Z}^4\,( 243 - 432\,{S_w}^2 + 320\,{S_w}^4 ) \,( t + u )  -   2\,{m_Z}^2\,( 320\,s^2\,{S_w}^4 - 27\,( -9 + 16\,{S_w}^2 ) \,{( t + u ) }^2 )       )  ) \,v^2\nonumber\\
H_5 &= -( s\,( t\,u\,( -4\,{m_Z}^4\,( 81 - 144\,{S_w}^2 + 160\,{S_w}^4 )  +  ( -243 + 432\,{S_w}^2 - 384\,{S_w}^4 ) \,t^2 \nonumber\\
& + 2\,{( 9 - 8\,{S_w}^2 ) }^2\,t\,u - 3\,( 81 - 144\,{S_w}^2 + 128\,{S_w}^4 ) \,u^2 + 4\,{m_Z}^2\,( 81 - 144\,{S_w}^2 + 160\,{S_w}^4 ) \,( t + u )  ) \nonumber\\
& + 3\,{m_t}^2\,( 81 - 144\,{S_w}^2 + 128\,{S_w}^4 ) \, ( t^3 + u^3 + {m_Z}^4\,( t + u )  - 2\,{m_Z}^2\,( t^2 + u^2 )  )  ) \,v^2 ) \nonumber\\
H_6 &=-4\,( -243\,t^4 + 432\,{S_w}^2\,t^4 - 384\,{S_w}^4\,t^4 - 512\,{S_w}^4\,t^3\,u - 640\,{S_w}^4\,t^2\,u^2 - 512\,{S_w}^4\,t\,u^3\nonumber\\
& - 243\,u^4 + 432\,{S_w}^2\,u^4 - 384\,{S_w}^4\,u^4 - {m_Z}^4\,( ( 243 - 432\,{S_w}^2 + 704\,{S_w}^4 ) \,t^2 + 1024\,{S_w}^4\,t\,u \nonumber\\
& + ( 243 - 432\,{S_w}^2 + 704\,{S_w}^4 ) \,u^2 )  + 2\,{m_Z}^2\,( ( 243 - 432\,{S_w}^2 + 448\,{S_w}^4 ) \,t^3 + 768\,{S_w}^4\,t^2\,u \nonumber\\
& +  768\,{S_w}^4\,t\,u^2 + ( 243 - 432\,{S_w}^2 + 448\,{S_w}^4 ) \,u^3 )  + {m_t}^2\,( ( 243 - 432\,{S_w}^2 + 384\,{S_w}^4 ) \,t^3 \nonumber\\
& + 320\,{S_w}^4\,t^2\,u +  320\,{S_w}^4\,t\,u^2 + 3\,( 81 - 144\,{S_w}^2 + 128\,{S_w}^4 ) \,u^3 \nonumber\\
& + {m_Z}^4\,( 243 - 432\,{S_w}^2 + 704\,{S_w}^4 ) \,( t + u )  -  2\,{m_Z}^2\,( ( 243 - 432\,{S_w}^2 + 448\,{S_w}^4 ) \,t^2 \nonumber\\
& + 512\,{S_w}^4\,t\,u + ( 243 - 432\,{S_w}^2 + 448\,{S_w}^4 ) \,u^2 )  )  ) \,v^4 \nonumber\\
H_7 &= -2\,( t^2\,u^2\,( ( 243 - 432\,{S_w}^2 + 704\,{S_w}^4 ) \,t^2 + 1024\,{S_w}^4\,t\,u +
       ( 243 - 432\,{S_w}^2 + 704\,{S_w}^4 ) \,u^2 ) \nonumber\\
&  +    192\,{m_t}^6\,{S_w}^4\,u\,( {m_Z}^4 - 2\,{m_Z}^2\,t + t\,( t + u )  )  +    2\,{m_Z}^2\,t\,u\,( 256\,s^3\,{S_w}^4 - 192\,s^2\,{S_w}^4\,u \nonumber\\
& - 192\,s\,{S_w}^4\,u^2 +       27\,( -9 + 16\,{S_w}^2 ) \,t\,u\,( t + u )  )  +    64\,{m_t}^4\,{S_w}^4\,( 6\,{m_Z}^2\,t\,u\,( 2\,s + u ) \nonumber\\
& +       3\,{m_Z}^4\,( 2\,s^2 - 2\,s\,u + u^2 )  - 2\,t\,u\,( 3\,t^2 + 2\,t\,u + 3\,u^2 )  )  +    2\,{m_Z}^4\,( 96\,s^4\,{S_w}^4 + 128\,s^3\,{S_w}^4\,u \nonumber\\
& + 320\,s^2\,{S_w}^4\,u^2 +       384\,s\,{S_w}^4\,u^3 + 3\,u^2\,( ( 81 - 144\,{S_w}^2 ) \,t^2 + 64\,{S_w}^4\,u^2 )  )    \nonumber\\
& - {m_t}^2\,( 2\,{m_Z}^2\,t\,u\,        ( 832\,s^2\,{S_w}^4 + ( 243 - 432\,{S_w}^2 ) \,t^2 +          3\,{( 9 - 8\,{S_w}^2 ) }^2\,u^2 ) \nonumber\\
& +       t\,u\,( ( -243 + 432\,{S_w}^2 - 384\,{S_w}^4 ) \,t^3 + 448\,{S_w}^4\,t^2\,u +          448\,{S_w}^4\,t\,u^2 \nonumber\\
& - 3\,( 81 - 144\,{S_w}^2 + 128\,{S_w}^4 ) \,u^3 )  +       {m_Z}^4\,( 576\,s^3\,{S_w}^4 + 64\,s^2\,{S_w}^4\,u + 576\,s\,{S_w}^4\,u^2 \nonumber\\
& +          3\,u\,( 9\,( -9 + 16\,{S_w}^2 ) \,t^2 + 9\,( -9 + 16\,{S_w}^2 ) \,t\,u +             256\,{S_w}^4\,u^2 )  )  )  ) \nonumber
\end{align}
}

{\small
\begin{align}
H_8 &= -8\,m_t\,( -2\,{m_Z}^2\,t\,u\,( ( 243 - 432\,{S_w}^2 + 64\,{S_w}^4 ) \,t^2 +       4\,( 243 - 432\,{S_w}^2 + 128\,{S_w}^4 ) \,t\,u \nonumber\\
& +       ( 243 - 432\,{S_w}^2 + 64\,{S_w}^4 ) \,u^2 )  -    t\,u\,( ( -243 + 432\,{S_w}^2 - 384\,{S_w}^4 ) \,t^3 \nonumber\\
& +       2\,( -243 + 432\,{S_w}^2 + 32\,{S_w}^4 ) \,t^2\,u +       2\,( -243 + 432\,{S_w}^2 + 32\,{S_w}^4 ) \,t\,u^2 \nonumber\\
& -       3\,( 81 - 144\,{S_w}^2 + 128\,{S_w}^4 ) \,u^3 )  -    128\,{m_t}^2\,{S_w}^4\,t\,u\,( 4\,{m_Z}^4 + 3\,t^2 - 2\,t\,u + 3\,u^2 -       4\,{m_Z}^2\,( t + u )  )  \nonumber\\
& +    192\,{m_t}^4\,{S_w}^4\,( -4\,{m_Z}^2\,t\,u + {m_Z}^4\,( t + u )  +       t\,u\,( t + u )  ) \nonumber\\
& - {m_Z}^4\,     ( -729\,t\,u\,( t + u )  + 1296\,{S_w}^2\,t\,u\,( t + u )  +       64\,{S_w}^4\,( 3\,t^3 - 8\,t^2\,u - 8\,t\,u^2 + 3\,u^3 )  )  ) \,v \nonumber\\
H_9 &= -4\,( -( t\,u\,( ( 243 - 432\,{S_w}^2 + 448\,{S_w}^4 ) \,t^3 + 768\,{S_w}^4\,t^2\,u +
         768\,{S_w}^4\,t\,u^2 \nonumber\\
& + ( 243 - 432\,{S_w}^2 + 448\,{S_w}^4 ) \,u^3 )  )  +    3\,{m_t}^2\,t\,u\,( {m_Z}^4\,( 243 - 432\,{S_w}^2 + 256\,{S_w}^4 ) \nonumber\\
& -       4\,{m_Z}^2\,{( 9 - 8\,{S_w}^2 ) }^2\,( t + u )  +       2\,{( 9 - 8\,{S_w}^2 ) }^2\,( t^2 + u^2 )  )  +    64\,{m_t}^4\,{S_w}^4\,( 4\,{m_Z}^4\,( t + u ) \nonumber\\
& + 4\,t\,u\,( t + u )  -       {m_Z}^2\,( t^2 + 14\,t\,u + u^2 )  )  -    {m_Z}^4\,( -243\,s\,t\,u + 432\,s\,{S_w}^2\,t\,u \nonumber\\
& +       64\,{S_w}^4\,( 4\,t^3 + 15\,t^2\,u + 15\,t\,u^2 + 4\,u^3 )  )  +    2\,{m_Z}^2\,( 243\,t\,u\,( t^2 + u^2 )  - 432\,{S_w}^2\,t\,u\,( t^2 + u^2 )\nonumber\\
&  +       32\,{S_w}^4\,( t^4 + 22\,t^3\,u + 30\,t^2\,u^2 + 22\,t\,u^3 + u^4 )  )  ) \,v^2 \nonumber\\
H_{10} &= 24\,m_t\,( -2\,{m_Z}^2\,t\,u\,( 128\,{m_t}^4\,{S_w}^4 +       3\,( 27 - 48\,{S_w}^2 + 64\,{S_w}^4 ) \,t^2 + 256\,{S_w}^4\,t\,u \nonumber\\
& +       3\,( 27 - 48\,{S_w}^2 + 64\,{S_w}^4 ) \,u^2 - 256\,{m_t}^2\,{S_w}^4\,( t + u )       )  + t\,u\,( ( 81 - 144\,{S_w}^2 + 128\,{S_w}^4 ) \,t^3 \nonumber\\
& + 192\,{S_w}^4\,t^2\,u  +       192\,{S_w}^4\,t\,u^2 + ( 81 - 144\,{S_w}^2 + 128\,{S_w}^4 ) \,u^3 +       64\,{m_t}^4\,{S_w}^4\,( t + u )  \nonumber\\
& - 128\,{m_t}^2\,{S_w}^4\,{( t + u ) }^2 )       \nonumber\\
& + {m_Z}^4\,( t + u ) \,( 64\,s^2\,{S_w}^4 - 64\,s\,{S_w}^4\,u +       u\,( 64\,{m_t}^2\,{S_w}^4 + 81\,t - 144\,{S_w}^2\,t - 64\,{S_w}^4\,u )  )  ) \,  v \nonumber\\
H_{11} &= -4\,( t\,u\,( ( 243 - 432\,{S_w}^2 + 448\,{S_w}^4 ) \,t^3 + 768\,{S_w}^4\,t^2\,u +       768\,{S_w}^4\,t\,u^2 \nonumber\\
& + ( 243 - 432\,{S_w}^2 + 448\,{S_w}^4 ) \,u^3 +       256\,{m_t}^4\,{S_w}^4\,( t + u )  - 512\,{m_t}^2\,{S_w}^4\,{( t + u ) }^2       ) \nonumber\\
& + {m_Z}^4\,( t + u ) \,( 256\,s^2\,{S_w}^4 - 192\,s\,{S_w}^4\,u +       3\,u\,( 64\,{m_t}^2\,{S_w}^4 + 81\,t - 144\,{S_w}^2\,t - 64\,{S_w}^4\,u )  ) \nonumber\\
& -    2\,{m_Z}^2\,( 243\,t\,u\,( t^2 + u^2 )  - 432\,{S_w}^2\,t\,u\,( t^2 + u^2 )  +       32\,{m_t}^4\,{S_w}^4\,( t^2 + 14\,t\,u + u^2 ) \nonumber\\
& -       64\,{m_t}^2\,{S_w}^4\,( t^3 + 15\,t^2\,u + 15\,t\,u^2 + u^3 )  +       32\,{S_w}^4\,( t^4 + 22\,t^3\,u + 30\,t^2\,u^2 + 22\,t\,u^3 + u^4 )  )  ) \,v^2 \nonumber\\
H_{12} &= -8\,s\,( {m_t}^2\,{( 9 - 8\,{S_w}^2 ) }^2\,( {m_Z}^2 - t ) \,     ( {m_Z}^2 - u ) \,( t + u )  +    t\,u\,( 4\,{m_Z}^4\,( 81 - 144\,{S_w}^2 + 160\,{S_w}^4 ) \nonumber\\
& +       ( 243 - 432\,{S_w}^2 + 384\,{S_w}^4 ) \,t^2 - 2\,{( 9 - 8\,{S_w}^2 ) }^2\,t\,u +       3\,( 81 - 144\,{S_w}^2 + 128\,{S_w}^4 ) \,u^2 \nonumber\\
& -       4\,{m_Z}^2\,( 81 - 144\,{S_w}^2 + 160\,{S_w}^4 ) \,( t + u )  )  ) \,v^2 \nonumber\\
H_{13} &= -16\,m_t\,( -( t\,u\,( ( -243 + 432\,{S_w}^2 + 64\,{S_w}^4 ) \,t^2 +         512\,{S_w}^4\,t\,u \nonumber\\
& + ( -243 + 432\,{S_w}^2 + 64\,{S_w}^4 ) \,u^2 )  )  +    64\,{m_t}^2\,{S_w}^4\,( 4\,{m_Z}^4\,( t + u )  + 4\,t\,u\,( t + u ) \nonumber\\
& -       {m_Z}^2\,( t^2 + 14\,t\,u + u^2 )  )  -    2\,{m_Z}^4\,( -243\,t\,u + 432\,{S_w}^2\,t\,u + 64\,{S_w}^4\,( 2\,t^2 + t\,u + 2\,u^2 )  ) \nonumber\\
&  +    2\,{m_Z}^2\,( -243\,t\,u\,( t + u )  + 432\,{S_w}^2\,t\,u\,( t + u )  +       32\,{S_w}^4\,( t^3 + 9\,t^2\,u + 9\,t\,u^2 + u^3 )  )  ) \,v^3
\end{align}
}

{\bf Process (2)}

\begin{align}
\frac{d\sigma^Z}{dt} & = \frac{e^2}{20736\,\pi\, C_w\,S_w\,(m_Z^2-t)^2\, s^2\,\Lambda^4}\biggl\{H_1\,|\alpha^Z_{ut}|^2+\,H_2\,|\alpha^Z_{tu}|^2\, +\,H_3\,|\beta^Z_{ut}|^2\,\Bigr]\nonumber\\
& +\,H_4 \, |\beta^Z_{tu}|^2\,+\,H_5 \,\Bigl[|\eta|^2\,+\,|\bar\eta|^2\,-2\,Re(\eta\, \bar\eta^*)\Bigr] \,+\,H_6 \, |\theta|^2 +\,H_7 \, Re(\alpha^Z_{ut}\, \alpha^Z_{tu}) \nonumber\\
& + H_8 \, Im(\alpha^Z_{ut}\, \beta^Z_{tu}) \, + \, H_9 \, Re(\alpha^Z_{ut}\, \theta^*) \, + \, H_{10} \, Im(\alpha^Z_{tu}\, \beta^{Z*}_{tu})  \, + \, H_{11} \, Re(\alpha^Z_{tu}\, \theta) \nonumber\\
& + \, H_{12} \Bigl[ Re(\beta^Z_{ut}\, \eta^*) \,- \,Re(\beta^Z_{ut}\, \bar \eta^*) \Bigr]\, + \, H_{13} \, Im(\beta^Z_{tu}\, \theta) \biggr\}
\end{align}
with coefficients
{\small
\begin{align}
H_1 &=  t\,( -256\,{m_t}^6\,{S_w}^4 +    256\,{m_t}^4\,{S_w}^4\,( s + t + u )  +    t\,( 64\,s^2\,{S_w}^4 + {( 9 - 8\,{S_w}^2 ) }^2\,u^2 ) \nonumber\\
& -    {m_t}^2\,( 64\,s^2\,{S_w}^4 + 256\,s\,{S_w}^4\,t +       {( 9 - 8\,{S_w}^2 ) }^2\,u\,( 4\,t + u )  )  )\nonumber\\
H_2 &=  -( ( {m_t}^2 - t ) \,t\,    ( 64\,s^2\,{S_w}^4 + {( 9 - 8\,{S_w}^2 ) }^2\,u^2 )  )\nonumber\\
H_3 &=  -16\,t\,( s\,( -81 + 144\,{S_w}^2 - 128\,{S_w}^4 ) \,u +    {m_t}^2\,( s\,{( 9 - 8\,{S_w}^2 ) }^2 + 64\,{S_w}^4\,u )    ) \,v^2\nonumber\\
H_4 &=  16\,t\,( -64\,{m_t}^2\,{S_w}^4\,( s + u )  +    u\,( 128\,s\,{S_w}^4 + 9\,( -9 + 16\,{S_w}^2 ) \,( t + u )       )  ) \,v^2\nonumber\\
H_5 &=  -( s\,( 81 - 144\,{S_w}^2 + 128\,{S_w}^4 ) \,    ( {m_t}^2 - t ) \,u\,v^2 )\nonumber\\
H_6 &=  -4\,( -64\,s^2\,{S_w}^4 - {( 9 - 8\,{S_w}^2 ) }^2\,u^2 +    {m_t}^2\,( 64\,s\,{S_w}^4 + {( 9 - 8\,{S_w}^2 ) }^2\,u )    ) \,v^4\nonumber\\
H_7 &=  -2\,t\,( {m_t}^2 + t ) \,( 64\,s^2\,{S_w}^4 +    {( 9 - 8\,{S_w}^2 ) }^2\,u^2 )\nonumber\\
H_8 &=  -8\,m_t\,t\,( 64\,{m_t}^4\,{S_w}^4 -    128\,{m_t}^2\,{S_w}^4\,u + 81\,u\,( 2\,t + u )  -    144\,{S_w}^2\,u\,( 2\,t + u ) \nonumber\\
& -    64\,{S_w}^4\,( t^2 - 2\,t\,u - 2\,u^2 )  ) \,v\nonumber\\
H_9 &= -4\,t\,( -64\,s^2\,{S_w}^4 - {( 9 - 8\,{S_w}^2 ) }^2\,u^2 +    {m_t}^2\,( 128\,s\,{S_w}^4 + 2\,{( 9 - 8\,{S_w}^2 ) }^2\,u       )  ) \,v^2 \nonumber\\
H_{10} &= 8\,m_t\,t\,( 64\,s^2\,{S_w}^4 + {( 9 - 8\,{S_w}^2 ) }^2\,u^2 )    \,v \nonumber\\
H_{11} &= -4\,t\,( 64\,s^2\,{S_w}^4 + {( 9 - 8\,{S_w}^2 ) }^2\,u^2 ) \,v^2 \nonumber\\
H_{12} &= -8\,s\,( 81 - 144\,{S_w}^2 + 128\,{S_w}^4 ) \,t\,u\,v^2\nonumber\\
H_{13} &= -16\,m_t\,t\,( 64\,{m_t}^2\,{S_w}^4 - 64\,{S_w}^4\,t + 81\,u -    144\,{S_w}^2\,u ) \,v^3
\end{align}
}

{\bf Process (3)}

\begin{align}
\frac{d\sigma^Z}{dt} & = \frac{e^2}{62208\,\pi\, C_w\,S_w\,(m_Z^2-t)^2\,  (m_Z^2-s)^2\, s^2\,\Lambda^4}\biggl\{H_1\,|\alpha^Z_{ut}|^2+\,H_2\,|\alpha^Z_{tu}|^2\, +\,H_3\,|\beta^Z_{ut}|^2\,\Bigr]\nonumber\\
& +\,H_4 \, |\beta^Z_{tu}|^2\,+\,H_5 \,\Bigl[|\eta|^2\,+\,|\bar\eta|^2\,-2\,Re(\eta\, \bar\eta^*)\Bigr] \,+\,H_6 \, |\theta|^2 +\,H_7 \, Re(\alpha^Z_{ut}\, \alpha^Z_{tu}) \nonumber\\
& + H_8 \, Im(\alpha^Z_{ut}\, \beta^Z_{tu}) \, + \, H_9 \, Re(\alpha^Z_{ut}\, \theta^*) \, + \, H_{10} \, Im(\alpha^Z_{tu}\, \beta^{Z*}_{tu})  \, + \, H_{11} \, Re(\alpha^Z_{tu}\, \theta) \nonumber\\
& + \, H_{12} \Bigl[ Re(\beta^Z_{ut}\, \eta^*) \,- \,Re(\beta^Z_{ut}\, \bar \eta^*) \Bigr]\, + \, H_{13} \, Im(\beta^Z_{tu}\, \theta) \biggr\}
\end{align}
with coefficients
{\small
\begin{align}
H_1 &= s^2\,t^2\,( s^2\,( 243 - 432\,{S_w}^2 + 448\,{S_w}^4 )  +     512\,s\,{S_w}^4\,t + ( 243 - 432\,{S_w}^2 + 448\,{S_w}^4 ) \,t^2 )
 \nonumber\\
&  - 192\,{m_t}^6\,{S_w}^4\,   ( -2\,{m_Z}^2\,s\,t + s\,t\,( s + t )  +     {m_Z}^4\,( 5\,t + 4\,u )  )  +  2\,{m_Z}^2\,s\,t\,( 27\,s^2\,( -9 + 16\,{S_w}^2 ) \,t \nonumber\\
& +     27\,s\,( -9 + 16\,{S_w}^2 ) \,t^2 +     64\,{S_w}^4\,u\,( -3\,t^2 - 3\,t\,u + 2\,u^2 )  )  +  2\,{m_Z}^4\,( 27\,s^2\,( 9 - 16\,{S_w}^2 ) \,t^2 \nonumber\\
& +     32\,{S_w}^4\,( 6\,t^4 + 12\,t^3\,u + 14\,t^2\,u^2 + 8\,t\,u^3 + 3\,u^4 )  ) \nonumber\\
& +  64\,{m_t}^4\,{S_w}^4\,( 6\,s^3\,t +     {m_Z}^4\,( 21\,t^2 + 62\,t\,u + 24\,u^2 )  +     2\,s\,t\,( 3\,t^2 + {m_Z}^2\,( 15\,t + 2\,u )  )  ) \nonumber\\
& +  {m_t}^2\,( -( s\,t\,( s^3\,           ( 243 - 432\,{S_w}^2 + 384\,{S_w}^4 )  +          4\,s^2\,( 243 - 432\,{S_w}^2 + 208\,{S_w}^4 ) \,t \nonumber\\
& +          4\,s\,( 243 - 432\,{S_w}^2 + 208\,{S_w}^4 ) \,t^2 +          3\,( 81 - 144\,{S_w}^2 + 128\,{S_w}^4 ) \,t^3 )  ) \nonumber\\
& -     2\,{m_Z}^2\,s\,t\,( 27\,s^2\,( -9 + 16\,{S_w}^2 )  +        216\,s\,( -9 + 16\,{S_w}^2 ) \,t +        3\,( -81 + 144\,{S_w}^2 + 320\,{S_w}^4 ) \,t^2 \nonumber\\
& + 768\,{S_w}^4\,t\,u +        64\,{S_w}^4\,u^2 )  + {m_Z}^4\,      ( 135\,s^2\,( -9 + 16\,{S_w}^2 ) \,t +        135\,s\,( -9 + 16\,{S_w}^2 ) \,t^2 \nonumber\\
& -        64\,{S_w}^4\,( 12\,t^3 + 59\,t^2\,u + 55\,t\,u^2 + 15\,u^3 )  )  ) \nonumber\\
H_2 &= s^2\,t^2\,( s^2\,( 243 - 432\,{S_w}^2 + 448\,{S_w}^4 )  +     512\,s\,{S_w}^4\,t + ( 243 - 432\,{S_w}^2 + 448\,{S_w}^4 ) \,t^2 ) \nonumber\\
&  - 192\,{m_t}^6\,{S_w}^4\,t\,   ( {m_Z}^4 - 2\,{m_Z}^2\,s + s\,( s + t )  )  +  2\,{m_Z}^2\,s\,t\,( 27\,s^2\,( -9 + 16\,{S_w}^2 ) \,t \nonumber\\
& +     27\,s\,( -9 + 16\,{S_w}^2 ) \,t^2 +     64\,{S_w}^4\,u\,( -3\,t^2 - 3\,t\,u + 2\,u^2 )  )  +  2\,{m_Z}^4\,( 27\,s^2\,( 9 - 16\,{S_w}^2 ) \,t^2 \nonumber\\
& +     32\,{S_w}^4\,( 6\,t^4 + 12\,t^3\,u + 14\,t^2\,u^2 + 8\,t\,u^3 + 3\,u^4 )  )  +  64\,{m_t}^4\,{S_w}^4\,t\,   ( 6\,s^3 + 16\,s^2\,t \nonumber\\
& + 3\,{m_Z}^4\,( 3\,t + 2\,u )  +     6\,s\,( t^2 - {m_Z}^2\,( 3\,t + 2\,u )  )  )  +  {m_t}^2\,( -( s\,t\,( s^3\,           ( 243 - 432\,{S_w}^2 + 384\,{S_w}^4 ) \nonumber\\
& + 1088\,s^2\,{S_w}^4\,t +          1088\,s\,{S_w}^4\,t^2 + 3\,( 81 - 144\,{S_w}^2 + 128\,{S_w}^4 ) \,           t^3 )  )  + 2\,{m_Z}^2\,s\,t\,      ( s^2\,( 243 - 432\,{S_w}^2 ) \nonumber\\
& +        9\,( 27 - 48\,{S_w}^2 + 64\,{S_w}^4 ) \,t^2 + 768\,{S_w}^4\,t\,u +        448\,{S_w}^4\,u^2 )  + {m_Z}^4\,      ( 27\,s^2\,( -9 + 16\,{S_w}^2 ) \,t \nonumber\\
& +        27\,s\,( -9 + 16\,{S_w}^2 ) \,t^2 -        64\,{S_w}^4\,( 12\,t^3 + 15\,t^2\,u + 11\,t\,u^2 + 3\,u^3 )  )  ) \nonumber
\end{align}
}

{\small
\begin{align}
H_3 &= -16\,( 192\,{m_t}^6\,s\,{S_w}^4\,t +    192\,{m_t}^4\,{S_w}^4\,     ( -2\,{m_Z}^2\,s\,t + {m_Z}^4\,( 2\,t + u )  -       s\,t\,( 2\,t + 3\,u )  ) \nonumber\\
&  -    t\,u\,( 8\,{m_Z}^2\,s\,( 81 - 144\,{S_w}^2 + 112\,{S_w}^4 ) \,u +       s\,( 4\,( 81 - 144\,{S_w}^2 + 160\,{S_w}^4 ) \,t^2 \nonumber\\
& +          4\,( 81 - 144\,{S_w}^2 + 160\,{S_w}^4 ) \,t\,u +          3\,( 81 - 144\,{S_w}^2 + 128\,{S_w}^4 ) \,u^2 )  -       8\,{m_Z}^4\,( 9\,s\,( -9 + 16\,{S_w}^2 ) \nonumber\\
& +          112\,{S_w}^4\,( t + u )  )  )  +    {m_t}^2\,( 4\,{m_Z}^2\,s\,        ( -81 + 144\,{S_w}^2 + 128\,{S_w}^4 ) \,t\,u \nonumber\\
& -       {m_Z}^4\,( -243\,( s + t ) \,u + 432\,{S_w}^2\,( s + t ) \,u +          64\,{S_w}^4\,( 6\,t^2 + 20\,t\,u + 3\,u^2 )  ) \nonumber\\
&  +       s\,t\,( 81\,u\,( 4\,t + 3\,u )  - 144\,{S_w}^2\,u\,( 4\,t + 3\,u )  +          128\,{S_w}^4\,( 3\,t^2 + 8\,t\,u + 6\,u^2 )  )  )  ) \,v^2 \nonumber\\
H_4 &= 16\,( -192\,{m_t}^6\,s\,{S_w}^4\,t -    6\,{m_t}^4\,( -( {m_Z}^2\,s\,{( 9 - 8\,{S_w}^2 ) }^2\,t )           + 32\,{m_Z}^4\,{S_w}^4\,( 2\,t + u ) \nonumber\\
& -       32\,s\,{S_w}^4\,t\,( 2\,t + 3\,u )  )  +    t\,u\,( 2\,{m_Z}^2\,s\,( 243 - 432\,{S_w}^2 + 448\,{S_w}^4 ) \,u +       s\,( s^2\,( 243 - 432\,{S_w}^2 ) \nonumber\\
& +          ( 243 - 432\,{S_w}^2 + 640\,{S_w}^4 ) \,t^2 + 640\,{S_w}^4\,t\,u +          384\,{S_w}^4\,u^2 )  -       2\,{m_Z}^4\,( 27\,s\,( -9 + 16\,{S_w}^2 ) \nonumber\\
& +          448\,{S_w}^4\,( t + u )  )  )  +    {m_t}^2\,( -4\,{m_Z}^2\,s\,        ( 243 - 432\,{S_w}^2 + 128\,{S_w}^4 ) \,t\,u +       s\,t\,( 27\,s^2\,( -9 + 16\,{S_w}^2 ) \nonumber\\
& +          ( -243 + 432\,{S_w}^2 - 384\,{S_w}^4 ) \,t^2 -          1024\,{S_w}^4\,t\,u - 768\,{S_w}^4\,u^2 ) \nonumber\\
& +       2\,{m_Z}^4\,( 27\,s\,( -9 + 16\,{S_w}^2 ) \,t +          32\,{S_w}^4\,( 6\,t^2 + 20\,t\,u + 3\,u^2 )  )  )  ) \,v^2 \nonumber\\
H_5 &= -( ( -( s\,t\,( 8\,{m_Z}^4\,            ( 81 - 144\,{S_w}^2 + 112\,{S_w}^4 )  +           s^2\,( 243 - 432\,{S_w}^2 + 384\,{S_w}^4 )  +           2\,s\,{( 9 - 8\,{S_w}^2 ) }^2\,t \nonumber\\
& +           3\,( 81 - 144\,{S_w}^2 + 128\,{S_w}^4 ) \,t^2 -           8\,{m_Z}^2\,( 81 - 144\,{S_w}^2 + 112\,{S_w}^4 ) \,            ( s + t )  )  ) \nonumber\\
& +      3\,{m_t}^2\,( 81 - 144\,{S_w}^2 + 128\,{S_w}^4 ) \,       ( s^3 + t^3 + {m_Z}^4\,( s + t )  -         2\,{m_Z}^2\,( s^2 + t^2 )  )  ) \,u\,v^2 ) \nonumber\\
H_6 &= 4\,( 243\,s^4 - 432\,s^4\,{S_w}^2 + 384\,s^4\,{S_w}^4 + 256\,s^3\,{S_w}^4\,t +    128\,s^2\,{S_w}^4\,t^2 + 256\,s\,{S_w}^4\,t^3 \nonumber\\
& + 243\,t^4 - 432\,{S_w}^2\,t^4 +    384\,{S_w}^4\,t^4 - 2\,{m_Z}^2\,     ( s^3\,( 243 - 432\,{S_w}^2 + 320\,{S_w}^4 )  +       384\,s^2\,{S_w}^4\,t \nonumber\\
& + 384\,s\,{S_w}^4\,t^2 +       ( 243 - 432\,{S_w}^2 + 320\,{S_w}^4 ) \,t^3 )  +    {m_Z}^4\,( s^2\,( 243 - 432\,{S_w}^2 ) \nonumber\\
& +       ( 243 - 432\,{S_w}^2 + 384\,{S_w}^4 ) \,t^2 + 384\,{S_w}^4\,t\,u +       448\,{S_w}^4\,u^2 )  + {m_t}^2\,     ( s^3\,( -243 + 432\,{S_w}^2 - 384\,{S_w}^4 ) \nonumber\\
& -       64\,s^2\,{S_w}^4\,t - 64\,s\,{S_w}^4\,t^2 -       3\,( 81 - 144\,{S_w}^2 + 128\,{S_w}^4 ) \,t^3 +       2\,{m_Z}^2\,( s^2\,( 243 - 432\,{S_w}^2 + 320\,{S_w}^4 ) \nonumber\\
& +          256\,s\,{S_w}^4\,t + ( 243 - 432\,{S_w}^2 + 320\,{S_w}^4 ) \,t^2          )  + {m_Z}^4\,( 27\,s\,( -9 + 16\,{S_w}^2 ) \nonumber\\
& +          ( -243 + 432\,{S_w}^2 - 384\,{S_w}^4 ) \,t - 448\,{S_w}^4\,u )           )  ) \,v^4 \nonumber\\
H_7 &= -2\,( s^2\,t^2\,( s^2\,( 243 - 432\,{S_w}^2 + 448\,{S_w}^4 )  +       512\,s\,{S_w}^4\,t + ( 243 - 432\,{S_w}^2 + 448\,{S_w}^4 ) \,t^2       ) \nonumber\\
& + 192\,{m_t}^6\,{S_w}^4\,t\,     ( {m_Z}^4 - 2\,{m_Z}^2\,s + s\,( s + t )  )  +    2\,{m_Z}^2\,s\,t\,( 27\,s^2\,( -9 + 16\,{S_w}^2 ) \,t +       27\,s\,( -9 + 16\,{S_w}^2 ) \,t^2 \nonumber\\
& +       64\,{S_w}^4\,u\,( -3\,t^2 - 3\,t\,u + 2\,u^2 )  )  +    2\,{m_Z}^4\,( 27\,s^2\,( 9 - 16\,{S_w}^2 ) \,t^2 \nonumber\\
& +       32\,{S_w}^4\,( 6\,t^4 + 12\,t^3\,u + 14\,t^2\,u^2 + 8\,t\,u^3 + 3\,u^4 )  )  -    64\,{m_t}^4\,{S_w}^4\,( 6\,s^3\,t + 8\,s^2\,t^2 \nonumber\\
& -       3\,{m_Z}^4\,( t^2 - 2\,t\,u + 2\,u^2 )  +       6\,s\,t\,( t^2 - {m_Z}^2\,( t + 2\,u )  )  )  -    {m_t}^2\,( -( s\,t\,( s^3\,             ( 243 - 432\,{S_w}^2 + 384\,{S_w}^4 ) \nonumber\\
& + 64\,s^2\,{S_w}^4\,t +            64\,s\,{S_w}^4\,t^2 + 3\,( 81 - 144\,{S_w}^2 + 128\,{S_w}^4 ) \,             t^3 )  )  + 2\,{m_Z}^2\,s\,t\,        ( s^2\,( 243 - 432\,{S_w}^2 ) \nonumber\\
& +          3\,{( 9 - 8\,{S_w}^2 ) }^2\,t^2 + 704\,{S_w}^4\,u^2 )  +       {m_Z}^4\,( 27\,s^2\,( -9 + 16\,{S_w}^2 ) \,t +          27\,s\,( -9 + 16\,{S_w}^2 ) \,t^2 \nonumber\\
& +          64\,{S_w}^4\,( 12\,t^3 + 9\,t^2\,u + 5\,t\,u^2 + 9\,u^3 )  )  )  ) \nonumber
\end{align}
}

{\small
\begin{align}
H_8 &= -8\,m_t\,( 192\,{m_t}^6\,s\,{S_w}^4\,t +    s\,t\,u\,( 27\,s^2\,( -9 + 16\,{S_w}^2 )  +       27\,s\,( -9 + 16\,{S_w}^2 ) \,t \nonumber\\
& +       ( -243 + 432\,{S_w}^2 - 704\,{S_w}^4 ) \,t^2 - 704\,{S_w}^4\,t\,u -       384\,{S_w}^4\,u^2 )  + 2\,{m_Z}^2\,s\,t\,     ( 27\,s^2\,( -9 + 16\,{S_w}^2 ) \nonumber\\
& +       108\,s\,( -9 + 16\,{S_w}^2 ) \,t +       ( -243 + 432\,{S_w}^2 + 384\,{S_w}^4 ) \,t^2 + 384\,{S_w}^4\,t\,u -       320\,{S_w}^4\,u^2 ) \nonumber\\
& + {m_t}^2\,s\,t\,     ( s^2\,( 243 - 432\,{S_w}^2 )  +       {m_Z}^4\,( 729 - 1296\,{S_w}^2 - 1024\,{S_w}^4 )  +       27\,s\,( 9 - 16\,{S_w}^2 ) \,t + 243\,t^2 \nonumber\\
& - 432\,{S_w}^2\,t^2 +       192\,{S_w}^4\,t^2 - 256\,{m_Z}^2\,{S_w}^4\,( 3\,t - u )  +       896\,{S_w}^4\,t\,u + 768\,{S_w}^4\,u^2 ) \nonumber\\
& +
    192\,{m_t}^4\,{S_w}^4\,     ( -2\,{m_Z}^2\,s\,t + {m_Z}^4\,( s + t )  - s\,t\,( t + 3\,u )       ) \nonumber\\
& + {m_Z}^4\,( -192\,s^3\,{S_w}^4 + 1024\,s^2\,{S_w}^4\,t -       192\,{S_w}^4\,t^3 + s\,t\,( 1024\,{S_w}^4\,t - 729\,u + 1296\,{S_w}^2\,u          )  )  ) \,v \nonumber\\
H_9 &= -4\,( -( s\,t\,( s^3\,( 243 - 432\,{S_w}^2 + 320\,{S_w}^4 )  +         384\,s^2\,{S_w}^4\,t + 384\,s\,{S_w}^4\,t^2 \nonumber\\
& +         ( 243 - 432\,{S_w}^2 + 320\,{S_w}^4 ) \,t^3 )  )  -    2\,{m_Z}^2\,( 32\,s^4\,{S_w}^4 +       s^3\,( -243 + 432\,{S_w}^2 - 448\,{S_w}^4 ) \,t \nonumber\\
& -       576\,s^2\,{S_w}^4\,t^2 + s\,( -243 + 432\,{S_w}^2 - 448\,{S_w}^4 ) \,        t^3 + 32\,{S_w}^4\,t^4 ) \nonumber\\
& +    3\,{m_t}^2\,s\,t\,( {m_Z}^4\,        ( 243 - 432\,{S_w}^2 + 256\,{S_w}^4 )  -       4\,{m_Z}^2\,{( 9 - 8\,{S_w}^2 ) }^2\,( s + t )  +       2\,{( 9 - 8\,{S_w}^2 ) }^2\,( s^2 + t^2 )  ) \nonumber\\
& +    64\,{m_t}^4\,{S_w}^4\,( 2\,{m_Z}^4\,( s + t )  +       2\,s\,t\,( s + t )  + {m_Z}^2\,( s^2 - 10\,s\,t + t^2 )  ) \nonumber\\
& -    {m_Z}^4\,( 128\,s^3\,{S_w}^4 + 576\,s^2\,{S_w}^4\,t +       128\,{S_w}^4\,t^3 + 9\,s\,t\,( 64\,{S_w}^4\,t - 27\,u + 48\,{S_w}^2\,u          )  )  ) \,v^2 \nonumber\\
H_{10} &= 24\,m_t\,( 64\,{m_t}^6\,s\,{S_w}^4\,t +    2\,{m_Z}^2\,s\,t\,( 9\,s^2\,( -9 + 16\,{S_w}^2 )  +       ( -81 + 144\,{S_w}^2 - 128\,{S_w}^4 ) \,t^2 \nonumber\\
& - 128\,{S_w}^4\,t\,u -       192\,{S_w}^4\,u^2 )  + s\,t\,u\,     ( 9\,s^2\,( -9 + 16\,{S_w}^2 )  + 9\,s\,( 9 - 16\,{S_w}^2 ) \,t \nonumber\\
& -       3\,( 27 - 48\,{S_w}^2 + 64\,{S_w}^4 ) \,t^2 - 192\,{S_w}^4\,t\,u -       128\,{S_w}^4\,u^2 ) \nonumber\\
& + 64\,{m_t}^4\,{S_w}^4\,t\,     ( {m_Z}^4 - 2\,{m_Z}^2\,s - 3\,s\,( t + u )  )  +    {m_Z}^4\,u\,( 9\,s\,( -9 + 16\,{S_w}^2 ) \,t +       64\,{S_w}^4\,( t^2 + t\,u - u^2 )  ) \nonumber\\
& +    {m_t}^2\,( 256\,{m_Z}^2\,s\,{S_w}^4\,t\,( t + u )  +       s\,t\,( s^2\,( 81 - 144\,{S_w}^2 )  +          9\,s\,( -9 + 16\,{S_w}^2 ) \,t \nonumber\\
& +          3\,( 27 - 48\,{S_w}^2 + 64\,{S_w}^4 ) \,t^2 + 384\,{S_w}^4\,t\,u +          256\,{S_w}^4\,u^2 ) \nonumber\\
& +       {m_Z}^4\,( 9\,s\,( 9 - 16\,{S_w}^2 ) \,t -          64\,{S_w}^4\,( t^2 + 2\,t\,u - u^2 )  )  )  ) \,v \nonumber\\
H_{11} &= -4\,( 192\,{m_t}^6\,s\,{S_w}^4\,t +    2\,{m_Z}^2\,( 32\,s^4\,{S_w}^4 +       s^3\,( -243 + 432\,{S_w}^2 - 448\,{S_w}^4 ) \,t -       576\,s^2\,{S_w}^4\,t^2 \nonumber\\
& + s\,( -243 + 432\,{S_w}^2 - 448\,{S_w}^4 ) \,        t^3 + 32\,{S_w}^4\,t^4 )  +    s\,t\,u\,( 27\,s^2\,( -9 + 16\,{S_w}^2 ) \nonumber\\
& +       27\,s\,( 9 - 16\,{S_w}^2 ) \,t -       9\,( 27 - 48\,{S_w}^2 + 64\,{S_w}^4 ) \,t^2 - 576\,{S_w}^4\,t\,u -       320\,{S_w}^4\,u^2 ) \nonumber\\
& + 64\,{m_t}^4\,{S_w}^4\,     ( 3\,{m_Z}^4\,t + {m_Z}^2\,( s^2 - 10\,s\,t + t^2 )  -       9\,s\,t\,( t + u )  )  +    {m_Z}^4\,u\,( 27\,s\,( -9 + 16\,{S_w}^2 ) \,t \nonumber\\
& +       64\,{S_w}^4\,( 3\,t^2 + 3\,t\,u - 2\,u^2 )  )  +    {m_t}^2\,( -128\,{m_Z}^2\,{S_w}^4\,        ( s^3 - 9\,s^2\,t - 9\,s\,t^2 + t^3 ) \nonumber\\
& +       s\,t\,( s^2\,( 243 - 432\,{S_w}^2 )  +          27\,s\,( -9 + 16\,{S_w}^2 ) \,t +          9\,( 27 - 48\,{S_w}^2 + 64\,{S_w}^4 ) \,t^2 \nonumber\\
& + 1152\,{S_w}^4\,t\,u +          704\,{S_w}^4\,u^2 )  +       {m_Z}^4\,( 27\,s\,( 9 - 16\,{S_w}^2 ) \,t -          64\,{S_w}^4\,( 3\,t^2 + 6\,t\,u - 2\,u^2 )  )  )  ) \,v^2 \nonumber
\end{align}
}

{\small
\begin{align}
H_{12} &= 8\,u\,( -( {m_t}^4\,{m_Z}^2\,{( 9 - 8\,{S_w}^2 ) }^2\,       ( s + t )  )  - s\,t\,( 8\,{m_Z}^4\,        ( 81 - 144\,{S_w}^2 + 112\,{S_w}^4 ) \nonumber\\
& +       s^2\,( 243 - 432\,{S_w}^2 + 384\,{S_w}^4 )  +       2\,s\,{( 9 - 8\,{S_w}^2 ) }^2\,t +       3\,( 81 - 144\,{S_w}^2 + 128\,{S_w}^4 ) \,t^2 \nonumber\\
& +       8\,{m_Z}^2\,( 81 - 144\,{S_w}^2 + 112\,{S_w}^4 ) \,u )  +    {m_t}^2\,( {m_Z}^4\,{( 9 - 8\,{S_w}^2 ) }^2\,( s + t )  +       s\,{( 9 - 8\,{S_w}^2 ) }^2\,t\,( s + t ) \nonumber\\
& +       {m_Z}^2\,( {( 9 - 8\,{S_w}^2 ) }^2\,t\,u +          s\,( 8\,( 81 - 144\,{S_w}^2 + 112\,{S_w}^4 ) \,t +             {( 9 - 8\,{S_w}^2 ) }^2\,u )  )  )  ) \,v^2\nonumber\\
H_{13} &= 16\,m_t\,( s\,t\,( s^2\,( -243 + 432\,{S_w}^2 - 64\,{S_w}^4 )  +       256\,s\,{S_w}^4\,t + ( -243 + 432\,{S_w}^2 - 64\,{S_w}^4 ) \,t^2       ) \nonumber\\
& + 64\,{m_t}^4\,{m_Z}^2\,{S_w}^4\,( 6\,t - u )  +    2\,{m_Z}^4\,( 27\,s\,( -9 + 16\,{S_w}^2 ) \,t +       64\,{S_w}^4\,( 3\,t^2 + 3\,t\,u + u^2 )  ) \nonumber\\
& -    2\,{m_Z}^2\,u\,( 27\,s\,( 9 - 16\,{S_w}^2 ) \,t +       32\,{S_w}^4\,( 6\,t^2 + 6\,t\,u + u^2 )  )  -    2\,{m_t}^2\,( 64\,s\,{S_w}^4\,t\,( s + t ) \nonumber\\
& +       64\,{m_Z}^4\,{S_w}^4\,( 3\,t + u )  +       {m_Z}^2\,( 27\,s\,( -9 + 16\,{S_w}^2 ) \,t +          64\,{S_w}^4\,( 3\,t^2 - u^2 )  )  )  ) \,v^3
\end{align}
}
where $C_w$ and $S_w$ are the sine and cosine of Weinberg angle, $m_Z$ the mass of the Z boson.

The interference strong-electroweak (gluon+photon) FCNC contribution is given by the following expression

{\bf Process (1)}

\begin{align}
\frac{d\sigma^{S\,\gamma}}{dt} & = \frac{e\, g_3}{27\,\pi\, s^2\,\Lambda^4}\biggl\{s\,\left( -4\,{m_t}^2 + s \right)\,Re(\alpha^{S}_{ut}\, \alpha^{\gamma *}_{ut})\nonumber\\
& -s^2\Bigl[ Re(\alpha^{S}_{ut}\, \alpha^{\gamma }_{tu})\, + \, Re(\alpha^{S}_{tu}\, \alpha^{\gamma }_{ut})  \, - \, Re(\alpha^{S}_{tu}\, \alpha^{\gamma *}_{tu}) \Bigr]  -8\,m_t\,s\,v \Bigr[Im(\alpha^{S}_{ut}\, \beta^{\gamma}_{tu}) \, + \, Im(\beta^{S}_{tu}\, \alpha^{\gamma}_{ut}) \Bigr] \nonumber\\
& -16\,s\,v^2 \Bigl[ Re(\beta^{S}_{ut}\, \beta^{\gamma *}_{ut})\, + \, Re(\beta^{S}_{tu}\, \beta^{\gamma *}_{tu})  \Bigr]
\end{align}

{\bf Process (3)}

\begin{align}
\frac{d\sigma^{S\,\gamma}}{dt} & = \frac{e\, g_3}{27\,\pi\, s^2\,\Lambda^4}\biggl\{( 4\,{m_t}^2 - u) \,u \,Re(\alpha^{S}_{ut}\, \alpha^{\gamma *}_{ut})\nonumber\\
&+ u^2\Bigl[ Re(\alpha^{S}_{ut}\, \alpha^{\gamma }_{tu})\, + \, Re(\alpha^{S}_{tu}\, \alpha^{\gamma }_{ut})  \, - \, Re(\alpha^{S}_{tu}\, \alpha^{\gamma *}_{tu}) \Bigr] \, +\, 8\,m_t\,u\,v \Bigr[Im(\alpha^{S}_{ut}\, \beta^{\gamma}_{tu}) \, + \, Im(\beta^{S}_{tu}\, \alpha^{\gamma}_{ut}) \Bigr] \nonumber\\
&+ 16\,u\,v^2 \Bigl[ Re(\beta^{S}_{ut}\, \beta^{\gamma *}_{ut})\, + \, Re(\beta^{S}_{tu}\, \beta^{\gamma *}_{tu})  \Bigr]
\end{align}


The interference strong-electroweak (gluon+Z boson) FCNC contribution is given by

{\bf Process (1)}

\begin{align}
\frac{d\sigma^{S\, Z}}{dt} & = \frac{e \, g_3}{162\,\pi\, C_w\,(m_Z^2-t)^2\,  (m_Z^2-u)^2\, s^2\,\Lambda^4}\biggl\{FH_1\,Re(\alpha^S_{ut}\, \alpha^{Z *}_{ut}) \nonumber\\
&+ \,FH_2\Bigl[Re(\alpha^S_{ut}\, \alpha^{Z}_{tu})\, + \, Re(\alpha^S_{tu}\, \alpha^{Z}_{ut}) \, - \,Re(\alpha^S_{tu}\, \alpha^{Z *}_{tu})\Bigr] +\,FH_3\Bigl[ Im(\alpha^S_{ut}\, \beta^{Z}_{tu}) \, + \, Im(\beta^S_{tu}\, \alpha^{Z}_{ut})\Bigr]\nonumber\\
& +\,FH_4 \, Re(\alpha^S_{ut}\, \theta^*)\,+\,FH_5 \,Re(\alpha^S_{tu}\, \theta)\, + \,FH_6 \, Re(\beta^S_{ut}\, \beta^{Z *}_{ut}) +\,FH_7 \Bigl[Re(\beta^S_{ut}\, \eta^*) \, - \, Re(\beta^S_{ut}\, \bar\eta^*)\Bigr] \nonumber\\
& + FH_8 \, Re(\beta^S_{tu}\, \beta^{Z *}_{tu}) \, + \, FH_9 \, Im(\beta^S_{tu}\, \theta) \biggr\}
\end{align}
with coefficients
\begin{align}
FH_1 &= -2\,\left( 4\,{m_t}^2 - s \right) \,s\,S_w\,\left( -2\,t\,u + {m_Z}^2\,\left( t + u \right)  \right) \nonumber\\
FH_2 &= -2\,s^2\,S_w\,\left( -2\,t\,u + {m_Z}^2\,\left( t + u \right)  \right) \nonumber\\
FH_3 &= -16\,m_t\,s\,S_w\,\left( -2\,t\,u + {m_Z}^2\,\left( t + u \right)  \right) \,v \nonumber\\
FH_4 &= -4\,\left( 2\,{m_t}^2 - s \right) \,s\,S_w\,\left( 2\,{m_Z}^2 - t - u \right) \,v^2 \nonumber\\
FH_5 &= -4\,s^2\,S_w\,\left( 2\,{m_Z}^2 - t - u \right) \,v^2 \nonumber\\
FH_6 &= \frac{-4\,s}{S_w}\,\left( -9 + 8\,{S_w}^2 \right) \,\left( -2\,t\,u + {m_Z}^2\,\left( t + u \right)  \right) \,v^2 \nonumber\\
FH_7 &= - \frac{s}{S_w}\,\left( -9 + 8\,{S_w}^2 \right) \,\left( {m_Z}^2\,s +
        {m_t}^2\,\left( {m_Z}^2 - t - u \right)  + 2\,t\,u \right) \,v^2  \nonumber\\
FH_8 &= -32\,s\,S_w\,\left( -2\,t\,u + {m_Z}^2\,\left( t + u \right)  \right) \,v^2 \nonumber\\
FH_9 &= -16\,m_t\,s\,S_w\,\left( 2\,{m_Z}^2 - t - u \right) \,v^3
\end{align}

{\bf Process (3)}

\begin{align}
\frac{d\sigma^{S\, Z}}{dt} & = \frac{e \, g_3}{81\,\pi\, C_w\,(m_Z^2-t)^2\,  (m_Z^2-s)^2\, s^2\,\Lambda^4}\biggl\{FH_1\,Re(\alpha^S_{ut}\, \alpha^{Z *}_{ut}) \nonumber\\
&+ \,FH_2\Bigl[Re(\alpha^S_{ut}\, \alpha^{Z}_{tu})\, + \, Re(\alpha^S_{tu}\, \alpha^{Z}_{ut}) \, - \,Re(\alpha^S_{tu}\, \alpha^{Z *}_{tu})\Bigr] +\,FH_3\Bigl[ Im(\alpha^S_{ut}\, \beta^{Z}_{tu}) \, + \, Im(\beta^S_{tu}\, \alpha^{Z}_{ut})\Bigr]\nonumber\\
& +\,FH_4 \, Re(\alpha^S_{ut}\, \theta^*)\,+\,FH_5 \,Re(\alpha^S_{tu}\, \theta)\, + \,FH_6 \, Re(\beta^S_{ut}\, \beta^{Z *}_{ut}) +\,FH_7 \Bigl[Re(\beta^S_{ut}\, \eta^*) \, - \, Re(\beta^S_{ut}\, \bar\eta^*)\Bigr] \nonumber\\
& + FH_8 \, Re(\beta^S_{tu}\, \beta^{Z *}_{tu}) \, + \, FH_9 \, Im(\beta^S_{tu}\, \theta) \biggr\}
\end{align}
with coefficients
\begin{align}
FH_1 &= S_w\,\left( 4\,{m_t}^2 - u \right) \,u\,
  \left( {m_t}^2\,{m_Z}^2 - 2\,s\,t - {m_Z}^2\,u \right) \nonumber\\
FH_2 &= S_w\,u^2\,\left( {m_t}^2\,{m_Z}^2 - 2\,s\,t - {m_Z}^2\,u \right) \nonumber\\
FH_3 &= 8\,m_t\,S_w\,u\,\left( {m_t}^2\,{m_Z}^2 - 2\,s\,t - {m_Z}^2\,u
    \right) \,v \nonumber\\
FH_4 &= -2\,S_w\,\left( 2\,{m_t}^2 - u \right) \,
  \left( {m_t}^2 - 2\,{m_Z}^2 - u \right) \,u\,v^2 \nonumber\\
FH_5 &= -2\,S_w\,\left( {m_t}^2 - 2\,{m_Z}^2 - u \right) \,u^2\,v^2 \nonumber\\
FH_6 &=  \frac{-2\,\left( -9 + 8\,{S_w}^2 \right) \,u\,
    \left( -\left( {m_t}^2\,{m_Z}^2 \right)  + 2\,s\,t + {m_Z}^2\,u \right) \,v^2}{
    S_w} \nonumber\\
FH_7 &=  \frac{\left( -9 + 8\,{S_w}^2 \right) \,u\,
    \left( {m_t}^2\,\left( {m_Z}^2 - s - t \right)  + 2\,s\,t + {m_Z}^2\,u \right) \,v^2}
    {2\,S_w} \nonumber\\
FH_8 &= -16\,S_w\,u\,\left( -\left( {m_t}^2\,{m_Z}^2 \right)  + 2\,s\,t +
    {m_Z}^2\,u \right) \,v^2 \nonumber\\
FH_9 &= -8\,m_t\,S_w\,\left( {m_t}^2 - 2\,{m_Z}^2 - u \right) \,u\,v^3
\end{align}
%


The interference electroweak-electroweak (photon+Z boson) FCNC contribution is given by

{\bf Process (1)}

\begin{align}
\frac{d\sigma^{\gamma \,Z}}{dt} & = \frac{e^2}{2592\,\pi\, C_w\,S_w\,(m_Z^2-t)^2\,  (m_Z^2-u)^2\, s^2\,\Lambda^4}\biggl\{GH_1\,Re(\alpha^\gamma_{ut} \, \alpha^Z_{ut})\, \nonumber\\
&+ \,GH_2\Bigl[Re(\alpha^\gamma_{ut} \, \alpha^Z_{tu}) \, + \, Re(\alpha^\gamma_{tu} \, \alpha^Z_{ut})\Bigr] \, +\,GH_3\Bigl[Im(\alpha^\gamma_{ut} \, \beta^Z_{tu}) \, + \, Im(\beta^\gamma_{tu} \, \alpha^Z_{ut})\Bigr]\nonumber\\
& +\,GH_4 \, Re(\alpha^\gamma_{ut} \, \theta)\,+\,GH_5 \,Re(\alpha^\gamma_{tu} \, \alpha^{Z\,*}_{tu}) \,+\,GH_6\Bigl[Im(\alpha^\gamma_{tu} \, \beta^{Z\,*}_{tu}) \, + \, Im(\beta^\gamma_{tu} \, \alpha^{Z\,*}_{tu})\Bigr]\nonumber\\
& +\,GH_7 \, Re(\alpha^\gamma_{tu} \, \theta)  + GH_8 \, Re(\beta^\gamma_{ut} \, \beta^{Z\,*}_{ut}) \, + \, GH_9 \Bigl[Re(\beta^\gamma_{ut} \, \eta^{Z\,*}) \, - \, Re(\beta^\gamma_{ut} \, \bar\eta^{Z\,*})\Bigr] \nonumber\\
&+ \, GH_{10} \, Re(\beta^\gamma_{tu} \, \beta^{Z\,*}_{tu})  \, + \, GH_{11} \, Im(\beta^\gamma_{tu} \, \theta)
\end{align}
with coefficients
{\small
\begin{align}
GH_1 &= -48\,{m_t}^4\,{S_w}^2\,{( t - u ) }^2 +  24\,{m_t}^6\,{S_w}^2\,( -2\,{m_Z}^2 + t + u )  +  t\,u\,( ( 27 - 88\,{S_w}^2 ) \,t^2 - 128\,{S_w}^2\,t\,u \nonumber\\
& + ( 27 - 88\,{S_w}^2 ) \,u^2     )  + {m_Z}^2\,( t + u ) \,( -27\,t\,u + 8\,{S_w}^2\,( 4\,t^2 + 11\,t\,u + 4\,u^2 )  )\nonumber\\
& + {m_t}^2\,( {m_Z}^2\,( ( 27 - 8\,{S_w}^2 ) \,t^2 +        8\,( 27 - 20\,{S_w}^2 ) \,t\,u + ( 27 - 8\,{S_w}^2 ) \,u^2 ) \nonumber\\
& +     ( t + u ) \,( ( -27 + 48\,{S_w}^2 ) \,t^2 - ( 81 + 8\,{S_w}^2 ) \,t\,u +        3\,( -9 + 16\,{S_w}^2 ) \,u^2 )  ) \nonumber\\
GH_2 &= -24\,{m_t}^6\,{S_w}^2\,( {m_Z}^2 - t - u )  +  t\,u\,( ( -27 + 88\,{S_w}^2 ) \,t^2 + 128\,{S_w}^2\,t\,u + ( -27 + 88\,{S_w}^2 ) \,u^2     ) \nonumber\\
& + {m_Z}^2\,( 32\,s^3\,{S_w}^2 - 24\,s^2\,{S_w}^2\,u - 24\,s\,{S_w}^2\,u^2 +     27\,t\,u\,( t + u )  )  + 8\,{m_t}^4\,{S_w}^2\,   ( 3\,{m_Z}^2\,( 2\,s + u ) \nonumber\\
& - 2\,( 3\,t^2 + 2\,t\,u + 3\,u^2 )  )  -  {m_t}^2\,( ( 27 - 48\,{S_w}^2 ) \,t^3 + 56\,{S_w}^2\,t^2\,u + 56\,{S_w}^2\,t\,u^2 \nonumber\\
& +     3\,( 9 - 16\,{S_w}^2 ) \,u^3 + {m_Z}^2\,      ( 104\,s^2\,{S_w}^2 - 27\,t^2 + 3\,( -9 + 8\,{S_w}^2 ) \,u^2 )  ) \nonumber\\
GH_3 &= -4\,m_t\,( 24\,{m_t}^4\,{S_w}^2\,( 2\,{m_Z}^2 - t - u )  +    {m_Z}^2\,( ( -27 + 8\,{S_w}^2 ) \,t^2 + 4\,( -27 + 16\,{S_w}^2 ) \,t\,u \nonumber\\
& +       ( -27 + 8\,{S_w}^2 ) \,u^2 )  -    ( t + u ) \,( ( -27 + 48\,{S_w}^2 ) \,t^2 - ( 27 + 56\,{S_w}^2 ) \,t\,u \nonumber\\
& +       3\,( -9 + 16\,{S_w}^2 ) \,u^2 )  -    16\,{m_t}^2\,{S_w}^2\,( -3\,t^2 + 2\,t\,u - 3\,u^2 + 2\,{m_Z}^2\,( t + u )  )  ) \,v \nonumber\\
GH_4 &= -2\,( 32\,{m_t}^4\,{S_w}^2\,( 2\,{m_Z}^2 - t - u )  +    {m_Z}^2\,( ( 27 - 88\,{S_w}^2 ) \,t^2 - 128\,{S_w}^2\,t\,u +       ( 27 - 88\,{S_w}^2 ) \,u^2 ) \nonumber\\
& +    ( t + u ) \,( ( -27 + 56\,{S_w}^2 ) \,t^2 + ( 27 + 40\,{S_w}^2 ) \,t\,u +       ( -27 + 56\,{S_w}^2 ) \,u^2 ) \nonumber\\
& +    6\,{m_t}^2\,( -9 + 8\,{S_w}^2 ) \,( -t^2 - u^2 + {m_Z}^2\,( t + u )  )  ) \,  v^2 \nonumber\\
GH_5 &= -24\,{m_t}^6\,{S_w}^2\,( {m_Z}^2 - t - u )  +  t\,u\,( ( 27 - 88\,{S_w}^2 ) \,t^2 - 128\,{S_w}^2\,t\,u + ( 27 - 88\,{S_w}^2 ) \,u^2     ) \nonumber\\
& + {m_Z}^2\,( -32\,s^3\,{S_w}^2 + 24\,s^2\,{S_w}^2\,u + 24\,s\,{S_w}^2\,u^2 -     27\,t\,u\,( t + u )  )  + 8\,{m_t}^4\,{S_w}^2\,   ( 3\,{m_Z}^2\,( 2\,s + 3\,u ) \nonumber\\
& - 2\,( 3\,t^2 + 10\,t\,u + 3\,u^2 )  )  +  {m_t}^2\,( ( t + u ) \,( ( -27 + 48\,{S_w}^2 ) \,t^2 +        ( 27 + 152\,{S_w}^2 ) \,t\,u \nonumber\\
& + 3\,( -9 + 16\,{S_w}^2 ) \,u^2 ) +     {m_Z}^2\,( -40\,s^2\,{S_w}^2 - 96\,s\,{S_w}^2\,u +        9\,( 3\,t^2 + ( 3 - 8\,{S_w}^2 ) \,u^2 )  )  ) \nonumber\\
GH_6 &= 12\,m_t\,( 8\,{m_t}^4\,{S_w}^2\,( 2\,{m_Z}^2 - t - u )  -    16\,{m_t}^2\,{S_w}^2\,( 2\,{m_Z}^2 - t - u ) \,( t + u ) \nonumber\\
& +    {m_Z}^2\,( 3\,( -3 + 8\,{S_w}^2 ) \,t^2 + 32\,{S_w}^2\,t\,u +       3\,( -3 + 8\,{S_w}^2 ) \,u^2 ) \nonumber\\
& -    ( t + u ) \,( ( -9 + 16\,{S_w}^2 ) \,t^2 + ( 9 + 8\,{S_w}^2 ) \,t\,u +       ( -9 + 16\,{S_w}^2 ) \,u^2 )  ) \,v \nonumber
\end{align}
}

{\small
\begin{align}
GH_7 &= -2\,( 32\,{m_t}^4\,{S_w}^2\,( 2\,{m_Z}^2 - t - u )  -    64\,{m_t}^2\,{S_w}^2\,( 2\,{m_Z}^2 - t - u ) \,( t + u ) \nonumber\\
& -    ( t + u ) \,( ( -27 + 56\,{S_w}^2 ) \,t^2 + ( 27 + 40\,{S_w}^2 ) \,t\,u +       ( -27 + 56\,{S_w}^2 ) \,u^2 ) \nonumber\\
& +    {m_Z}^2\,( ( -27 + 88\,{S_w}^2 ) \,t^2 + 128\,{S_w}^2\,t\,u +       ( -27 + 88\,{S_w}^2 ) \,u^2 )  ) \,v^2 \nonumber\\
GH_8 &= 16\,( 2\,{m_Z}^2\,s^2\,( 9 - 20\,{S_w}^2 )  + 27\,s\,t^2 + 48\,{S_w}^2\,t^3 -    24\,{m_t}^4\,{S_w}^2\,( {m_Z}^2 - t - u )  - 18\,s\,t\,u \nonumber\\
& + 32\,{S_w}^2\,t^2\,u + 27\,s\,u^2 +    32\,{S_w}^2\,t\,u^2 + 48\,{S_w}^2\,u^3 +    {m_t}^2\,( 4\,{m_Z}^2\,s\,( 9 + 4\,{S_w}^2 )  - 27\,s\,( t + u ) \nonumber\\
& -  16\,{S_w}^2\,( 3\,t^2 + 2\,t\,u + 3\,u^2 )  )  ) \,v^2 \nonumber\\
GH_9 &= 4\,s\,( 2\,{m_Z}^2\,s\,( -9 + 20\,{S_w}^2 )  - 27\,t^2 + 48\,{S_w}^2\,t^2 + 18\,t\,u -
    16\,{S_w}^2\,t\,u - 27\,u^2 + 48\,{S_w}^2\,u^2 \nonumber\\
& +    {m_t}^2\,( {m_Z}^2\,( 36 - 56\,{S_w}^2 )  +       ( -9 + 8\,{S_w}^2 ) \,( t + u )  )  ) \,v^2 \nonumber\\
GH_{10} &= -16\,( 24\,{m_t}^4\,{S_w}^2\,( {m_Z}^2 - t - u )  -    16\,{m_t}^2\,{S_w}^2\,( {m_Z}^2\,s - 3\,t^2 - 2\,t\,u - 3\,u^2 ) \nonumber\\
&  -    ( t + u ) \,( ( -27 + 48\,{S_w}^2 ) \,t^2 - 16\,{S_w}^2\,t\,u +       3\,( -9 + 16\,{S_w}^2 ) \,u^2 ) \nonumber\\
& +    {m_Z}^2\,( 40\,s^2\,{S_w}^2 - 27\,{( t + u ) }^2 )  ) \,v^2 \nonumber\\
GH_{11} &= -8\,m_t\,( 27\,t^2 + 8\,{S_w}^2\,t^2 + 64\,{S_w}^2\,t\,u + 27\,u^2 + 8\,{S_w}^2\,u^2 \nonumber\\
& +    {m_Z}^2\,( 40\,s\,{S_w}^2 - 27\,( t + u )  )  +    8\,{m_t}^2\,{S_w}^2\,( 3\,{m_Z}^2 - 4\,( t + u )  )  ) \,v^3
\end{align}
}

{\bf Process (2)}

\begin{align}
\frac{d\sigma^{\gamma \,Z}}{dt} & = \frac{e^2}{864\,\pi\, C_w\,S_w\,(m_Z^2-t)^2\, s^2\,\Lambda^4}\biggl\{GH_1\,Re(\alpha^\gamma_{ut} \, \alpha^Z_{ut})\, \nonumber\\
&+ \,GH_2\Bigl[Re(\alpha^\gamma_{ut} \, \alpha^Z_{tu}) \, + \, Re(\alpha^\gamma_{tu} \, \alpha^Z_{ut})\Bigr] \, +\,GH_3\Bigl[Im(\alpha^\gamma_{ut} \, \beta^Z_{tu}) \, + \, Im(\beta^\gamma_{tu} \, \alpha^Z_{ut})\Bigr]\nonumber\\
& +\,GH_4 \, Re(\alpha^\gamma_{ut} \, \theta)\,+\,GH_5 \,Re(\alpha^\gamma_{tu} \, \alpha^{Z\,*}_{tu}) \,+\,GH_6\Bigl[Im(\alpha^\gamma_{tu} \, \beta^{Z\,*}_{tu}) \, + \, Im(\beta^\gamma_{tu} \, \alpha^{Z\,*}_{tu})\Bigr]\nonumber\\
& +\,GH_7 \, Re(\alpha^\gamma_{tu} \, \theta)  + GH_8 \, Re(\beta^\gamma_{ut} \, \beta^{Z\,*}_{ut}) \, + \, GH_9 \Bigl[Re(\beta^\gamma_{ut} \, \eta^{Z\,*}) \, - \, Re(\beta^\gamma_{ut} \, \bar\eta^{Z\,*})\Bigr] \nonumber\\
&+ \, GH_{10} \, Re(\beta^\gamma_{tu} \, \beta^{Z\,*}_{tu})  \, + \, GH_{11} \, Im(\beta^\gamma_{tu} \, \theta)
\end{align}
with coefficients
{\small
\begin{align}
GH_1 &= -32\,{m_t}^6\,{S_w}^2 + 32\,{m_t}^4\,{S_w}^2\,( s + t + u )  +  t\,( 8\,s^2\,{S_w}^2 + ( -9 + 8\,{S_w}^2 ) \,u^2 ) \nonumber\\
& -
  {m_t}^2\,( 8\,s^2\,{S_w}^2 + 32\,s\,{S_w}^2\,t +     ( -9 + 8\,{S_w}^2 ) \,u\,( 4\,t + u )  ) \nonumber\\
GH_2 &= -( ( {m_t}^2 + t ) \,    ( 8\,s^2\,{S_w}^2 + ( -9 + 8\,{S_w}^2 ) \,u^2 )  ) \nonumber\\
GH_3 &= -4\,m_t\,( 8\,{m_t}^4\,{S_w}^2 -    16\,{m_t}^2\,{S_w}^2\,u - 9\,u\,( 2\,t + u )  -    8\,{S_w}^2\,( t^2 - 2\,t\,u - 2\,u^2 )  ) \,v\nonumber\\
GH_4 &=  -2\,( -8\,s^2\,{S_w}^2 + ( 9 - 8\,{S_w}^2 ) \,u^2 +    {m_t}^2\,( 16\,s\,{S_w}^2 - 18\,u + 16\,{S_w}^2\,u )  ) \,v^2\nonumber\\
GH_5 &= -( ( {m_t}^2 - t ) \,    ( 8\,s^2\,{S_w}^2 + ( -9 + 8\,{S_w}^2 ) \,u^2 )  ) \nonumber\\
GH_6 &=  4\,m_t\,( 8\,s^2\,{S_w}^2 + ( -9 + 8\,{S_w}^2 ) \,u^2 ) \,v\nonumber\\
GH_7 &=  -2\,( 8\,s^2\,{S_w}^2 + ( -9 + 8\,{S_w}^2 ) \,u^2 ) \,v^2\nonumber\\
GH_8 &=  -16\,( s\,( 9 - 16\,{S_w}^2 ) \,u +    {m_t}^2\,( s\,( -9 + 8\,{S_w}^2 )  + 8\,{S_w}^2\,u )    ) \,v^2\nonumber\\
GH_9 &=  4\,s\,( 9 - 16\,{S_w}^2 ) \,u\,v^2\nonumber\\
GH_{10} &=  -16\,( 8\,{m_t}^2\,{S_w}^2\,( s + u )  -    u\,( 16\,s\,{S_w}^2 + 9\,( t + u )  )  ) \,v^2\nonumber\\
GH_{11} &= -8\,m_t\,( 8\,{m_t}^2\,{S_w}^2 - 8\,{S_w}^2\,t - 9\,u ) \,v^3
\end{align}
}

{\bf Process (3)}

\begin{align}
\frac{d\sigma^{\gamma \,Z}}{dt} & = \frac{e^2}{2592\,\pi\, C_w\,S_w\,(m_Z^2-t)^2\,  (m_Z^2-s)^2\, s^2\,\Lambda^4}\biggl\{GH_1\,Re(\alpha^\gamma_{ut} \, \alpha^Z_{ut})\, \nonumber\\
&+ \,GH_2\Bigl[Re(\alpha^\gamma_{ut} \, \alpha^Z_{tu}) \, + \, Re(\alpha^\gamma_{tu} \, \alpha^Z_{ut})\Bigr] \, +\,GH_3\Bigl[Im(\alpha^\gamma_{ut} \, \beta^Z_{tu}) \, + \, Im(\beta^\gamma_{tu} \, \alpha^Z_{ut})\Bigr]\nonumber\\
& +\,GH_4 \, Re(\alpha^\gamma_{ut} \, \theta)\,+\,GH_5 \,Re(\alpha^\gamma_{tu} \, \alpha^{Z\,*}_{tu}) \,+\,GH_6\Bigl[Im(\alpha^\gamma_{tu} \, \beta^{Z\,*}_{tu}) \, + \, Im(\beta^\gamma_{tu} \, \alpha^{Z\,*}_{tu})\Bigr]\nonumber\\
& +\,GH_7 \, Re(\alpha^\gamma_{tu} \, \theta)  + GH_8 \, Re(\beta^\gamma_{ut} \, \beta^{Z\,*}_{ut}) \, + \, GH_9 \Bigl[Re(\beta^\gamma_{ut} \, \eta^{Z\,*}) \, - \, Re(\beta^\gamma_{ut} \, \bar\eta^{Z\,*})\Bigr] \nonumber\\
&+ \, GH_{10} \, Re(\beta^\gamma_{tu} \, \beta^{Z\,*}_{tu})  \, + \, GH_{11} \, Im(\beta^\gamma_{tu} \, \theta)
\end{align}
with coefficients
{\small
\begin{align}
GH_1 &= -24\,{m_t}^6\,{S_w}^2\,( {m_Z}^2 - s - t )  +  s\,t\,( s^2\,( 27 - 56\,{S_w}^2 )  - 64\,s\,{S_w}^2\,t +     ( 27 - 56\,{S_w}^2 ) \,t^2 )\nonumber\\
&  -  8\,{m_t}^4\,{S_w}^2\,( 6\,s^2 + 6\,t^2 +     {m_Z}^2\,( 15\,t + 2\,u )  ) \nonumber\\
& +  {m_Z}^2\,( -27\,s^2\,t - 27\,s\,t^2 +     8\,{S_w}^2\,u\,( 3\,t^2 + 3\,t\,u - 2\,u^2 )  )  +  {m_t}^2\,( ( s + t ) \,      ( s^2\,( -27 + 48\,{S_w}^2 ) \nonumber\\
& + s\,( -81 + 56\,{S_w}^2 ) \,t +        3\,( -9 + 16\,{S_w}^2 ) \,t^2 ) \nonumber\\
& +     {m_Z}^2\,( 27\,s^2 + 216\,s\,t + 3\,( 9 + 40\,{S_w}^2 ) \,t^2  +        96\,{S_w}^2\,t\,u + 8\,{S_w}^2\,u^2 )  ) \nonumber\\
GH_2 &= -24\,{m_t}^6\,{S_w}^2\,( {m_Z}^2 - s - t )  +  s\,t\,( s^2\,( -27 + 56\,{S_w}^2 )  + 64\,s\,{S_w}^2\,t +     ( -27 + 56\,{S_w}^2 ) \,t^2 ) \nonumber\\
& -  8\,{m_t}^4\,{S_w}^2\,( 6\,s^2 + 8\,s\,t + 6\,t^2 -     3\,{m_Z}^2\,( t + 2\,u )  ) \nonumber\\
&  +
  {m_Z}^2\,( 27\,s^2\,t + 27\,s\,t^2 +     8\,{S_w}^2\,u\,( -3\,t^2 - 3\,t\,u + 2\,u^2 )  )  +  {m_t}^2\,( ( s + t ) \,      ( s^2\,( -27 + 48\,{S_w}^2 ) \nonumber\\
& + s\,( 27 - 40\,{S_w}^2 ) \,t +        3\,( -9 + 16\,{S_w}^2 ) \,t^2 )  +     {m_Z}^2\,( 27\,s^2 + ( 27 - 24\,{S_w}^2 ) \,t^2 -        88\,{S_w}^2\,u^2 )  ) \nonumber\\
GH_3 &= 4\,m_t\,( 24\,{m_t}^6\,{S_w}^2 -    24\,{m_t}^4\,{S_w}^2\,( {m_Z}^2 + t + 3\,u )  +    u\,( 27\,s^2 + 27\,s\,t + ( 27 - 88\,{S_w}^2 ) \,t^2 \nonumber\\
& - 88\,{S_w}^2\,t\,u -       48\,{S_w}^2\,u^2 )  + {m_Z}^2\,     ( 27\,s^2 + 108\,s\,t + ( 27 + 48\,{S_w}^2 ) \,t^2 + 48\,{S_w}^2\,t\,u -       40\,{S_w}^2\,u^2 ) \nonumber\\
& + {m_t}^2\,     ( -27\,s^2 - 27\,s\,t - 27\,t^2 + 24\,{S_w}^2\,t^2 + 112\,{S_w}^2\,t\,u +       96\,{S_w}^2\,u^2 + 16\,{m_Z}^2\,{S_w}^2\,( -3\,t + u )  )    ) \,v \nonumber\\
GH_4 &= 2\,( -( ( s + t ) \,( s^2\,( -27 + 40\,{S_w}^2 )  +         s\,( 27 + 8\,{S_w}^2 ) \,t + ( -27 + 40\,{S_w}^2 ) \,t^2 )       ) \nonumber\\
& + 8\,{m_t}^4\,{S_w}^2\,     ( -3\,{m_Z}^2 + 2\,( s + t )  )  +    {m_Z}^2\,( -27\,s^2 + ( -27 + 48\,{S_w}^2 ) \,t^2 +       48\,{S_w}^2\,t\,u + 56\,{S_w}^2\,u^2 ) \nonumber\\
& +    2\,{m_t}^2\,( 3\,( -9 + 8\,{S_w}^2 ) \,( s^2 + t^2 )  +       {m_Z}^2\,( 27\,s + 27\,t - 24\,{S_w}^2\,t - 32\,{S_w}^2\,u )  )    ) \,v^2 \nonumber\\
GH_5 &=  -24\,{m_t}^6\,{S_w}^2\,( {m_Z}^2 - s - t )  +  s\,t\,( s^2\,( 27 - 56\,{S_w}^2 )  - 64\,s\,{S_w}^2\,t +     ( 27 - 56\,{S_w}^2 ) \,t^2 ) \nonumber\\
& -  8\,{m_t}^4\,{S_w}^2\,( 6\,s^2 + 16\,s\,t + 6\,t^2 -     3\,{m_Z}^2\,( 3\,t + 2\,u )  ) \nonumber\\
&  +  {m_Z}^2\,( -27\,s^2\,t - 27\,s\,t^2 +     8\,{S_w}^2\,u\,( 3\,t^2 + 3\,t\,u - 2\,u^2 )  ) \nonumber\\
&  +  {m_t}^2\,( ( s + t ) \,      ( s^2\,( -27 + 48\,{S_w}^2 )  + s\,( 27 + 88\,{S_w}^2 ) \,t +        3\,( -9 + 16\,{S_w}^2 ) \,t^2 ) \nonumber\\
& +     {m_Z}^2\,( 27\,s^2 + ( 27 - 72\,{S_w}^2 ) \,t^2 -        96\,{S_w}^2\,t\,u - 56\,{S_w}^2\,u^2 )  ) \nonumber\\
GH_6 &= 12\,m_t\,( -8\,{m_t}^6\,{S_w}^2 +    u\,( -9\,s^2 + 9\,s\,t + 3\,( -3 + 8\,{S_w}^2 ) \,t^2 + 24\,{S_w}^2\,t\,u +       16\,{S_w}^2\,u^2 ) \nonumber\\
& + {m_Z}^2\,     ( -9\,s^2 + ( -9 + 16\,{S_w}^2 ) \,t^2 + 16\,{S_w}^2\,t\,u +       24\,{S_w}^2\,u^2 )  + 8\,{m_t}^4\,{S_w}^2\,     ( {m_Z}^2 + 3\,( t + u )  ) \nonumber\\
& +    {m_t}^2\,( 9\,s^2 - 9\,s\,t + 9\,t^2 - 24\,{S_w}^2\,t^2 - 48\,{S_w}^2\,t\,u -       32\,{S_w}^2\,u^2 - 16\,{m_Z}^2\,{S_w}^2\,( t + u )  )  ) \,  v \nonumber
\end{align}
}

{\small
\begin{align}
GH_7 &= 2\,( 24\,{m_t}^6\,{S_w}^2 +    {m_Z}^2\,( 27\,s^2 + ( 27 - 48\,{S_w}^2 ) \,t^2 -       48\,{S_w}^2\,t\,u - 56\,{S_w}^2\,u^2 ) \nonumber\\
&  -    u\,( -27\,s^2 + 27\,s\,t + 9\,( -3 + 8\,{S_w}^2 ) \,t^2 + 72\,{S_w}^2\,t\,u +       40\,{S_w}^2\,u^2 )  - 24\,{m_t}^4\,{S_w}^2\,     ( {m_Z}^2 + 3\,( t + u )  ) \nonumber\\
& +    {m_t}^2\,( -27\,s^2 + 27\,s\,t - 27\,t^2 + 72\,{S_w}^2\,t^2 +       144\,{S_w}^2\,t\,u + 88\,{S_w}^2\,u^2 +       48\,{m_Z}^2\,{S_w}^2\,( t + u )  )  ) \,v^2 \nonumber\\
GH_8 &= 16\,( 24\,{m_t}^6\,{S_w}^2 -    24\,{m_t}^4\,{S_w}^2\,( {m_Z}^2 + 2\,t + 3\,u )  +    u\,( ( 36 - 80\,{S_w}^2 ) \,t^2 + 4\,( 9 - 20\,{S_w}^2 ) \,t\,u \nonumber\\
& +       u\,( {m_Z}^2\,( 36 - 56\,{S_w}^2 )  + 27\,u - 48\,{S_w}^2\,u          )  ) \nonumber\\
& + {m_t}^2\,     ( 9\,( 2\,{m_Z}^2 - 4\,t - 3\,u ) \,u +       16\,{S_w}^2\,( 3\,t^2 + 8\,t\,u + 2\,u\,( {m_Z}^2 + 3\,u )  )  )    ) \,v^2 \nonumber\\
GH_9 &=  4\,u\,( s^2\,( -27 + 48\,{S_w}^2 )  + 2\,s\,( -9 + 8\,{S_w}^2 ) \,t -    27\,t^2 + 48\,{S_w}^2\,t^2 \nonumber\\
& + {m_t}^2\,     ( {m_Z}^2\,( 18 - 40\,{S_w}^2 )  -       ( -9 + 8\,{S_w}^2 ) \,( s + t )  )  - 36\,{m_Z}^2\,u +    56\,{m_Z}^2\,{S_w}^2\,u ) \,v^2 \nonumber\\
GH_{10} &= 16\,( 24\,{m_t}^6\,{S_w}^2  -    3\,{m_t}^4\,( {m_Z}^2\,( -9 + 8\,{S_w}^2 )  +       8\,{S_w}^2\,( 2\,t + 3\,u )  )  +    u\,( 27\,s^2 + ( 27 - 80\,{S_w}^2 ) \,t^2 \nonumber\\
& - 80\,{S_w}^2\,t\,u +       u\,( {m_Z}^2\,( 27 - 56\,{S_w}^2 )  - 48\,{S_w}^2\,u )       ) \nonumber\\
& + {m_t}^2\,( -27\,s^2 + ( -27 + 48\,{S_w}^2 ) \,t^2 +       128\,{S_w}^2\,t\,u + 2\,u\,( {m_Z}^2\,( -27 + 16\,{S_w}^2 )  +          48\,{S_w}^2\,u )  )  ) \,v^2 \nonumber\\
GH_{11} &= -8\,m_t\,( s^2\,( 27 - 8\,{S_w}^2 )  + 32\,s\,{S_w}^2\,t + 27\,t^2 -    8\,{S_w}^2\,t^2 \nonumber\\
& + {m_t}^2\,     ( {m_Z}^2\,( -27 + 24\,{S_w}^2 )  -       16\,{S_w}^2\,( s + t )  )  + 27\,{m_Z}^2\,u +    8\,{m_Z}^2\,{S_w}^2\,u ) \,v^3
\end{align}
}

Finally, interference between the SM diagrams with the anomalous one is given by

{\bf Process (6)}

\begin{align}
\frac{d\sigma}{dt} & = \frac{e\,g_W^2}{1728\,\pi\, C_w\, S_w\, (m_Z^2-s)\, (t-m_W^2)\, u\,s^2\,\Lambda^2}\biggl\{-16\,m_t\,u\,v\,V_{td}\,V_{ud}\,Im(\beta^S_{ut}) \nonumber\\
& +\,2\,m_t\,u\,v\,V_{td}\,V_{ud}\,Im(\beta^\gamma_{ut})\, +\,4\,m_t\,s\,( 9 + 2\,{S_w}^2 ) \,u\,v\,V_{td}\,V_{ud}\,Re(\beta^Z_{ut})\nonumber\\
& +\,m_t\,( 9 + 2\,{S_w}^2 ) \,t\,u\,v\,V_{td}\,V_{ud} \Bigl[ Re(\eta)\, -\, Re(\bar\eta) \Bigr] \biggr\}
\end{align}
where $m_W$ is the W boson mass, $V_{td}$ and $V_{ud}$ are elements of the CKM matrix.

The other process in table~\ref{tab:ffchannel} are deduct in the following way: processes (4) and (5) have expressions very similar to these for process (2), with the substitution $s\leftrightarrow t$ and $s\leftrightarrow u,$ respectively and processes (7) and (8) have expressions very similar to these for process (6), with the substitution $t\leftrightarrow u$ and $t\leftrightarrow s,$ respectively.

\newpage

\section{Cross section expression for the $t\bar t$ FCNC production process} \label{sec:ttbar}

The expression for the $t\bar t$ FCNC cross section is also given by three terms, as eq.~\ref{eq:blabla}. The strong FCNC contribution is given by eq.~\ref{eq:tototo}.

\begin{equation}
\frac{d\sigma}{dt}=\frac{d\sigma^S}{dt}+\frac{d\sigma^{\gamma}}{dt}+\frac{d\sigma^{Z}}{dt}.\label{eq:blabla}
\end{equation}

\begin{align}
\frac{d\sigma^S}{dt} & = \frac{4}{2916\,\pi\, C^2_w \,(s-m_Z^2) t\, u\,s^3\,\Lambda^4}\biggl\{F_1\,|\alpha^S_{ut}|^2+\,F_2\,|\alpha^S_{tu}|^2\, +\,F_3\,|\beta^S_{ut}|^2\,+\, F_4\, |\beta^S_{tu}|^2\nonumber\\
& +\,F_5 \, Re(\alpha^S_{ut}\, \alpha^S_{tu})\,+\,F_6 \, Im(\alpha^S_{ut}\, \beta^{S}_{tu})\,+\,F_7 \, Im(\alpha^S_{tu}\, \beta^{S*}_{tu}) \biggr\}\label{eq:tototo}
\end{align}
where $F_i$ function are given by
\begin{align}
F_1 &= 9\,{C_w}^2\,( 8\,e^2 - 3\,{g_3}^2 ) \,( {m_Z}^2 - s ) \,   ( {m_t}^8 + 2\,{m_t}^6\,t + 6\,{m_t}^4\,t^2 + t^2\,u^2 -     2\,{m_t}^2\,t^2\,( t + 4\,u )  ) \nonumber\\
& -  8\,e^2\,s\,( 4\,{m_t}^8\,{S_w}^2 +     {m_t}^6\,( -9 + 8\,{S_w}^2 ) \,t +     6\,{m_t}^4\,( -3 + 4\,{S_w}^2 ) \,t^2 + 4\,{S_w}^2\,t^2\,u^2 \nonumber\\
& +     {m_t}^2\,t^2\,( ( 9 - 8\,{S_w}^2 ) \,t + 18\,u - 32\,{S_w}^2\,u )     ) \nonumber\\
F_2 &= ( 9\,{C_w}^2\,( 8\,e^2 - 3\,{g_3}^2 ) \,( {m_Z}^2 - s )  -    32\,e^2\,s\,{S_w}^2 ) \,{( {m_t}^4 - t\,u ) }^2 \nonumber\\
F_3 &= 8\,( 9\,{C_w}^2\,( 8\,e^2 - 3\,{g_3}^2 ) \,( {m_Z}^2 - s ) \,     ( 2\,{m_t}^6 - 3\,{m_t}^4\,t + {m_t}^2\,s\,t + t^2\,( 2\,s + t )  )\nonumber\\
&  - 4\,e^2\,s\,( 16\,{m_t}^6\,{S_w}^2 -      3\,{m_t}^4\,( 3 + 8\,{S_w}^2 ) \,t \nonumber\\
& +       2\,{m_t}^2\,( 4\,s\,{S_w}^2 - 9\,t ) \,t +      t^2\,( 16\,s\,{S_w}^2 + 9\,t + 8\,{S_w}^2\,t + 18\,u )  )  ) \,v^2 \nonumber\\
F_4 &= \frac{1}{{S_w}^2} -4\,( 54\,{C_w}^2\,{g_3}^2\,( {m_Z}^2 - s ) \,{S_w}^2\,       ( 2\,{m_t}^6 - 3\,{m_t}^4\,t + {m_t}^2\,s\,t + t^2\,( 2\,s + t )  )\nonumber\\
& + e^2\,( {m_t}^6\,( -288\,{C_w}^2\,{m_Z}^2\,{S_w}^2 +            s\,( 162 + 288\,( -1 + {C_w}^2 ) \,{S_w}^2 + 128\,{S_w}^4 )            ) \nonumber\\
& - 6\,{m_t}^4\,( -72\,{C_w}^2\,{m_Z}^2\,{S_w}^2 +            s\,( 27 + 36\,( -1 + 2\,{C_w}^2 ) \,{S_w}^2 + 32\,{S_w}^4 )            ) \,t \nonumber\\
& + {m_t}^2\,s\,t\,          ( -144\,{C_w}^2\,( {m_Z}^2 - s ) \,{S_w}^2 +            64\,s\,{S_w}^4 + 81\,t - 288\,{S_w}^2\,t - 81\,u + 144\,{S_w}^2\,u ) \nonumber\\
& +    8\,{S_w}^2\,t^2\,( -18\,{C_w}^2\,( {m_Z}^2 - s ) \,      ( 2\,s + t )  + s\,( 16\,s\,{S_w}^2 + 9\,t + 8\,{S_w}^2\,t + 18\,u )            )  )  ) \,v^2 \nonumber
\end{align}
\begin{align}
F_5 &= 2\,( 9\,{C_w}^2\,( 8\,e^2 - 3\,{g_3}^2 ) \,( {m_Z}^2 - s )  -    32\,e^2\,s\,{S_w}^2 ) \,( {m_t}^4 - t\,u ) \,  ( {m_t}^4 - 2\,{m_t}^2\,t + t\,u ) \nonumber\\
F_6 &=    8\,m_t\,( 9\,{C_w}^2\,( 8\,e^2 - 3\,{g_3}^2 ) \,     ( {m_Z}^2 - s ) \,( {m_t}^6 + 3\,{m_t}^2\,t^2 -       t^2\,( t + 3\,u )  ) \nonumber\\
& - 4\,e^2\,s\,     ( 8\,{m_t}^6\,{S_w}^2 - 9\,{m_t}^4\,t +       6\,{m_t}^2\,( -3 + 4\,{S_w}^2 ) \,t^2 \nonumber\\
& +       t^2\,( ( 9 - 8\,{S_w}^2 ) \,t + 6\,( 3 - 4\,{S_w}^2 ) \,u )       )  ) \,v \nonumber\\
F_7 &= -8\,m_t\,( 9\,{C_w}^2\,( 8\,e^2 - 3\,{g_3}^2 ) \,     ( {m_Z}^2 - s )  - 32\,e^2\,s\,{S_w}^2 ) \,( {m_t}^2 - t ) \,  ( {m_t}^4 - t\,u ) \,v
\end{align}

The weak (photon) FCNC contribution is given by eq.~\ref{eq:tototo2}.

\begin{align}
\frac{d\sigma^\gamma}{dt} & = \frac{1}{3888\,\pi\, C^2_w \,(s-m_Z^2) t\, u\,s^3\,\Lambda^4}\biggl\{G_1\,|\alpha^S_{ut}|^2+\,G_2\,|\alpha^S_{tu}|^2\, +\,G_3\,|\beta^S_{ut}|^2\,+\, G_4\, |\beta^S_{tu}|^2\nonumber\\
& +\,G_5 \, Re(\alpha^S_{ut}\, \alpha^S_{tu})\,+\,G_6 \, Im(\alpha^S_{ut}\, \beta^{S}_{tu})\,+\,G_7 \, Im(\alpha^S_{tu}\, \beta^{S*}_{tu}) \biggr\}\label{eq:tototo2}
\end{align}
where $G_i$ function are given by
\begin{align}
G_1 &= 2\,( 9\,{C_w}^2\,( e^2 + 3\,{g_3}^2 ) \,( {m_Z}^2 - s ) \,     ( {m_t}^8 + 2\,{m_t}^6\,t + 6\,{m_t}^4\,t^2 + t^2\,u^2  -       2\,{m_t}^2\,t^2\,( t + 4\,u )  )\nonumber\\
&  -    e^2\,s\,( 4\,{m_t}^8\,{S_w}^2 +       {m_t}^6\,( -9 + 8\,{S_w}^2 ) \,t +       6\,{m_t}^4\,( -3 + 4\,{S_w}^2 ) \,t^2 + 4\,{S_w}^2\,t^2\,u^2 \nonumber\\
& +       {m_t}^2\,t^2\,( ( 9 - 8\,{S_w}^2 ) \,t + 18\,u - 32\,{S_w}^2\,u )       )  )\nonumber\\
G_2 &= 2\,( 9\,{C_w}^2\,( e^2 + 3\,{g_3}^2 ) \,( {m_Z}^2 - s )  -    4\,e^2\,s\,{S_w}^2 ) \,{( {m_t}^4 - t\,u ) }^2\nonumber\\
G_3 &= 8\,( 18\,{C_w}^2\,( e^2 + 3\,{g_3}^2 ) \,( {m_Z}^2 - s ) \,     ( 2\,{m_t}^6 - 3\,{m_t}^4\,t + {m_t}^2\,s\,t + t^2\,( 2\,s + t )  )\nonumber\\
&  - e^2\,s\,( 16\,{m_t}^6\,{S_w}^2 -       3\,{m_t}^4\,( 3 + 8\,{S_w}^2 ) \,t +       2\,{m_t}^2\,( 4\,s\,{S_w}^2 - 9\,t ) \,t \nonumber\\
& +       t^2\,( 16\,s\,{S_w}^2 + 9\,t + 8\,{S_w}^2\,t + 18\,u )  )  ) \,v^2 \nonumber\\
G_4 &= - \frac{1}{{S_w}^2}  ( -432\,{C_w}^2\,{g_3}^2\,( {m_Z}^2 - s ) \,{S_w}^2\,         ( 2\,{m_t}^6 - 3\,{m_t}^4\,t + {m_t}^2\,s\,t + t^2\,( 2\,s + t )           ) \nonumber\\
& + e^2\,( {m_t}^6\,            ( -288\,{C_w}^2\,{m_Z}^2\,{S_w}^2 +              s\,( 162 + 288\,( -1 + {C_w}^2 ) \,{S_w}^2 + 128\,{S_w}^4 )              ) \nonumber\\
& - 6\,{m_t}^4\,( -72\,{C_w}^2\,{m_Z}^2\,{S_w}^2 +              s\,( 27 + 36\,( -1 + 2\,{C_w}^2 ) \,{S_w}^2 + 32\,{S_w}^4 )              ) \,t \nonumber\\
& + {m_t}^2\,s\,t\,            ( -144\,{C_w}^2\,( {m_Z}^2 - s ) \,{S_w}^2 +              64\,s\,{S_w}^4 + 81\,t - 288\,{S_w}^2\,t - 81\,u + 144\,{S_w}^2\,u ) \nonumber\\
& +           8\,{S_w}^2\,t^2\,( -18\,{C_w}^2\,( {m_Z}^2 - s ) \,               ( 2\,s + t )  + s\,( 16\,s\,{S_w}^2 + 9\,t + 8\,{S_w}^2\,t + 18\,u )              )  )  ) \,v^2 \nonumber
\end{align}
\begin{align}
G_5 &=  4\,( 9\,{C_w}^2\,( e^2 + 3\,{g_3}^2 ) \,( {m_Z}^2 - s )  -    4\,e^2\,s\,{S_w}^2 ) \,( {m_t}^4 - t\,u ) \,  ( {m_t}^4 - 2\,{m_t}^2\,t + t\,u )\nonumber\\
G_6 &= 8\,m_t\,( 18\,{C_w}^2\,( e^2 + 3\,{g_3}^2 ) \,     ( {m_Z}^2 - s ) \,( {m_t}^6 + 3\,{m_t}^2\,t^2 -       t^2\,( t + 3\,u )  ) \nonumber\\
&  + e^2\,s\,     ( -8\,{m_t}^6\,{S_w}^2 + 9\,{m_t}^4\,t -       6\,{m_t}^2\,( -3 + 4\,{S_w}^2 ) \,t^2 \nonumber\\
& +       t^2\,( ( -9 + 8\,{S_w}^2 ) \,t + 6\,( -3 + 4\,{S_w}^2 ) \,u )       )  ) \,v \nonumber\\
G_7 &= -16\,m_t\,( 9\,{C_w}^2\,( e^2 + 3\,{g_3}^2 ) \,     ( {m_Z}^2 - s )  - 4\,e^2\,s\,{S_w}^2 ) \,( {m_t}^2 - t ) \,  ( {m_t}^4 - t\,u ) \,v
\end{align}

The weak (Z boson) FCNC contribution is given by eq.~\ref{eq:tototo3}.
\begin{align}
\frac{d\sigma^Z}{dt} & = \frac{1}{3888\,\pi\, C^2_w\,(m_Z^2-t)\,  (s-m_Z^2)\, s^3\,\Lambda^4}\biggl\{H_1\,|\alpha^Z_{ut}|^2+\,H_2\,|\alpha^Z_{tu}|^2\, +\,H_3\,|\beta^Z_{ut}|^2\, +\,H_4 \, |\beta^Z_{tu}|^2\,\nonumber\\
&+\,H_5 \,|\eta|^2\,+\,H_6 \,|\bar\eta|^2\,+\,H_7 \, |\theta|^2\,+\,H_8 \, Re(\alpha^Z_{ut}\, \alpha^Z_{tu})\, + H_9 \, Im(\alpha^Z_{ut}\, \beta^Z_{tu})\, + \, H_{10} \, Re(\alpha^Z_{ut}\, \theta^*) \,\nonumber\\
&+ \, H_{11} \, Im(\alpha^Z_{tu}\, \beta^{Z*}_{tu}) \, + \, H_{12} \, Re(\alpha^Z_{tu}\, \theta)\,+ \, H_{13}  Re(\beta^Z_{ut}\, \eta^*) +\, H_{14}\,Re(\beta^Z_{ut}\, \bar \eta^*)\, \nonumber\\
&+ \, H_{15} \, Im(\beta^Z_{tu}\, \theta)\, + \, H_{16}\,Re(\eta\, \bar\eta^*) \biggr\}\label{eq:tototo3}
\end{align}
where $H_i$ function are given by
\begin{align}
H_1 &= -2\,( 9\,{C_w}^2\,( e^2 + 3\,{g_3}^2 ) \,( {m_Z}^2 - s ) \,     ( {m_t}^8 + 2\,{m_t}^6\,t + 6\,{m_t}^4\,t^2 + t^2\,u^2 -       2\,{m_t}^2\,t^2\,( t + 4\,u )  ) \nonumber\\
& -    e^2\,s\,( 4\,{m_t}^8\,{S_w}^2 +       {m_t}^6\,( -9 + 8\,{S_w}^2 ) \,t +       6\,{m_t}^4\,( -3 + 4\,{S_w}^2 ) \,t^2 + 4\,{S_w}^2\,t^2\,u^2 \nonumber\\
& +       {m_t}^2\,t^2\,( ( 9 - 8\,{S_w}^2 ) \,t + 18\,u - 32\,{S_w}^2\,u )       )  ) \nonumber\\
H_2 &= -2\,( 9\,{C_w}^2\,( e^2 + 3\,{g_3}^2 ) \,( {m_Z}^2 - s )  -    4\,e^2\,s\,{S_w}^2 ) \,{( {m_t}^4 - t\,u ) }^2 \nonumber\\
H_3 &= \frac{1}{{S_w}^2}( -432\,{C_w}^2\,{g_3}^2\,( {m_Z}^2 - s ) \,{S_w}^2\,       ( 2\,{m_t}^6 - 3\,{m_t}^4\,t + {m_t}^2\,s\,t + t^2\,( 2\,s + t )  )\nonumber\\
&  + e^2\,( {m_t}^6\,( -288\,{C_w}^2\,{m_Z}^2\,{S_w}^2 +    s\,( 162 + 288\,( -1 + {C_w}^2 ) \,{S_w}^2 + 128\,{S_w}^4 )          ) \nonumber\\
& - 6\,{m_t}^4\,( -72\,{C_w}^2\,{m_Z}^2\,{S_w}^2 +         s\,( 27 + 36\,( -1 + 2\,{C_w}^2 ) \,{S_w}^2 + 32\,{S_w}^4 )          ) \,t \nonumber\\
& + {m_t}^2\,s\,t\,          ( -144\,{C_w}^2\,( {m_Z}^2 - s ) \,{S_w}^2 +            64\,s\,{S_w}^4 + 81\,t - 288\,{S_w}^2\,t - 81\,u + 144\,{S_w}^2\,u ) \nonumber\\
& +         8\,{S_w}^2\,t^2\,( -18\,{C_w}^2\,( {m_Z}^2 - s ) \,             ( 2\,s + t )  + s\,( 16\,s\,{S_w}^2 + 9\,t + 8\,{S_w}^2\,t + 18\,u )            )  )  ) \,v^2 \nonumber\\
H_4 &= -8\,( 18\,{C_w}^2\,( e^2 + 3\,{g_3}^2 ) \,( {m_Z}^2 - s ) \,     ( 2\,{m_t}^6 - 3\,{m_t}^4\,t + {m_t}^2\,s\,t + t^2\,( 2\,s + t )  )\nonumber\\
&   - e^2\,s\,( 16\,{m_t}^6\,{S_w}^2 -       3\,{m_t}^4\,( 3 + 8\,{S_w}^2 ) \,t +       2\,{m_t}^2\,( 4\,s\,{S_w}^2 - 9\,t ) \,t \nonumber\\
& +       t^2\,( 16\,s\,{S_w}^2 + 9\,t + 8\,{S_w}^2\,t + 18\,u )  )  ) \,v^2 \nonumber
\end{align}
\begin{align}
H_5 &= -\frac{1}{16\,{m_Z}^2\,{S_w}^2}( ( {m_t}^4 + 2\,{m_t}^2\,t + t^2 - 2\,{m_Z}^2\,( t + u )  )\nonumber\\
 & \times        \,( 432\,{C_w}^2\,{g_3}^2\,( {m_Z}^2 - s ) \,{S_w}^2\,         ( 3\,{m_t}^4 + t^2 - {m_t}^2\,( 3\,t + u )  ) \nonumber\\
 & +        e^2\,( 24\,{m_t}^4\,{S_w}^2\,            ( 18\,{C_w}^2\,( {m_Z}^2 - s )  +              s\,( 15 - 8\,{S_w}^2 )  )  +           8\,{S_w}^2\,( 18\,{C_w}^2\,( {m_Z}^2 - s ) \nonumber\\
 & +              s\,( 9 - 8\,{S_w}^2 )  ) \,t^2 +           {m_t}^2\,( -81\,s^2 - 144\,{C_w}^2\,{m_Z}^2\,{S_w}^2\,               ( 3\,t + u ) \nonumber\\
 & + 16\,s\,{S_w}^2\,               ( 3\,( -6 + 9\,{C_w}^2 + 4\,{S_w}^2 ) \,t +                 ( -9 + 9\,{C_w}^2 + 4\,{S_w}^2 ) \,u )  )  )  ) \,v^2      )  \nonumber\\
H_6 &= -\frac{1}{16\,{m_Z}^2\,{S_w}^2} ( ( {m_t}^4 + 2\,{m_Z}^2\,s - 2\,{m_t}^2\,t + t^2 ) \,\nonumber\\
&      ( 432\,{C_w}^2\,{g_3}^2\,( {m_Z}^2 - s ) \,{S_w}^2\,         ( 3\,{m_t}^4 + t^2 - {m_t}^2\,( 3\,t + u )  ) \nonumber\\
&  +        e^2\,( 24\,{m_t}^4\,{S_w}^2\,            ( 18\,{C_w}^2\,( {m_Z}^2 - s )  +              s\,( 15 - 8\,{S_w}^2 )  )  +           8\,{S_w}^2\,( 18\,{C_w}^2\,( {m_Z}^2 - s ) \nonumber\\
& +              s\,( 9 - 8\,{S_w}^2 )  ) \,t^2 +           {m_t}^2\,( -81\,s^2 - 144\,{C_w}^2\,{m_Z}^2\,{S_w}^2\,               ( 3\,t + u ) \nonumber\\
& + 16\,s\,{S_w}^2\,               ( 3\,( -6 + 9\,{C_w}^2 + 4\,{S_w}^2 ) \,t +                 ( -9 + 9\,{C_w}^2 + 4\,{S_w}^2 ) \,u )  )  )  ) \,v^2      ) \nonumber\\
H_7 &= \frac{-2}{{m_Z}^2}\,( 18\,{C_w}^2\,( e^2 + 3\,{g_3}^2 ) \,       ( {m_t}^6\,( {m_Z}^2 - 3\,s )  +         2\,{m_Z}^2\,( {m_Z}^2 - s ) \,u^2 \nonumber\\
& +         {m_t}^4\,( 6\,{m_Z}^4 + s\,( 3\,t + u )  -            {m_Z}^2\,( 3\,s + 2\,u )  )  -         {m_t}^2\,( s\,t^2 + 2\,{m_Z}^4\,( t + 3\,u ) \nonumber\\
& +            {m_Z}^2\,( s^2 - 6\,s\,u - u^2 )  )  )  +      e^2\,s\,( {m_t}^6\,( 9 - 24\,{S_w}^2 )  -         16\,{m_Z}^2\,{S_w}^2\,u^2 \nonumber\\
& +         {m_t}^4\,( -48\,{m_Z}^2\,{S_w}^2 +            6\,( -3 + 4\,{S_w}^2 ) \,t + 8\,{S_w}^2\,u )  +         {m_t}^2\,( ( 9 - 8\,{S_w}^2 ) \,t^2 \nonumber\\
& +            {m_Z}^2\,( 18\,s + 16\,{S_w}^2\,( t + 3\,u )  )  )  )      ) \,v^4 \nonumber\\
H_8 &=  -4\,( 9\,{C_w}^2\,( e^2 + 3\,{g_3}^2 ) \,( {m_Z}^2 - s )  -    4\,e^2\,s\,{S_w}^2 ) \,( {m_t}^4 - t\,u ) \,  ( {m_t}^4 - 2\,{m_t}^2\,t + t\,u )\nonumber\\
H_9 &= -8\,m_t\,( 18\,{C_w}^2\,( e^2 + 3\,{g_3}^2 ) \,     ( {m_Z}^2 - s ) \,( {m_t}^6 + 3\,{m_t}^2\,t^2 -       t^2\,( t + 3\,u )  ) \nonumber\\
& + e^2\,s\,     ( -8\,{m_t}^6\,{S_w}^2 + 9\,{m_t}^4\,t -       6\,{m_t}^2\,( -3 + 4\,{S_w}^2 ) \,t^2 \nonumber\\
& +       t^2\,( ( -9 + 8\,{S_w}^2 ) \,t + 6\,( -3 + 4\,{S_w}^2 ) \,u )       )  ) \,v \nonumber\\
H_{10} &= -4\,( 18\,{C_w}^2\,( e^2 + 3\,{g_3}^2 ) \,( {m_Z}^2 - s ) \,     ( 2\,{m_t}^6 + 3\,{m_t}^4\,t + t\,u^2 - {m_t}^2\,t\,( t + 5\,u )  )\nonumber\\
&  + e^2\,s\,( {m_t}^6\,( 9 - 16\,{S_w}^2 )  -       6\,{m_t}^4\,( -3 + 4\,{S_w}^2 ) \,t - 8\,{S_w}^2\,t\,u^2 \nonumber\\
& +       {m_t}^2\,t\,( -9\,t + 8\,{S_w}^2\,t - 18\,u + 40\,{S_w}^2\,u )  )    ) \,v^2 \nonumber\\
H_{11} &= 16\,m_t\,( 9\,{C_w}^2\,( e^2 + 3\,{g_3}^2 ) \,     ( {m_Z}^2 - s )  - 4\,e^2\,s\,{S_w}^2 ) \,( {m_t}^2 - t ) \,  ( {m_t}^4 - t\,u ) \,v \nonumber\\
H_{12} &= 8\,( 9\,{C_w}^2\,( e^2 + 3\,{g_3}^2 ) \,( {m_Z}^2 - s )  -    4\,e^2\,s\,{S_w}^2 ) \,( {m_t}^2 - u ) \,( {m_t}^4 - t\,u ) \,
  v^2\nonumber\\
\end{align}
\begin{align}
H_{13} &= \frac{-1}{2\,{S_w}^2}( ( {m_t}^2 - t ) \,      ( -432\,{C_w}^2\,{g_3}^2\,( {m_Z}^2 - s ) \,{S_w}^2\,         ( {m_t}^4 - {m_t}^2\,( 3\,t + u )  + t\,( t + 2\,u )  ) \nonumber\\
& +        e^2\,( -8\,{m_t}^4\,{S_w}^2\,            ( 18\,{C_w}^2\,( {m_Z}^2 - s )  +              s\,( 27 - 8\,{S_w}^2 )  )  -           8\,{S_w}^2\,( 18\,{C_w}^2\,( {m_Z}^2 - s ) \nonumber\\
& +              s\,( 9 - 8\,{S_w}^2 )  ) \,t\,( t + 2\,u )  +           {m_t}^2\,( 81\,s^2 + 144\,{C_w}^2\,{m_Z}^2\,{S_w}^2\,               ( 3\,t + u ) \nonumber\\
& - 16\,s\,{S_w}^2\,               ( 3\,( -6 + 9\,{C_w}^2 + 4\,{S_w}^2 ) \,t +                 ( -9 + 9\,{C_w}^2 + 4\,{S_w}^2 ) \,u )  )  )  ) \,v^2      )  \nonumber\\
H_{14} &= \frac{1}    {2\,{S_w}^2}( {m_t}^2 - t ) \,( -432\,{C_w}^2\,{g_3}^2\,       ( {m_Z}^2 - s ) \,{S_w}^2\,       ( {m_t}^4 - {m_t}^2\,( 3\,t + u )  + t\,( t + 2\,u )  ) \nonumber\\
& +      e^2\,( -8\,{m_t}^4\,{S_w}^2\,          ( 18\,{C_w}^2\,( {m_Z}^2 - s )  +            s\,( 27 - 8\,{S_w}^2 )  )  -         8\,{S_w}^2\,( 18\,{C_w}^2\,( {m_Z}^2 - s ) \nonumber\\
& +            s\,( 9 - 8\,{S_w}^2 )  ) \,t\,( t + 2\,u )  +         {m_t}^2\,( 81\,s^2 + 144\,{C_w}^2\,{m_Z}^2\,{S_w}^2\,             ( 3\,t + u ) \nonumber\\
& - 16\,s\,{S_w}^2\,             ( 3\,( -6 + 9\,{C_w}^2 + 4\,{S_w}^2 ) \,t +               ( -9 + 9\,{C_w}^2 + 4\,{S_w}^2 ) \,u )  )  )  ) \,v^2 \nonumber\\
H_{15} &= -8\,m_t\,( 18\,{C_w}^2\,( e^2 + 3\,{g_3}^2 ) \,     ( {m_Z}^2 - s ) \,( {m_t}^4 + {m_t}^2\,( 3\,t + u )  -       t\,( t + 4\,u )  ) \nonumber\\
& + e^2\,s\,     ( {m_t}^4\,( 9 - 8\,{S_w}^2 )  -       2\,{m_t}^2\,( 3\,( -3 + 4\,{S_w}^2 ) \,t + 4\,{S_w}^2\,u ) \nonumber\\
& +       t\,( -9\,t + 8\,{S_w}^2\,t - 18\,u + 32\,{S_w}^2\,u )  )  ) \,v^3\nonumber\\
H_{16} &= \frac{1}    {8\,{m_Z}^2\,{S_w}^2}( {m_t}^4 + 2\,{m_t}^2\,{m_Z}^2 - t^2 - 2\,{m_Z}^2\,u ) \,\nonumber\\
&\times    ( 432\,{C_w}^2\,{g_3}^2\,( {m_Z}^2 - s ) \,{S_w}^2\,       ( 3\,{m_t}^4 + t^2 - {m_t}^2\,( 3\,t + u )  ) \nonumber\\
& +      e^2\,( 24\,{m_t}^4\,{S_w}^2\,          ( 18\,{C_w}^2\,( {m_Z}^2 - s )  +            s\,( 15 - 8\,{S_w}^2 )  )  +         8\,{S_w}^2\,( 18\,{C_w}^2\,( {m_Z}^2 - s ) \nonumber\\
& +            s\,( 9 - 8\,{S_w}^2 )  ) \,t^2 +         {m_t}^2\,( -81\,s^2 - 144\,{C_w}^2\,{m_Z}^2\,{S_w}^2\,             ( 3\,t + u ) \nonumber\\
& + 16\,s\,{S_w}^2\,             ( 3\,( -6 + 9\,{C_w}^2 + 4\,{S_w}^2 ) \,t +               ( -9 + 9\,{C_w}^2 + 4\,{S_w}^2 ) \,u )  )  )  ) \,v^2
\end{align}

\end{subappendices}

\chapter{Lepton flavour violating processes in the effective Lagrangian approach}\label{cha:lepton}

In the previous chapter we studied the dimension six operators responsible for the flavour violations in the production of the top quark. In the leptonic sector this study has some features that distinguish and somewhat simplify the hadronic sector. The first aspect concerns the experimental evidence. The neutrinos oscillation gives us evidence that neutrinos have nonzero mass and, in this case, the flavour violation in the charged leptonic becomes reality. On the other hand, there is no evidence about this violation between the charged leptons. In fact, we have many experimental results which set stringent limits on the extent of flavour violation that may occur. Because of the very tiny neutrino mass expected, the flavour violation between charged lepton must be much suppressed. Nevertheless, even with all known experimental constraints it is possible that signal of LFV may be observed at the ILC, as we will see, tacking advantage of the large luminosities planned. Even without ILC functioning the lepton collider great luminosity and the signal's simplicity gives us stringent limits for LFV and great possibilities of obtaining even more stringent limits.

We will now deal with LFV in a very similar way to what we have done in the production of quark top with FCNC.

\section{FLV effective operators}
\label{sec:eff}

We are interested in those ${\cal L}^{(6)}$ operators
that give rise to LFV. Throughout this we will use $l_h$ to
represent a heavy lepton and $l_l$ denotes a light one (whose mass
we consider zero). In processes where a tau lepton is present, both
the muon and the electron will be taken to be massless. If a given
process only involves muons and electrons, then the electron mass
will be set to zero, but the muon mass will be kept. Whenever the
lepton's mass has no bearing on the result we will use $l$ for all
massless leptons, and drop the generation index.

The effective operators that will be important for our studies fall
in three categories:
\begin{enumerate}
\item those that generate flavour-violating vertices of the form $Z
\,l_h\, l_l$ and $\gamma \,l_h \,l_l$ (and also, for some operators,
vertices like $\gamma\,\gamma \,l_h \,l_l$); these operators always
involve gauge fields, either explicitly or in the form of covariant
derivatives;
\item four-fermion operators, involving only leptonic spinors;
\item and a type of operator that involves only scalar and fermionic
fields that will roughly correspond to a wave function
renormalization of the fermion fields.
\end{enumerate}

\subsection{Effective operators generating  $Z \,l_h\, l_l$ and $\gamma \,l_h \,l_l$ vertices}

There are five tree-level dimension 6 effective operators that can
generate a new $Z \,l_h\, l_l$ interaction. This means that these
interactions are compatible with SM symmetries at tree level. Again
following the notations of~\cite{Buchmuller:1985jz} we write the
first two operators as
\begin{align}
{\cal O}_{D_e}&
=\frac{\eta^{R}_{ij}}{\Lambda^2}\,\left(\bar{\ell}^i_L\,
D^{\mu}\,e^j_R\right)\, D_{\mu} \phi, \nonumber \\
{\cal O}_{\bar{D}_e} & = \frac{\eta^{L}_{ij}}{\Lambda^2}\,\left(
D^{\mu} \bar{\ell}^i_L\, \,e^j_R\right)\, D_{\mu} \phi  .
\label{eq:op1}
\end{align}
The coefficients $\eta^{R(L)}_{ij}$ are complex dimensionless
couplings and the $(i,j)$ are flavour indices. For flavour violation
to occur, these indices must differ. $\ell^i_L$ is a left-handed
$SU(2)_{L}$ doublet, $e^j_R$ is a right-handed $U(1)_{Y}$ singlet,
$\phi$ is the Higgs scalar $SU(2)$ doublet.
%
%
There is no $\gamma\, l_h\, l_l$ interaction stemming from these
terms, although one may obtain contributions to vertices involving
also a Higgs field, such as $\gamma\, \phi\, l_h\, l_l$ and $Z\,
\phi\, l_h\, l_l$.

The remaining three operators that contribute to the vertices $Z
\,l_h\, l_l$ but not to $\gamma\, l_h\, l_l$ are given by
\begin{align}
{\cal O}_{\phi e} & =
i\frac{\theta^{R}_{ij}}{\Lambda^2}\,\left(\phi^{\dagger} D_{\mu}
\phi \right)\left(\bar{e}^i_R \, \gamma^{\mu} \,e^j_R \, \right),\nonumber  \\
{\cal O}^{(1)}_{\phi \ell} & =
i\frac{\theta^{L(1)}_{ij}}{\Lambda^2}\,\left(\phi^{\dagger} D_{\mu}
\phi \right)\left(\bar{\ell}^i_L \, \gamma^{\mu} \,\ell^j_L \,
\right) ,\nonumber\\
{\cal O}^{(3)}_{\phi \ell} & =
i\frac{\theta^{L(3)}_{ij}}{\Lambda^2}\,\left(\phi^{\dagger} D_{\mu}
\tau_{I} \phi \right)\left(\bar{\ell}^i_L \, \gamma^{\mu} \tau^{I}
\,\ell^j_L \, \right).
 \label{eq:op2}
\end{align}
Again, $\theta^{R}_{ij}$ and $\theta^{L(1),(3)}_{ij}$ are complex
dimensionless couplings, and the contributions to $Z \,l_h\, l_l$
arise when both scalar fields acquire a vev $v$. Because the
covariant derivatives act on those same fields and the SM Higgs has
no coupling to the photon, there are no contributions to $\gamma
\,l_h\, l_l$ from these operators. There are however five dimension
six operators that contribute to both the $Z \,l_h\, l_l$ and
$\gamma\, l_h\, l_l$ vertices and are only present at the one-loop
level. They are given by
\begin{align}
{\cal O}_{eB} &=
\;\;i\,\frac{\alpha^{B\,R}_{ij}}{\Lambda^2}\,\left(\bar{e}^i_R\,
\gamma^\mu\,D^\nu\,e^j_R\right)\,B_{\mu\nu}  ,\nonumber \\
{\cal O}_{\ell B} &=
\;\;i\,\frac{\alpha^{B\,L}_{ij}}{\Lambda^2}\,\left(\bar{\ell}^i_L\,
\gamma^\mu\,D^\nu\,\ell^j_L\right)\,B_{\mu\nu}  , \nonumber \\
{\cal O}_{eB\phi} &=
\;\;\frac{\beta^{B}_{ij}}{\Lambda^2}\,\left(\bar{\ell}^i_L\,
\sigma^{\mu\nu}\,e^j_R\right)\, \phi\,B_{\mu\nu}  ,\nonumber \\
{\cal O}_{\ell W} & =
\;\;i\,\frac{\alpha^{W\,L}_{ij}}{\Lambda^2}\,\left(\bar{\ell}^i_L \,
\tau_{I} \,  \gamma^\mu\,D^\nu\,\ell^j_L\right)\,W^{I}_{\mu\nu} ,\nonumber\\
{\cal O}_{eW\phi} &
=\frac{\beta^{W}_{ij}}{\Lambda^2}\,\left(\bar{\ell}^i_L\, \,
\tau_{I} \, \sigma^{\mu\nu}\,e^j_R\right)\, \phi\,W^{I}_{\mu\nu}
.\label{eq:op4-5}
\end{align}
$\alpha_{ij}$ and $\beta_{ij}$ are complex dimensionless couplings,
$B_{\mu\nu}$ and $W^{I}_{\mu\nu}$ are the usual $U(1)_Y$ and
$SU(2)_L$ field tensors, respectively. These tensors ``contain''
both the photon and Z boson fields, through the well-known Weinberg
rotation. Thus they contribute to both $Z \,l_h\, l_l$ and $\gamma\,
l_h\, l_l$ when we consider the partial derivative of $D^\mu$ in the
equations~\eqref{eq:op4-5} or when we replace the
Higgs field $\phi$ by its vev $v$ in them. We will return to this
point in section~\ref{sec:gamma}.

\subsection{Four-Fermion effective operators producing an $e\,e\, l_h l_l$
contact interaction}
Because we are specifically interested in studying the phenomenology
of the ILC, we will only consider four-fermion operators where two
of the spinors involved correspond to the colliding
electrons/positrons of that collider. Another spinor will correspond
to a heavy lepton, $l_h$. There are four relevant types of
four-fermion operators that contribute to $e^+\,e^- \to l_h\, l_l$,
\begin{align}
{\cal O}_{\ell \ell}^{(1)} & = \frac{\kappa_{\ell \ell}^{(1)}}{2}
\left( \bar\ell_L \gamma_\mu \ell_L \right ) \left(\bar \ell_L
\gamma^\mu \ell_L \right),\nonumber\\
{\cal O}_{\ell \ell}^{(3)} & = \frac{\kappa_{\ell \ell}^{(3)}}{2}
\left( \bar\ell_L \gamma_\mu \tau^I \ell_L \right ) \left(\bar
\ell_L \gamma^\mu \tau^I \ell_L \right), \nonumber \\
{\cal O}_{e e} & = \frac{\kappa_{e e}}{2} \left( \bar e_R \gamma_\mu
e_R \right ) \left(\bar e_R \gamma^\mu e_R \right)  ,\nonumber\\
{\cal O}_{\ell e} & = \kappa_{\ell e} \left( \bar \ell_L e_R \right
) \left(\bar e_R \ell_L \right). \label{eq:ff}
\end{align}
Again, all couplings in these operators are, in general,
complex. As we have done with the previous operators, we should now
consider all possible ``placements'' of the $l_h$ spinor, and
consider different couplings for each of them. But that would lead
to an unmanageable number of fermionic operators, all with the same
Lorentz structure but differing simply in the location of the heavy
lepton spinor. Thus we will simplify our approach and define only
one coupling constant for each type of operator. An exception is the
operator ${\cal O}_{\ell e} = \kappa_{\ell e} \left( \bar \ell_L e_R
\right ) \left(\bar e_R \ell_L \right)$, which corresponds to an
interaction between a right-handed current and a left-handed one.
Depending on where we place the $l_h$ spinor, then, we might have
two different effective operators. For example, if we consider the
operators that would contribute to $e^+\,e^- \to \tau^-\, e^+$, the
two possibilities are
\begin{align}
{\cal O}_{\tau e} & = \kappa_{\tau e}\,
\left(\bar{\ell}^\tau_L\,\gamma_R\,e_R\right) \left(\bar e_R
\,\gamma_L\, \ell^e_L \right) ,\nonumber\\
{\cal O}_{e \tau}  & = \kappa_{e
\tau}\,\left(\bar{\ell}^e_L\,\gamma_R\,e_R\right) \left(\bar \tau_R
\,\gamma_L\, \ell^e_L \right) \;\;\;,
\end{align}
where $\ell^e$ and $\ell^\tau$ are the leptonic doublets from the
first and third generations, respectively.
 As we see, we find
two different Lorentz structures depending on where we ``insert''
the $\tau$ spinor. Therefore we define two different couplings, each
corresponding to the two possible flavour-violating
interactions.\label{par:lep2}

It will be simpler, however, to parameterize the four-fermion
effective Lagrangian built with the operators above in the manner of
ref.~\cite{BarShalom:1999iy}. For the $e^+e^- l_h l_l$ interaction,
we have
\begin{equation}
{\cal L}_{ee l_l l_h}  =  {1\over\Lambda^2} \sum_{i,j=L,R} \biggl[
V_{ij}^s \left({\bar e} \gamma_\mu \gamma_i e \right) \left( \bar
l_h \gamma^\mu \gamma_j l_l \right) + S_{ij}^s \left( {\bar e}
\gamma_i e \right) \left( \bar l_h \gamma_j l_l \right)  \biggr]
\;\;\; .
\end{equation}
The vector-like ($V_{ij}$) and scalar-like ($S_{ij}$) couplings may
be expressed in terms of the coefficients of the four four-fermion
operators written in eq.~\eqref{eq:ff}~\footnote{The tensor
operators were eliminated using Fierz transformations. A tensor
exchange can thus be hidden in the vector and scalar operators} in
the following manner:

\begin{align}
V_{LL} & = \frac{1}{2} \left( \kappa_{\ell \ell}^{(1)} -
\kappa_{\ell\ell}^{(3)} \right),\nonumber\\
V_{RR} & = \frac{1}{2} \kappa_{e e}  ,\nonumber\\
V_{LR} & = 0,\nonumber\\
V_{RL} & = 0, \nonumber\\
S_{RR} & = 0 ,\nonumber\\
S_{LL} & = 0,\nonumber\\
S_{LR} & = \kappa_{\ell e}^L ,\nonumber\\
S_{RL} & = \kappa_{\ell e}^R .
\end{align}

\subsection{Effective operators generating an $l_h \,l_l$ mixing}

There is a special kind of interaction that corresponds to a
wave-function renormalization, which has its origin in the operator
\begin{align}
{\cal O}_{e \ell \phi} &=
\frac{\delta_{ij}}{\Lambda^2}\,\left(\phi^{\dagger} \phi
\right)\left(\bar{\ell}^i_L\, \,e^j_R \, \phi \right) \;\; ,
\label{eq:delt}
\end{align}
where  $\delta_{ij}$ are complex dimensionless couplings. After
spontaneous symmetry breaking the neutral component of the field
$\phi$ acquires a vev ($\phi_0\,\rightarrow\,\phi_0\,+\,v$, with
$v\,=\, 246/\sqrt{2}$ GeV) and a dimension three operator is
generated which is a flavour violating self-energy like term. In
other words, it mixes, at the level of the propagator, the leptons
of different families. We consider these operators here for
completeness, even though we will show that they have no impact in
the phenomenology whatsoever.

\section{The complete Lagrangian}
\label{sec:lag}
The complete effective Lagrangian can now be written as a function
of the operators defined in the previous section
\begin{align}
{\cal L}\;\; = &\;\; {\cal L}_{ee l_l l_h}\;+\;{\cal O}_{D_e}\;+\;
{\cal O}_{\bar{D}_e}\;+\; {\cal O}_{\phi e}\;+\; {\cal
O}^{(1)}_{\phi
\ell}\;+\; {\cal O}^{(3)}_{\phi \ell}\;+\; \nonumber \\
& \;\; {\cal O}_{e \ell \phi}\;+\;{\cal O}_{eB} \;+\;{\cal O}_{\ell
B}\;+\;{\cal O}_{\ell W}\;+\;{\cal O}_{eB\phi}\;+\;{\cal
O}_{eW\phi}\;+\; \mbox{h.c.} \, \, \, .
\end{align}
This Lagrangian describes new vertices of the form $\gamma
\,\bar{l_h}\, l_l$, $Z \,\bar{l_h}\, l_l$, $\bar{e} \,e
\,\bar{l_h}\, l_l$, $\bar{l_h} l_l$ (and many others) and all of
their charge conjugate vertices. We will also consider an analogous
Lagrangian with flavour indices exchanged - in other words, we will
consider couplings of the form $\eta_{hl}$ and $\eta_{lh}$, for
instance - except for the four-fermion Lagrangian, as was explained
in the previous section. Rather than write the Feynman rules for
these anomalous vertices and start the calculation of all LFV decay
widths and cross sections, we shall use all experimental and
theoretical constraints to reduce as much as possible the number of
independent couplings. After imposing these constraints we will
write the Feynman rules for the remaining Lagrangian and proceed
with the calculation.

\subsection{The constraints from $l_h \rightarrow l_l \,\gamma$}
\label{sec:gamma}

Some of the operators presented in the previous section can be
immediately discarded due to the very stringent experimental bounds
which exist for the decays $\tau \rightarrow \mu\, \gamma$, $\tau
\rightarrow e\, \gamma$ and $\mu \rightarrow e\, \gamma$. The
argument is as follows: all the operators in
eqs.~\eqref{eq:op4-5} contribute to both
$\gamma\,l_h\,l_l$ and $Z\,l_h\,l_l$ interactions, due to the
presence of the gauge fields $B_\mu$ and $W^3_\mu$ in the field
tensors $B_{\mu\nu}$ and $W_{\mu\nu}$ that compose them. Then we can
write, for instance, an operator $\cal{O}_{\ell \gamma}$, given by
\begin{equation}
{\cal O}_{\ell\gamma}\,=
\;\;i\,\frac{\alpha^{\gamma\,L}_{ij}}{\Lambda^2}\,\left(\bar{\ell}^i_L\,
\gamma^\mu\,\partial^\nu\,\ell^j_L\right)\,F_{\mu\nu}
\end{equation}
where $F_{\mu\nu}$ is the usual electromagnetic tensor. This
operator was constructed from both $\cal{O}_{\ell B}$ and ${\cal
O}_{\ell W}$, and the new effective coupling
$\alpha^{\gamma\,L}_{ij}$ is related to $\alpha^{B\,L}_{ij}$ and
$\alpha^{W\,L}_{ij}$ through the Weinberg angle $\theta_W$ by
\begin{equation}
\alpha^{\gamma\, L}_{i j}\;=\;\cos\theta_W \, \alpha^{B\,L}_{i j} -
\sin\theta_W \, \alpha^{W\, L}_{i j} \;\;\; . \label{eq:wein1}
\end{equation}
Following the same exact procedure we can also obtain an operator
${\cal O}_{\ell Z}$, with coupling constant given by
\begin{equation}
\alpha^{Z\, L}_{i j}\;=\;-\sin\theta_W \, \alpha^{B\,L}_{i j} -
\cos\theta_W \, \alpha^{W\, L}_{i j} \;\;\; .
\end{equation}
New operators with photon and Z interactions appear from the
remaining terms, with coupling constants given by
\begin{align}
\alpha^{\gamma\, R}_{i j}\;=\;\cos\theta_W \, \alpha^{B\,R}_{i j} &
& \alpha^{Z\, R}_{i j}\;=\;-\sin\theta_W \, \alpha^{B\,R}_{i j}
\nonumber \\
\beta^{\gamma\, R}_{i j}\;=\;\cos\theta_W \, \beta^{B}_{i j} & &
\beta^{Z\, R}_{i j}\;=\;-\sin\theta_W \, \beta^{B}_{i j} \nonumber \\
\beta^{\gamma\, L}_{i j}\;=\;\cos\theta_W \, \beta^{B}_{i j} -
\sin\theta_W \, \beta^{W}_{i j} & & \beta^{Z\, L}_{i
j}\;=\;-\sin\theta_W \, \beta^{B}_{i j} - \cos\theta_W \,
\beta^{W}_{i j} \;\;\; . \label{eq:wein3}
\end{align}

\begin{figure}[!htbp]
  \begin{center}
  \includegraphics[scale=1.0]{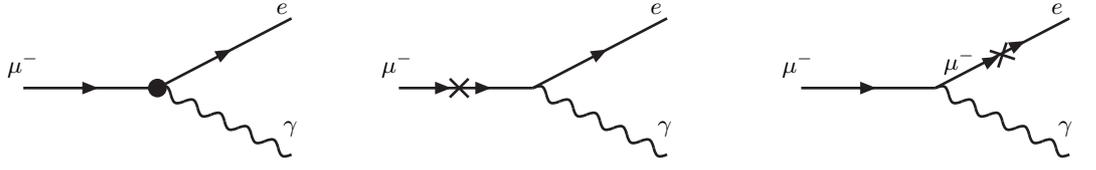}
    \caption{$\mu \rightarrow e \gamma$ with effective anomalous vertices involving the couplings $\alpha$, $\beta$
and $\delta$.}
    \label{fig:decayegamma}
  \end{center}
\end{figure}

It is a simple matter to obtain the Feynman rules for the
$\gamma\,l_h\,l_l$ interactions from the Lagrangian.
%
 In figure~\ref{fig:decayegamma} we present the Feynman diagrams for
the decay $\mu \rightarrow e \gamma$ (in fact, for any decay of the
type $l_h \rightarrow l_l \gamma$) with vertices containing the
effective couplings $\alpha$, $\beta$ and $\delta$.
 Interestingly,
the $\delta$ contributions cancel out, already at the level of the
amplitude\footnote{This cancellation occurs even if we consider the
case where all the leptons have masses.}.
 The
calculation of the remaining diagram is quite simple and gives us
the following expression for the width of the anomalous decay $l_h
\rightarrow l_l \gamma$ in terms of the $\alpha$ and $\beta$
couplings:
\begin{eqnarray}
\Gamma(l_h\rightarrow l_l\gamma)&=&\frac{m_h^3}{64 \pi
\Lambda^4}\biggl[
 m_h^2(|\alpha^{\gamma R}_{lh}+\alpha^{{\gamma R}*}_{hl}|^2+|\alpha^{\gamma L}_{lh}+
 \alpha^{{\gamma L}*}_{hl}|^2) + 16v^2(|\beta^{\gamma}_{lh}|^2+|\beta^{\gamma}_{hl}|^2)\nonumber\\
&+&  8m_hv\: \mbox{Im}(\alpha^{\gamma
R}_{hl}\beta^{\gamma}_{hl}-\alpha^{\gamma R}_{lh}\beta^{\gamma
*}_{hl}-\alpha^{\gamma L}_{lh}\beta^{\gamma *}_{lh}+\alpha^{\gamma L}_{hl}\beta^{\gamma}_{lh})
\biggr] \, \, . \label{eq:wid}
\end{eqnarray}
So, for the decay $\mu \rightarrow e \gamma$, using the data
from~\cite{Yao:2006px}, we get (with $\Lambda$ expressed in TeV)
\begin{eqnarray}
\mbox{BR}(\mu\rightarrow e \gamma)&=& \frac{0.22}{\Lambda^4} \biggl[
(|\alpha^{\gamma R}_{e \mu}+\alpha^{{\gamma R}*}_{\mu
e}|^2+|\alpha^{\gamma L}_{e \mu}+\alpha^{{\gamma L}*}_{\mu e}|^2) +
4.3 \times 10^{7}(|\beta^{\gamma}_{e \mu}|^2+|\beta^{\gamma}_{\mu
e}|^2)\nonumber\\
&+&  1.3 \times 10^{4} \: \mbox{Im}(\alpha^{\gamma R}_{\mu
e}\beta^{\gamma}_{\mu e}-\alpha^{\gamma R}_{e \mu}\beta^{\gamma
*}_{\mu e}-\alpha^{\gamma L}_{e \mu}\beta^{\gamma *}_{e
\mu}+\alpha^{\gamma L}_{\mu e}\beta^{\gamma}_{e \mu}) \biggr] \quad
\, .
\end{eqnarray}
Now, all decays $l_h \rightarrow l_l \gamma$ are severely
constrained by experiment, especially in the case of $\mu
\rightarrow e \gamma$ but also in $\tau \rightarrow e \gamma$ and
$\tau \rightarrow \mu \gamma$. To obtain a crude constraint on the
couplings, we can use the experimental constraint BR$(\mu
\rightarrow e \gamma) < 1.2 \times 10^{-11}$~\cite{Yao:2006px} and
set all couplings but one to zero. With this procedure we get the
approximate bound
\begin{equation}
\frac{|\alpha_{e \mu}^{\gamma\,L,R}|}{\Lambda^2} \, \leq \, 7.4
\times 10^{-6}\;\;\; \mbox{TeV}^{-2}
\end{equation}
and identical bounds for the $\alpha_{\mu e}^{\gamma\,L,R}$
couplings. The constraints on the $\beta$ constants are roughly four
orders of magnitude smaller. Using the same procedure for the two
remaining LFV processes we get
\begin{align}
\frac{|\alpha_{e \tau}^{\gamma\,L,R}|}{\Lambda^2} & \leq
1.6\times 10^{-3} \;\;\; \mbox{TeV}^{-2}\\
\frac{|\alpha_{\mu \tau}^{\gamma\,L,R}|}{\Lambda^2} & \leq  1.3
\times 10^{-3} \;\;\; \mbox{TeV}^{-2}
\end{align}
with the $\beta$ couplings even more constrained in their values.

\begin{figure}[!htbp]
  \begin{center}
  \includegraphics[scale=0.7]{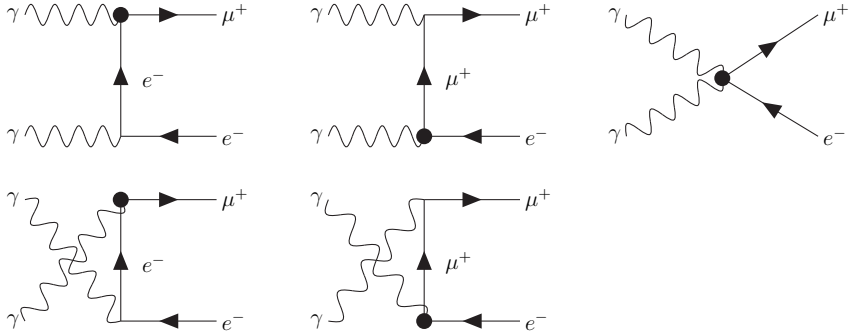}
    \caption{Feynman diagrams for the $\gamma\,\gamma\,\rightarrow\,\mu^+\, e^-$ process.}
    \label{fig:ggll}
  \end{center}
\end{figure}

The experimental bounds on the various branching ratios are so
stringent that they pretty much curtail any possibility of these
anomalous operators having any observable effect on any experiences
performed at the ILC. To see this, let us consider the
flavour-violating reaction $\gamma\,\gamma\,\rightarrow\,l_h\,l_l$,
which in principle could occur at the ILC~\cite{Cannoni:2005gy}.
There are five Feynman diagrams (figure~\ref{fig:ggll}) involving
the $\{\alpha\,,\,\beta\}$ couplings that contribute to this
process.
There are also three diagrams involving the $\delta$ couplings of
eq.~\eqref{eq:delt}, but their contributions (once again) cancel at
the level of the amplitude. The calculation of the cross section for
this process is laborious but unremarkable. The end result, however,
is extremely interesting. The cross section is found to be
\begin{equation}
\frac{d\sigma(\gamma\,\gamma\,\rightarrow \,l_h\,l_l)}{d t} \;
\;=\;\; -\,\frac{4\pi\alpha\,F_{\gamma\gamma}}{{m_h}^3\,s\,
    {( {m_h}^2 - t ) }^2\,t\,{( {m_h}^2 - u ) }^2\,u}\;\Gamma(l_h\,\rightarrow
    \,l_l\,\gamma)\;\;\;,
\label{eq:ggl}
\end{equation}
with a function $F_{\gamma\gamma}$ given by
\begin{align}
F_{\gamma\gamma} &=\;\;\;
      m_h^{10}\,( t + u )- 12 \,m_h^8\,t\,u +
      m_h^6\,( t + u )\,( t^2 + 13\,t\,u + u^2 )  - m_h^4\,t\,u\,( 7\,t^2 + 24\,t\,u + 7\,u^2 )\nonumber \\
      & \;\;\; \;\;\;+\,12\,m_h^2\,t^2\,u^2\,( t + u ) - 6\,t^3\,u^3 \;\;\; .
      \label{eq:fgg}
\end{align}
The remarkable thing about eq.~\eqref{eq:ggl} is the proportionality
of the (differential) cross section to the width of the anomalous
decay $l_h\,\rightarrow\,l_l\,\gamma$, which is to say (modulus the
total width of $l_h$, which is well known), to its branching ratio.
A similar result had been obtained for gluonic flavour-changing
vertices in refs.~\cite{Ferreira:2005dr,Ferreira:2006xe}. Because
the allowed branching ratios for the $l_h\,\rightarrow\,l_l\,\gamma$
are so constrained, the predicted cross sections for the ILC are
extremely small. We have
\begin{align}
\sigma(\gamma\,\gamma\,\rightarrow \,\mu^-\,e^+)\;\sim&
\;\;\;10^{-8}\,\times\,\mbox{BR}(\mu\,\rightarrow \,e\,\gamma) \;\;\mbox{pb} \nonumber \\
\sigma(\gamma\,\gamma\,\rightarrow \,\tau^-\,\mu^+)\;\sim&
\;\;\;10^{-5}\,\times\,\mbox{BR}(\tau\,\rightarrow \,\mu\,\gamma) \;\;\mbox{pb}  \nonumber \\
\sigma(\gamma\,\gamma\,\rightarrow \,\tau^-\,e^+)\;\sim&
\;\;\;10^{-5}\,\times\,\mbox{BR}(\tau\,\rightarrow \,e\,\gamma)
\;\;\mbox{pb} \;\;\; , \label{eq:crl}
\end{align}
with $\sqrt{s}\,=\,1$ TeV. With the current branching ratios of the
order of $10^{-12}$ for the muon decay and $10^{-7}$ for the tau
ones, it becomes obvious that these reactions would have
unobservable cross sections.

Our conclusion is thus that the $\alpha_{ij}^{\gamma}$ and
$\beta_{ij}^{\gamma}$ couplings are too small to produce observable
signals in foreseeable collider experiments. However, both
$\{\alpha^{\gamma}_{i j}\,,\,\beta^{\gamma}_{i j}\}$ and
$\{\alpha^{Z}_{i j}\,,\,\beta^{Z}_{i j}\}$ are written in terms of
the original $\{\alpha^{B,W}_{i j}\,,\,\beta^{B,W}_{i j}\}$
couplings, via coefficients (sine and cosine of $\theta_W$) of order
1. Hence, unless there was some bizarre unnatural cancellation, the
couplings $\{\gamma\,,\,Z\}$ and $\{B\,,\,W\}$ should be of the same
order of magnitude. Since we have no reason to assume such a
cancellation, we come to the conclusion that the $\alpha$ and
$\beta$ couplings are simply too small to be considered interesting.
They will have no bearing whatsoever on anomalous LFV interactions
mediated by the $Z$ boson. From now on, we will simply consider them
to be zero, which means that there will not be any anomalous
vertices of the form $\gamma\ell_i\ell_j.$

\subsection{A set of free parameters}

In the previous section we have presented the complete set of
operators that give contributions to the flavour violating processes
$e^+ e^- \rightarrow l_h l_l$. However, these operators are not,
{\em a priori}, all independent. It can be shown that (see
refs.~\cite{Buchmuller:1985jz,Ferreira:2005dr,Grzadkowski:2003tf,Ferreira:2006in}
for details), for instance, there is a relation between operators of
the types ${\cal O}_{eB\phi}$ and ${\cal O}_{eB}$ and some of the
four-fermion operators, modulo a total derivative. These relations
between operators appear when one uses the fermionic equations of
motion, along with integration by parts. They could be used to
discard operators whose coupling constants are $\alpha$ and $\beta$,
or some of the four-fermion operators. We used this argument to
present the results in
refs.~\cite{Ferreira:2005dr,Ferreira:2006xe,Ferreira:2006in} in a
more simplified fashion. However, in the present circumstances, we
already discarded the $\alpha$ and $\beta$ operators due to the size
of their contributions to physical processes being extremely limited
by the existing bounds on flavour-violating leptonic decays with a
photon. Since we already threw away these two sets of operators, we
are not entitled to use the equations of motion to attempt to
eliminate another.

Notice also that in most of the work that was done with the
effective Lagrangian approach one replaces, at the level of the
amplitude, operators of the type ${\cal O}_{D_e}$ by operators of
the type ${\cal O}_{eZ\phi}$ by using Gordon identities. In fact, it
can be shown that the following relation holds for free fermionic
fields,
\begin{equation}
\bar{e}^i_L\, \partial^{\mu}\,e^j_R\,  = \, m_j \, \bar{e}^i_L\,
\gamma^{\mu} \,e^j_R - \bar{e}^i_L\, \sigma^{\mu\alpha}\
\partial^{\alpha}\,e^j_R \, \, .
\end{equation}
Notice that the use of Gordon identities is not the same thing as
using the field's equations of motion to eliminate operators: in the
latter case, one proves that different operators are related to one
another and use those conditions to choose among them; in the
former, all we are doing is rewriting the amplitude in a different
form. And in our case, this procedure does not bring any
simplification.

Finally, using the equations of motion, a relation can be
established between operators ${\cal O}_{D_e}$ and ${\cal
O}_{\bar{D}_e}$, namely
\begin{equation}
{\cal O}_{D_e} + {\cal O}_{\bar{D}_e} + \left(\bar{\ell}_L \,e_R
\right) \, \left[ \Gamma_e^{\dagger} \bar{e}_R \,\ell_L + \Gamma_u
\bar{q}_L \epsilon \,u_R  +  \Gamma_d^{\dagger} \bar{d}_R \, q_L
\right] \, = \, 0 \;\;\;,
\end{equation}
where the $\Gamma_e$ coefficients are the leptonic Yukawa couplings
and $\epsilon$ the bidimensional Levi-Civita tensor. We see that the
relationship between these two operators involves four-fermion terms
as well. This relation means we can choose between one of the two
operators ${\cal O}_{D_e}$ and ${\cal O}_{\bar{D}_e}$, given that
the four-fermion operators appearing in this expression have already
been considered by us. This means that only one of the $\eta_{ij}^R$
and $\eta_{ij}^L$ couplings will appear in the calculation. We chose
the first one and will drop, from this point onwards, the
superscript ``$R$''. Also, after expanding the operators of
eq.~\eqref{eq:op2}, we see that the $\theta$ couplings always appear
in the same combinations. We therefore define two new couplings,
$\theta_R$ and $\theta_L$, as
\begin{align}
\theta_R & =  \theta_{lh}^R + \theta_{hl}^{R*} ,\\
\theta_L & =  \theta_{lh}^{L(1)} +
\theta_{hl}^{L(1)*}-\theta_{lh}^{L(3)} - \theta_{hl}^{L(3)*} .
\end{align}

\section{Decay widths}\label{sec:dec}

As we said before, all LFV processes are severely constrained by
experimental data. Now that we have settled on a set of anomalous
effective operators, we should first consider what is the effect of
those operators on leptonic LFV decays. The existing data severely
constrains two types of decay: a heavy lepton decaying into three
light ones, $l_h\,\rightarrow\,l\,l\,l$, such as
$\tau^-\,\rightarrow\,e^-\,e^+\,e^-$, and decays of the Z boson to
two different leptons, $Z\,\rightarrow\,l_h\,l_l$ (such as
$Z\,\rightarrow\,\tau^+\,e^-$). Flavour-violating processes
involving neutrinos in the final state (such as, say,
$Z\,\rightarrow\,\nu_\tau\,\bar{\nu}_e$) are not constrained by
experimental data, as they are indistinguishable from the ``normal''
processes.

\begin{figure}[!htbp]
  \begin{center}
      \includegraphics[scale=1.0]{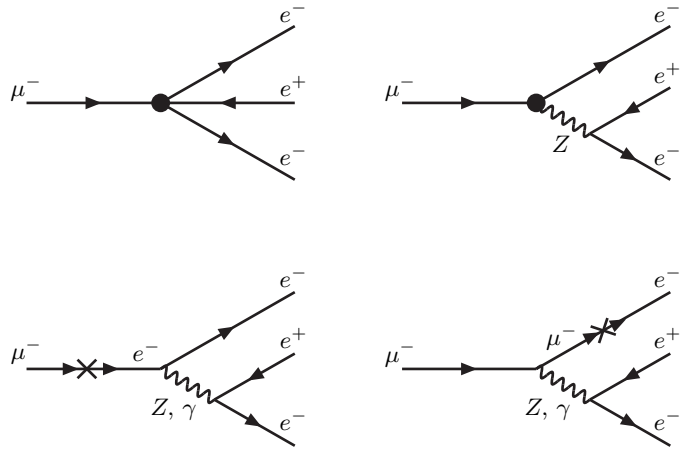}
    \caption{Feynman diagrams for the decay $\mu^- \,\rightarrow \,e^- \,e^+ \,e^-$.}
    \label{fig:decayZ}
  \end{center}
\end{figure}

\begin{figure}[!htbp]
  \begin{center}
    \includegraphics[scale=0.7]{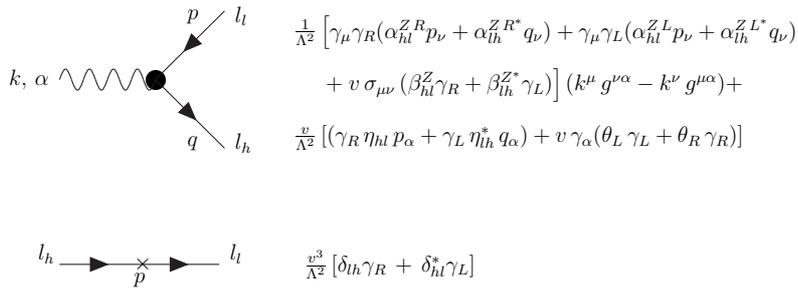}
    \caption{Feynman rules for anomalous $Z \, \bar{l_h} \, l_l$ and
$\bar{l_l} \, l_h$ vertices.}
    \label{fig:feynman}
  \end{center}
\end{figure}

For the 3-lepton decay, there are three distinct contributions,
whose Feynman diagrams are shown in figure~\ref{fig:decayZ}
 for the
particular case of $\mu^- \,\rightarrow \,e^- \,e^+ \,e^-$. As
before, the contributions involving the $\delta$ operators cancel at
the level of the amplitude and have absolutely no effect on the
physics. Using the Feynman rules in figure~\ref{fig:feynman}
 and the
four-fermion Lagrangian we can determine the expression for the
decay $l_h \rightarrow l l l$. Remember that $l$ stands for a
massless lepton whatever its flavour is.
The decay width obtained is the sum of three terms, to wit
\begin{equation}
\Gamma (l_h \,\rightarrow\, l\, l\, l)\;=\;\Gamma_{4f}(l_h
\,\rightarrow\, l\, l\, l)\,+\,\Gamma_Z(l_h \,\rightarrow\, l\, l\,
l)\,+\,\Gamma_{int}(l_h \,\rightarrow\, l\, l\, l)\;\;\; ,
\end{equation}
where $\Gamma_{4f}$ contains the contributions from the four-fermion
graph in figure~\ref{fig:decayZ}, $\Gamma_Z$ those from the Feynman
diagram with a Z boson and $\Gamma_{int}$ the interference between
both diagrams. A simple calculation yields
\begin{align}
\Gamma_{4f} (l_h \,\rightarrow\, l\, l\, l) & =  \frac{m_h^5}{6144
\, \pi^3\, \Lambda^4}
\,\left[|S_{LR}|^2+|S_{RL}|^2+4(|V_{LL}|^2+|V_{RR}|^2)\right]
\nonumber \vspace{0.3cm} \\
\Gamma_Z (l_h \,\rightarrow\, l\, l\, l) & = \frac{(g_A^2+g_V^2)
v^2}{768\, M_z^4 \,\pi^3 \,\Lambda^4} \, \left\{ \left(
|\theta_L|^2+|\theta_R|^2 \right) v^2 m_h^5 + \frac{1}{2} \mbox{Re}
\left(\eta_{lh} \theta_L^* +  \eta_{hl} \theta_R^* \right) v\, m_h^6
\right.
\nonumber \vspace{0.3cm} \\
&\left.  + \frac{m_h^7}{10 M_z^2} \left[ (|\eta_{lh}|^2  +
|\eta_{hl}|^2) M_z^2 + 6 (|\theta_L|^2  +  |\theta_R|^2)\, v^2
\right] \right\} \nonumber \\
\Gamma_{int} (l_h \,\rightarrow\, l\, l\, l) & = \frac{v^2 \,
m_h^5}{768 M_z^2 \pi^3 \Lambda^4} \,  \left[ \left( 1+ \frac{3
m_h^2}{10 M_z^2} \right) \biggl\{  \left( g_V +g_A \right) \mbox{Re}
\left( \theta_L V_{LL}^* \right)   \right.
\nonumber \vspace{0.3cm} \\
& +    \left( g_V -g_A \right) \mbox{Re} \left( \theta_R V_{RR}^*
\right) \biggr\}  - \frac{m_h}{4v} \left( 1+ \frac{m_h^2}{5 M_z^2}
\right) \biggl\{ \left( g_V +g_A \right) \mbox{Re}
\left( \eta_{lh} V_{RR}^*  \right)   \nonumber \vspace{0.3cm} \\
& + \left( g_V -g_A \right) \mbox{Re} \left( \eta_{hl} V_{LL}
\right) \biggr\} \biggr]  \;\;\; . \label{eq:expa}
\end{align}
where
\begin{align}
g_V & = -\,\frac{e}{\sin\theta_W\,\cos\theta_W}\left(-\frac{1}{4} +
\sin\theta_W^2\right),\\
g_A &= \frac{e}{4\sin\theta_W\,\cos\theta_W}
\end{align}
and $e$ is the elementary electric charge. An important remark about
these results: they are not, in fact, the {\em exact} expressions
for the decay widths. The full expressions for $\Gamma_Z (l_h
\,\rightarrow\, l\, l\, l)$ and $\Gamma_{int} (l_h \,\rightarrow\,
l\, l\, l)$ are actually the sum of a logarithmic term and a
polynomial one. However, it so happens that the first four terms of
the Taylor expansion of the logarithm in $m_h/M_z$ cancel the
polynomial exactly. The expressions of eq.~\eqref{eq:expa} are
therefore the first surviving terms of that Taylor expansion, and
constitute an excellent approximation to the exact result, and one
that is much easier to deal with numerically (the cancellation
mentioned poses a real problem in numerical calculations).

As for the LFV decays of the Z-boson, there is an extensive
literature on this
subject~\cite{Delepine:2001di,*Illana:2000ic,*FloresTlalpa:2001sp,*Perez:2003ad}.
There are, of course, no four-fermion contributions to this decay
width, and a simple calculation provides us the following
expression:
\begin{align}
\Gamma (Z \,\rightarrow\, l_h\, l_l) & = \frac{(M_z^2-m_h^2)^2
\,v^2}{128\, M_z^5 \,\pi \,\Lambda^4} \, \left[ (M_z^2-m_h^2)^2
(|\eta_{hl}|^2+|\eta_{lh}|^2)
\right. \nonumber\\
& + 4\, (m_h^2+ 2 M_z^2) v^2 (|\theta_{L}|^2+|\theta_{R}|^2) +
\left. \,4\, m_h\, (m_h^2-M_z^2)\, v \,\mbox{Re} \left( \theta_L
\eta_{hl}+ \theta_R \eta_{lh}^* \right) \right] .
\end{align}

\section{Cross Sections}
\label{sec:cross}
\begin{figure}[!htbp]
  \begin{center}
  \includegraphics[scale=1.0]{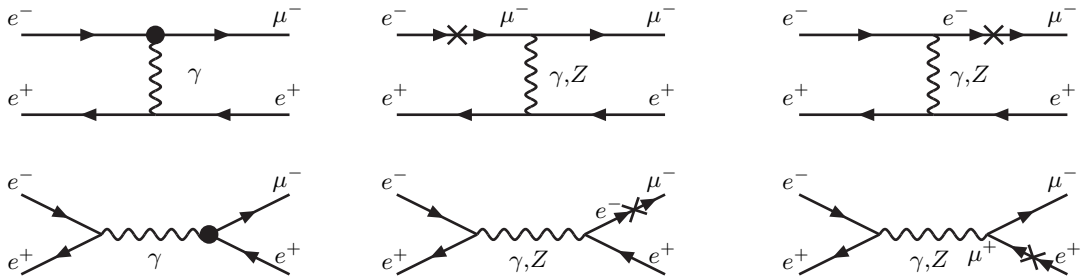}
    \caption{Feynman diagrams describing the process $e^+ e^- \rightarrow \mu^- e^+$}
    \label{fig:cross}
  \end{center}
\end{figure}

\begin{figure}[!htbp]
  \begin{center}
  \includegraphics[scale=1.2]{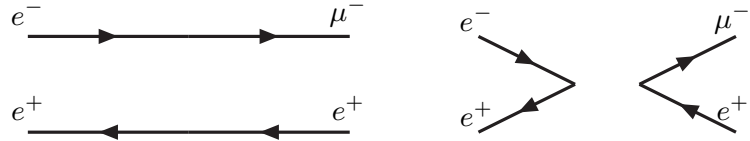}
    \caption{Interpretation of the four-fermion terms
             contributing to the process $e^+ e^- \rightarrow \mu^- e^+$
             in terms of currents; notice the analog of a $t$ channel
             and an $s$ one.}
    \label{fig:cross4f}
  \end{center}
\end{figure}

In this section we will present expressions for the cross sections
of various LFV processes that may occur at the ILC. There are three
such processes, namely:
\begin{enumerate}
\item $e^+ e^- \rightarrow \mu^- e^+;$
\item $e^+ e^- \rightarrow \tau^- e^+;$
\item $e^+ e^- \rightarrow \tau^- \mu^+,$
\end{enumerate}
The respective charge-conjugates must be included, as well. We have
calculated all cross sections keeping both final state masses.
However, given the energies involved, the contributions to the cross
sections which arise from the lepton masses are extremely small, and
setting them to zero is an excellent approximation. We thus present
all formulae with zero leptonic masses, as they are much simpler
than the complete expressions. In figures~\ref{fig:cross}
and~\ref{fig:cross4f} we present all diagrams that contribute to the
process $e^+ e^- \rightarrow \mu^- e^+.$
A brief word about our conventions. There are two types of LFV
production cross sections, corresponding to different sets of
Feynman diagrams. In the case of process (1), we see from
figures~\ref{fig:cross} and~\ref{fig:cross4f} that the reaction
can proceed through both a $t$-channel and an $s$-channel - this is
obvious for the diagrams involving the exchange of a photon or a $Z$
boson. For the four-fermion channels less so, but
figure~\ref{fig:cross4f} illustrates the $t$ and $s$-channel
analogy. Depending on the ``location'' of the incoming electron
spinor in the operators of eq.~\eqref{eq:ff}, we can interpret those
operators as two fermionic currents interacting with one another,
that interaction is obviously analog to the two different channels.
Process (2) has diagrams identical to those of process (1). Process
(3), however, can only occur through the $s$ channel - that is
obvious once one realizes that for process (3) there is no positron
in the final state. In fact for process (3) there are only
``s-channel'' contributions from the four-fermion operators.

A simple way of condensing the different four-fermion cross sections
into a single expression is to adopt the following convention: we
will include indices ``s'' and ``t'' in the four-fermion couplings.
If we are interested in the cross sections for processes (1) and (2)
- which occur through both $s$ and $t$ channels - then all ``s''
couplings will be equal to the ``t'' ones. If we wish to obtain the
cross section for process (3) (which only has $s$ channels) we must
simply set all couplings with a ``t'' index to zero. We have further
considered the likely possibility that in the ILC one may be able to
polarize the beams of incoming electrons and
positrons~\cite{MoortgatPick:2005cw}. Thus, $\sigma_{IJ}$ represents
the polarized cross section for an $I$ polarized electron and a $J$
polarized positron, with $\{I\,,\,J\}\,=\,\{R\,,\,L\}$ - that is,
beams with a right-handed polarization or a left-handed one. The
explicit expressions for the four-fermion differential cross
sections are then given by
\begin{align}
\frac{d \sigma_{LL}}{dt} \, = & \, \frac{1}{16 \pi s^2 \Lambda^4}
\left[ 4 u^2 |V_{LL}^s + V_{LL}^t|^2 + t^2 (4|V_{LR}^s|^2 +
|S_{RL}^t|^2)\right]
\nonumber \\
\frac{d \sigma_{RR}}{dt} \, = & \, \frac{1}{16 \pi s^2 \Lambda^4}
\left[ 4
u^2 |V_{RR}^s + V_{RR}^t|^2 +  t^2 (4|V_{RL}^s|^2 + |S_{LR}^t|^2) \right] \nonumber \\
\frac{d \sigma_{LR}}{dt} \, = & \, \frac{1}{16 \pi s^2 \Lambda^4}
\left[ u^2 |S_{LL}^s + S_{LL}^t|^2 + s^2 (4|V_{RL}^t|^2 +
|S_{LR}^s|^2) \right]
\nonumber \\
\frac{d \sigma_{RL}}{dt} \, = & \, \frac{1}{16 \pi s^2 \Lambda^4}
\left[ u^2 |S_{RR}^s + S_{RR}^t|^2 + s^2 (4|V_{LR}^t|^2 +
|S_{RL}^s|^2) \right]
\label{eq:sig4f}
\end{align}
See Appendix~\ref{sec:cross2} for the full calculation. The unpolarized cross
section is obviously the averaged sum over the four terms of
eq.~\eqref{eq:sig4f}. To re-emphasize, the four-fermion cross
section for processes (1) and (2) is obtained from this expression
by setting all ``s'' couplings equal to the ``t'' ones; and to
obtain the cross section for process (3) one must simply set all
``t'' couplings to zero.
The total cross sections for each of the processes are then given by
\begin{align}
\sigma^{(1,2)}(e^- e^+ \rightarrow l_h e^+) &=\;\; \sigma_Z^{(1,2)}
\,+\, \sigma_{4f}^{(1,2)}\,+\, \sigma_{int}^{(1,2)}
\nonumber \\
\sigma^{(3)}(e^- e^+ \rightarrow \tau^- \mu^+) &=\;\; \sigma_Z^{(3)}
\,+\, \sigma_{4f}^{(3)} \,+\, \sigma_{int}^{(3)}\;\;\; ,
\end{align}
where $\sigma_Z$ is the cross section involving only the anomalous
$Z$ interactions of figure~\ref{fig:feynman}, $\sigma_{4f}$ the
four-fermion cross section - whose calculation we already explained
- and $\sigma_{int}$ the interference between both of these. The
$\delta$ couplings also present in figure~\ref{fig:feynman} end up
not contributing at all to the physical cross sections, once again.
For completeness, then, the remaining terms in the differential
cross section for processes (1) and (2) are given by
\begin{align}
\frac{d\sigma_{Z}^{(1,2)}}{dt}\ & =  \frac{- v^2}{32\,\pi
\,\Lambda^4\, {\left( M_z^2 - s \right) }^2\,s^2\,{\left(
M_z^2 - t \right) }^2}\nonumber\\
& \times v^2 \, \left[F_1 (g_A,g_V) \, |\theta_L|^2 + F_1 (g_A,-g_V)
|\theta_R|^2 \right] \, + \, F_2 (g_A,g_V) \, |\eta_{lh}|^2 + F_2
(g_A,-g_V) |\eta_{hl}|^2
\end{align}
with
\begin{align}
F_1 (g_A,g_V) & = 2\,\left\{( g_A + g_V)^2 \left[ s\, t \, (2 \,
M_z^4+2\, u \, M_z^2+s^2+t^2) - u \, M_z^2 \, (u\, M_z^2+2\, s^2+2
\, t^2) \right]\right.
\nonumber \\
& + \left. 2 \, ( g_A^2 + g_V^2) u \, (t^3+s^3) + (g_A - g_V)^2 s \,
u^2 \, t \right\} \nonumber \\
F_2 (g_A,g_V) & = \, - t \, u \, s \, \left[ ( g_A^2 + g_V^2)(3 \,
M_z^4+3\, u \, M_z^2+s^2+t^2+ s \, t)\right.\nonumber\\
& \left. + 2 \, g_A \, g_V \, \left( M_z^2 - s \right) \,\left(
M_z^2 - t \right) \right]\;\;\; .
\end{align}
The interference term is  given by
\begin{align}
\frac{d\sigma_{int}^{(1,2)}}{dt} & = \frac{(t-s) \, v^2 }{16\,\pi
\,\Lambda^4\,
   \left( M_z^2 - s \right) \,s^2\,\left( M_z^2 - t \right)}
\nonumber \\
& \times \biggl[  \left(  g_A \,
         \mbox{Re} \left( \theta_L S_{LR}^* - \theta_R S_{RL}^* \right)   +
          g_V \, \mbox{Re} \left( \theta_L S_{LR}^* + \theta_R
S_{RL}^* \right)    \right) \,
        \left( s\,t + \left( - M_z^2 + s + t \right) \,u \right)
\nonumber \\
&       +\,4\, ( g_A - g_V) \,(s+t) \,u \, \mbox{Re} \left( \theta_L
V_{LL}^* \right) -
     4\,( g_A + g_V) \,(s+t) \,u  \, \mbox{Re} \left( \theta_R
V_{RR}^* \right) \biggr]
\end{align}
For process (3), we have
\begin{align}
\frac{d\sigma_{Z}^{(3)}}{dt} & = \frac{- v^2 }{32\,\pi \,\Lambda^4\,
   {\left(  M_z^2 - s \right) }^2\,s^2} \biggl\{ 2\, v^2 \left[
   (g_A^2+g_V^2)\, (2tu-s^2) \, (|\theta_L|^2+|\theta_R|^2) \right.
\nonumber \\
& \left. + \, 2 \, g_A \, g_V \,s (t-u)\, (|\theta_L|^2-
|\theta_R|^2) \right] - (g_A^2+g_V^2) \, t \, u \, s \,
(|\eta_{hl}|^2+|\eta_{lh}|^2) \biggr\}
\end{align}
and finally, the interference terms are
\begin{equation}
\frac{d\sigma_{int}^{(3)}}{dt}=\frac{(s+t) \, u \, v^2 }{8\,\pi
\,\Lambda^4\, \left( M_z^2 - s \right) \,s^2} \biggl[ ( g_V - g_A)
\, Re \left( \theta_L V_{LL}^* \right) +
   ( g_V + g_A)  \, \mbox{Re} \left( \theta_R
V_{RR}^* \right) \biggr]\;\;\; .
\label{eq:int}
\end{equation}
At this point we must remark on the different energy behavior that
these various terms have. Once integrated in $t$, the four-fermion
terms grow linearly with $s$, whereas those arising from the
anomalous $Z$ couplings have a much smoother evolution with $s$ -
whereas the first ones diverge as $s\,\rightarrow\,\infty$, the
second ones tend to zero. See Appendix~\ref{sec:cross2} for the expressions
of the integrated cross sections. This could be interpreted as a
clear dominance of the four-fermion terms over the remaining
anomalous couplings. However, we must remember that we are working
in a non-renormalizable formalism. We know, from the beginning, that
these operators only offer a reasonable description of high-energy
physics up to a given scale, of the order of $\Lambda$. The
dominance of the four-fermion cross section must therefore be
carefully considered - it may simply happen, as there is nothing
preventing it, that the four-fermion couplings of eq.~\eqref{eq:ff}
are much smaller in size than the $Z$ boson ones of
figure~\ref{fig:feynman}.

As we saw, the $\delta$ couplings end up not contributing to either
decay widths or cross sections (and this is true regardless of
whether the light leptons are considered massless or not). As we
mentioned before, their inclusion could be interpreted as an
on-shell renormalization of the leptonic propagators. On that light,
their cancellation suggests that the effective operator formalism is
equivalent to an on-shell renormalization scheme. This is further
supported by the fact that the list of effective operators of
ref.~\cite{Buchmuller:1985jz} was obtained by using the fields'
equations of motion to simplify several terms. However, we must
mention that at least in some Feynman diagrams (some of those
contributing to $\gamma\,\gamma\,\rightarrow\,l_h\,l_l$, for
instance), the ``$\delta$-insertions'' were made in {\em internal}
fermionic lines, so that this cancellation is not altogether
obvious.

\subsection{Asymmetries}
In a collider with polarized beams, asymmetries can play a major
role in the determination of flavour violating couplings. A great
advantage of using these observables is that, as will soon become
obvious, all dependence on the scale of unknown physics, $\Lambda$,
vanishes due to their definition. There is a strong possibility that
the ILC could have both beams polarized, therefore allowing a number
of different possibilities for the polarization of each beam, and
consequently for the asymmetries that could be measured. For a more
detailed study see~\cite{MoortgatPick:2005cw}. A particulary
appealing situation is found when the contributions from the $Z$
boson anomalous couplings are not significant when compared with the
four-fermion ones. In this case the study of asymmetries would allow
us, in principle, to determine each four-fermion coupling
individually. We will now concentrate on one of the most feasible
scenarios, which is to have a polarized electron beam and an
unpolarized positron beam. We will take both the right-handed and
left-handed polarizations to be 100\%, which is obviously what
is expected to occur (recent studies show that a 90 \% polarization
is attainable)~\cite{MoortgatPick:2005cw}. The differential cross
sections for left-handed ($P_{e^-}=-1$) and right-handed
($P_{e^-}=+1$) polarized electrons are
\begin{align}
\frac{d \sigma_{L}}{dt} \, = & \, \frac{1}{16 \pi s^2 \Lambda^4}
\left( 4 u^2 |V_{LL}^s + V_{LL}^t|^2 + t^2 |S_{RL}^t|^2 + s^2
|S_{LR}^s|^2 \right)
\nonumber \\
\frac{d \sigma_{R}}{dt} \, = & \, \frac{1}{16 \pi s^2 \Lambda^4}
\left( 4 u^2 |V_{RR}^s + V_{RR}^t|^2 +  t^2 |S_{LR}^t|^2 + s^2
|S_{RL}^s|^2\right) \, \, .
\end{align}
Two forward-backward asymmetries for the left-handed and
right-handed polarized cross sections can now be defined as
\begin{equation}
A_{FB,L (R)} \, = \, \frac{\int_{0}^{\pi/2} d \sigma_{L (R)}
(\theta) \, - \, \int_{\pi/2}^{\pi} d \sigma_{L
(R)}(\theta)}{\sigma_{L (R)}}
\end{equation}
and we can also define a left-right asymmetry, given by
\begin{equation}
A_{LR} \, = \, \frac{\sigma_L -
\sigma_R}{\sigma_{L}+\sigma_{R}}\;\;\; ,
\end{equation}
where $\sigma_{L (R)}$ is the total cross section for a left-handed
(right-handed) polarized electron beam. Note that we have assumed
that the polarization of the final state particles is not measured.
Otherwise we could get even more information by building an
asymmetry related to the measured final state polarizations. Using
the expressions on Appendix~\ref{sec:cross2} it is simple to find, for these
asymmetries, the following expressions:
\begin{equation}
A_{FB,L} \, = \, \frac{12 \, |V_{LL}^s + V_{LL}^t|^2 - 3 \,
|S_{RL}^t|^2}{16 \, |V_{LL}^s + V_{LL}^t|^2 + 4 \, |S_{RL}^t|^2+ 12
\, |S_{LR}^s|^2}
\label{eq:afbl}
\end{equation}
and
\begin{equation}
A_{FB,R} \, = \, \frac{12 \, |V_{RR}^s + V_{RR}^t|^2 - 3 \,
|S_{LR}^t|^2}{16 \, |V_{RR}^s + V_{RR}^t|^2 + 4 \, |S_{LR}^t|^2+ 12
\, |S_{RL}^s|^2}  \, \, .
\label{eq:afbr}
\end{equation}
Finally, the left-right asymmetry reads
\begin{equation}
A_{LR} \, = \, \frac{|V_{LL}^s + V_{LL}^t|^2 - |V_{RR}^s +
V_{RR}^t|^2+ |S_{RL}^t|^2 - |S_{LR}^t|^2 + 3 \,
(|S_{LR}^s|^2-|S_{RL}^s|^2)}{|V_{LL}^s + V_{LL}^t|^2 + |V_{RR}^s +
V_{RR}^t|^2+ |S_{RL}^t|^2 + |S_{LR}^t|^2 + 3 \,
(|S_{LR}^s|^2+|S_{RL}^s|^2)} \, ,
\label{eq:alr}
\end{equation}
which has no dependence on $\Lambda$. Notice that all of these
expressions assume an unpolarized positron beam, and a completely
polarized electron beam, either left- or right-handed. If the
electron beam is not perfectly polarized, but instead has a
percentage of polarization $P_{e^-}$, we can still write
\begin{equation}
\sigma_{P_{e^-}} \, =   \, \sigma_0 \left[ 1-P_{e^-} A_{LR} \right]
\end{equation}
with $\sigma_0 \, = \, (\sigma_{L}+\sigma_{R})/4$. So if in reality
we only have access at the ILC to beams with +80 \% (- 80 \%)
polarization  we could still use them to determine $\sigma_0$ and
$A_{LR}$. If we had access to a positron polarized beam, we could
then write a similar expression for the cross section obtained from
the polarized positrons. Notice that $A_{LR}$ would be different -
the indices left and right would then refer to the positron and not
to the electron.

The most interesting possibility is, of course, when both beams are
polarized, with different percentages, $P_{e^-}$ and $P_{e^+}$. We
could then perform experiments where the four different combinations
of beam polarizations were used. The resulting cross section would
be
\begin{align}
\sigma_{P_{e^-}P_{e^+}} \, = &  \, \frac{1}{4} \biggl[
(1+P_{e^-})(1+P_{e^+}) \sigma_{RR}\, + \, (1-P_{e^-})(1-P_{e^+})
\sigma_{LL}+
\nonumber \\
& (1+P_{e^-})(1 - P_{e^+})  \sigma_{RL} \, + \,
(1-P_{e^-})(1+P_{e^+}) \sigma_{LR} \biggr]\;\;\; .
\end{align}
As such, we would be able to determine $\sigma_{RR}$, $\sigma_{LL}$,
$\sigma_{RL}$ and $\sigma_{LR}$ - and consequently each of the four
four-fermion couplings, $V_{LL}$, $V_{RR}$, $S_{RL}$ and $S_{LR}$.

\section{Results and Discussion}
\label{sec:res}
In the previous sections we computed cross sections and decay widths
for several flavour-violating processes. We will now consider the
possibility of their observation at the ILC. To do so we will use
one set of the proposed parameters \cite{MoortgatPick:2005cw} for
the ILC, i.e., a center-of mass energy of $\sqrt{s} = 1$ TeV and an
integrated luminosity of ${\cal L} = 1$ ab$^{-1}$. At this point we
remark that, other than the experimental constraints on the
flavour-violating decay widths computed in sec.~\ref{sec1:lepton}
(see table~\ref{tab:dec} in the page~\pageref{tab:dec}), we have no
bounds on the values of the anomalous couplings.
The range of values chosen for each of the coupling constants was
$10^{-4} \leq |a/\Lambda^2| \leq 10^{-1}$, where $a$ stands for a
generic coupling and $\Lambda$ is in TeV. For $a \approx 1$ the
scale of new physics can be as large as 100 TeV. This means that if
the scale for LFV is much larger than 100 TeV, it will not be probed
at the ILC unless the values of coupling constants are unusually
large. The asymmetry plots are not affected by this choice as
explain before.

\begin{figure}[!htbp]
  \begin{center}
    \includegraphics[scale=0.7]{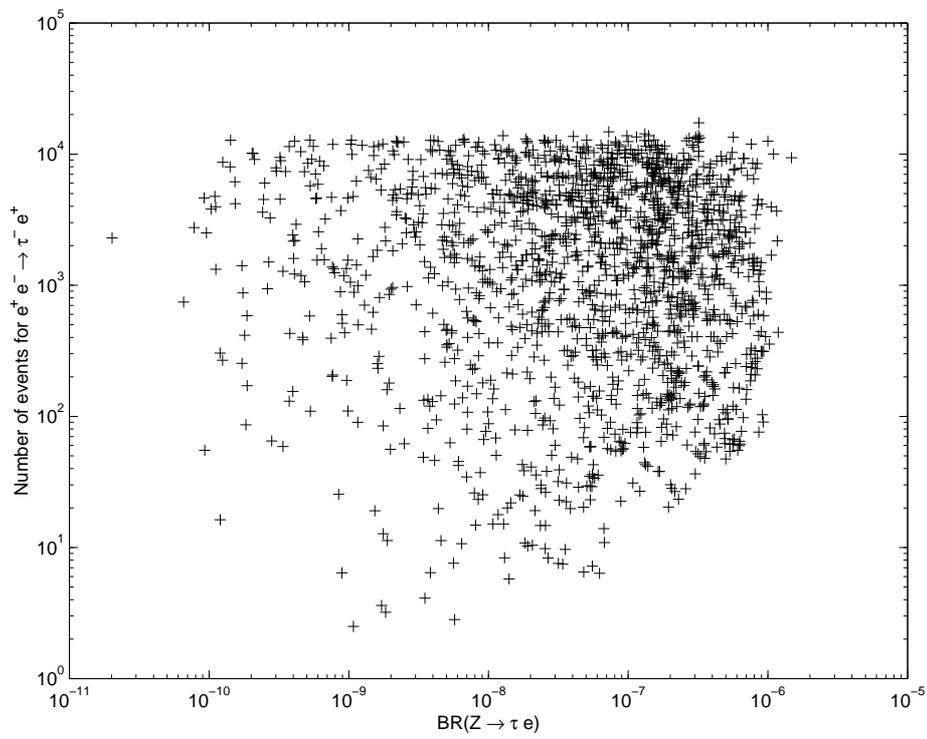}
    \caption{Number of expected events at the ILC for the reaction $e^+\,e^-\,\rightarrow\,\tau^-\,
    e^+$, with a center-of-mass energy of 1 TeV and a total luminosity of 1 ab$^{-1}$.}
    \label{fig:plot2}
  \end{center}
\end{figure}

We will therefore generate random values for all anomalous couplings
(four-fermion and $Z$ alike), and discard those combinations of
values of the couplings for which the several branching ratios we
computed earlier are larger than the corresponding experimental
upper bounds from table~\ref{tab:dec}. This procedure allows for the
possibility that one set of anomalous couplings (the $Z$ or
four-fermion ones) might be much larger than the other. When an
acceptable combination of values is found, it is used in
expressions~\eqref{eq:sig4f}-~\eqref{eq:int} to compute the value of
the flavour-violating cross section. In figure~\ref{fig:plot2} we
plot the number of events expected at the ILC for the process
$e^+\,e^-\,\rightarrow\,\tau^-\,e^+$, in terms of the branching
ratio BR$(Z\,\rightarrow\,\tau\,e)$. To obtain the points shown in
this graph, we demanded that the values of the effective couplings
were such that all of the branching ratios for the decays of the
$\tau$ lepton into three light leptons and
BR$(Z\,\rightarrow\,\tau\,\mu)$ were smaller than the experimental
upper bounds on those quantities shown in table~\ref{tab:dec}. We
observe that, even for fairly small values of the $\tau$
flavour-violating decays ($10^{-9}- 10^{-6}$), there is the
possibility of a large number of events for the anomalous cross
section.

\begin{figure}[!htbp]
  \begin{center}
    \includegraphics[scale=0.7]{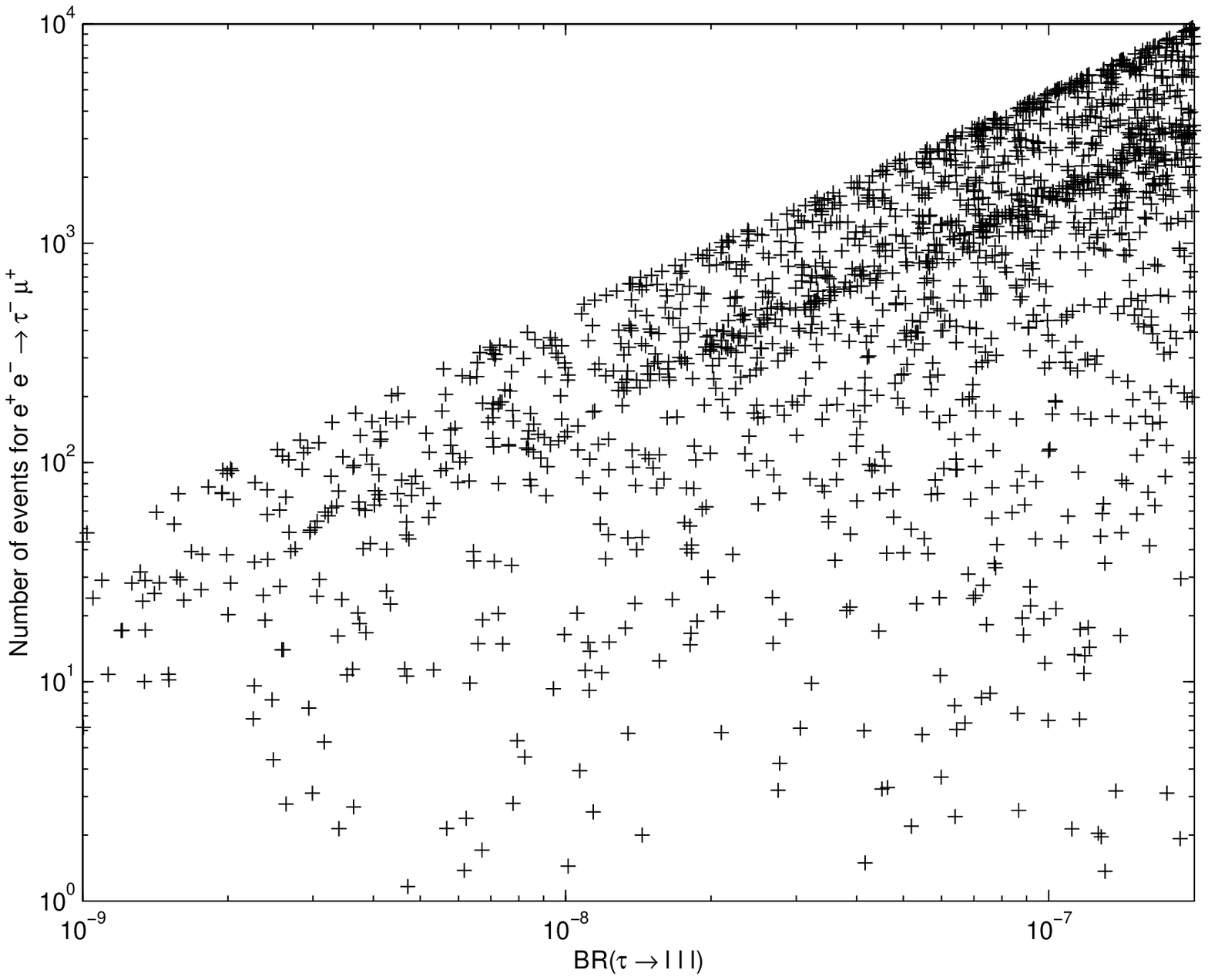}
    \caption{Number of expected events at the ILC for the reaction $e^+\,e^-\,\rightarrow\,\tau^-\,
    \mu^+$, with a center-of-mass energy of 1 TeV and a total luminosity of 1 ab$^{-1}$.}
    \label{fig:plot3}
  \end{center}
\end{figure}

\begin{figure}[!htbp]
  \begin{center}
    \includegraphics[scale=0.7]{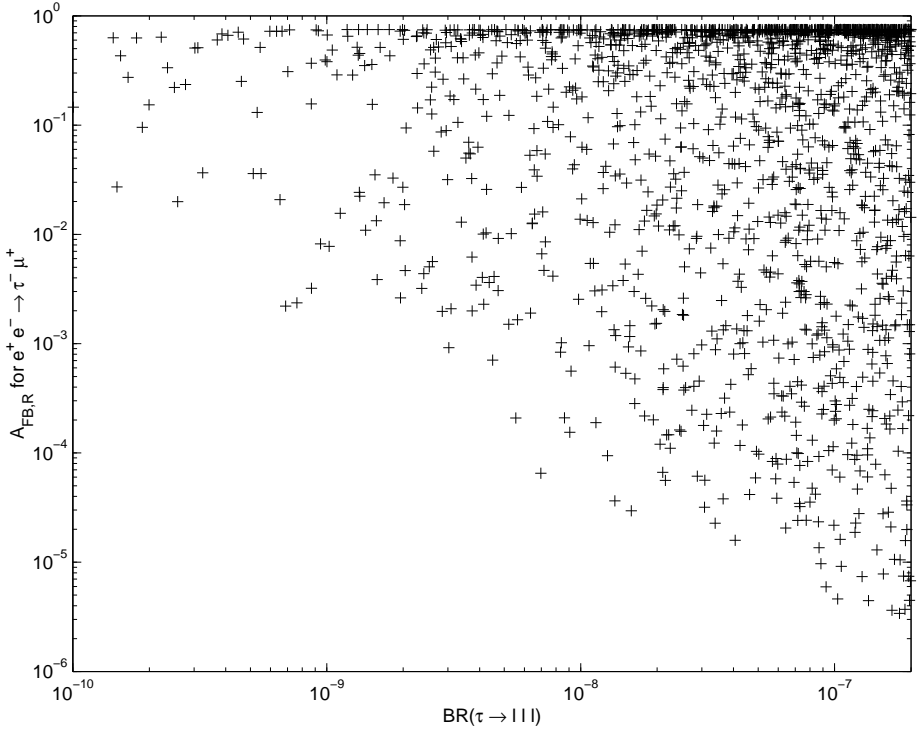}
    \caption{$A_{FB,R}$ asymmetry for the process $e^+\,e^-\,\rightarrow\,
    \tau^-\,\mu^+$ versus $BR(\tau\,\rightarrow\,l\,l\,l)$}
    \label{fig:asy}
  \end{center}
\end{figure}

By following the opposite procedure -- requiring first that the
branching ratios BR$(Z\,\rightarrow\,\tau\,e)$ and
BR$(Z\,\rightarrow\,\tau\,\mu)$ be according to the experimental
values, and letting BR$(\tau\,\rightarrow\,l\,l\,l)$ free, where $l$
is either an electron or a muon -- we obtain the plot shown in
figure~\ref{fig:plot3}. This time we analyse the process
$e^+\,e^-\,\rightarrow\,\tau^-\, \mu^+$, but a similar plot is found
for $e^+\,e^-\,\rightarrow\,\tau^-\, e^+$. The number of events
rises sharply with increasing branching ratio of $\tau$ into three
leptons. It is possible to discern a thin ``band'' of events in the
middle of points of figure~\ref{fig:plot3}, rising linearly with
BR$(\tau\,\rightarrow\,l\,l\,l)$. This ``band'' corresponds to
events for which the four-fermion couplings are dominant over $Z$
events. In that case, they dominate both
BR$(\tau\,\rightarrow\,l\,l\,l)$ and
$\sigma(e^+\,e^-\,\rightarrow\,\tau^-\, \mu^+)$, and the larger
 one is, the larger the other will be -- which explains the linear
growth of this subset of points in the plot of
figure~\ref{fig:plot3}. This ``isolated'' contribution from the
four-fermion terms is not visible in figure~\ref{fig:plot2} since
the branching ratios of the $Z$ decays are independent of those same
couplings. Finally and for completeness, in figure~\ref{fig:asy} we
show the values of the asymmetry
coefficient $A_{FB,R}$ defined in~\eqref{eq:afbr}, for the process
$e^+\,e^-\,\rightarrow\,\tau^-\, \mu^+$, versus the three-lepton
decay of the $\tau$. A similar plot is obtained for the asymmetry
$A_{FB,L}$. We observe a fairly uniform dependence on the branching
ratio $BR(\tau\,\rightarrow\,l\,l\,l)$, which is to say, on the
values of the four-fermion couplings.\label{par:lep3}

\begin{figure}[!htbp]
  \begin{center}
    \includegraphics[scale=0.8]{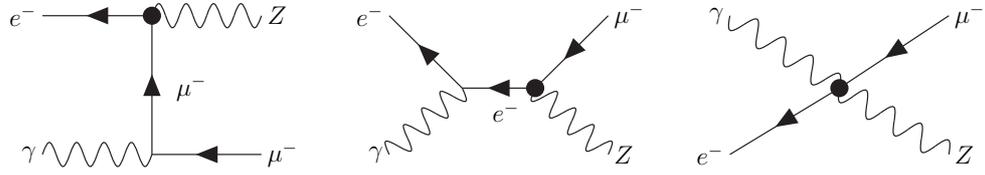}
    \caption{Feynaman diagrams to $e^-\,\gamma\,\rightarrow\,\mu^-\, Z.$}
    \label{fig:gemuZ}
  \end{center}
\end{figure}

Finally, we also considered another possible process of LFV, namely
$\gamma\,e^-\,\rightarrow\,\mu^-\,Z$. There are three Feynman
diagrams contributing to this process, according the
figure~\ref{fig:gemuZ},
  one of which involving a quartic vertex which emerges from the
effective operators of eqs.~\eqref{eq:op1} and~\eqref{eq:op2}.
 This process might occur at
the ILC, if we consider the almost-collinear photons emitted by the
colliding leptons, well described by the so-called equivalent photon
approximation (EPA)~\cite{Brodsky:1971ud,*Gordon:1994wu}. An
estimate of the cross section for this process, however, showed it
to be much lower than the remaining ones we considered in this
thesis. This is due to the EPA introducing an extra electromagnetic
coupling constant into the cross section, and also to the fact that
the final state of this process includes at least three particles
(one of the beam particles ``survives'' the interaction) -- thus
there is, compared to the other processes which have only leptons in
the final state, an additional phase space suppression. Notice,
however, that an optional upgrade for the ILC is to have $e\,\gamma$
collisions, with center-of-mass energies and luminosities similar to
those of the $e^+\, e^-$ mode, so this cross section might become
important.

The flavour-violating channels are experimentally interesting, as
they present a final state with an extremely clear signal, which can
be easily identified. The argument is that the final state will
always present two very energetic leptons of different flavour, more
to the point, an electron and a muon. LFV can be seen in one of the
three channels $e^+\, e^-\rightarrow\mu^-\, e^+,$ $e^+\,
e^-\rightarrow\tau^-\, e^+,$ $e^+\, e^-\rightarrow\tau^-\, \mu^+,$
and charge conjugate channels. The first channel is the best one,
with the two leptons back to back and almost free of backgrounds.
For the other production processes, we may ``select'' the decays of
the tau that best suit our purposes: for the second we should take
the tau decay $\tau^-\rightarrow\mu^-\bar\nu_\mu\nu_\tau,$ and for
the third process, $\tau^-\rightarrow e^-\bar\nu_e\nu_\tau.$ The
branching ratios for both of these tau decays are around 17\%, so
the loss of signal is affordable. The conclusion is that, for every
lepton flavour-violating process, one can always end up with a final
state with an electron and a muon. If the ILC detectors have superb
detection performances for these particles, then the odds of
observing violation of the leptonic number at the ILC, if those
processes do exist, seem reasonable.

Clearly, our prediction that significant numbers of anomalous events
may be produced at the ILC needs to be further investigated
including the effects of a real detector. Notice also that due to
{\it beamstrahlung} effects which reduce the effective beam energy,
the total LFVevent rates might be reduced, specially in the case of
the four-fermion cross sections, which increase with $s.$ Also, one
must take into account the many different backgrounds that could
mask our signal. And the fact that, even in the best case scenario,
only a few thousand events are produced with an integrated
luminosity of $1\; ab^{-1},$ could limit the signal-to-background
ratio.
  We can show that with some very simple cuts most of the background
can be eliminated. Because of the weaker experimental constraints on
processes involving $\tau$ leptons, the most promising LFV reactions
at the linear collider are $\mu\, e$ and $\mu\,\tau$ production. For
illustrative purposes we will study the backgrounds to the LFV
process $e^+\, e^-\rightarrow\tau^+\, e^-\rightarrow\mu^+\,
e^-\,\nu_\mu\bar\nu_\tau.$ The main source of background to this
process are $e^+\, e^-\rightarrow\mu^+\, e^-\,\nu_\mu\bar\nu_e$ and
$e^+\, e^-\rightarrow\tau^+\, \tau^-\rightarrow\mu^+\,
e^-\,\nu_\mu\bar\nu_e\,\nu_\tau\,\bar\nu_\tau.$ The cross section to
the background process $e^+\, e^-\rightarrow\mu^+\,
e^-\,\nu_\mu\bar\nu_e$ was calculated using
WPHACT~\cite{Accomando:2002sz,*Accomando:1996es} and confirmed using
RACOONWW~\cite{Denner:2002cg}. The cross section for the remaining
background was evaluated using PYTHIA~\cite{Sjostrand:2000wi}. In
$e^+\, e^-\rightarrow\tau^+\, e^-\rightarrow\mu^+\,
e^-\,\nu_\mu\bar\nu_\tau$ the electron is produced in a two body
final state. Therefore its energy is approximately half of the
center-of-mass energy. Furthermore, if $\theta_e$ is the angle
between the electron and the beam, then the transverse momentum of
the electron is $p_T=\sqrt{s}/2\,\sin\theta_e.$ This means that a
cut in $\theta_e$ implies a cut in $p_T.$  The main contribution to
this cross section comes from the fourfermion interaction. There are
no propagators involved and consequently the dependence in
$\theta_e$ (and in $p_T$) is very mild. This can be seen from the
expression~\ref{eq:4ftot} in the Appendix~\ref{ape:lep1}. Making all coupling
constants $V_{ij}$ and $S_{ij}$ equal, it can be shown that a
$10^{\circ}$ cut will reduce the cross section by 2\% while a
$60^{\circ}$ cut will reduce it only by 58\%. In
Table~\ref{tab:tab2} we show the cross sections for the signal and
for the backgrounds as a function of a cut in $\theta_e$ and a
corresponding cut in $p_T.$
\begin{table}[!htb]
\begin{center}
\begin{tabular}{lcccccc}
\hline
 Cut in $\theta_e$ (degrees) & 10 & 20 & 30 & 40 & 50 & 60\\
\hline
$e^+\,e^-\rightarrow\tau^+\,e^-\rightarrow\mu^+\,e^-\,\nu_\mu\,\bar\nu_\tau$
& 4.9 & 4.6 & 4.1 & 3.5 & 2.8 & 2.1\\
$e^+\,e^-\rightarrow\mu^+\,e^-\,\nu_\mu\,\bar\nu_e$ & 68.2 & 26.3 &
10.8 & 4.4 & 1.6 & 0.5 \\
$e^+\,e^-\rightarrow\tau^+\,\tau^-\rightarrow\mu^+\,e^-\,\nu_\mu\,\bar\nu_e\,\nu_\tau\,\bar\nu_\tau$
& 1.3 & 0.8 & 0.3 & 0.2 & 0.06 & 0.01\\
\hline
\end{tabular}
\caption{Cross section (in $fbarn)$ for the LFV signal and most
relevant backgrounds to that process for several values of the angle
cut between the outgoing electron and the beam axis.}
\label{tab:tab2}
\end{center}
\end{table}
For the signal we start with a cross section of $5\; fbarn$ when no
cuts are applied. Because of the mild dependence on $\theta_e$, a
cut of 60 degrees will make the signal well above background. A
further cut on the energy of the electron could be applied, say $E_e
> 300\; GeV.$ This would not affect the signal but will reduce the
background even further. All calculations were performed at tree
level with initial state radiation and final state radiation turned
off. Another possibility for background reduction would be to use
the polarization of the beams, a method known to be very efficient.
Notice, however, that this procedure might affect the extraction of
four-fermion couplings from polarized beam experiments -- if the
signal is observed only for certain combinations of beam
polarizations, it could happen that only certain couplings, or
combinations thereof, can be measured.

Finally, some comments on the dependence of these
results vis-a-vis expected improvements on the measurements
of the LFV branching ratios of tab~\ref{tab:dec}.
Could it be that future experiments would tighten the constraints so
much that there was no room available for discovery? Tau physics at
BABAR and BELLE has provided the best limits so far on LFV involving
the $\tau$ lepton. The combined results from BABAR and BELLE on
$\tau\rightarrow l\gamma.$ are now reaching the level of $10^{-8}$
and will be close to just a few $10^{-8}$ by 2008
\cite{Banerjee:2007rj}. More important to us are the decays
$\tau\rightarrow lll,$ due to the constraints imposed on the four
fermion operators. The latest results on $Br(\tau\rightarrow lll)$
from BABAR and BELLE are of the order of $10^{-7},$ with less than
$100\; fb^{-1}$ of data analysed. A value of the order of a few
$10^{-8}$ is expected when all data is taken into account
\cite{Salvatore:2006}. Other planned experiments like MEG or
SINDRUM2 (see \cite{nicolo:2005}) will provide much more precise
results for both $\mu\rightarrow e\gamma$ and $\mu e$ conversion,
respectively.\label{par:tira1} However, those results will not constrain any further the four-fermion couplings.
 The current limit $Br(\mu\rightarrow e e
e) < 10^{-12}$ at $90\%$ CL \cite{Wilson:1998} already excludes the
possibility of finding LFV in the $\mu e e e$ coupling. This limit
will be improved by the Sundrum experiment (see \cite{nicolo:2005}).
Another possibility is the GigaZ option for the ILC, which probably
would be earlier than an energy upgrade to $1 T
eV.$\label{par:tira2} Again, the limits on the LFV branching ratios of the $Z$ boson would be improved \cite{Bellgardt:1987du} but the bounds on the four-fermion couplings would not be affected.
 Lastly, LFV searches will also take
place at the LHC. Preparatory studies on the LFV decay
$\tau\rightarrow\mu\mu\mu$ are being conducted by CMS
\cite{Mori:2007zza}, ATLAS  \textcolor[gray]{0.3}{[31]} and also by
LHCb \cite{Shapkin:2007zz}. During the initial low luminosity runs
$(10-30\; fb^{-1}/year)$ for 2008-2009, searches for this decay may
be possible. So far the limits predicted are only slightly better
than the known limits from the $B$-factories.\label{par:tira3} Therefore, in the foreseeable future, the constraints on the four fermion $\tau$ couplings arising from the branching ratios of table I could go down one order of magnitude, to the order of $10^{-8}.$ Accordingly, and repeating the calculations that led to figs.~\ref{fig:plot2} and~\ref{fig:plot3}, the maximum number of events expected at the ILC also goes down by one order of magnitude, to about 1000 events. Given the discussion on backgrounds above, we expect that detection of LFV at the ILC would still be possible, although harder.

\newpage


\addcontentsline{toc}{section}{{\bf Appendix 4}}

\begin{subappendices}
\section{Single top production via gamma-gamma
collisions}\label{ape:lep1}

In section~\ref{sec:gamma} we argued that the couplings
corresponding to the operators of
eqs.~\eqref{eq:op4-1}--~\eqref{eq:op4-5} were extremely limited in
size by the existing experimental data for the branching ratios of
the decays $l_h\,\rightarrow\,l_l\,\gamma$. In fact, we even showed
that the cross sections for the processes
$\gamma\,\gamma\,\rightarrow\,l_h\,l_l$, eq.~\eqref{eq:ggl}, were
directly proportional to those branching ratios, and their values at
the ILC were predicted to be exceedingly small. It is easy to
understand, though, that we has defined operators analogous to those
of eqs.~\eqref{eq:op4-5} for quarks instead of
leptons. In particular, we considered flavour-changing operators
involving the top quark, which would describe decays such as
$t\,\rightarrow\,u\,\gamma$ or $t\,\rightarrow\,c\,\gamma$ - and
these decay widths have not yet been measured.
The total top quark width being also a lot
larger than the tau's or the muon's, it seems possible that the
cross section for single top production via flavour-violating
photon-photon interactions may well present us with observable
values.

The corresponding calculation (eq.~\ref{eq:widW}) is altogether identical to the one we
presented for the leptonic case. We find an expression for the width
of the anomalous decay $t\,\rightarrow\,q\,\gamma$ nearly identical
to that of eq.~\eqref{eq:wid}.
 Likewise, considering that the top quark's
charge is $2/3$ and the quarks have three colour degrees of freedom,
we may rewrite the analog of eq.~\eqref{eq:ggl} as
\begin{equation}
\frac{d\sigma(\gamma\,\gamma\,\rightarrow \,t\,\bar{q})}{d t} \;
\;=\;\; -\,\frac{16\pi\alpha\,F_{\gamma\gamma}}{3\,{m_t}^3\,s\,
    {({m_t}^2 - t ) }^2\,t\,{( {m_t}^2 - u ) }^2\,u}\;\Gamma(t\,\rightarrow
    \,q\,\gamma)\;\;\;,
    \label{eq:ggtq}
\end{equation}
with $F_{\gamma\gamma}$ given by an expression identical to
eq.~\eqref{eq:fgg}, with the substitution
$m_h\,\leftrightarrow\,m_t$. With a top total width of about 1.42
GeV and for $\sqrt{s}$ equal to 1 TeV, this expression can be
integrated in $t$ (with a $p_T$ cut of 10 GeV on the final state
particles, to prevent any collinear singularities) and the total
cross section estimated to be of the order
\begin{align}
\sigma(\gamma\,\gamma\,\rightarrow \,t\,\bar{q})\;\sim&
\;\;\;90\,\times\,\mbox{BR}(t\,\rightarrow \,q\,\gamma)
\;\;\mbox{pb} \;\;\; .
\end{align}
We see a considerable difference vis-a-vis the predicted leptonic
cross sections, from eqs.~\eqref{eq:crl} - this one is much larger.
To pass from the photon-photon cross section to an electron-positron
process, we apply the standard procedure: use the equivalent photon
approximation~\cite{Brodsky:1971ud,*Gordon:1994wu} to provide us
with the probability of an electron/positron with energy $E$
radiating photons with a fraction $x$ of E and integrate
eq.~\eqref{eq:ggtq} over $x$. For recent studies of photon-photon
collisions at the ILC, see for instance~\cite{Cao:2003vf}. The
numerical result we found for the single top production cross
section is
\begin{equation}
\sigma(e^+\,e^-\,\rightarrow \,e^+\,e^-\,t\,\bar{q})\;\;=
\;\;1.08\,\times\,\mbox{BR}(t\,\rightarrow \,q\,\gamma)
\;\;\mbox{pb} \;\;\; . \label{eq:ggtop}
\end{equation}
For an integrated luminosity of about 1 ab$^{-1}$, this gives us
about one event observed at the ILC for branching ratios of
$t\,\rightarrow \,q\,\gamma$ near the maximum of its theoretical
predictions~\footnote{Obtained in several models, such the two-Higgs
doublet model or R-parity violationg SUSY
theories~\cite{Luke:1993cy,*Atwood:1996vj,*Yang:1997dk,*Guasch:1999jp,*Delepine:2004hr,*Liu:2004qw}.},
$\sim\,10^{-6}$. Clearly, this result means that this process should
not be observed at the ILC, even in the best case scenario. However,
in the event of non-observation, eq.~\eqref{eq:ggtop} could be
useful to impose an indirect limit on the branching ratio
$\mbox{BR}(t\,\rightarrow \,q\,\gamma)$. Several authors have
studied single top production in $e^+\,e^-$
collisions~\cite{BarShalom:1999iy,Han:1998yr,*Gouz:1998rk,*Hikasa:1984um,*Cao:2002si}.
For gamma-gamma reactions, single top production at the ILC in the
framework of the effective operator formalism may has been studied
in~\cite{Jiang:1998gm,*Yu:1999tw,*Abraham:1997zf}, and for specific
models, such as SUSY and technicolor, in ref.~\cite{Cao:2003vf}.

\newpage

\section{Total cross section expressions}\label{sec:cross2}

We write the amplitude for the four-fermion cross sections in two
parts. One for the $s$ channel and the other one for the $t$
channel. In doing so we are generalizing the four-fermion Lagrangian
which for a gauge theory has equal couplings for both $s$ and $t$
channels. For the $s$ channel the amplitude reads
\begin{equation}
T_{ij}^s \, = \, \frac{1}{\Lambda^4} \left[ V_{ij}^s \, (\bar{v}_e
\gamma_{\alpha} \gamma_i u_e) (\bar{u}_{l_h} \gamma^{\alpha}
\gamma_j v_{l_l}) \, + \, \, S_{ij}^s \, (\bar{v}_e \gamma_i u_e)
(\bar{u}_{l_h} \gamma_j v_{l_l}) \right]
\end{equation}
while for the t channel we have
\begin{equation}
T_{ij}^t \, = \,  -\,\frac{1}{\Lambda^4} \left[ V_{ij}^t \,
(\bar{u}_{l_h} \gamma_{\alpha} \gamma_i u_e) (\bar{v}_e
\gamma^{\alpha} \gamma_j v_e) \, + \, \, S_{ij}^t \, (\bar{u}_{l_h}
\gamma_i u_e) (\bar{v}_e \gamma_j v_e) \right]
\end{equation}
with $i,j = L,R$. With these definitions we can write
\begin{align}
|T (e_L^- e_L^+ \rightarrow {l_h}_L^- {l_l}_L^+)|^2 \, = &
\,\frac{1}{\Lambda^4} (4 u^2 |V_{LL}^s + V_{LL}^t|^2)
\nonumber \\
|T (e_R^- e_R^+ \rightarrow {l_h}_R^- {l_l}_R^+)|^2 \, = & \,
\frac{1}{\Lambda^4} (4 u^2
|V_{RR}^s + V_{RR}^t|^2) \nonumber \\
|T (e_L^- e_L^+ \rightarrow {l_h}_R^- {l_l}_R^+)|^2 \, = & \,
\frac{1}{\Lambda^4}\,t^2 \,|S_{RL}^t|^2
\nonumber \\
|T (e_R^- e_R^+ \rightarrow {l_h}_L^- {l_l}_L^+)|^2 \, = & \,
\frac{1}{\Lambda^4} t^2 \,|S_{LR}^t|^2
\nonumber \\
|T (e_L^- e_R^+ \rightarrow {l_h}_L^- {l_l}_R^+)|^2 \, = & \,
\frac{1}{\Lambda^4} \, s^2 \,|S_{LR}^s|^2
\nonumber \\
|T (e_R^- e_L^+ \rightarrow {l_h}_R^- {l_l}_L^+)|^2 \, = & \,
\frac{1}{\Lambda^4} \, s^2 \, |S_{RL}^s|^2
\end{align}
and to obtain the expressions when only the $t$ or $s$ channels are
present, you just have to set the $s$ couplings or the $t$
couplings, respectively, equal to zero. $u$, $t$ and $s$ are the
Mandelstam variables defined in the usual way.

The cross sections for polarized electron and positron beams with no
detection of the polarization of the final state particles were
given in eq.~\eqref{eq:sig4f}. The International Linear Collider
will have a definite degree of polarization that will depend on the
final design of the machine. For longitudinal polarized beams the
cross section can be written as
\begin{align}
\frac{d \sigma_{P_{e^-}P_{e^+}}}{dt} \, = &  \, \frac{1}{4} \biggl[
(1+P_{e^-})(1+P_{e^+}) \frac{d \sigma_{RR}}{dt} \, + \,
(1-P_{e^-})(1-P_{e^+}) \frac{d \sigma_{LL}}{dt}+
\nonumber \\
& (1+P_{e^-})(1 - P_{e^+}) \frac{d \sigma_{RL}}{dt} \, + \,
(1-P_{e^-})(1+P_{e^+}) \frac{d \sigma_{LR}}{dt} \biggr]
\end{align}
where $\sigma_{RL}$ corresponds to a cross section where the
electron beam is completely right-handed polarized ($P_{e^-}=+1$)
and the positron beam is completely left-handed polarized
($P_{e^+}=-1$). This reduces to the usual averaging over spins in
the case of totally unpolarized beams. For the general expression
for polarized beams, as well as a study on all the advantages of
using those beams, see~\cite{MoortgatPick:2005cw}.

In the main text we presented expressions for the differential cross
sections. For completeness we now present the formulae for the total
cross sections. For the four-fermion case, the expressions have a
very simple dependence on the $p_T$ cut one might wish to apply, so
we exhibit it. The quantity $x=\sqrt{1-4p_T^2/s}$, with $p_T$ being
the value of the minimum transverse momentum for the heaviest
lepton, gives us an immediate way of obtaining these cross sections
with a cut on the $p_T$ of the final particles. The total cross
section is obviously the sum over all polarized ones, which gives us
\begin{equation}
\sigma \, =  \, \frac{s \, x \, (3+x^2)}{768 \pi \Lambda^4} \left( 4
\, |V_{LL}^s + V_{LL}^t|^2 + |S_{RL}^t|^2  + 4\,
 |V_{RR}^s + V_{RR}^t|^2 + |S_{LR}^t|^2 \right)
+ \, \frac{s \, x}{64 \pi} \left( |S_{LR}^s|^2 + |S_{RL}^s|^2
\right) \, . \label{eq:4ftot}
\end{equation}
As explained in the main text, the cross sections for processes
$(1,2)$ are obtained from eq.~\eqref{eq:4ftot} by setting all of the
``s'' couplings equal to the ``t'' ones, and, for process $(3)$, by
setting the ``t'' couplings to zero.

For the remaining cross section expressions we imposed no $p_T$ cut
on any of the final particles. The total cross section for the $Z$
couplings is given by, for processes $(1,2)$,
\begin{align}
\sigma_Z^{(1,2)} (e^- e^+ \rightarrow l_h l_l) \;=&\;\; \frac{
v^2}{192 \,\pi\,\Lambda^4\,M_z^2\, s^2\,(M_z^2-s)^2 \, (M_z^2+s) }
\left[ F_3 (g_A) |\eta_{lh}|^2 \right.
\nonumber \vspace{0.3cm} \\
& \; \left. + F_3 (-g_A) |\eta_{hl}|^2 + F_4 (g_A) |\theta_{L}|^2 +
F_4 (-g_A) |\theta_{R}|^2  \right] \, \, ,
\end{align}
with
\begin{align}
F_3 (g_A) & = 6 \,s\,M_z^2\,(M_z^4-s^2) \log \left(
\frac{M_z^2+s}{M_z^2}\right)\nonumber\\
&\times \left[ (g_A-g_V)^2 M_z^4 -2 (g_A^2+g_A
g_V+g_V^2) s M_z^2 - (g_A^2+g_V^2) s^2 \right]- s^2\,M_z^2\, (M_z^2+s)\nonumber\\
& \times  \left[ 6(g_A-g_V)^2 M_z^4 -3 (7g_A^2-2 g_A g_V+7 g_V^2) s
M_z^2 + 2 (7g_A^2+3 g_A g_V+7 g_V^2) s^2 \right] \nonumber
\end{align}
\begin{align}
F_4 (g_A) \;=&\;\; 48 v^2 (M_z^4-s^2) (M_z^2+s) M_z^4 \log \left(
\frac{M_z^2}{M_z^2+s}\right) (g_A-g_V)^2
\nonumber\\
&  +8 v^2 s \left[ 3 (g_A-g_V)^2 M_z^6 (2M_z^2 + s)- (5 g_A^2 - 18
g_A g_V+ 5 g_V^2) s^2 M_z^4
\right.\nonumber\\
& \left.  - 5 (g_A^2+g_V^2) s^3 M_z^2 + 3 (g_A^2+g_V^2) s^4 \right],
 \nonumber
\end{align}
with interference terms
\begin{align}
\sigma_{int}^{(1,2)} \;=&\;\; -\,\frac{ v^2}{48 \, \Lambda^4\,\pi
\,(s-M_z^2)\,s^2 } \biggr\{
\nonumber \vspace{0.3cm} \\
& s \biggr[ \left( g_A -g_V \right) \left\{
(12M_z^4+6sM_z^2-14s^2)\mbox{Re} \left( \theta_L V_{LL}^*
\right)-s^2\mbox{Re} \left( \theta_R S_{RL}^* \right) \right\}
\nonumber \vspace{0.3cm} \\
& \, \,  -\left( g_A +g_V \right) \left\{
(12M_z^4+6sM_z^2-14s^2)\mbox{Re} \left( \theta_R V_{RR}^*
\right)-s^2\mbox{Re} \left( \theta_L S_{LR}^* \right) \right\}
\biggl]
\nonumber \vspace{0.3cm} \\
& +\,3\,(M_z^2-s) \biggr[ \left( g_A -g_V \right) \left\{
(4M_z^4+8sM_z^2-4s^2)\mbox{Re} \left( \theta_L V_{LL}^*
\right)-s^2\mbox{Re} \left( \theta_R S_{RL}^* \right) \right\}
\nonumber \vspace{0.3cm} \\
& \, \,  -\left( g_A +g_V \right) \left\{
(4M_z^4+8sM_z^2-4s^2)\mbox{Re} \left( \theta_R V_{RR}^*
\right)-s^2\mbox{Re} \left( \theta_L S_{LR}^* \right) \right\}
\biggl]
\nonumber \vspace{0.3cm} \\
&\log \left( \frac{M_z^2}{M_z^2+s}   \right) \biggl\} \;\;\; .
\end{align}
Finally, for process $(3)$, we have
\begin{equation}
\sigma_Z^{(3)}  \;=\; \frac{\left( g_V^2 +g_A^2 \right) v^2
s}{192\,\pi\, \Lambda^4\,(M_z^2-s)^2 }  \left[ 8 \left(
|\theta_L|^2+|\theta_R|^2 \right) v^2 + \left(
|\eta_{hl}|^2+|\eta_{lh}|^2 \right) s \right] \, \, ,
\end{equation}
and
\begin{align}
\sigma_{int}^{(3)} \;= &\; \frac{s v^2}{24 \pi (s-M_z^2) \,
\Lambda^4} \left[  \left( g_V +g_A \right) \mbox{Re} \left(\theta_R
V_{RR}^* \right) + \left( g_V -g_A \right) \, \mbox{Re} \left(
\theta_L V_{LL}^*\right) \right] \;\;\; .
\end{align}

\newpage

\section{Numerical values for decay widths and cross sections}

We present here numerical values for the several decay widths and
cross sections given in the text. We have set, in the following
expressions, $\Lambda$ to $1\;TeV,$ the dependence in $\Lambda$
being trivially recovered if we wish a different value for it.
\begin{align}
\mbox{BR}_{4f} (\mu \rightarrow l l l) &=\;  2.3 \times 10^{-4}
(|S_{LR}|^2+|S_{RL}|^2+4(|V_{LL}|^2+|V_{RR}|^2)) \vspace{0.4cm}  \nonumber \\
\mbox{BR}_{4f} (\tau \rightarrow l l l) &=\; 4.0 \times 10^{-5}
(|S_{LR}|^2+|S_{RL}|^2+4(|V_{LL}|^2+|V_{RR}|^2))\vspace{0.4cm}\nonumber \\
\mbox{BR}_Z (\mu \rightarrow l l l) &=\; 8.2 \times 10^{-4} \left(
|\theta_L|^2+|\theta_R|^2 \right)  + 2.5 \times 10^{-7} \mbox{Re}
\left(
\eta_{lh} \theta_L^* +  \eta_{hl} \theta_R \right)\vspace{0.4cm}\nonumber \\
\mbox{BR}_Z (\tau \rightarrow l l l) &=\; 1.4 \times 10^{-4} \left(
|\theta_L|^2+|\theta_R|^2 \right)  + 7.3 \times 10^{-7} \mbox{Re}
\left( \eta_{lh} \theta_L^* +  \eta_{hl} \theta_R
\right)\vspace{0.4cm}\nonumber \\
\mbox{BR}_{int} (\mu \rightarrow l l l) &=\; -1.4 \times 10^{-3}
\mbox{Re} \left( \theta_L V_{LL}^*\right) + 1.1 \times 10^{-3}
\mbox{Re} \left( \theta_R V_{RR}^* \right) \vspace{0.4cm}
\nonumber  \\
& \;   +\,1.7 \times 10^{-7} Re \left( \eta_{lh} V_{RR}^*  \right)
-2.1 \times 10^{-7} \mbox{Re} \left( \eta_{hl}
V_{LL} \right)\vspace{0.4cm}\nonumber \\
\mbox{BR}_{int} (\tau \rightarrow l l l) &=\; -2.4 \times 10^{-4} Re
\left( \theta_L V_{LL}^* \right) + 1.9 \times 10^{-4} \mbox{Re}
\left( \theta_R V_{RR}^* \right) \vspace{0.4cm}
\nonumber  \\
 & \; +\, 4.8 \times 10^{-7} Re \left( \eta_{lh}
V_{RR}^*  \right)  -6.0 \times 10^{-7} \mbox{Re} \left( \eta_{hl}
V_{LL} \right)\;\;\; .
\end{align}

\begin{align}
\mbox{BR} (Z \rightarrow l l)     &=\;   2.3 \times 10^{-5}
(|\eta_{hl}|^2+|\eta_{lh}|^2) + 6.7 \times 10^{-4}
(|\theta_{L}|^2+|\theta_{R}|^2) \nonumber \\
\mbox{BR} (Z \rightarrow \mu l)     &=\;  Br (Z \rightarrow l l)-
2.0 \times 10^{-7} \mbox{Re} \left( \theta_L \eta_{hl}+ \theta_R
\eta_{lh}^* \right) \nonumber \\
\mbox{BR} (Z \rightarrow \tau l)     &=\;   Br (Z \rightarrow l l) -
2.4 \times 10^{-6} \mbox{Re} \left( \theta_L \eta_{hl}+ \theta_R
\eta_{lh}^* \right) \;\;\;.
\end{align}

For the cross sections, taking $\sqrt{s}= 1$ TeV  and imposing a cut
of 10 GeV on the $p_T$ of the particles in the final state, we have
(in picobarn):
\begin{align}
\sigma_{4f}^{(1,2)} (e^- e^+ \rightarrow l l)  &=\; 2.58 \left(
|S_{LR}|^2+|S_{RL}|^2 \right) + 10.33 \left( |V_{LL}|^2+|V_{RR}|^2
 \right)\nonumber \\
\sigma_{4f}^{(3)} (e^- e^+ \rightarrow l l)  &=\; 1.94 \left(
|S_{LR}|^2+|S_{RL}|^2 \right) + 2.58 \left(
|V_{LL}|^2+|V_{RR}|^2 \right)\nonumber \\
\sigma_Z^{(1,2)} (e^- e^+ \rightarrow l l) &=\; 1.0 \times 10^{-2}
|\eta_{lh}|^2 +9.7 \times 10^{-3} |\eta_{hl}|^2 + 5.7 \times 10^{-2}
\left( \theta_{L}|^2 +  |\theta_{R}|^2  \right)\nonumber \\
\sigma_Z^{(3)} (e^- e^+ \rightarrow l l)  &=\; 1.6 \times 10^{-4}
\left( |\theta_L|^2+|\theta_R|^2 \right) + 6.7 \times 10^{-4} \left(
|\eta_{hl}|^2+|\eta_{lh}|^2 \right)\nonumber \\
\sigma_{int}^{(1,2)} (e^- e^+ \rightarrow l l) &=\; 0.70 \,
\mbox{Re} \left( \theta_L V_{LL}^* \right) + 0.19 \, \mbox{Re}
\left( \theta_L S_{RL}^* \right) - 0.56 \, \mbox{Re} \left( \theta_R
V_{RR}^* \right) - 0.24 \, \mbox{Re}
\left( \theta_R S_{LR}^* \right)\nonumber \\
\sigma_{int}^{(3)} (e^- e^+ \rightarrow l l)  &= \; - 2.6 \times
10^{-2} \, \mbox{Re} \left( \theta_R V_{RR}^* \right) +3.2 \times
10^{-2} \,  \mbox{Re} \left( \theta_L V_{LL}^* \right) \;\;\; .
\end{align}

\end{subappendices}




\bibliography{thesis.bbl}

\bibliographystyle{h-physrev}

\end{document}